\documentclass[12pt,letterpaper,twoside]{report}
\usepackage{graphicx}
\usepackage{fancyhdr}
\usepackage{latexsym}

\def\bg{\begin{eqnarray}}
\def\nd{\end{eqnarray}}

\newcommand{\be}{\begin{eqnarray}}
\newcommand{\ee}{\end{eqnarray}}
\newcommand{\ben}{\begin{eqnarray*}}
\newcommand{\een}{\end{eqnarray*}}

\newcommand{\bcent}{\begin{center}}
\newcommand{\ecent}{\end{center}}
\newcommand{\benum}{\begin{enumerate}}
\newcommand{\eenum}{\end{enumerate}}
\newcommand{\bdesc}{\begin{description}}
\newcommand{\edesc}{\end{description}}
\newcommand{\bitem}{\begin{itemize}}
\newcommand{\eitem}{\end{itemize}}
\newcommand{\bquote}{\begin{quote}}
\newcommand{\equote}{\end{quote}}
\newcommand{\bhalfp}{\begin{minipage}{0.45\textwidth}}
\newcommand{\ehalfp}{\end{minipage}}
\newcommand{\bhead}{\begin{center}\bf \Large}
\newcommand{\ehead}{\end{center}\bigskip}
\newcommand{\non}{\nonumber\\}

 \newcommand{\calD}{{\cal D}}

 \newcommand{\calG}{{\cal G}}
 \newcommand{\calH}{{\cal H}}

 \newcommand{\ave}[1]{\langle {#1} \rangle}

\newcommand{\umax}{u_{\rm max}}

\newcommand{\xmax}{x_{\rm max}}
\newcommand{\ymax}{y_{\rm max}}

\def\be{\begin{equation}}
\def\ee{\end{equation}}
\def\ba{\begin{eqnarray}}
\def\ea{\end{eqnarray}}

\newcommand{\roughly}[1]{\mathrel{\raise.3ex\hbox{$#1$\kern-0.85em
\lower1ex\hbox{$\sim$}}}}

\def\2pi{\left(2\pi\right)}

\def\beq{\begin{equation}}
\def\eeq{\end{equation}}
\def\bg{\begin{eqnarray}}
\def\nd{\end{eqnarray}}
\def\bea{\begin{eqnarray}}
\def\eea{\end{eqnarray}}

\def\D3{\overline{\mbox{D3}}}

\begin{document}

\begin{titlepage}
\begin{center}
\rule{2in}{0in}\\
\vspace{50.2pt}
{\huge \bf From string theory  \\
 to \\
 \vspace{18pt}
 large N QCD}\\
\vspace{48pt}
{\Large \bf Mohammed Shahpur Mia }\\
\vspace{48pt}
{\large Department of Physics}\\
\vspace{36pt}
{\large McGill University}\\
{\large Montreal, Quebec, Canada.}\\
{\large August, 2010.}\\
\vspace{36pt}
A thesis submitted to McGill University\\
in partial fulfillment of the requirements of the degree of Doctor of Philosophy\\
\vspace{36pt}
Copyright \copyright\ mohammed, 2010
\end{center}
\end{titlepage}

\rule{2in}{0in}
\thispagestyle{empty}
\newpage

\pagenumbering{roman}
\section*{\centering \LARGE \bf Abstract}
\addcontentsline{toc}{section}{\bf Abstract}
We propose the dual gravity of a non conformal gauge theory which has logarithmic running of couplings in the IR but becomes almost conformal
in the far UV. The theory has matter in fundamental representation, non-zero temperature and under a cascade of Seiberg dualities, can be
described in terms of gauge groups of lower and lower rank. We outline the procedure of holographic renormalization and propose a mechanism to
UV complete the gauge theory by modifying the dual geometry at large radial distances. As an example, we construct the brane configuration and
sources required to
attach a Klebanov-Witten type geometry at large $r$ to a Klebanov-Strassler type geometry at small $r$. Using the supergravity description for
the dual geometry, we compute thermal mass  of a fundamental `quark' in our theory along with drag and diffusion coefficients of the gauge theory
plasma. We compute
the stress tensor of the gauge theory and formulate the wake a probe leaves behind as it traverses the medium. Transport coefficient 
shear viscosity $\eta$ and its ratio to entropy $\eta/s$ are calculated and finally we show how confinement of `quarks' at
large separation can occur at low temperatures. We classify the most general dual geometry that  gives rise to linear confinement at low
temperatures and show how quarkonium states can melt at high temperatures to liberate `quarks'.          
\newpage

\section*{\centering \LARGE \bf R\'{e}sum\'{e}}
\addcontentsline{toc}{section}{\bf R\'{e}sum\'{e}}
    
    Nous proposons une th\'{e}orie de la gravit\'{e} qui correspond \`{a} une th\'{e}orie de jauge avec des
constantes de couplage \`{a} comportement logarithmique dans l'infrarouge et devenant
quasi-conforme dans l'ultraviolet profond. La th\'{e}orie contient la mati\`{e}re dans la
repr\'{e}sentation fondamentale, a une temp\'{e}rature non-nulle et peut-\^{e}tre d\'{e}crite en terme de
groupes de jauge de rangs de plus en plus petits par une cascade de dualit\'{e}s de Seiberg.
Nous discutons bri\`{e}vement du processus de renormalisation holographique et proposons
un m\'{e}canisme pour compl\'{e}ter la th\'{e}orie de jauge dans l'ultraviolet. Comme exemple,
nous construisons une configuration de membranes et de sources qui joignent une
g\'{e}om\'{e}trie de Klebanov-Witten \`{a} grand r \`{a} une g\'{e}om\'{e}trie de Klebanov-Strassler \`{a} petit r.
En utilisant la supergravit\'{e} pour d\'{e}crire la g\'{e}om\'{e}trie duale, nous calculons la masse
thermique d'un quark fondamental ainsi que les coefficients de tra\^{i}n\'{e}e et de diffusion du
plasma de la th\'{e}orie de jauge. Nous \'{e}valuons le tenseur de stress de la th\'{e}orie de jauge et
quantifions le sillon laiss\'{e} par une sonde traversant le milieu. Nous calculons le rapport
de la viscosit\'{e} de cisaillement \`{a} la densit\'{e} d'entropie, $\eta/s$. Finalement, nous montrons
comment le g\'{e}om\'{e}trie duale la plus g\'{e}n\'{e}rale possible peut donner lieu \`{a} la fois au
confinement lin\'{e}aire de paires quark-antiquark \`{a} basse temp\'{e}rature et \`{a} la dissolution du
quarkonium \`{a} haute temp\'{e}rature.

\newpage
\section*{\centering \LARGE \bf Dedication}
\addcontentsline{toc}{section}{\bf Dedication}
To my brother Mashrur Mia. 
\newpage
\section*{\centering \LARGE \bf Acknowledgements}
\addcontentsline{toc}{section}{\bf Acknowledgements}
First and foremost, I would like to thank my advisor Professor Charles Gale for believing in me and being there when it mattered the most. I
would not be a graduate student if it wasn't for Charles.   
The research for my doctoral studies would not have been
possible without the guidance of Professor Keshav Dasgupta. The effort and hours he
spent on a daily basis explaining the intricacies of string theory to someone like me with the minimal knowledge and that too for two and a half years, 
is truly inspiring. I cannot imagine the completion of this work without Keshav. I am in debt to Professor Sangyong Jeon, for the training he has given me in the last six years, to question, to be
critical and most importantly, to be more of a physicist.

I would not be a scientist, if it wasn't for my mom and dad, who brought me to this world and taught me how to think and raised me to who I
am.
Everything that is good and creative in me is from them. I have to thank my brother Sa'ad for being my moral guru, the guardian he has always 
been and for enduring all my irregularities, making our apartment our home. I'm truly blessed to have my better half Roshni to be there 
always, I would be totally lost 
without her. I have to thank my office mates Rhiannon, Aaron, Paul and Simon, my colleagues  Alisha, Qin, Nima and Hoi for all the discussions on
physics and non-physics topics that keep us going. Thanks to Andrew Frey and Alejandra for always answering my questions with the best insight
into the subject. 

My cousins and family in Montreal, Choyon, Ayon, Ruben bhaiya, Juthi bhaabi, Evan, Salek bhai, Sani- was always there to make Montreal my 
home. Thanks Yuri for lending your ears, painting my face and kicking me out to freeze. I'm grateful to all my brothers and sisters in McGill- 
Sayem, Takshed, 
 Khalid, Saquib bhai, Rafa Apu, Sadaf Apu, Rafi bhai, Ruhan bhai and Mehdi bhai- you have always managed to 
entertain and put a smile on my face. 

Finally, I thank Allah, the almighty, for everything.                          
\newpage

\section*{\centering \LARGE \bf Statement of originality}
\addcontentsline{toc}{section}{\bf Statement of originality}
The thesis is based on my research done in collaboration with Keshav Dasgupta, Charles Gale and Sangyong Jeon and the
results presented constitute original work that appeared  in the following articles:
\begin{itemize}
\item {\bf M.~Mia}, K.~Dasgupta, C.~Gale and S.~Jeon,
  ``{\it Five Easy Pieces: The Dynamics of Quarks in Strongly Coupled Plasmas}'',
  Nucl.\ Phys.\  B {\bf 839}, 187 (2010), 107pp,
  [arXiv:0902.1540 [hep-th]].
\item {\bf M.~Mia}, K.~Dasgupta, C.~Gale and S.~Jeon,
  ``{\it Toward Large N Thermal QCD from Dual Gravity: The Heavy Quarkonium
  Potential}'',
  Phys.\ Rev.\  D {\bf 82}, 026004 (2010), 28pp, 
  [arXiv:1004.0387 [hep-th]].
\item  {\bf M.~Mia}, K.~Dasgupta, C.~Gale and S.~Jeon,
  ``{\it Heavy Quarkonium Melting in Large N Thermal QCD}'', in press for Phys. Lett. B (2010),\\ 
  doi:10.1016/j.physletb.2003.10.071, 15pp,  
  arXiv:1006.0055 [hep-th].
\item {\bf M.~Mia} and C.~Gale,
  ``{\it Jet quenching and the gravity dual}'',
  Nucl.\ Phys.\  A {\bf 830}, 303C (2009), 4pp, 
  [arXiv:0907.4699 [hep-ph]].
\item  {\bf M.~Mia}, K.~Dasgupta, C.~Gale and S.~Jeon,
  ``{\it Quark dynamics, thermal QCD, and the gravity dual}'',
  Nucl.\ Phys.\  A {\bf 820}, 107C (2009), 4pp.
\end{itemize} 
All the original calculations presented here were performed by me except the ones concerning arbitrary UV completions, the
explicit form of the fluxes and the proof of melting of the potential, which were done by my collaborators.   
\newpage

\tableofcontents
\listoftables
\addcontentsline{toc}{section}{\bf List of Tables}
\listoffigures
\addcontentsline{toc}{section}{\bf List of Figures}
\thispagestyle{plain}

\pagestyle{headings}
\renewcommand{\chaptermark}[1]{\markboth{\hskip24pt Chapter \thechapter\hskip12pt #1\hfill}{}}
\renewcommand{\sectionmark}[1]{\markright{\hfill\thesection\hskip12pt #1\hskip24pt}{}}

\chapter{Introduction}
Since the discovery of a unified description of strong, weak and electromagnetic interaction back in the 1970's, much
activity was centered around its consequences. The foundation of the unification lies in the principle
of local gauge symmetry where interactions between particles are mediated by gauge fields. Through the mechanism of
spontaneous symmetry breaking as in the Higgs model \cite{Higgs1}-\cite{Englert:1964et},  Weinberg \cite{Weinberg} and Salam
\cite{Salam} showed that
at low energies only 
photons and neutrinos remain massless while vector bosons responsible for weak interaction 
acquire mass. In the far UV  the vector bosons become massless and  
strong and weak interactions become long range- just like electromagnetic interaction. Thus at high energies
symmetry between strong, weak and electromagnetic force gets restored \cite{Salam}\cite{Weinberg}. 
\pagenumbering{arabic}

On the other hand phase transition
is a process which changes the symmetry of the system and before or after the phase transition, symmetry is
broken \cite{Linde}. This suggests the existence of various phases in the theory of elementary particles. In particular, highly dense
matter under extreme temperature and pressure should exhibit various phases as temperature and density is altered. By
heating nuclear matter up to extreme temperature and pressure, one expects to create a new phase of matter, a plasma of
quarks and gluons where quarks become massless with  strong, weak and electromagnetic interaction becoming long range,
restoring the symmetry between them. The estimates for the critical temperature $T_c$ for phase transition in
renormalizable gauge theories were first
made by \cite{Weinberg-2}-\cite{Linde-1} and there is an extensive literature on the phase structure of quark matter 
along with studies of critical temperature \cite{MSc1}-\cite{MScX}.

These theoretical expectations for a new phase of matter known as the Quark Gluon Plasma (QGP) led to the
Relativistic Heavy Ion Collider (RHIC) program where heavy nuclei are made to collide at relativistic velocities. The matter formed in the
early stages of the collision has indeed been identified not as a collection of color neutral hadrons, 
but a new state of dense matter \cite{RHICwp1}-\cite{RHICwp4}. The relativistic fluid created at RHIC cannot be described in terms of
hadronic degrees of freedom and one may conclude that the energies reached by the experiments give rise to a plasma of quarks and gluons i.e.
QGP. But what are the properties of this fluid? In particular, is the coupling between the constituents of the fluid strong 
or weak? The answer to
this is crucial as it will determine the theoretical tools to study the fluid. At weak coupling, perturbative methods can be useful whereas at
strong coupling, effective  or lattice field theory need to be applied. 

To address this issue, first observe that  
one of the key characteristics of a fluid is how it responds to pressure variation and 
flows to equilibrium state. At the heavy ion collider, the overlapping region between two heavy nuclei is of elliptic shape with one
axis longer than the other- thus creating anisotropic pressure. The fluid will of course try to equilibrate to spherical symmetric
configuration and this leads to elliptic flow. This flow as observed in the experiments at RHIC \cite{RHIC-Eflow-1}-\cite{RHIC-Eflow-4} is 
well described by ideal hydrodynamics which assumes that the QGP fluid is strongly coupled. This means perturbative field theory techniques
are not useful to analyze this strongly coupled system and one usually studies the theory on the lattice or using effective field theory
. While the former becomes  quite challenging with lattice simulations limited by computational ability of the numerical method, the latter
relies on effective Lagrangians which are only approximations and could lead to incomplete results. 

But these are not the only tools to
describe a strongly coupled field theory.  One may apply the principle of holography which relates the Hilbert space of a 
gauge theory with that of a theory of gravity, to study strongly coupled field theories. The key observation was made by Maldacena in the late 1990's while studying anti de Sitter
black hole solutions and how they arise from D brane configurations \cite{Maldacena-1}. He conjectured that strongly coupled ${\cal N}=4$  supersymmetric conformal field theory is dual 
to a theory of weakly
coupled gravitons describing $AdS_5\times S_5$ geometry, where $AdS_5$ is the five dimensional anti de Sitter space and $S_5$ is the five 
sphere. This duality known as the AdS/CFT correspondence opened a whole new avenue to study strongly coupled quantum field theories using
weakly coupled classical gravity and is only a part of a more general correspondence between gauge theory and geometry. 

In this thesis we extend
the AdS/CFT correspondence to incorporate renormalization group flow in the gauge theory and study the system at strong coupling using weakly
coupled dual gravity. We propose
the dual geometry for a strongly coupled gauge theory which has logarithmic running of couplings in the far IR but becomes almost conformal in
the far UV. The gauge theory has matter in fundamental representation and has non-zero temperature- which is achieved by introducing a black hole
in the dual geometry. Using the weakly coupled gravity description for the strongly coupled field theory, one can easily compute expectation
values of various gauge theory operators that are extremely difficult to obtain using conventional techniques. By computing 
quantum correlation functions using dual classical action,  one can analyze the thermodynamic properties and learn about the kinematics of 
 strongly coupled gauge theory plasma.  But how does the rich structure of quantum field theory get exact description in terms of a simple classical
 theory of gravity? To be more precise, what are the limits of validity of the correspondence? 
 
 It turns out that for a field theory at strong coupling to have a description in terms of weakly coupled gravity, the number of colors $N$ in
 the gauge theory must be very large. In the large $N$ limit, it seems that the field theory has a classical description in terms of dual
 gravity where quantum correlation functions can be exactly calculated. At first glance this simplification might appear to be quite
 `accidental' but a careful analysis can shed light on the issue. The answer may be linked to an observation made by  't Hooft who suggested 
 that the number of colors $N$ in
 a gauge theory can be thought of as an expansion parameter \cite{'tHooft}. There are planar and non planar diagrams that contribute to the
 propagator and in the limit $N\rightarrow \infty$, only planar diagrams survive. This is because  non planar diagrams are
 suppressed by ${\cal O}(1/N^k),k\ge 1$,
 relative to planar ones \cite{'tHooft}\cite{Witten-planar}, resulting in a drastic simplification of the propagator. While at finite $N$
 propagators and subsequently S matrix elements are extremely difficult to compute due to the large number of non planar diagrams at various
 loop order, in the large $N$ limit, the theory has a rather simple description. This simplification from a field theory analysis is
 completely consistent with our dual gravity description where theory can be described by classical gravity.        
                       
The thesis is organized as follows: In chapter two, after a brief discussion on AdS/CFT correspondence, 
we describe in some detail gauge theories that arise from various brane configurations and then present their dual 
geometries. We outline the procedure of holographic renormalization of non conformal gauge theories which have dual gravity
description and then discuss how to UV complete the theory by attaching geometries. In the final section of chapter two, we
give a specific example of such UV completion by considering localized sources which allow one to attach asymptotic AdS
geometry at large $r$, to a Klebanov-Strassler type geometry at small $r$. Using the dual geometries of chapter two, in 
chapter three we compute various gauge theory
quantities crucial in analyzing a plasma. We compute the thermal mass of `quark' in fundamental representation, the drag it
experiences as it moves through the plasma and the wake it leaves behind as it traverses the medium. We also compute the
momentum broadening of a fast moving jet along with transport coefficients
such as shear viscosity and its ratio to entropy. Finally we show how confinement can be achieved using dual gravity and
propose the most general dual geometry that can realize linear confinement at large distances. We conclude the thesis with a
brief summary of our results along with future directions that can be explored.

\chapter{The Gauge/Gravity Correspondence}
 
 Quantum field theory with gauge symmetry has been extremely successful in describing the dynamics and collective
 excitations of highly energetic
particles at very short distance scales. On the other hand, Einstein's theory of gravity has been crucial in 
understanding the dynamics of massive objects separated by large distances under the force of gravity. But what is the
connection between these two seemingly distinct theories at opposite distance scales? General relativity
dictates that any stress energy will couple to gravitons and thus given a field theory one can always compute the gravitons
sourced by the stress tensor of the field theory. Thus given a gauge theory, we can always compute the corresponding
gravitons and there is a natural equivalence between  energies of the gravitons with that of the fields. 
Note the coupling of field theory with gravity is very weak compared to other gauge couplings (such as
strong interaction coupling $\alpha_s$ or the electromagnetic or weak interaction coupling) up to the energy scale relevant for current experiments. 
Nevertheless there is a naturalness in this correspondence between gauge theory and the geometry it sources, all because of
the coupling of the two.

In string theory, a more remarkable and a little less obvious duality arises. Open string excitations are described by a
quantum field theory which has scalar and vector fields. In addition, closed string excitations are described by
quantum fields which are tensors of rank two and higher \cite{Witten-DBExc}\cite{stringtext1}\cite{DBPrimer}. Thus open string quantization can give rise to gauge
theories with fermions and spin one bosons whereas closed string quantization can incorporate gravitons. 
In general there will be interactions between open and closed string modes but it turns out that when the interaction is
weak, there is indication that the Hilbert space of the open string modes and the Hilbert space of the closed string modes are
identical. Thus the conjecture \cite{Maldacena-1} that  open string field theory is dual to closed string field theory in the limit 
where the modes decouple! This remarkable duality is one the primary motivations for proposing a general correspondence between a
gauge theory and  theory of gravity - in the limit they decouple from each other. In the following sections we will explore
this gauge gravity duality, first for conformal field theories and then for gauge theories with running couplings.

\section{Conformal Field Theory and AdS Geometry}
Gauge theories naturally arise when one studies the excitations of branes. In particular excitations of open strings ending
on D branes can be described by supersymmetric Yang-Mills multiplet with  vector, spinor and scalar fields \cite{Witten-DBExc}. 
If we consider strings ending on a single 
D3 brane where the D3 brane is embedded in ten dimensional flat space time, then the gauge group is $U(1)$. The D3 brane fills up four dimensional Minkowski
space and can move in six dimensional space with coordinates $x^j,j=4,..,9$. 
its  world volume has
coordinates $x^i, i=0,1,2,3$ and the massless gauge bosons $A_l(x^i),i,l=0,..,3$ live on
the D3 brane world volume. There are six scalar fields $\phi_j(x^i),j=4,..,9$
and they describe the oscillation of the position of the D3 brane. 

On the other hand
if we consider $N$ number of parallel D3 branes, then two ends of a string can lie on different branes  or on the same brane. There are
$N^2$ choices to place the  endpoints which gives $N^2$ number of vectors. Thus the excitations of the strings stretching 
between and ending on the branes is described by a $U(N)$ gauge theory \cite{Witten-DBExc} \cite{Klebanov-review} with vector, scalars 
and spinor fields. 
Note that one
takes the massless modes of the string excitations which means strings have vanishing length and   imply that we 
have coincident branes. Thus excitations of $N$ 
coincident D3 branes give rise to $U(N)$ gauge theory living in four dimensional flat space time. It also turns out that the
gauge coupling does not run and we have a conformal field theory with $U(N)$ gauge group in four dimensions. 

The total action for the collection of branes embedded in the ten dimensional space time can be written as \cite{Magoo}
\bg
{\cal S}={\cal S}_{\rm branes}+ {\cal S}_{\rm gravity}+ {\cal S}_{\rm int} 
\nd 
where ${\cal S}_{\rm branes}$ is the action for theory on the branes, ${\cal S}_{\rm gravity}$ is the action for the
background geometry on which the branes have been embedded and ${\cal S}_{\rm int}$ is the interaction of the brane theory
with gravity i.e the the interactions of the gauge fields with gravitons. At low energy, the coupling of the gauge theory with the gravitons is negligible and we can ignore the interaction term to obtain a gauge theory living in 
flat space and a theory of free low energy gravitons. 

Let us analyze the system of branes embedded in this geometry  using  supergravity which is the low energy limit of string
theory. 
The supergravity solution incorporating the back reaction of the flux sourced by the branes was computed back in
the 90's \cite{Hor-Stro}. The metric reads \cite{Klebanov-review}  
\bg \label{Sugra-1}
ds^2&=&\frac{1}{\sqrt{H}}\left(-\tilde{G}(r)dt^2+d\overrightarrow{x}^2\right)+\frac{\sqrt{H}}{\tilde{G}(r)} dr^2+r^2 \sqrt{H} d\Omega_{5}^2\nonumber\\
H&=&1+L^4/r^4, ~~\tilde{G}(r)=1-\frac{\tilde{r}_0^4}{r^4},~~L^4=4\pi g_sN \alpha'^2\nonumber\\
\nd     
where $d\Omega_5^2$ is the metric of five dimensional sphere $S^5$, $g_s$ is the string coupling and $\alpha'=l_s^2$ is the
string scale. Here horizon is located at $r=\tilde{r}_0$ and extremal limit is achieved by taking $\tilde{r}_0=0$. In the
extremal limit, 
we see that we have an horizon at $r=0$ and for the near horizon limit, $r\ll L$,
the above metric takes the following form
\bg \label{AdS-metric}
ds^2&=&\frac{r^2}{L^2}\left(-dt^2+d\overrightarrow{x}^2\right)+\frac{L^2}{r^2} dr^2+L^2 d\Omega_{5}^2
\nd 
 which is the metric of $AdS_5\times S_5$, where the AdS throat radius given by $L$.  For an observer located at the boundary $r=\infty$, 
 gravitons from the near horizon region appear to be of low energy as they are red-shifted. As the energy of the gravitons
 should be measured by an observer at the boundary, low energy modes consist of the gravitons coming from the near horizon
 region i.e. $AdS_5\times S_5$ geometry and low energy gravitons from the bulk. 
 
 Thus we have two descriptions of the same system of brane configuration at low energy; one in terms of  branes and low energy
 gravitons, the other in terms of $AdS_5\times S_5$ geometry and low energy gravitons.  We can identify the two
 descriptions 
 and conclude that the brane theory described by ${\cal S}_{\rm branes}$ has an equivalent description in terms
 of gravitons of $AdS_5\times S_5$ geometry. Thus the gauge theory on the branes which is a Conformal field theory in 
 flat four dimensional space time is dual to Anti de-Sitter geometry \cite{Maldacena-1}. This duality is known as the
 AdS/CFT correspondence. 
 \begin{figure}[htb]\label{cone}
		\begin{center}
\includegraphics[height=8cm, width=12cm]{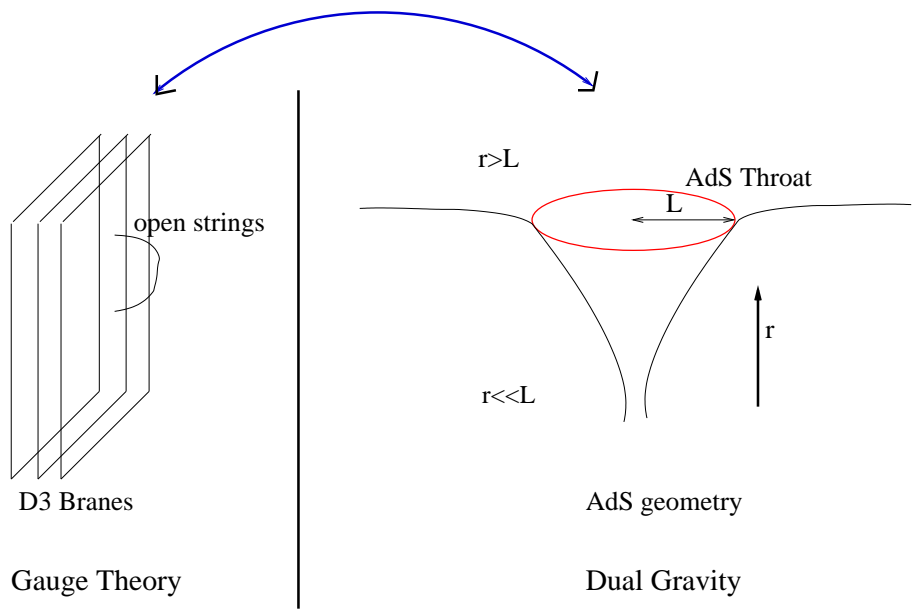}
		\caption{{Open string gauge theory from D3 branes  and its dual geometry. }}

		\end{center}
		\end{figure}

 One may ask at what regime of couplings constant space is this duality valid? To answer this, first note that  the string coupling $g_s$ which describes the coupling between the strings and thus
 the coupling between the gauge fields, can be identified with the Yang-Mills coupling $g_{\rm YM}^2=4\pi g_s$.
 On the other hand the t' Hooft coupling for $U(N)$ gauge theory is $\lambda=g_{\rm YM}^2N$ and thus  
 the AdS throat radius $L^4=g_{\rm YM}^2N \alpha'^2=\lambda \alpha'^2$  in $\alpha'$ units is the 't Hooft coupling. 
 The supergravity solution in (\ref{Sugra-1},\ref{AdS-metric}) is only valid when curvature of space is small. Observing
 that the 
 Ricci scalar for $AdS_5$
 and $S_5$ are of ${\cal O}(L^{-2})$, we conclude that for the solution in (\ref{Sugra-1},\ref{AdS-metric}) to be valid, 
 we need $L$ to be very large. 
 Keeping in mind that length is measured in string units $\alpha'$, this means $g_s N=\lambda/4\pi\gg 1$. We also want 
 to keep $g_s$ small so we can ignore effects of string loops in the gravity
 action. The only way to keep $g_s\ll 1$ and still have $g_s N\gg 1$ is to take $N\rightarrow \infty$. 
 Thus in these limits, we
 have large $N$ strongly coupled gauge theory with 't Hooft coupling $\lambda\gg 1$ dual to Anti de-Sitter geometry with small
 curvature. Hence strongly coupled Conformal field theory is dual to weakly coupled Anti de-Sitter gravity.

\section{Non Conformal Field Theory and Dual Geometry}
The duality between conformal field theory and AdS geometry is a particular example of the Holographic Principle mapping
Hilbert spaces of  gauge theories  with that of gravity. By studying brane excitations in various background geometries one
can attempt to generalize the AdS/CFT correspondence to include theories with running couplings. In the following sections
we first analyze the gauge theories that arise from excitations of branes placed in geometries with conical singularity and
then describe their weakly coupled dual gravity.

\subsection{Gauge Theory from Brane Configuration}
We will consider branes placed in geometries with conifolds and study their excitations. Before going into the brane setup,
 we briefly discuss the conifold geometry. For details, consult \cite{O18}-\cite{4}
 
Consider a six dimensional cone with base $T^{1,1}$ and radial coordinate $r$. 
This cone is a manifold with conic singularity i.e a conifold which is a solution
to the Einstein equations in vacuum  and it has the metric 
\bg \label{conemt}
ds^2_{6}=dr^2+r^2 ds^2_{T^{1,1}}
\nd
The metric of the base $T^{1,1}$ is given by 
\bg \label{mt11}
ds^2_{T^{1,1}}=\frac{1}{9}\left[d\psi +\sum_{i=1}^2 {\rm cos}\theta_i d\phi_i\right]^2
+\frac{1}{6}\sum_{i=1}^2\left[d\theta_i^2 +{\rm sin}^2\theta_1 d\phi_i^2\right]
\nd
where $\psi$ is an angular coordinate ranging from $0$ to $4\pi$ and $(\theta_1,\phi_1)$ , $(\theta_2,\phi_2)$ parameterizes
the two  two-spheres $S^2$'s. The form of the metric makes it clear that $T^{1,1}$ is a $U(1)$ bundle over $S^2\times
S^2$ and has the topology of $S^2\times S^3$.

Recall that a two dimensional cone embedded in three dimensional space 
has the familiar embedding equation 
\bg
x^2+y^2=z^2\nonumber
\nd         
where the base of the cone is a circle $S^1$ with radius $z^2=r^2$. The six dimensional conifold is a generalization to
higher dimensions and has the embedding  
\bg \label{con-1}
z_1z_2-z_3z_4=0
\nd
in $C^4$, where $z_i$ are complex coordinates given by \cite{O18}
\bg
z_1&=& r^{3/2} e^{i/2\left(\psi-\phi_1-\phi_2\right)}{\rm sin}\left(\theta_1/2\right) {\rm
sin}\left(\theta_2/2\right)\nonumber\\
z_2&=& r^{3/2} e^{i/2\left(\psi+\phi_1+\phi_2\right)}{\rm cos}\left(\theta_1/2\right) {\rm
cos}\left(\theta_2/2\right)\nonumber\\
z_3&=& r^{3/2} e^{i/2\left(\psi+\phi_1-\phi_2\right)}{\rm cos}\left(\theta_1/2\right) {\rm
sin}\left(\theta_2/2\right)\nonumber\\
z_4&=& r^{3/2} e^{i/2\left(\psi-\phi_1+\phi_2\right)}{\rm sin}\left(\theta_1/2\right) {\rm
cos}\left(\theta_2/2\right)
\nd

To see the symmetries of the space, we can parameterize the conifold with another set of complex coordinates $w_i$ given by
\bg
&&z_1=w_1+iw_2, ~~~~~~~~ z_2=w_1-iw_2\nonumber\\
&&z_3=-w_3+iw_4,~~~~~~ z_4=-w_3-iw_4
\nd
In terms of $w_i$ coordinates, the conifold  equation (\ref{con-1}) becomes
\bg \label{t11a}
\sum w_i^2=0
\nd
On the other hand, the base $T^{1,1}$ is the intersection of the cone with surface 
\bg \label{t11b}
\sum |w_i|^2=r^3
\nd
We note from (\ref{t11a}), (\ref{t11b}) that $T^{1,1}$ is invariant  under rotation of the four $w_i$ coordinates, that is
under the group $SO(4)\simeq SU(2)\times SU(2)$ and an overall phase rotation. Thus the symmetry group of $T^{1,1}$ is
$SU(2)\times SU(2)\times U(1)$. This will come in handy shortly. 

With the understanding of the symmetries of our conifold geometry, consider embedding $N$ D3 branes in ten dimensional manifold
with the metric 
\bg
ds_{10}^2=-dt^2+d\overrightarrow{x}^2+ds_6^2
\nd
where $ds_6^2$ is given by (\ref{conemt}). That is we have four dimensional Minkowski space along with the six dimensional conifold. The D3 branes live in the flat four
dimensional space and and are placed at the tip of the conifold at fixed radial location $r=0$, as shown in Fig. {\bf 2.2}.
\begin{figure}[htb]\label{cone}
		\begin{center}
\includegraphics[height=6cm, width=10cm]{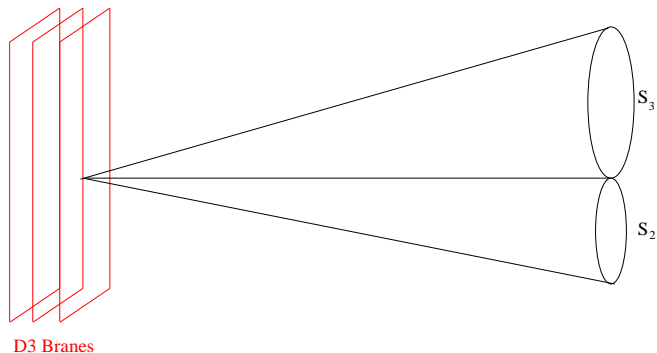}
		\caption{{$N$ D3 branes placed on conifold singularity.}}

		\end{center}
		\end{figure}
   
The excitations of the massless open strings ending on these D3 branes are described by 
gauge fields and complex matter fields $A_i,B_i$, $i=1,2$ which
transform as bi-fundamental fields under the gauge group $SU(N)\times SU(N)$. Note that the  matter fields $A_1, A_2$ transform 
under global $SU(2)$ and so do $B_1,B_2$ under another $SU(2)$ and we also have global $U(1)$ phase rotation. Thus we have
$SU(2)\times SU(2)\times U(1)$ global symmetry, which is also the symmetry of the conifold! This is not surprising as these
fields describe motion of the D3 branes \cite{Witten-DBExc} and  the branes move in the
conifold direction. Thus the fields $A_i, B_i$ are really coordinates-ordinates of the conifold and can be written as 
\bg
&&z_1=A_1B_1, ~~~~~ z_2=A_2B_2\nonumber\\
&&z_3=A_1B_2, ~~~~~ z_4=A_2B_1
\nd

Having analyzed the global symmetry, we would like to understand the origin of the gauge group $SU(N)\times SU(N)$ and the
bi-fundamental nature of the matter fields. The nature of the gauge group becomes clear once we analyze the T dual setup of
the Type IIB brane configuration in conifold geometry. We will be brief on what follows and for detailed discussion on T dual of Type
IIB which is Type IIA brane constructions, please consult
\cite{Hanany}-\cite{Park}. 
\begin{figure}[htb]\label{cone}
		\begin{center}
\includegraphics[height=6cm, width=12cm]{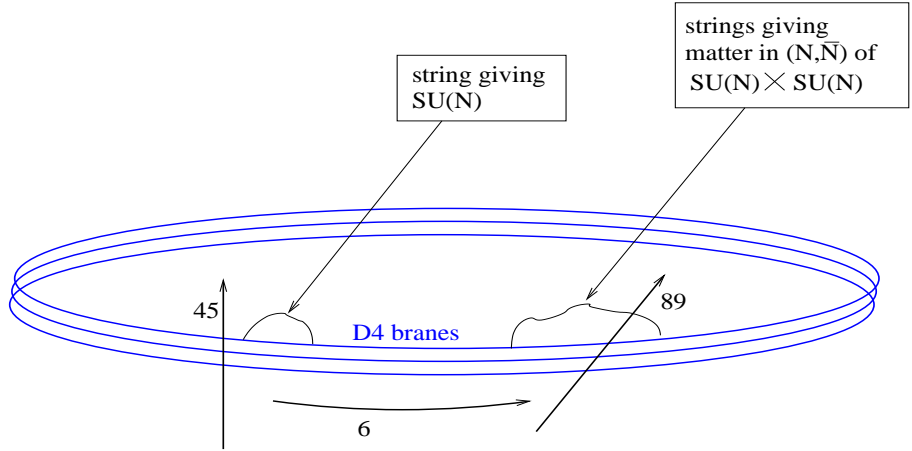}
		\caption{{T dual of KW model.}}

		\end{center}
		\end{figure}
   
Under T
duality the cone becomes two intersecting NS5 branes extending along the 012345 and 012389 directions, while the D3 
branes become D4 branes along 01236 directions  \cite{keshav-1}. Here 0123 are the four Minkowski directions , 6 is the
radial direction and 45789 are the angular directions. The T dual setup is shown in Fig {\bf 2.3} where we have suppressed 0123
and the 7 direction. As D4 branes can end on NS5 branes, they get divided into two branches
between the NS5 branes along the 6 direction as shows in Fig {\bf 2.3}. On each branch there are $N$ D4 branes giving rise 
to $SU(N)$ gauge group. But there are also strings with one end on a D4 brane on one side of the NS5 and the other 
on the other side NS5 brane, as in Fig {\bf 2.3}. These strings give rise to matter fields which transform as $(N,\bar{N})$
of  $SU(N)\times SU(N)$ gauge group. Thus D3 branes at the tip of the conifold gives $SU(N)\times
SU(N)$ gauge theory and this is known as the Klebanov-Witten (KW) model \cite{KW}.

With the understanding of the bi-fundamental nature of the gauge group, we now analyze the various couplings of the KW model. 
There are three couplings
namely the gauge couplings $g_1,g_2$ corresponding to the gauge group $SU(N)\times SU(N)$ and the coupling $h$ to quartic
super potential \cite{KW,Strassler:2005qs}. The beta functions for the couplings which govern how they change with
energy scale $\mu$ is given by

\bg\label{kwbeta}
\beta_{g_1} = - {g_1^3 N\over 16\pi^2} \left({1+2\gamma_0\over 1-{g_1^2N\over 8\pi^2}}\right), ~~~
\beta_{g_2} = - {g_2^3 N\over 16\pi^2} 
\left({1+2\gamma_0\over 1-{g_2^2N\over 8\pi^2}}\right), ~~~ \beta_\eta = \eta(1+2\gamma_0)
\nd
where $\eta = h\mu$ is the dimensionless coupling of the theory. Observe that  
all the fields in the theory have the same anomalous dimension 
$\gamma_0(g_1, g_2, h)$ \cite{KW, Strassler:2005qs} and the three beta functions vanish exactly when
\bg\label{betazero1}
\gamma_0(g_1, g_2, h) ~ = ~ -{1\over 2}
\nd   
which is one equation for the three couplings. 
\begin{figure}[htb]\label{rgsurface1}
		\begin{center}
\includegraphics[height=6cm]{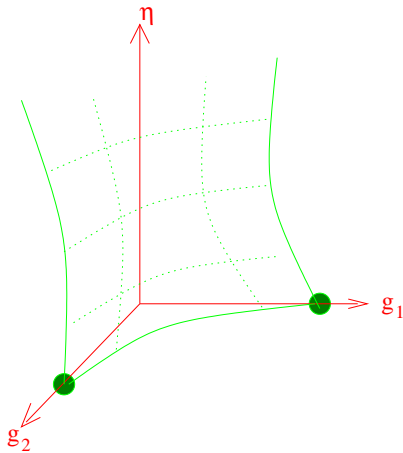}
		\caption{{The two-dimensional RG surface in the Klebanov-Witten theory.}}
		\end{center}
		\end{figure} 
The couplings flow with scale $\mu$ and once they reach a value such that 
$(g_1,g_2,h)$ satisfy equation
(\ref{betazero1}), there is no flow in the theory.  The theory reaches a fixed point and all the couplings stay at that 
value as energy scale is changed thereafter i.e. the field theory becomes conformal. As equation (\ref{betazero1}) is an equation of a  surface in three
dimensions, the fixed points in this theory form a {\it two-dimensional}
surface in the three-dimensional space of couplings (see Fig {\bf 2.4} for details). 
The  flow of the couplings with scale is of course the Renormalization Group (RG) flow as illustrated in 
Fig {\bf 2.5} below.
\begin{figure}[htb]\label{rgflow1}
		\begin{center}
	\includegraphics[height=6cm]{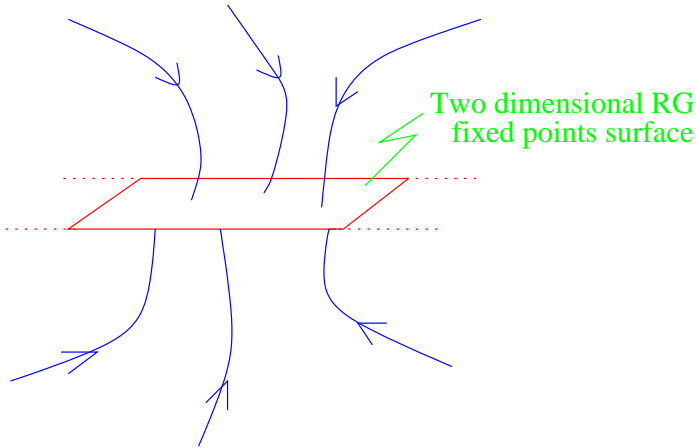}
				\caption{{ The typical RG flows in the Klebanov-Witten theory.}}
		\end{center}
		\end{figure}
Since the sign of the two beta functions are negative any arbitrary flow in the coupling constant space 
brings us to the fixed point surface. Thus the brane setup in Fig. {\bf 2.2} leads to a conformal field theory with gauge 
group $SU(N)\times SU(N)$.

Now consider embedding  $M$ D5
branes in the conifold geometry with D3 branes where the D5 branes wrap the two cycle $S^2$ of the conifold base and extend 
in four Minkowski directions. The D5 branes will
slide down to the tip at r=0 and  we have the brane setup of Fig.{\bf 2.6}.
\begin{figure}[htb]\label{cone}
		\begin{center}
\includegraphics[height=6cm, width=10cm]{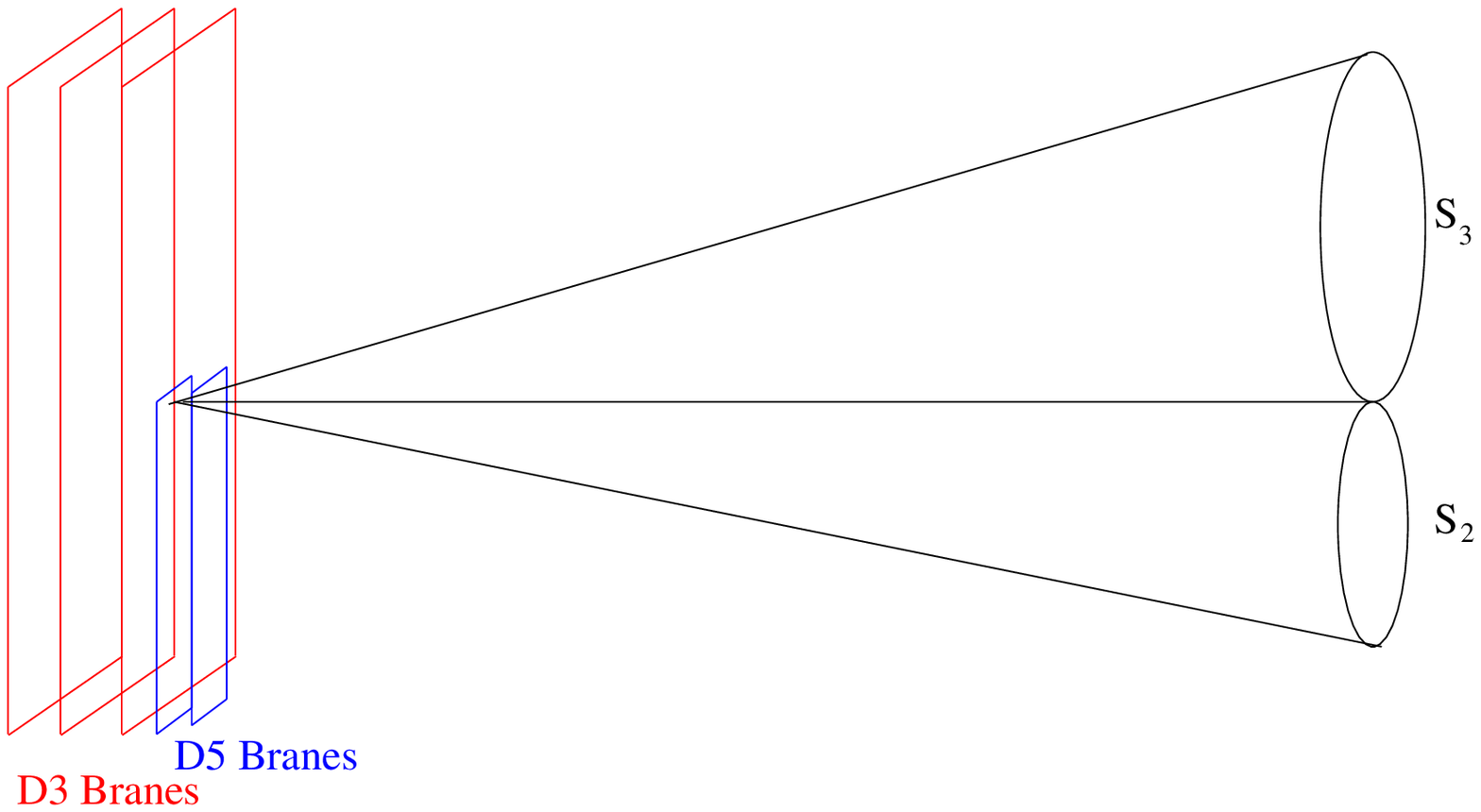}
		\caption{{D3-D5 system placed at conifold singularity.}}

		\end{center}
		\end{figure}
 We again have bi-fundamental matter fields
$A_i,B_i$ with global symmetry group $SU(2)\times SU(2)\times U(1)$. But now due to the additional D5 branes wrapping one 
of the $S^2$'s, the gauge group is $SU(N+M)\times SU(N)$. 

The bi-fundamental nature of the fields and the different size the
gauge groups become clear when we go to
the T dual picture as shown in fig. {\bf 2.7}. The M D5 branes which wrap only one of the $S^2$'s (albeit of vanishing
radius) become D4 branes 
which stretch between the NS5 branes 
only on one branch and thus
contribute to only one of the gauge groups \cite{keshav-2}. This effectively gives $N+M$ D4 branes stacked in one branch between the NS5
branes where the other branch has $N$ D4 branes. This gives rise to the group $SU(N+M)\times SU(N)$. 
\begin{figure}[htb]\label{cone}
		\begin{center}
\includegraphics[height=6cm, width=12cm]{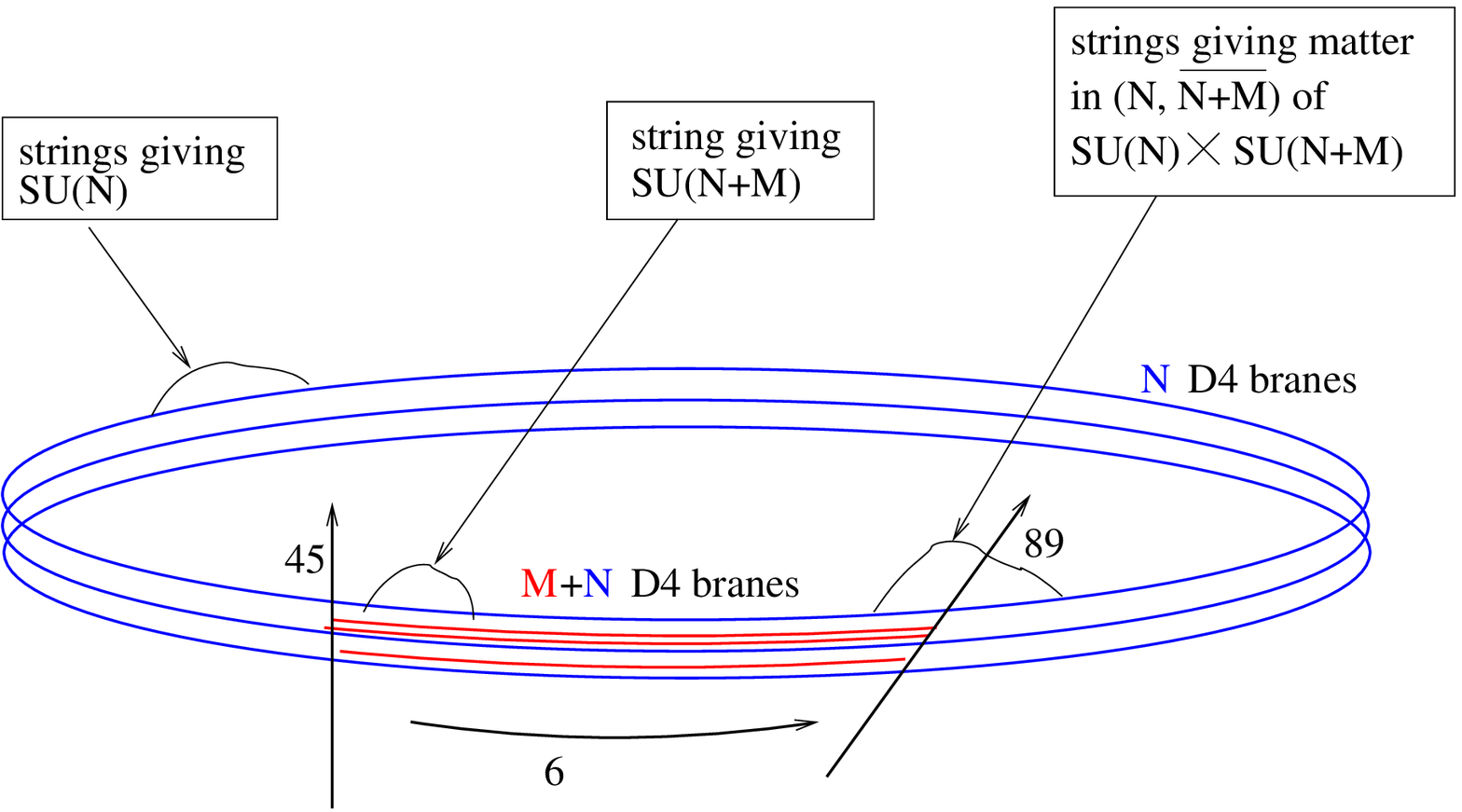}
		\caption{{T dual of KS model.}}

		\end{center}
		\end{figure}

As $M\neq 0$ we can
define  $kM\equiv
N+M $ and then the gauge theory is  $SU(kM) \times 
SU((k-1)M)$. The gauge couplings are now $g_k, g_{k-1}$ for the two gauge groups respectively and we also have $\eta$ to be 
the  dimensionless coupling of the quartic  super potential. The three beta functions now are \cite{1, Strassler:2005qs}:
\bg\label{betaksnow}
&&\beta_k = -{g_k^3 kM\over 16\pi^2}\left[ {(1+2\gamma_0) + {2\over k}(1-\gamma_0)\over 1-{g_k^2 kM\over 8\pi^2}}\right],
~~~~~ \beta_\eta = \eta(1+2\gamma_0)\nonumber\\
&&\beta_{k-1} = -{g_{k-1}^3 (k-1)M\over 16\pi^2}
\left[ {(1+2\gamma_0) - {2\over k-1}(1-\gamma_0)\over 1-{g_{k-1}^2 (k-1)M\over 8\pi^2}}\right]
\nd
from which we see that they differ from (\ref{kwbeta}) by ${\cal O}(1/k)$ factors. Note that
there is  
no point in the coupling constant space where all the three beta functions  vanish exactly and thus the theory 
has no conformal fixed points (except when all couplings are zero and we have free theory at {\it all scales}). 
\begin{figure}[htb]\label{cone}
		\begin{center}
\includegraphics[height=8cm, width=10cm]{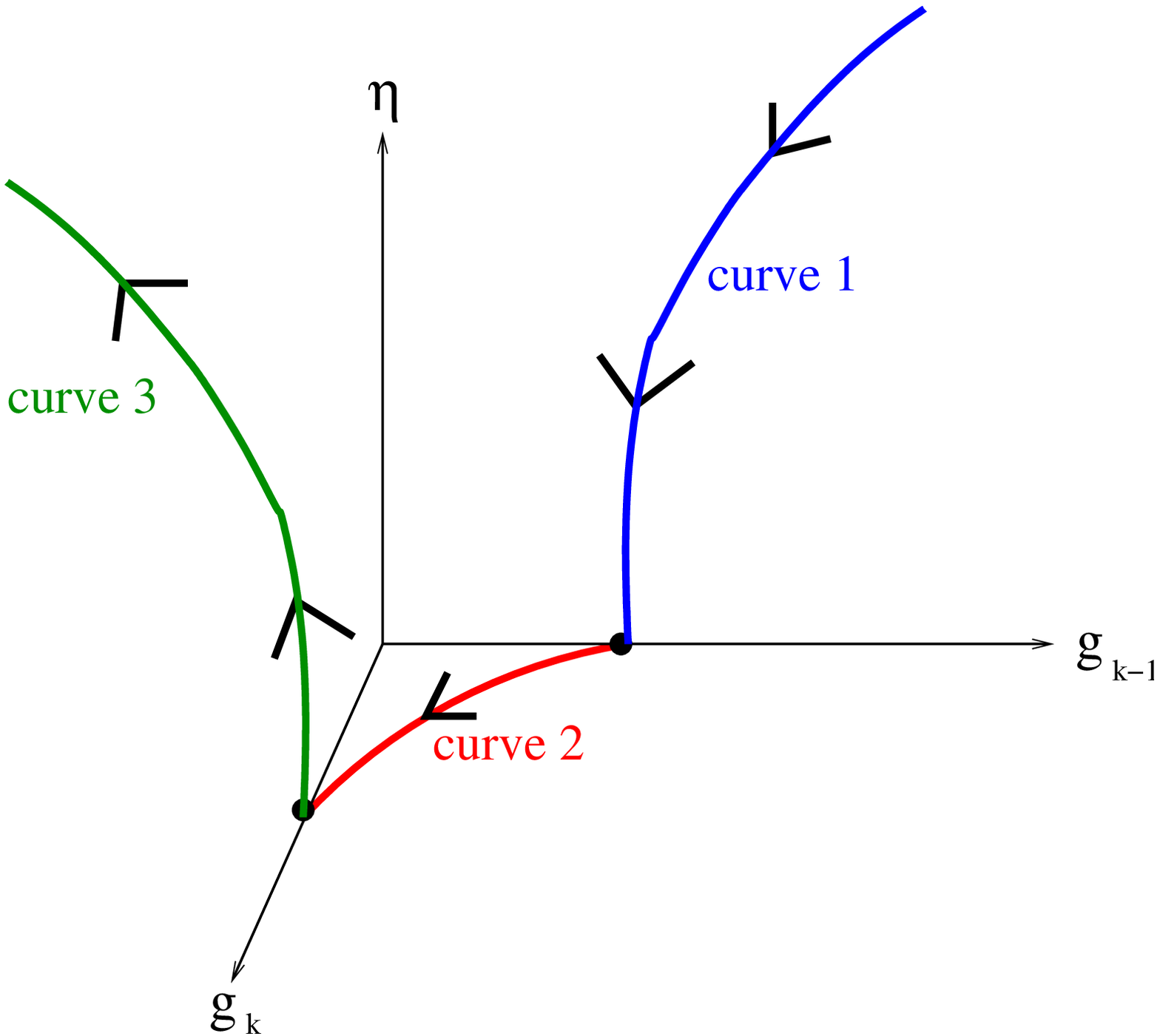}
		\caption{{RG flow of KS model.}}

		\end{center}
		\end{figure}

Solutions to (\ref{betaksnow}) along with boundary conditions determine the RG flow of theory and the flow in the space of 
coupling constants is depicted by the arrows in Fig {\bf 2.8}. From (\ref{betaksnow}) we see that the gauge couplings run
logarithmically with scale $\mu$ and when both $g_k, g_{k-1}$ are not zero, the difference between them grows with scale. 
This is the Klebanov-Strassler (KS) model \cite{1}. 
As the QCD coupling $\alpha_s$ also runs logarithmically with scale, the gauge theory we obtained above could have common
features with QCD. Furthermore, as we shall see in the next section, this gauge theory has a dual gravity description and 
we may learn about strongly coupled QCD by analyzing the dual geometry of the brane setup of Fig {\bf 2.6}. But
before making connections to QCD, lets first analyze the  RG flow of KS
model. 

The RG flow  incorporates a Seiberg duality cascade, where under a series of dualities  higher rank gauge groups have
equivalent description in terms of  lower rank groups. Here we will briefly review the Seiberg duality cascade and its realization
in the KS model. For a comprehensive review please consult \cite{Strassler:2005qs, Seiberg-1}.

Seiberg duality states that strongly coupled  
$SU({\cal N})$
gauge theory  with $N_f$ flavors is dual to weakly coupled $SU(N_f-{\cal N})$ gauge theory with $N_f$ flavors. 
Recall that we have $SU(N+M)\times SU(N)$ gauge group and the $SU(N+M)$ branch has $2N$
effective flavors while the $SU(N)$ branch has $2(N+M)$ flavors \cite{1,Strassler:2005qs}. To see this, consider the fields
$A_1,A_2$ and note that each of them  have color indices under two color groups $SU(N+M)$ and $SU(N)$. If one fixes 
the color under one of the group, say $SU(N+M)$, then the color indices for the group $SU(N)$ can be thought of as flavor
indices. Then for the $SU(N+M)$ color symmetry group, the $SU(N)$ group appears as flavor symmetry group. As we have two
fields $A_1$ and $A_2$, this means that the $SU(N+M)$ branch has $2N$ effective flavors. The same argument shows that
$SU(N)$ group has $2(N+M)$ effective flavors.  

Now consider the flow on the $\eta=0$ plane given by  curve 2 of Fig
{\bf 2.8}. The coupling $g_{k-1}$ corresponding to gauge group $SU(N)$ shrinks while $g_k$ corresponding to $SU(N+M)$ grows 
as scale is changed from UV to IR. . At the end point of curve 
2, the $SU(N+M)$ gauge group is strongly coupled with $2N$ effective flavors and thus it is dual to weakly coupled
$SU(2N-(N+M))=SU(N-M)$ gauge theory under Seiberg duality. 
\begin{figure}[htb]\label{cone}
		\begin{center}
\includegraphics[height=8cm, width=14cm]{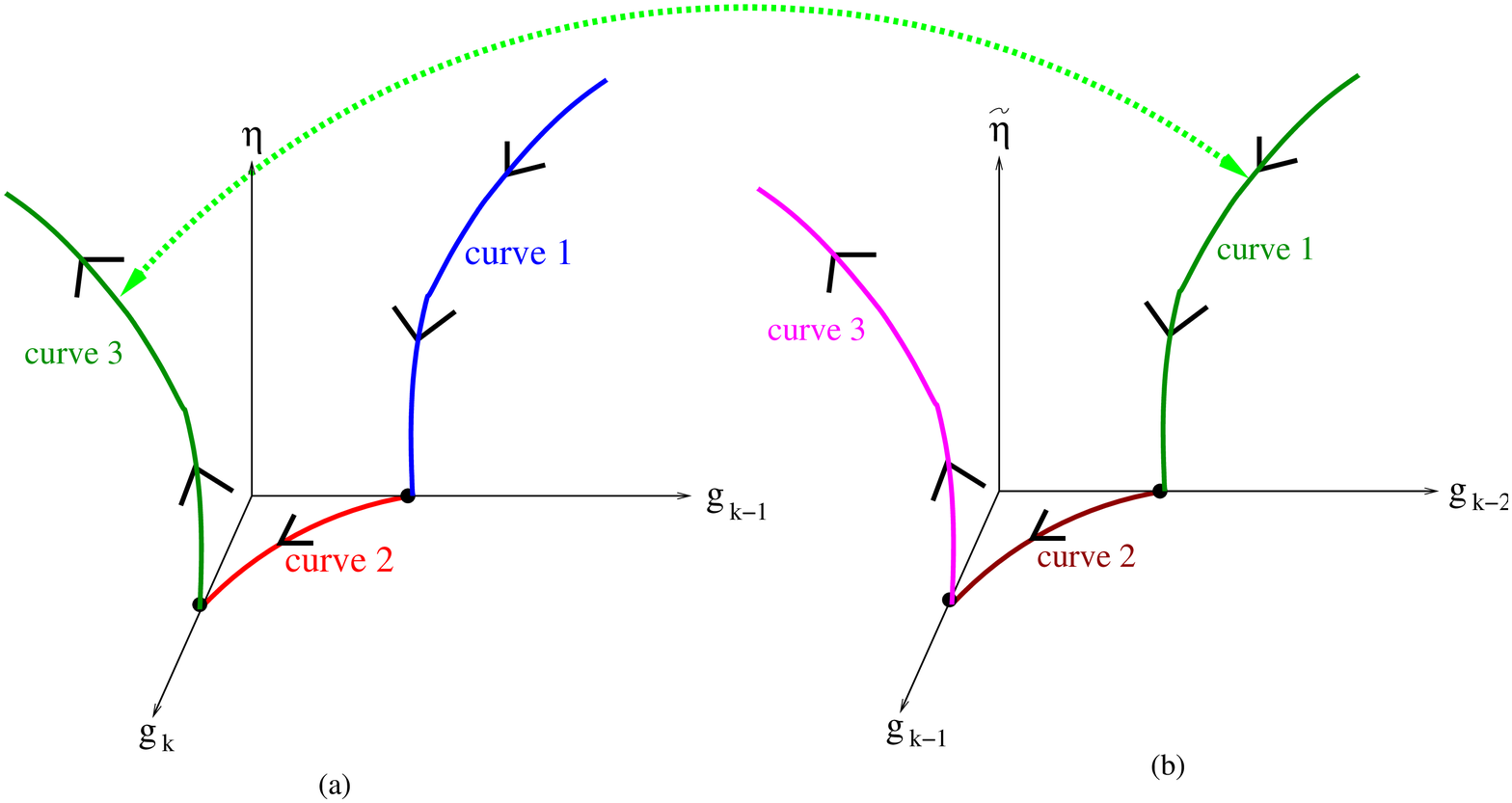}
		\caption{RG flow of  SU(N+M)$\times$ SU(N)along with its dual SU(N-M)$\times$SU(N)}
		\end{center}
		\end{figure}
This means the end point of curve 2 has two equivalent
descriptions; one in
terms of $SU(N+M)$ strongly coupled gauge theory with $2N$ flavors and the other in terms of $SU(N-M)$ weakly coupled gauge
theory with $2N$ flavors. Thus $SU(N+M)\times SU(N)$ gauge theory is dual to $SU(N-M)\times SU(N)$ gauge theory. 
But $N-M=(k-2)M$ and thus we can draw the endpoint of curve 2 on another coupling constant space with couplings
$g_{k-1},g_{k-2}$ as shown in Fig {\bf 2.9}. In fact the RG flow of  curve 3 can be identified with that of curve 1 in the 
space of new couplings $g_{k-1},g_{k-2}$ and new quartic coupling $\tilde{\eta}$. 
The RG flow of the new couplings
$g_{k-1}, g_{k-2}, \tilde{\eta}$ can be obtained by replacing $k$ with $k-1$ and $\eta$ by
$\tilde{\eta}$ in (\ref{betaksnow}). Solving the RG equations, one observe that when $g_k$ of $SU(N+M)$ sector grows, 
$g_{k-2}$ of $SU(N-M)$ shrinks, so
the flows of curve 3 in Fig {\bf 2.9(a)} and of curve 1 in Fig {\bf 2.9(b)} are in opposite direction. 
This also means $\eta\sim 1/\tilde{\eta}$ and the identification of curve 3 with curve 1 is shown in Fig {\bf 2.9} is justified.
 
 Denoting $N-M=\tilde{N}$, we now have $SU(\tilde{N}+M)\times SU(\tilde{N})$ and
repeating the same arguments as before we can perform another Seiberg duality to
obtain  the $SU(\tilde{N}-M)\times SU(\tilde{N})$ gauge group. At each step of the duality, the effective $\bar{N}$ (defined
from gauge group $SU(\bar{N}+M)\times SU(\bar{N})$  is 
reduced by $M$ units and we have a cascade of
dualities known as the Seiberg duality cascade. The
reduction continues until we reach a point when one of the groups has zero size and we end up with 
$SU(0)\times SU(M)=SU(M)$ gauge group. 
\begin{figure}[htb]\label{cone}
		\begin{center}
\includegraphics[height=8cm, width=12cm]{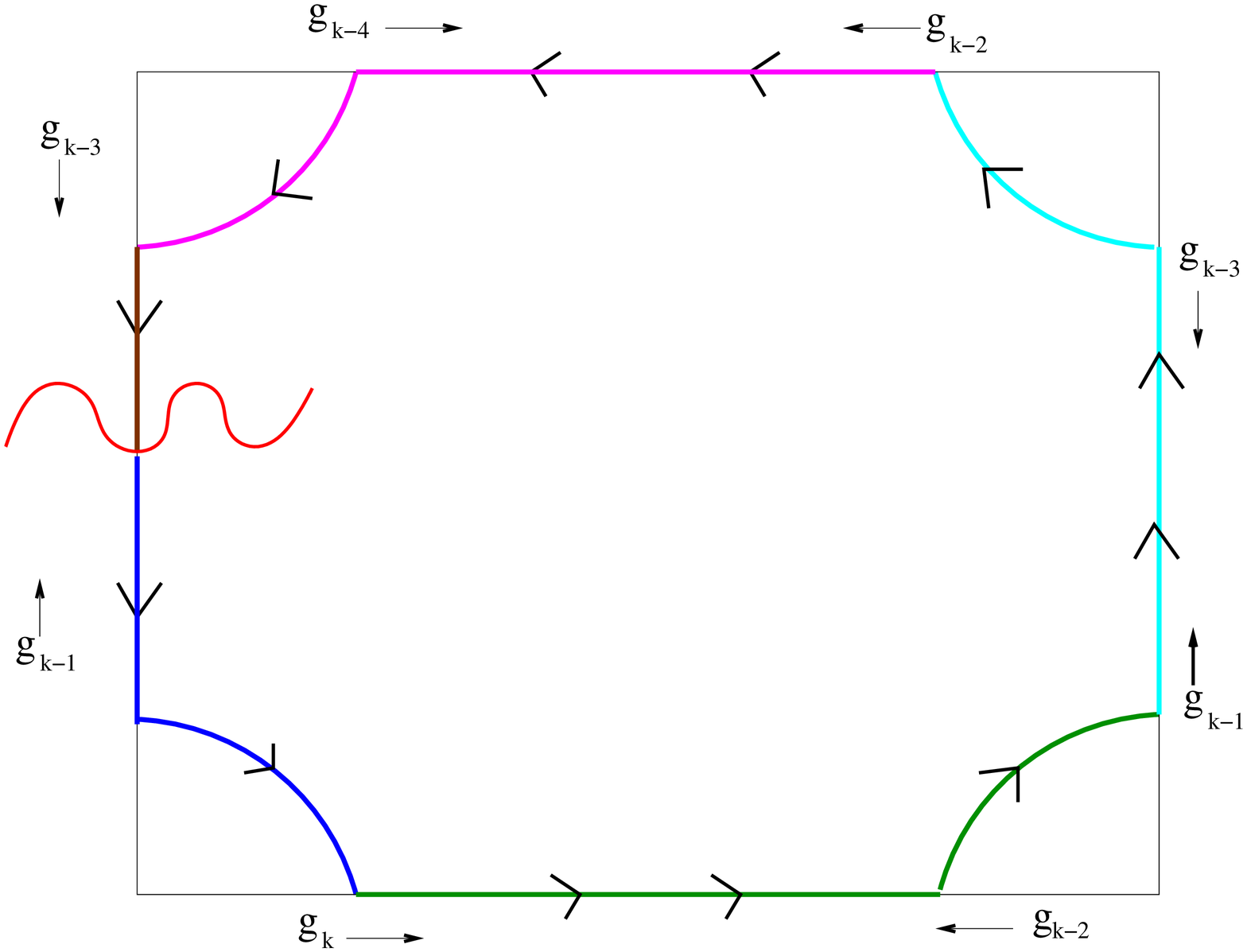}
		\caption{RG flow of the gauge theory as it cascades under Seiberg duality. }
		\end{center}
		\end{figure}

By suppressing the quartic coupling $\eta$,$\tilde{\eta}$, we can draw the RG flows of the couplings
$(g_{k},g_{k-1},...,g_{k-n})$ on the 
same plot as depicted on Fig {\bf 2.10}. Each line of the boundary is the axis for coupling $g_{\tilde{k}}$ and its Seiberg
dual $g_{\tilde{k}-2}$. This depiction is indeed valid as when $g_{\tilde{k}}$ grows and becomes strong, its dual $g_{\tilde{k}-2}$ 
shrinks and becomes weak and vice-versa. Intersecting lines describe gauge couplings $(g_{\tilde{k}},g_{\tilde{k}-1})$ and 
we move from one pair of couplings to the other with
the flow. The cut in the Fig {\bf 2.10} indicates that the flow do not take us back to the original coupling $g_{k-1}$,
rather the flow takes us to another coupling $g_{k-n}$ and we move from one sheet of coupling constant space to another.  

Thus we see that in the KS model there are multiple equivalent descriptions of the same gauge theory and each
description is related to the other by Seiberg duality. As we go from UV to IR, we can describe the gauge theory with lower
and lower rank groups. In the language of Wilsonian RG flow, we are integrating out UV modes and obtaining effective
Lagrangians for lower and lower rank gauge groups in the IR. If we keep all the relevant, irrelevant and marginal operators 
as we integrate out, physical observables of course should not change with the RG flow. But if we leave out certain 
operators to obtain the effective Lagrangians, the flow will result in changes in physical observables. We will have more to
say about this subtlety in the coming sections.

In summary, the gauge theory in the KS model has a very rich structure with the rank of the gauge group growing in the 
UV. In fact the far UV of gauge theory is Seiberg dual to an infinite rank gauge group while the far IR is a rather simple $SU(M)$
gauge theory. But we still do not have matter in the fundamental representation and to make connection with QCD, we need to amend
the KS brane setup.

In order to
obtain fundamental matter, one introduces D7 branes. The D7 branes can be embedded in various ways
\cite{4,Karch-katz,Sakai-Son,gtpapers,cotrone,cotrone2,kuperstein} and in
particular the D7 brane world volume will depend on the background geometry along with the boundary conditions.
In Ouyang's model \cite{4} seven branes are embedded via the following equation (see also \cite{sully}):
\bg \label{seven} 
z\equiv r^{3\over 2} {\rm exp}\left[i(\psi-\phi_1 - \phi_2)\over 2\right] {\rm sin}~{\theta_1 \over 2} 
~{\rm sin}~{\theta_2 \over 2} = \mu
\nd
where $\mu$ is a complex quantity. In the limit where $\mu \to 0$, the seven branes 
are oriented along  two branches:
\begin{eqnarray}\label{branches} \nonumber 
&&{\rm Branch ~1}: ~~\theta_1 = 0, ~~\phi_1 = 0 \\
&&{\rm Branch ~2}: ~~ \theta_2 = 0, ~~\phi_2 = 0
\end{eqnarray}
{}From the above observe that the seven branes in branch 1 wrap a four cycle ($\theta_2, \phi_2$) and 
($\psi, r$) in the internal space and is stretched along the space time directions ($t, x, y, z$). Similarly in 
 seven branes on branch 2 would wrap a four-cycle ($\theta_1, \phi_1, r, \psi$). 

With the above embedding one needs to check whether Gauss's law is violated. 
As the seven branes wrap a non-compact four cycle filling the entire $r$ direction, the field lines sourced by the branes
extend only in the compact directions. If we draw a Gaussian surface, all the field lines that go out of the surface come
back into the surface as the space is compact. This means there cannot be any net charge due to the seven branes.  
This paradox can  be resolved by allowing the seven brane to wrap a topologically trivial 
cycle so that it can end {\it abruptly} at some $r = r_{\rm min}$ when the embedding is (\ref{seven}). 
Thus the 
D7 brane extend in the radial  direction with $r_{\rm min}<r<\infty$ which is similar to the seven brane configuration
of \cite{Karch-katz,kuperstein}.  Of course there are $N_f$ D7 branes which can overlap and the final configuration of branes is sketched in
Fig {\bf 2.11}. In section {\bf 2.4} we will discuss modification of the embedding (\ref{seven}) which will be more relevant
in describing large $N$ QCD.

\begin{figure}[htb]\label{cone1}
		\begin{center}
\includegraphics[height=6cm, width=10cm]{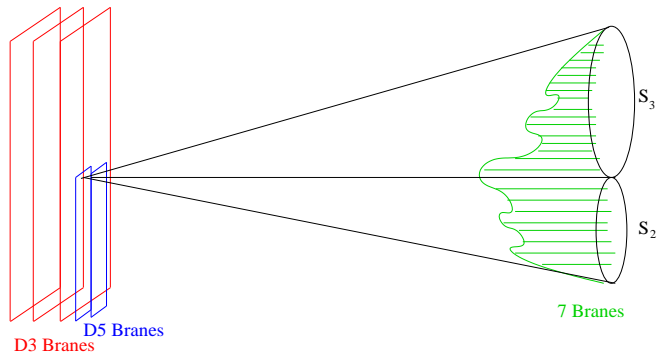}
		\caption{{D3-D5 branes along with D7 branes  embedded in conifold geometry.}}
		\end{center}
		\end{figure}

In addition to the bi-fundamental fields $A_i,B_i$, introduction of the D7 branes give rise to flavor symmetry group $SU(N_f)\times SU(N_f)$ and  matter fields
$q,\tilde{q},Q,\tilde{Q}$
which transform as fundamental under the gauge group $SU(N+M)\times SU(N)$ \cite{4}. The bi-fundamental nature of the
fields follows from the same T duality argument as before and the beta functions now take the form:

\bg\label{betaksnow1}
&&\beta_k = -{g_k^3 kM\over 16\pi^2}\left[ {(1+2\gamma_0) + {2\over k}(1-\gamma_0)\over 1-{g_k^2 kM\over
8\pi^2}}+\frac{N_f(1-2\gamma_{q})}{kM}\right],
~~~~~ \beta_\eta = \eta(1+2\gamma_0)\nonumber\\
&&\beta_{k-1} = -{g_{k-1}^3 (k-1)M\over 16\pi^2}
\left[ {(1+2\gamma_0) - {2\over k-1}(1-\gamma_0)\over 1-{g_{k-1}^2 (k-1)M\over 8\pi^2}}+\frac{N_f(1-2\gamma_{q})}{(k-1)M}\right]
\nd
where $\gamma_q$ is the anomalous dimension of the $q$ field. Again we have logarithmic running of the couplings but also
matter which transforms as fundamental of the gauge group. In Table {\bf 2.1}, we list the various
matter fields and their representation under local and global symmetry groups for the brane setup in Fig {\bf 2.11}.

\vspace{1cm}
\begin{table}
\begin{tabular}{|l|p{1.5in}|p{1.3in}|p{1.2in}|} \hline
\em Field &$SU(N+M)\times SU(N)$ &$SU(N_f)\times SU(N_f)$ &$SU(2)\times SU(2)$\\\hline
$q$ &({$\bf N+M$, 1})&({$\bf N_f$, 1})&({\bf 1,1})\\
$\tilde{q}$ &($\overline{\bf N+M}, 1$)&({1,$\bf N_f$})&({\bf 1,1})\\
$Q$ &(1,${\bf N+M}$)&($\overline{\bf N_f}$,1)&({\bf 1,1})\\
$\tilde{Q}$ &(1,$\overline{\bf N+M}$)&($1,\overline{\bf N_f}$)&({\bf 1,1}) \\
$A_{1,2}$ &(${\bf N+M}$,$\overline{\bf N+M}$ )&($\overline{\bf N_f}$,$\bf N_f$)&({\bf 2,1})\\
$B_{1,2}$ &($\overline{\bf N+M}$,${\bf N+M}$ )&($\bf N_f$,$\overline{\bf N_f}$)&({\bf 1,2})\\\hline
\end{tabular}
\caption{{The field content and their representation under symmetry groups.}}
\end{table}

\vspace{0.5 in}

The field theory here also cascades to lower and lower rank gauge groups under Seiberg duality. But now due to the presence
of fundamental flavor, the effective flavor for $SU(N+M)$ strongly coupled gauge group is $2N+N_f$ and its Seiberg dual weakly
coupled theory is $SU(2N+N_f-N-M)=SU(N-M+N_f)$. This means that at the $i$th step of the cascade, the effective $\bar{N}$
(again defined from the gauge group $SU(\bar{N}+M)\times SU(\bar{N})$)  is reduced by $(M-kN_f)$ units, with $k$ a natural number. This reduction
continues until we end up with just one group $SU(0)\times SU(M-jN_f)=SU(M-jN_f)$. Here $j$ is the number of dualities
performed starting with gauge group $SU(N+M)\times SU(N)$ at some UV scale. This brane setup in Fig {\bf 2.11} is the Ouyang
embedding of D7 branes on Klebanov-Strassler model and we will refer to it as the Ouyang-Klebanov-Strassler (OKS) model. 

In summary, 
for OKS model we have a field theory which is non conformal with matter in fundamental representation with $N_f$ flavors 
and the gauge couplings run logarithmically with scale. At some UV scale the gauge group has description in terms of 
$SU(N+M)\times SU(N)$ group while at the far IR we end up with $SU(M-jN_f)$ gauge group.

For the gauge theories arising from brane configurations in Fig {\bf 2.2}, {\bf 2.6} and {\bf 2.11}, we would like to compute
matrix elements  as they are physical quantities of interest.  
Note that it is the t'Hooft couplings $\lambda_k=(N+M)g_k^2(\Lambda)$ and $\lambda_{k-1}=Ng_{k-1}^2(\Lambda)$ that 
are relevant for computing propagators and scattering amplitudes.  At a scale when either of the two t'Hooft couplings
$\lambda_k,\lambda_{k-1}$ 
become large, perturbative methods fail and we need to resort to other techniques. One approach is to study the theory on 
the lattice while the other is to find the dual gravity. We will discuss the latter in the following section. 

\subsection{The Dual Gravity of the Brane Theory}
Having discussed in some detail the properties of  gauge theories that arise from branes in conifold geometry, we will now
analyze the same brane configurations at low energy using supergravity (SUGRA) approximation of string theory. 
Type IIB  supergravity action including local sources
 in ten dimensions  is \cite{DRS}\cite{GKP}:
\bg \label{action1}
S_{\rm total}&=&S_{\rm SUGRA}+S_{\rm loc}=\frac{1}{2\kappa^2_{10}}\int d^{10}x \sqrt{G}\left(R+\frac{\partial_\mu \bar{\tau}\partial^\mu\tau}{2|{\rm
Im}\tau|^2}-\frac{1}{2}|\widetilde{F}_5|^2-\frac{G_3 \cdot \bar{G}_3}{12 {\rm Im}\tau}\right)\nonumber\\
&+&
\int_{\Sigma_8} C_4\wedge R_{(2)}\wedge R_{(2)}+\frac{1}{8i\kappa_{10}^2}\int \frac{C_4\wedge G_3\wedge \bar{G}_3}{{\rm Im} \tau}+S_{\rm loc}
\nd
in Einstein frame, where $\tau=C_0+i e^{-\phi}$ is the axio-dilaton with $C_0$ being the axion and $\phi$ the dilaton field and  
$\widetilde{F}_5$ is the five-form flux sourced by the D3 branes. Here $G_3\equiv F_3-\tau
H_3$ with $F_3$ the RR three form flux sourced by D5 branes and $H_3=dB_2$  the NS-NS three form flux with $B_2$ being the
NS-NS two form. We also have
$C_4$ the four-form potential,   
$G=\sqrt{{\rm det}~G_{\mu\nu}}$ with $G_{\mu\nu}$
being the metric, $R_{(2)}$  the curvature two-form, and $S_{\rm loc}$ is the action for localized sources in the system
(i.e. D7 branes and other local sources that we may consider). The above action (\ref{action1}) is the most general
supergravity action one gets by placing branes in various geometries and we will consider the relevant terms for the
the brane setups of Fig {\bf 2.2, 2.6} and {\bf 2.11} separately. 

First consider the brane setup of Fig {\bf 2.2}. The $N$ D3 branes sources Ramond-Ramond (RR) five form flux in the supergravity action but
do not source three form flux and 
neither the axio-dilaton field. Minimizing the action (\ref{action1}) with  the only $\tilde{F}_5\neq 0$ being the source, we get the following 
metric \cite{KW} 

\bg \label{KWmetric}
ds^2&=&\frac{1}{\sqrt{H}}\left(-dt^2+d\overrightarrow{x}^2\right)+\sqrt{H} dr^2+r^2\sqrt{H}ds^2_{T^{1,1}}\nonumber\\
H&=&1+h,~~h=\frac{L^4}{r^4}, ~~L^4=4\pi g_sN \alpha'^2\nonumber\\
g_s F_5&=&d^4x\wedge dh^{-1}+\ast \left(d^4x \wedge dh^{-1}\right)
\nd 
where $ds^2_{T^{1,1}}$ is the metric of $T^{1,1}$ given by (\ref{mt11}), $\ast$ is the hodge star operator and we have taken the extremal limit.The horizon is
located at $r=0$ and the near horizon limit ($r\ll L$) of the above metric (\ref{KWmetric}) is that of $AdS_5\times T^{1,1}$
. Using the same arguments as in section {\bf 2.1}, the near horizon gravitons appear to be of low energy as measured by an observer at the
boundary $r=\infty$. On the other hand at low energy, the $SU(N)\times SU(N)$ gauge theory that lives on D3 brane world volume i.e on
four dimensional space, decouples from gravity. Thus we conclude that $SU(N)\times SU(N)$ gauge theory in four dimensional
flat space is dual to $AdS_5\times T^{1,1}$ geometry. 

But in what regime is this duality valid? For our supergravity solution to hold, we need $L\gg 1$ in
$\alpha'$ units, which means that $g_sN\gg 1$. As we want to ignore effects of string loops in supergravity action, we need
$g_s\ll 1$ and combined with the requirement $g_sN\gg 1$, this means $N\rightarrow \infty$. On the other hand the gauge
couplings $g_1,g_2$ corresponding to the gauge group $SU(N)\times SU(N)$ is related to the dilaton and $B_2$ appearing on
SUGRA action by \cite{KW}\cite{keshav-2}

\begin{eqnarray} \label{twocoup} \nonumber
&&\frac{8\pi^2}{g_1^2} = e^{-\Phi}\left[\pi  + \frac{1}{2\pi} \left(\int_{S^2} 
B_2\right)\right] \\  
&& \frac{8\pi^2}{g_2^2} = e^{-\Phi}\left[\pi  -
 \frac{1}{2\pi} \left(\int_{S^2} B_2\right)\right]  
\end{eqnarray}
 For brane configuration in Fig {\bf 2.2}, the dual
geometry has no $B_2$ and dilaton is constant with its value set by the string coupling i.e. $e^{-\phi}=1/g_s$ which  
gives $g_1^2=g_2^2=8\pi g_s$. Thus from dual
supergravity we conclude that
 the gauge couplings do not run with scale and we have conformal field theory with large t'Hooft coupling
$\lambda_1=\lambda_2=8\pi g_sN\gg 1$. This is exactly consistent with our analysis of the gauge theory of KW model in the previous
section  where we found
that the gauge couplings always reach conformal fixed point surface of Fig {\bf 2.5}. 

The field theories arising from the brane setups in Fig {\bf 2.2}, {\bf 2.6} and {\bf 2.11} are at zero temperature.
Introducing temperature in the field theory boils down to introducing black holes in the dual geometry. For the KW model,
metric for the dual geometry incorporating finite temperature of the field theory is given by  
\bg \label{KWmetricbh}
ds^2&=&-\frac{g}{\sqrt{h}} dt^2+\frac{1}{h}d\overrightarrow{x}^2+\frac{\sqrt{h}}{g} dr^2+r^2\sqrt{h}ds^2_{T^{1,1}}\nonumber\\
g&=&1-\frac{r_h^4}{r^4}
\nd 
where $r_h$ is the horizon of the black-hole. Note that the above metric (\ref{KWmetricbh}) can be obtained by minimizing
the SUGRA action (\ref{action1}) with only $\tilde{F}_5\neq 0$ and every other source zero. Thus  metrics in (\ref{KWmetric}) 
and (\ref{KWmetricbh}) can be obtained from the same action with the same sources.

Next for the brane setups in Fig {\bf 2.6} and {\bf 2.11}, the most general SUGRA solution takes the form \cite{FEP} 

\bg\label{bhmet}
ds^2 = {1\over \sqrt{h}}
\left(-g_1 dt^2+dx^2+dy^2+dz^2\right)+\sqrt{h}\Big[g_2^{-1}dr^2+r^2 d{\cal M}_5^2\Big]
\nd
where 
 $g_i$ are functions\footnote{They would in general be functions of ($r, \theta_i$). We will discuss this later.} 
that determine the presence of the black hole, $h$ is the $10d$ warp factor that could be a 
function of all the internal coordinates and $d{\cal M}_5^2$ is given by:
\begin{eqnarray}\label{bhmet2}\nonumber
d{\cal M}_5^2 = && h_1 (d\psi + {\rm cos}~\theta_1~d\phi_1 + {\rm cos}~\theta_2~d\phi_2)^2 + 
h_2 (d\theta_1^2 + {\rm sin}^2 \theta_1 ~d\phi_1^2) + \\ \nonumber 
&& + h_4 (h_3 d\theta_2^2 + {\rm sin}^2 \theta_2 ~d\phi_2^2) + h_5~{\rm cos}~\psi \left(d\theta_1 d\theta_2 - 
{\rm sin}~\theta_1 {\rm sin}~\theta_2 d\phi_1 d\phi_2\right) + \\ 
&& ~~~~~~~~~~~~~~~~~~~~~~~~ + h_5 ~{\rm sin}~\psi \left({\rm sin}~\theta_1~d\theta_2 d\phi_1 - 
{\rm sin}~\theta_2~d\theta_1 d\phi_2\right)
\end{eqnarray}
with $h_i$ being the six-dimensional warp factors. The advantage of writing the background in the above form 
is that it includes all possible deformations in the presence of seven branes, fluxes and other localized sources in the
theory. The difficulty however 
is that the equations for the warp factors $h_i$ are coupled higher order differential equations which do not 
have simple analytical solutions. The original KS solution without  black hole and seven branes is obtained in the limit 
\bg \label{ksagain} 
h_3 = g_i = 1, ~~ h_i = {\rm fixed} 
\nd

 In the presence of seven branes
we obtain:
\bg \label{resconi} 
h_5 = 0, ~~~ h_3 = 1, ~~~ h_4 - h_2 = a, ~~~ g_i = 1 \nd
which puts a seven brane in a resolved conifold background with $a$ = constant \cite{sully} 
using the so-called Ouyang embedding \cite{4}. For $a=0$ and the following choice of $h_i$,
\bg
h_1 = {1\over 9}, ~~~~~ h_2 = h_4 = {1\over 6}, ~~~~~ h_3 = 1 
\nd
 the metric takes the following form \cite{4}

\bg \label{Ouyangm}
ds^2 &=& {1\over \sqrt{h}}
\left(-dt^2+dx^2+dy^2+dz^2\right)+\sqrt{h}\Big[dr^2+r^2 ds^2_{\rm T^{1,1}}\Big]\nonumber\\
h&=&\frac{L^4}{r^4}
\Bigg[1+\frac{3g_sM^2}{2\pi N}{\rm log}r\left\{1+\frac{3g_sN_f}{2\pi}\left({\rm log}r +\frac{1}{2}\right)
+\frac{g_sN_f}{4\pi}{\rm log}\left({\rm sin}\frac{\theta_1}{2}
{\rm sin}\frac{\theta_2}{2}\right)\right\}\Bigg]\nonumber\\
\nd
which is Ouyang's solution \cite{4}. In the limit $N_f=0$, the above metric (\ref{Ouyangm}) reduces to the metric of 
Klebanov-Strassler model \cite{1}.

As mentioned above, a black hole could be inserted in this background by switching on a
non-trivial $g_i$. However a naive choice of fluxes in 
(\ref{resconi}) will break supersymmetry \cite{sully}. In general we will not restrict to  dual geometries which are 
supersymmetric even at zero temperature and in particular supersymmetry will be explicitly broken by the introduction of
black hole.    

 Once we introduce black hole, $g_i\neq 1$ and we do not expect $h_i$ to remain constant anymore. 
We also expect $M$ and $N_f$ in (\ref{Ouyangm}) to be given by some $M_{\rm eff}$ and $N_{f}^{\rm eff}$ respectively.
Our first approximation would then be to make 
the following ansatz for the $h_i, M_{\rm eff}$ and $N_{f}^{\rm eff}$:
\bg \label{hi}
&&h_1 = {1\over 9} + {\cal O}(g_s), ~~~~~ h_2 = h_4 = {1\over 6} + {\cal O}(g_s), ~~~~~ h_3 = 1 + {\cal O}(g_s)\\
&&M_{\rm eff} = M + \sum_{m\ge n} a_{mn} (g_sN_f)^m (g_sM)^n, ~~~~~ 
N_{f}^{\rm eff} = N_f + \sum_{m \ge n} b_{mn} (g_sN_f)^m (g_sM)^n\nonumber 
\nd
with $a_{mn}, b_{mn}$ could in principle be functions of the internal coordinates ($\psi,\phi_i,\theta_i$). 
Note that we have made $m \ge n$ in the above expansions because the precise limits for which our supergravity solution 
would be  valid are: 
\bg\label{prelim2} 
\left (g_s, ~g_s N_f, ~g^2_s M N_f, ~{g_s M^2\over N}\right) ~ \to ~ 0, ~~~~~
(g_s N, ~g_s M)~ \to ~ \infty
\nd  
These limits of the variables 
bring us closer to the Ouyang solution with  little squashing of the two-spheres. This also means 
that the warp factor $h$ in (\ref{bhmet}) can be written as \cite{FEP}:
\bg \label{hvalue}
h =\frac{L^4}{r^4}\Bigg[1+\frac{3g_sM_{\rm eff}^2}{2\pi N}{\rm log}r\left\{1+\frac{3g_sN^{\rm eff}_f}{2\pi}\left({\rm
log}r+\frac{1}{2}\right)+\frac{g_sN^{\rm eff}_f}{4\pi}{\rm log}\left({\rm sin}\frac{\theta_1}{2}
{\rm sin}\frac{\theta_2}{2}\right)\right\}\Bigg]\nonumber\\
\nd

\begin{figure}[htb]\label{cone}
		\begin{center}
\includegraphics[height=6cm, width=12cm]{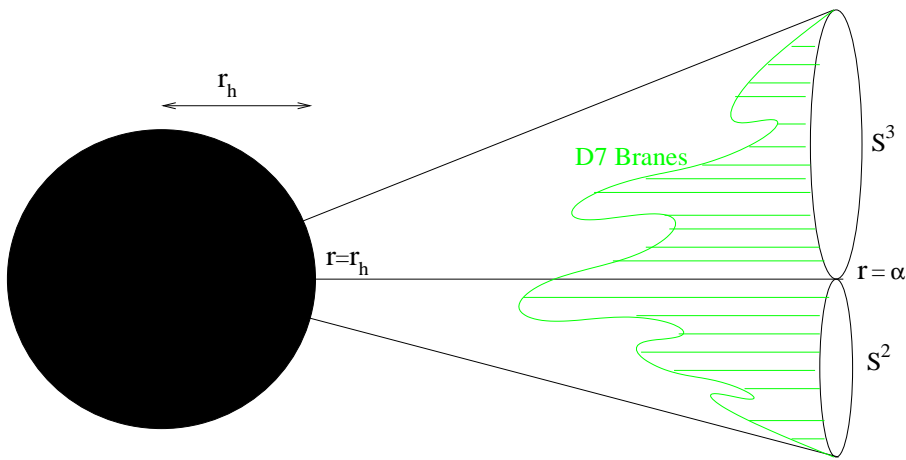}
		\caption{Dual geometry of Ouyang-Klebanov-Strassler model with black hole.  }
		\end{center}
		\end{figure}

Now for the black hole factors, what are the choices for $g_i$ that are consistent with taking back reactions of all the
sources in our setup?  
The Einstein's equations derived from the action (\ref{action1}) for the KS and OKS model (where sources are 
functions of
internal coordinates) fixes  $g_i$'s  to be functions of 
($r,\theta_1,\theta_2$) . Our ansatz therefore 
is \cite{FEP}:
\bg \label{grdef} 
g_1(r,\theta_1,\theta_2)= 1-\frac{r_h^4}{r^4} + {\cal O}(g^2_sM N_f), 
~~~~ g_2(r,\theta_1,\theta_2) = 1-\frac{r_h^4}{r^4} + {\cal O}(g^2_sM N_f)\nd
where $r_h$ is the horizon, and the ($\theta_1, \theta_2$) dependences come from the ${\cal O}(g^2_sM N_f)$ 
corrections. The resolution parameter $a$ is no longer a constant but a function of horizon radius and number of D5 and D7
branes i.e.  
$a = a(r_h) + {\cal O}(g_s^2 M N_f)$. In Fig {\bf 2.12} we
sketched the dual geometry of OKS model with a black hole. The geometry is a resolved deformed conifold with squashing of the two $S^2$'s and
as the local action for D7 branes appear explicitly in (\ref{action1}), we have sketched their embedding  in the dual
geometry. Note that D3 and D5 branes contribute to the geometry through the fluxes  but do not explicitly 
 enter into the action. Hence the these branes are not sketched in the dual geometry and only D7 branes form the localized
 sources.

But what are the fluxes in (\ref{action1}) that give rise to a metric of the form (\ref{bhmet}) with $g_i$ as in
(\ref{grdef})? The background RR three and five-form fluxes can be 
succinctly written as: 
\bg\label{h3f3}\nonumber
  H_3 &=& dr\wedge e_\psi\wedge(c_1\,d\theta_1+c_2\,d\theta_2) + dr\wedge(c_3\sin\theta_1\,d\theta_1\wedge d\phi_1-c_4\sin\theta_2\,d\theta_2\wedge d\phi_2)\\
  & & +\left(\frac{r^2+6a^2}{2r}\,c_1\sin\theta_2\,d\phi_2 
    -\frac{r}{2}\,c_2\sin\theta_1\,d\phi_1\right)\wedge d\theta_1\wedge d\theta_2\,,\nonumber\\
  \widetilde{F}_3 &=& -\frac{1}{g_s}\,dr\wedge e_\psi\wedge(c_1\sin\theta_1\,d\phi_1+c_2\sin\theta_2\,d\phi_2)\nonumber\\
  & &  +\frac{1}{g_s}\,e_\psi\wedge(c_5\sin\theta_1\,d\theta_1\wedge d\phi_1-c_6\sin\theta_2\,d\theta_2\wedge 
d\phi_2)\nonumber\\ 
  & & -\frac{1}{g_s}\,\sin\theta_1\sin\theta_2\left(\frac{r}{2}\,c_2 \,d\theta_1-\frac{r^2+6a^2}{2r}\,c_1\,d\theta_2 \right)
    \wedge d\phi_1\wedge d\phi_2\,.
\nd
where $H_3$ is closed and $\widetilde F_3 \equiv F_3 - C_0 H_3$, $C_0$ being the ten dimensional axion. 
The derivations of the 
coefficients appearing 
in (\ref{h3f3}) are rather involved and can be found in \cite{FEP} 
\footnote{Observe that the background 
EOMs cannot be trivially worked out by solving SUGRA EOMs with fluxes and seven branes sources. This is because, even if 
we know the energy momentum tensors for the fluxes, the energy momentum tensors for $N_f$ coincident 
seven branes are not 
known in the literature. In particular {\it non-abelian} Born-Infeld action for $N_f$ 
seven branes on a {\it curved} background is unknown. 
 In the absence of such direct approach, 
we use an alternative method to derive the EOMs. This method uses the ISD (imaginary self-duality) properties of the 
background fluxes and fields. Details on this  appears in \cite{4}\cite{sully} for the case without black hole 
. 
For the geometry with a black hole  we  found \cite{FEP} that one could find consistent solutions to EOMs using similar arguments. 
}. 
Here we just quote the 
results for a constant resolution parameter $a$: 
\bg\label{defc}\nonumber
  c_1 &=& \frac{g_s^2MN_f}{4\pi r(r^2+6a^2)^2}\,\big(72 a^4-3r^4-56a^2r^2\log r+a^2 r^2\log(r^2+9a^2)\big)\,
    \cot\frac{\theta_1}{2}\nonumber\\
c_2 &=& \frac{3g_s^2MN_f}{4\pi r^3}\,\big(r^2-9a^2\log(r^2+9a^2)\big)\, \cot\frac{\theta_2}{2}\\ \nonumber
  c_3 &=& \frac{3g_sM r}{r^2+9a^2}+\frac{g_s^2MN_f}{8\pi r(r^2+9a^2)}\,\Big[-36a^2-36 r^2\log a+34 r^2\log r\\ \nonumber
    & & \qquad\qquad\qquad\qquad +(10r^2+81a^2)\log(r^2+9a^2)+12r^2\log\left(\sin\frac{\theta_1}{2}\sin\frac{\theta_2}{2}\right)\Big]\\ \nonumber
c_4 &=& \frac{3g_sM(r^2+6a^2)}{\kappa r^3}+\frac{g_s^2MN_f}{8\pi\kappa r^3}\,\Big[18a^2-36(r^2+6a^2)\log a+(34 r^2+36a^2) 
    \log r\\ \nonumber 
  & & \qquad\qquad + (10 r^2+63a^2)\log(r^2+9a^2)+(12 r^2+72a^2)\log\left(\sin\frac{\theta_1}{2}\sin\frac{\theta_2}{2}\right)\Big] \\ \nonumber
c_5 &=& g_sM+\frac{g_s^2MN_f}{24\pi(r^2+6a^2)}\,\Big[18a^2-36(r^2+6a^2)\log a+8(2r^2-9a^2)\log r\\ \nonumber
  & & \qquad\qquad\qquad\qquad\qquad\qquad\qquad\qquad\qquad\qquad\;\;+(10r^2+63a^2)\log(r^2+9a^2)
    \Big]\\ \nonumber
  c_6 &=& g_sM+\frac{g_s^2MN_f}{24\pi r^2}\,\Big[-36a^2-36r^2\log a+16r^2\log r+(10r^2+81a^2)\log(r^2+9a^2)\Big]
\end{eqnarray}
with $\kappa = \frac{r^2 + 9a^2}{r^2 + 6a^2}$. All the above coefficients have further corrections that we will discuss
later. Finally,
this allows us to write the NS 2--form potential:
\begin{eqnarray}\label{btwo}
  B_2 &=& \left(b_1(r)\cot\frac{\theta_1}{2}\,d\theta_1+b_2(r)\cot\frac{\theta_2}{2}\,d\theta_2\right)\wedge e_\psi\\ \nonumber
  & + &\left[\frac{3g_s^2MN_f}{4\pi}\,\left(1+\log(r^2+9a^2)\right)\log\left(\sin\frac{\theta_1}{2}\sin\frac{\theta_2}{2}\right)
    +b_3(r)\right]\sin\theta_1\,d\theta_1\wedge d\phi_1\\ \nonumber 
  & - & \left[\frac{g_s^2MN_f}{12\pi r^2}\left(-36a^2+9r^2+16r^2\log r+r^2\log(r^2+9a^2)\right)
    \log\left(\sin\frac{\theta_1}{2}\sin\frac{\theta_2}{2}\right)+b_4(r)\right]\\ \nonumber
  & & \qquad\qquad \times \sin\theta_2\,d\theta_2\wedge d\phi_2
\end{eqnarray}
with the $r$-dependent functions 
\begin{eqnarray}\label{defb}\nonumber
  b_1(r) &=& \frac{g_S^2MN_f}{24\pi(r^2+6a^2)}\big(18a^2+(16r^2-72a^2)\log r+(r^2+9a^2)\log(r^2+9a^2)\big)\\
  b_2(r) &=& -\frac{3g_s^2MN_f}{8\pi r^2}\big(r^2+9a^2\big)\log(r^2+9a^2)
\end{eqnarray}
and $b_3(r)$ and $b_4(r)$ are given by the first order differential equations
\bg\label{gdas}
  b_3'(r) &=& \frac{3g_sMr}{r^2+9a^2} + \frac{g_s^2MN_f}{8\pi r(r^2+9a^2)}\Big[-36a^2-36a^2\log a
+34 r^2\log r\\ \nonumber
  & & \qquad\qquad\qquad\qquad\qquad\qquad+(10 r^2+81a^2)
    \log(r^2+9a^2)\Big]\nonumber\\
  b_4'(r) &=& -\frac{3g_sM(r^2+6a^2)}{\kappa r^3} - \frac{g_s^2MN_f}{8\pi\kappa r^3}\Big[18a^2-36(r^2+6a^2)\log a\\ \nonumber 
  & & \qquad\qquad\qquad+(34 r^2+36a^2)\log r +(10r^2+63a^2)\log(r^2+9a^2)\Big]
\nd

Putting back the forms of $c_i$ in (\ref{h3f3}) we can see how exactly the fluxes change with conifold coordinates, but
being quite involved, it is hard to compare with the Ouyang model. However
there exist an alternative way to rewrite 
the fluxes which would tell us exactly how the black hole modifies the original Ouyang setup. This can be 
presented 
in the following way:
\begin{eqnarray} \label{3form}
{\widetilde F}_3 & = & 2M {\bf A_1} \left(1 + {3g_sN_f\over 2\pi}~{\rm log}~r\right) ~e_\psi \wedge 
\frac{1}{2}\left({\rm sin}~\theta_1~ d\theta_1 \wedge d\phi_1-{\bf B_1}~{\rm sin}~\theta_2~ d\theta_2 \wedge
d\phi_2\right)\nonumber\\
&& -{3g_s MN_f\over 4\pi} {\bf A_2}~{dr\over r}\wedge e_\psi \wedge \left({\rm cot}~{\theta_2 \over 2}~{\rm sin}~\theta_2 ~d\phi_2 
- {\bf B_2}~ {\rm cot}~{\theta_1 \over 2}~{\rm sin}~\theta_1 ~d\phi_1\right)\nonumber \\
&& -{3g_s MN_f\over 8\pi}{\bf A_3} ~{\rm sin}~\theta_1 ~{\rm sin}~\theta_2 \left({\rm cot}~{\theta_2 \over 2}~d\theta_1 +
{\bf B_3}~ {\rm cot}~{\theta_1 \over 2}~d\theta_2\right)\wedge d\phi_1 \wedge d\phi_2\label{brend} \\
H_3 &=&  {6g_s {\bf A_4} M}\Bigg(1+\frac{9g_s N_f}{4\pi}~{\rm log}~r+\frac{g_s N_f}{2\pi} 
~{\rm log}~{\rm sin}\frac{\theta_1}{2}~
{\rm sin}\frac{\theta_2}{2}\Bigg)\frac{dr}{r}\nonumber \\
&& \wedge \frac{1}{2}\Bigg({\rm sin}~\theta_1~ d\theta_1 \wedge d\phi_1
- {\bf B_4}~{\rm sin}~\theta_2~ d\theta_2 \wedge d\phi_2\Bigg)
+ \frac{3g^2_s M N_f}{8\pi} {\bf A_5} \Bigg(\frac{dr}{r}\wedge e_\psi -\frac{1}{2}de_\psi \Bigg)\nonumber  \\
&& \hspace*{1.5cm} \wedge \Bigg({\rm cot}~\frac{\theta_2}{2}~d\theta_2 
-{\bf B_5}~{\rm cot}~\frac{\theta_1}{2} ~d\theta_1\Bigg)\nonumber
\end{eqnarray}
where we see that the background is exactly of the form presented in \cite{4} except that there are asymmetry factors ${\bf A_i}, {\bf B_i}$. These
asymmetry factors contain all the informations of the black hole etc in our background\footnote{One can easily see from 
these asymmetry factors that one of the two spheres is squashed. This squashing factor is of 
order ${\cal O}(g_s N_f)$ and therefore could have a perturbative expansion. Note also that although the resolution 
factor in the metric is hidden behind the horizon of the black hole the effect of this shows up in the fluxes. As far 
as we know, these details were first considered in \cite{FEP}}. 
To order ${\cal O}(g_sN_f)$ these 
asymmetry factors are given by:
\bg\label{asymmetry}
&& {\bf A_1} ~=~ 1 + {9g_s N_f \over 4\pi} \cdot {a^2\over r^2}\cdot (2 - 3~{\rm log}~r) + {\cal O}(a^2 g_s^2 N_f^2) \nonumber\\
&& {\bf B_2} ~=~ 1 + {36 a^2~{\rm log}~r \over r^3 + 18 a^2 r ~{\rm log}~r} + {\cal O}(a^2 g_s^2 N_f^2)\\
&& {\bf A_2} ~= ~1 + {18 a^2 \over r^2} \cdot {\rm log}~r + {\cal O}(a^2 g_s^2 N_f^2) \nonumber\\ 
&& {\bf B_1} ~=~ 1 + {81\over 2} \cdot 
{g_s N_f a^2 {\rm log}~r \over 4\pi r^2 + 9 g_s N_f a^2 (2 - 3~{\rm log}~r)} + {\cal O}(a^2 g_s^2 N_f^2)\nonumber\\ 
&& {\bf A_3} ~=~ 1 - {18 a^2 \over r^2}\cdot {\rm log}~r +  {\cal O}(a^2 g_s^2 N_f^2)\nonumber\\ 
&& {\bf B_3} ~ = ~ 1 + {36 a^2 {\rm log}~r \over r^2 - 18 a^2 {\rm log}~r} + {\cal O}(a^2 g_s^2 N_f^2)\nonumber\\ 
&& {\bf A_4} ~ = ~ 1 - {3a^2 \over r^2} + {\cal O}(a^2 g_s^2 N_f^2), ~~~~~ {\bf B_4} ~ = ~ 1 + {3g_s a^2 \over r^2 - 3 a^2} + {\cal O}(a^2 g_s^2 N_f^2)\nonumber\\ 
&& {\bf A_5} ~ = ~ 1 + {36 a^2 {\rm log}~r \over r} +  {\cal O}(a^2 g_s^2 N_f^2), ~~~~ 
{\bf B_5} ~ = ~ 1 + {72 a^2 {\rm log}~r \over r + 36 a^2 {\rm log}~r} + {\cal O}(a^2 g_s^2 N_f^2)\nonumber
\nd
These asymmetry factors tell us that corrections to the Ouyang background \cite{4} come from ${\cal O}(a^2/r^2)$ onwards. Thus to complete the  
picture all we now need are the values for the axio-dilaton $\tau$ and the five form $F_5$. If $z_1=\mu$ gives the location
of a single D7 brane as in (\ref{seven}), then from F theory one obtains $\tau\sim {\rm log}({z_1-\mu})$ \cite{4}\cite{senF} near
the D7 brane. In
section {\bf 2.4} we will discuss in some detail how one obtains this form of $\tau$. As $\tau=C_0+i e^{-\phi}$ we get the
following form for the dilaton and axion:

\bg\label{axfive}
&&e^{-\phi}=\frac{1}{g_s}-\frac{3N_f}{4\pi}{\rm log}r-\frac{N_f}{2\pi}{\rm log}\left({\rm sin}\frac{\theta_1}{2}{\rm
sin}\frac{\theta_2}{2}\right)+ {\cal O}(\mu,a)\nonumber\\
&&C_0 ~ = ~ {N_f \over 4\pi} (\psi - \phi_1 - \phi_2)+{\cal O}(\mu,a) \nonumber\\
&& F_5 ~ = ~ {1\over g_s} \left[ d^4 x \wedge d h^{-1} + \ast(d^4 x \wedge dh^{-1})\right]
\nd
where ${\cal O}(\mu,a)$ denotes all orders in $\mu$ and the resolution parameter $a$, $h$ is the ten dimensional warp factor discussed above. 
Thus combining (\ref{brend}) and (\ref{axfive}) our background can be written almost like the Ouyang background 
\cite{4} with deviations 
given by (\ref{asymmetry}).    

In order to extract temperature from the geometry, we look at the metric in (\ref{bhmet}) in the near horizon limit
$r\rightarrow r_h$. To be exact, we
really need to start from the ten dimensional supergravity action and then integrate out the 
internal directions to obtain a
five dimensional effective action. Minimization of that five dimensional effective action will give the five dimensional
effective metric which is the same as integrating the five dimensional metric over the internal directions. 
That is the five dimensional metric is given by $g_{\mu\nu}=\int d\theta_i d\phi_i d\psi G_{\mu\nu}(\theta_i,\phi_i,\psi)$
where $G_{\mu\nu}, \mu,\nu=0,..,4$ is the metric as in (\ref{Ouyangm}) and we only integrate over the internal coordinates
that $G_{\mu\nu}$ is a function of. We will denote the resulting warp factor for the five dimensional 
metric $g_{\mu\nu}$ as $h(r)$ for the ensuing analysis of temperature.  

Now, looking at the $r,t$ direction of the metric $g_{\mu\nu}$ and by change of variable, under the assumption that 
$\int d\theta_k d\phi_j d\psi ~g_i\approx g(r)$, $i,j,k=1,2$, 
we can define $\rho^2$ as: 
\bg \label{Temp1}
\rho^2=\frac{4\sqrt{h(r_h)}g(r)}{[g'(r_h)]^2}
\nd
so that the near horizon limit of five dimensional effective metric takes the following Rindler form: 
\bg \label{Temp2}
ds^2=-\rho^2 \frac{g'(r_h)^2}{4h(r_h)g(r_c)}dt_c^2+d\rho^2
\nd
where prime denotes differentiation with respect to $r$ and we only wrote the 
$r,t$ part of the metric in terms of new variable $\rho$ and $t_c \equiv \sqrt{g(r_c)}t$.  
The
reason behind rescaling time at fixed $r_c$ is that with this time coordinate $t_c$, the five dimensional metric induces a
four dimensional Minkowski metric at every $r_c$. 

Now the temperature observed by the field theory with time coordinate
$t_c$ can be extracted by writing the metric in (\ref{Temp2}) in the following form 
\bg \label{Temp3}
ds^2=-4\pi^2 T_c^2\rho^2 ~dt_c^2~ + ~ d\rho^2
\nd   
Thus comparing (\ref{Temp2}) and (\ref{Temp3}), we obtain the temperature $T_c$ as:
\bg \label{Temp4}
T_c~= ~ \frac{g'(r_h)}{4\pi\sqrt{h(r_h)g(r_c)}}
\nd   
In the limit where we have $g_1 = g_2 = g = 1-{r_h^4\over r^4}$, we can easily compute the corresponding temperature using
the above formula (\ref{Temp4}). This is given by:
\bg\label{Temp5}
T_c ~ = ~ {r_h \over \pi L^2} ~ + ~ {r_h^5\over 2\pi L^2 r_c^4} + \sum_{m,n,p} c_{mnp} {r_h^{m} {\rm log}^n r_h\over 
r_c^{p}}
~ \equiv  ~ {T}_b ~+~ {\cal O}(1/r_c)
\nd
where $L$ is defined earlier, $c_{mnp}$ is in general functions of ($g_s, M, N, N_f$), 
and ${T}_b > T_{\rm deconf}$ (where $T_{\rm deconf}$ is the deconfinement 
temperature) is the temperature at $r_c \to \infty$ i.e
\bg \label{bndtemp}
{T}_b ~\equiv~ T_{\rm boundary} ~=~ {g'(r_h)\over 4\pi \sqrt{h(r_h)}} , ~~ r_h ~\equiv~ F({T_b}) 
~\equiv ~{\cal T}
\nd
where $F(T_b)$ can be obtained by inverting the first equation above. We can do this exactly once we know the black hole 
factor $g(r_h)$ as well as the warp factor $h(r_h)$ to all orders in $g_sN_f, g_s M$. Here we identify the
 black hole horizon radius $r_h$ with what characteristic temperature $\cal T$ which is the only scale in the theory when
 there are no D7 branes.

With the knowledge of the dual gravity for the branes setups of KS and OKS models with temperature, 
one may ask whether Seiberg
duality cascade can be realized from supergravity. The answer lies in  identifying the gauge theory effective degrees of 
freedom using gravity. For AdS/CFT correspondence this identification is rather simple; the AdS throat radius $L$ is
directly related to the number of colors $N$ of the gauge theory by $N=\frac{L^4}{4\pi g_s \alpha'^2}$ as can be seen from
(\ref{Sugra-1}). Generalizing this result for the non-AdS/non-CFT models, one can immediately identify the effective degrees
of freedom of the gauge theory with the throat radius $\tilde{L}^4(r)\equiv r^4 h$ where $h$ is the warp factor appearing in
(\ref{bhmet}). This gives

\bg \label{Neff}
N_{\rm eff}(r)=\frac{\tilde{L}^4(r)}{V g_s \alpha'^2}=\frac{r^4 h}{V g_s \alpha'^2}
\nd      
where $V$ is a constant which depends on the five dimensional compact manifold ${\cal M}_5 $. For example, if ${\cal M}_5=T^{1,1}$, then
$V=\frac{27}{4}\pi$ \cite{KW}. 

\begin{figure}[htb]\label{cone}
		\begin{center}
\includegraphics[height=8cm, width=12cm]{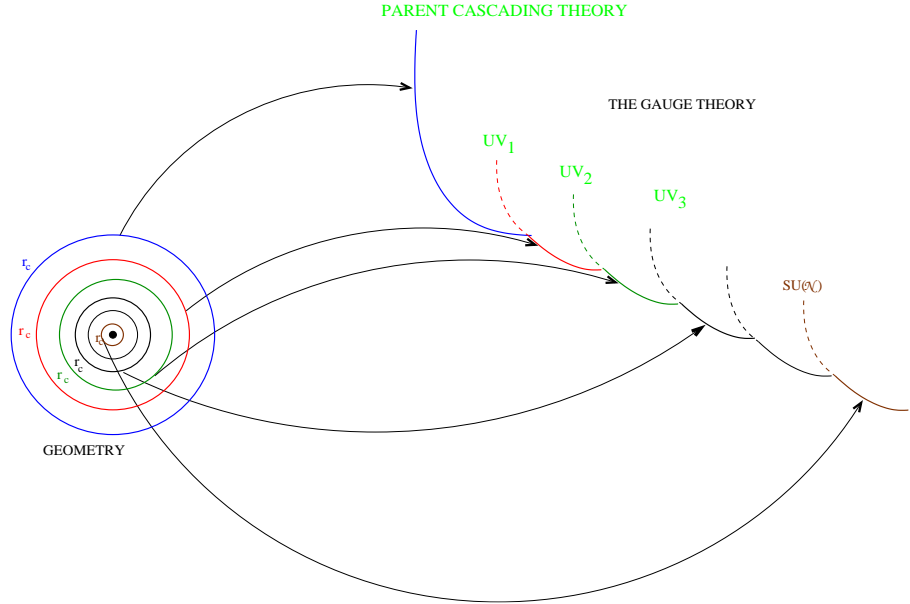}
		\caption{Seiberg duality cascade captured by dual geometry.  }
		\end{center}
		\end{figure}
Now with the warp factor $h$ given by (\ref{hvalue}), we see that $N_{\rm eff}$ grows with $r$ and this gives a flow of
effective degrees of freedom with changing $r$ of the dual gravity. More precisely one can introduce a cutoff $r_c$ in the
geometry  and evaluate the four dimensional dual  field theory on the boundary $r=r_c$. The bulk geometry obtained
this way extends from $r_h$ to $r_c$ and  the dual gauge theory has degrees of freedom $N_{\rm eff}(r_c)$.  
Note that the radial
coordinate of dual gravity is identified as the energy scale of the gauge theory, that is $r=\Lambda$ and the geometry
with cutoff $r_c$ is dual to effective field theory at scale $\Lambda_c$. Of course one has to be specially careful in obtaining
effective gauge theory Lagrangian from dual gravity. In particular one has to attach  geometries from $r=r_c$ to 
$r=\infty$ in order to account for  appropriate irrelevant and marginal operators that have been integrated out. This
procedure of attaching UV caps to geometries will be discussed in detail in the comings sections. 

As $N_{\rm eff}$ grows with $r$, we conclude from gravity that effective degrees of freedom grow in the UV and 
shrink in the IR, which is exactly
consistent with Seiberg duality cascade. At some UV scale $\Lambda_{\rm UV}$, we start with  $SU(N+M)\times SU(N)$ and as we
go to lower and lower energy scales, we can describe the theory with lower and lower rank gauge groups using a cascade of
Seiberg dualities. Thus supergravity captures the generic feature of Seiberg duality cascade and $N_{\rm eff}$ at some value of
of $r$, that is at some value of scale $\Lambda$ describes the overall degrees of freedom of the gauge theory. Fig {\bf
2.13} shows a simplified depiction of how an effective field theory at scale $\Lambda_c$ can be mapped to a geometry with
cutoff $r_c$. For more details consult \cite{Strassler:2005qs, FEP}.
\begin{figure}[htb]\label{cone}
		\begin{center}
\includegraphics[height=9cm, width=14cm]{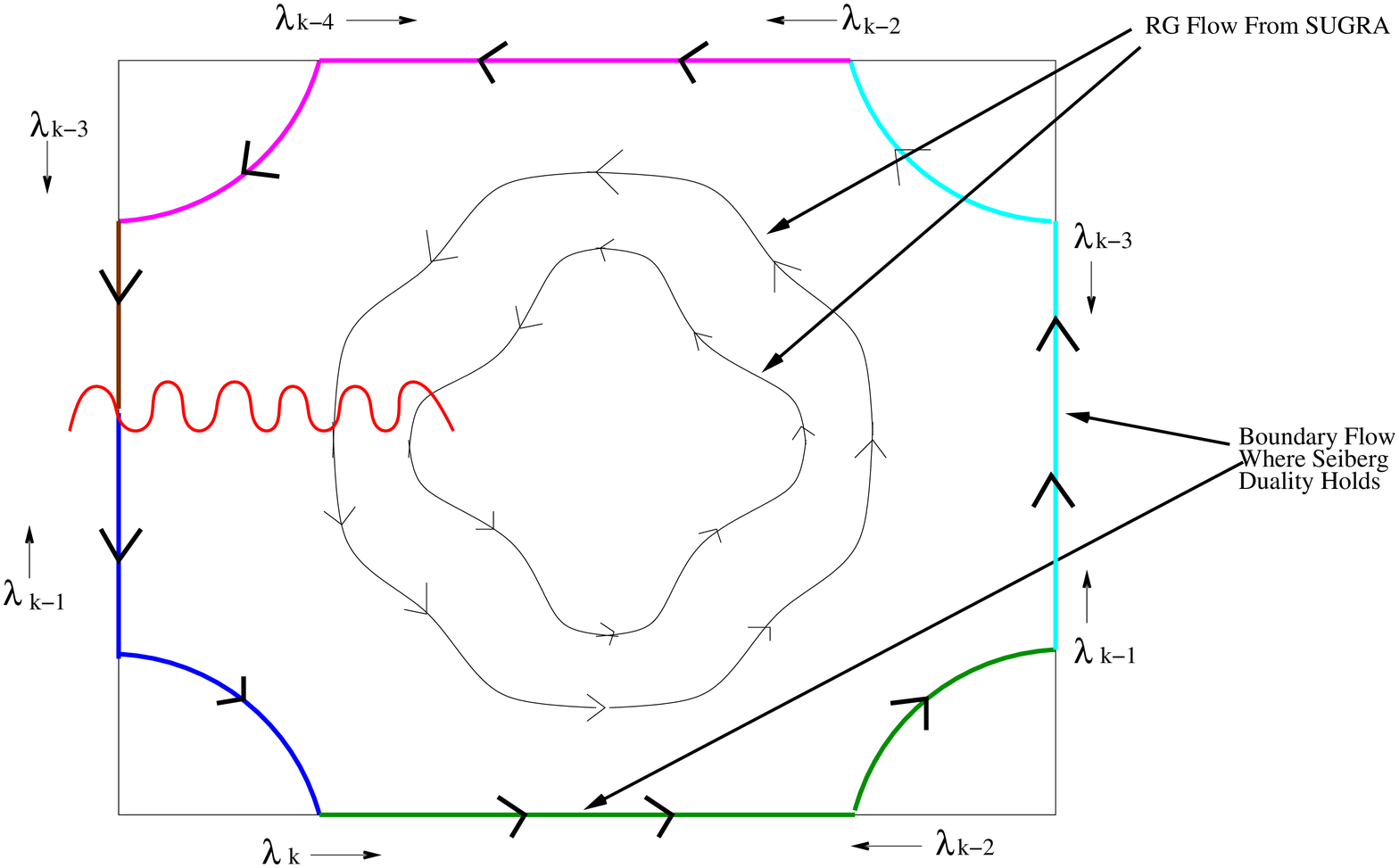}
		\caption{Effective field theories at various scales along with their dual geometry with cutoffs.  }
		\end{center}
		\end{figure}

Although generic features of the duality cascade is captured by gravity, there are crucial distinctions between the cascade
and the flow obtained from dual gravity. To describe a theory with its Seiberg dual, a key requirement is that the coupling
has to be either very strong or very weak. In the space of coupling constants, one must be at the boundary curve of Fig {\bf
2.14}. For  $N,M\sim {\cal O}(1)$, the t'Hooft couplings are roughly equal to the gauge couplings, that is $\lambda_{k}\sim
g_{k}$ so we can treat $g_{k}$ as the t'Hooft couplings. Then Seiberg duality can be performed at a scale when t'Hooft
coupling $\lambda_{k}$ of $SU(N+M)$ is very large and $\lambda_{k-1}$ of $SU(N)$ is very small. On the other hand
supergravity solution is valid when $N,M\gg 1$ and using (\ref{twocoup}) one obtains that both t'Hooft couplings
$\lambda_k,\lambda_{k-1}$ are very large. This means the SUGRA solution lies in the interior of coupling constant surface of
Fig {\bf 2.14}
and not on the boundary curve where Seiberg duality holds. The situation is depicted in Fig {\bf 2.14}. 

The regime where supergravity solution holds is precisely beyond the phase space where Seiberg duality holds. The flow
captured by supergravity is a smooth RG flow with continuously changing degrees of freedom rather than a step by step
reduction of gauge group by some units (for KS gauge theory by $M$ units) as described by the cascade. However supergravity may capture crucial features of the
gauge theory which are independent of smooth RG flow or patchy Seiberg duality cascade. 

   With a clear understanding of the gauge theories arising from the branes and their dual gravity, we would like to know how
to compute relevant gauge theory observables. In the following section we discuss the precise mapping between gauge theory
correlation functions and geometry.
 
\section{Gauge Theory Observables from Dual Gravity} 
In order to compute expectation values of certain observables  of the gauge theory, one first computes the partition
function. For a system at thermal equilibrium, various thermodynamic quantities such as free energy, pressure, entropy etc. 
can easily be obtained from the partition function. On the other hand propagators of quantum fields and subsequently
scattering amplitudes of particles can be easily evaluated by taking the functional derivative of the partition function with
respect to some appropriate sources. Thus once we know the partition function of a quantum field theory, we can essentially
determine its thermodynamic properties. 

As the Hilbert space of certain quantum field theories arising from brane excitations is contained in the Hilbert space of
the geometry sourced by the branes, there should be a one-to-one correspondence between the partition functions of the gauge
theory and the dual gravity. A precise mapping was proposed by Witten \cite{Witten-2} and subsequently by the authors of
\cite{Witten-2a}. Witten's proposal states that the partition function of
the strongly coupled quantum field theory should be identified with the partition function of the weakly coupled classical
gravity. In particular if we are interested in computing the expectation value of an operator $<{\cal O}>$ with a source
$\phi_0$, one can make the following identification of the partition function as a functional of the source $\phi_0$:
\bg \label{KS16}
{\cal Z}_{\rm gauge}[\phi_0] &~\equiv &~ \langle{\rm exp}\int_{M^4} \phi_0 {\cal O}\rangle
~=~ {\cal Z}_{\rm gravity}[\phi_0] \nonumber\\
&~ \equiv &  ~{\rm exp}(S_{\rm SUGRA}[\phi_0] + S_{\rm GH}[\phi_0] + S_{\rm counterterm}[\phi_0])
\nd
where $M^4$ is a Minkowski manifold, $S_{\rm GH}$ is the Gibbons-Hawking boundary term \cite{Gibbons-Hawking}, 
$\phi_0$ should be understood as a fluctuation over a given configuration of field
and $S_{\rm counterterm}$ is the 
counter-term action added to renormalize the action. We will briefly describe how each term is obtained from supergravity
action and their necessity. 

First observe that $\phi$ is a bulk field, which  is in  general a function of the 
coordinates of  the
geometry and $\phi_0$ is the boundary value of $\phi$. One integrates the classical ten dimensional
gravity action (\ref{action1}) over
the compact manifold ${\cal M}_5$ to obtain an effective action $S^{\rm eff}_5$ for a five dimensional manifold. 
For AdS/CFT correspondence the five dimensional manifold one obtains after integration over ${\cal M}_5=S_5$ is of course
$AdS_5$ geometry whereas for non-Ads geometries described in the previous section, one obtains modified AdS space. 
The bulk five dimensional geometry has a boundary at $r=\infty$ and the boundary describes a four dimensional manifold with
an induced Minkowski flat metric. Thus the boundary value of
$\phi$ 
\bg
\phi_0(t,x,y,z)\equiv \phi(r=\infty,t,x,y,z)
\nd
is a field which lives in four dimensional flat space and plays the role of source in the four dimensional gauge theory.

Now to obtain the gravity action as a functional of $\phi_0$, one  integrates  the five dimensional effective action $S^{\rm
eff}_5$ 
over the  radial coordinate $r$ and we are left with gravity action only being function of $\phi_0$. Given a boundary value
$\phi_0$, one can uniquely build the bulk field $\phi$ and write the bulk action for $\phi$ as a function of the boundary
value $\phi_0$. This was shown first by Witten \cite{Witten-2} for AdS bulk fields and for the geometries in
 \cite{FEP,LC} which are perturbations on AdS space, the same argument holds. 
 
 It turns out that in general for the gravity action $S[\phi]$ to be stationary under perturbation $\phi\rightarrow
 \phi+\delta\phi$, one needs to add surface terms which are known as Gibbons-Hawking terms \cite{Gibbons-Hawking}. On the
 other hand, after the radial integral is done, for some sources $\phi$, $S[\phi_0]$ becomes infinite and one needs to
 regularize the action to obtain finite expectation values for operators. The regularization requires addition of extra
 terms which cancel the infinities that appear in the gravity action and are denoted by $S_{\rm counterterm}$.  Keeping all this in mind, one takes the functional
 derivative of (\ref{KS16}) with respect to $\phi_0$ to obtain 
 
\bg \label{O}
<{\cal O}>&=&\frac{\delta {\cal Z}_{\rm gauge}}{\delta \phi_0}=\frac{\delta {\cal Z}_{\rm gravity}}{\delta
\phi_0}\nonumber\\
&\sim&{\rm exp}\left(S_{\rm total}[\phi_0]\right)\frac{\delta S_{\rm total}[\phi_0]}{\delta\phi_0}\arrowvert_{\phi_0=0} 
\nd 
where $S_{\rm total}[\phi_0]=S_{\rm SUGRA}[\phi_0] + S_{\rm GH}[\phi_0] + S_{\rm counterterm}[\phi_0]$.

Observe that in the usual AdS/CFT case we consider the action at the boundary to 
map it directly to the dual gauge theory side. For general gauge/gravity dualities, there are many 
possibilities of defining 
different gauge theories at the boundary depending on how we cut-off the geometry and add UV
caps. We will elaborate the addition of UV caps in section {\bf 2.3.2} and in the following section {\bf 2.3.1}, we briefly describe
renormalization of the supergravity action.

\subsection{Holographic Renormalization}
As already mentioned, supergravity action $S_{\rm sugra}[\phi_0]$, for some sources $\phi_0$, becomes infinity and one needs
to regularize the action. In particular if we want to compute the stress tensor $T^{pq}$, $p,q=0,1,2,3$ of a field theory in
four dimensional flat space time,
then the source $\phi_0$ is the metric $\eta_{pq}$, i.e the Minkowski metric. 
It turns out that if we naively integrate over the radial coordinate $r$ of $S^{\rm eff}_5$, the resulting action as a
function of $\eta_{pq}$ blows up.  

First observe that it is indeed possible to write the five dimensional supergravity action for AdS or modified AdS space as a
functional of flat four dimensional metric. This is because for both AdS and non-AdS space (KS and OKS background with
black hole), the five dimensional metric has
the following form:
\bg \label{AdSmetric}
ds^2&=&-\frac{g_1}{\sqrt{h}}dt^2+\frac{1}{\sqrt{h}}\left(dx^2+dy^2+dz^2\right)+\frac{\sqrt{h}}{g_2}dr^2\nonumber\\
&\equiv&g_{\mu\nu} dx^\mu dx^\nu
\nd 
 At the boundary $r=\infty$, $g_1\rightarrow 1$, and the induced metric is $\eta_{pq}/\sqrt{h(\infty)}$ which is a constant
 factor times the Minkowski metric. Thus $S^{\rm eff}_5[g_{\mu\nu}]$ can be written as a function of $\eta_{pq}$ once the
 radial integral is done. 
 
Now for AdS space, starting with (\ref{action1}) with only $\tilde{F}_5\neq 0$ being the source, 
integrating over the internal coordinates give the following action:
\bg \label{AdSaction}
S^{\rm eff}_{AdS_5}=\int d^4x \int_{r_h}^{r_c} dr \sqrt{g}\left(R_5+\Lambda\right)
\nd  
where $g={\rm det} g_{\mu\nu}$, $\Lambda$ is the cosmological constant and $r_c\rightarrow \infty$. Note that the above action is not only the action for AdS space
but also for asymptotic AdS spaces i.e. geometries  $G_{\mu\nu}$ such that 
$\lim_{r\rightarrow \infty} G_{\mu\nu}=g_{\mu\nu}$
with $g_{\mu\nu}$ given by (\ref{AdSmetric}) where $h=L^4/r^4$ \cite{kostas1}. 

After the $r$ integral is done, the result contains terms which diverge as $r_c\rightarrow \infty$. As the AdS or asymptotic
AdS metric is a power
series in $r$, the divergent terms  are precisely the
ones containing positive powers of $r_c$ and by subtracting these terms we get a finite result for the action. This
subtraction procedure can be done by introducing counter term $S_{\rm counter}$ to the action (\ref{AdSaction}) such that
they cancel the infinities coming from it. The renormalization scheme mentioned here was first proposed by Skenderis
 and for more details please consult \cite{kostas1}-\cite{kostas5}. 
 
With a finite renormalized action $S^{\rm ren}_{AdS_5}[\eta_{pq}]$, and using the equation of motion for the metric
$g_{\mu\nu}$, one can take the  derivative with respect to $\eta_{pq}$  to obtain the following result for the stress tensor
\cite{kostas3}
\bg
T^{pq}=\frac{1}{4\pi G_N}\left(g_{(4)pq}-\frac{1}{8}g_{(0)pq}\left[({\rm Tr}g_{(2)})^2-{\rm
Tr}g_{(2)^2}\right]-\frac{1}{2}(g_{(2)}^2)_{pq}+\frac{1}{4}g_{(2)pq}{\rm Tr}g_{(2)}\right)
\nd
where $G_N$ is the five dimensional Newton's constant and $g_{(l)}$ is defined through the bulk metric $G_{\mu\nu}$ the
following way:
\bg
G_{\mu\nu}dx^\mu dx^\nu=\frac{d\rho^2}{4\rho^2}+\frac{1}{\rho}\tilde{g}_{pq}(x^p,\rho)dx^p dx^q\nonumber\\
\tilde{g}(x,\rho)=g_{(0)}(x)+...+\rho^{d/2}g_{(d)}+..
\nd
where $\rho=1/r^2$. Thus $g_{(l)}$ are Taylor coefficients in the expansion of the four dimensional metric 
$\tilde{g}_{pq}(x^p,\rho)$ which induces  the four dimensional boundary metric $g_{(0)pq}$ at the boundary $\rho=0$.  
 
The story gets somewhat more involved for non-AdS geometries and we will briefly review the key points of holographic
renormalization of dual supergravity for non conformal field theories. Details of the analysis can be found in our work
\cite{FEP,LC}. 

The divergent terms appearing in the supergravity action are uniquely determined once we know the form of the warp factor $h$
in (\ref{bhmet}). Observe that the logarithms appearing in $h$ (\ref{hvalue}) results from the logarithmic running of the
dilaton field (\ref{axfive}) and that of NS-NS field $B_2$. The dilaton only behaves logarithmically near the location of
the D7 brane and its global behavior is determined by F theory. For large $r$ and away from any D7 branes, we expect the
dilaton to behave as a constant. In fact in section {\bf 2.4} we will analyze the precise running of the axio-dilaton field
for all $r$ where we will also see how the logarithmic running of NS-NS $B_2$ field can be modified to give finite value for
$B_2$ at large $r$. With $B_2$ and axio-dilaton $\tau$ behaving as logarithms  in the local neighborhood of some $r$ but
approaching constant finite values for large $r$, in \cite{FEP} we proposed the following form
of the warp factor:
\bg\label{logr}
h~=~ {L^4\over r^{4- \epsilon_1}} ~+~ {L^4\over r^{4- 2\epsilon_2}} ~-~ {2L^4\over r^{4- \epsilon_2}} ~+~ {L^4 \over 
r^{4-r^{\epsilon^2_2/2}}} 
~\equiv~ \sum_{\alpha=1}^4 {L^4_{(\alpha)}\over r^4_{(\alpha)}}
\nd
where $\epsilon_i, r_{(\alpha)}$ etc are defined as:
\bg\label{epde}
&&\epsilon_1 ~ = ~ {3g_s M^2\over 2\pi N} + {g_s^2 M^2 N_f\over 8\pi^2 N} + {3g_s^2 M^2 N_f \over 8\pi N} ~
{\rm log}\left({\rm sin}~{\theta_1\over 2} {\rm sin}~{\theta_2\over 2}\right),  
~~\epsilon_2 ~ = ~ {g_s M \over \pi}\sqrt{2N_f\over N}\nonumber\\
&& ~~~~~~~~~~~ r_{(\alpha)} = r^{1-\epsilon_{(\alpha)}}, ~~ \epsilon_{(1)} = {\epsilon_1\over 4}, ~~ 
\epsilon_{(2)} = {\epsilon_2\over 2}, ~~ \epsilon_{(3)} = {\epsilon_2\over 4}, ~~ \epsilon_{(4)} = 
{\epsilon^2_2\over 8}\nonumber\\
&& ~~~~~~~~~~~ r_{(\pm\alpha)} = r^{1\mp\epsilon_{(\alpha)}}, ~~~L_{(1)} = L_{(2)} = L_{(4)} = L^4, ~~~ L_{(3)} = -2L^4 
\nd
which makes sense because we can make $\epsilon_i$ to be very small.
Note that the choice of $\epsilon_i$ doesn't require us to have 
$g_s N_f$ small (although we consider it here). 
In fact we {\it can} have all ($N, M, N_f$) large but $\epsilon_i$ 
small. A simple way to achieve this would be to have the following scaling behaviors of 
($g_s, N, M, N_f$):
\bg\label{scalbet}
g_s ~\to ~\epsilon^\alpha, ~~~~~M ~\to ~\epsilon^{-\beta}, ~~~~~ 
N_f ~\to ~\epsilon^{-\kappa}, ~~~~~ N ~\to ~\epsilon^{-\gamma} 
\nd
where $\epsilon \to 0$ is  the tunable parameter. Therefore all we require to achieve that is to allow:
\bg\label{allow}
\alpha ~+~ \gamma ~>~ 2\beta ~+~ \kappa, ~~~~~~~ \alpha ~>~ \kappa, ~~~~~~~~ \gamma ~>~ \alpha
\nd 
where the last inequality can keep $g_s N_f$ small. Thus $g_s N, g_s M$ are very large, but $g_s, 
{g_s M^2 \over N}, g_s N_f^{}$ are all very small to justify our expansions (and the choice of 
supergravity background)\footnote{For example we can have $g_s$ going to zero as  $g_s \to \epsilon^{5/2}$ and 
($N, M_{}, N_f$) going to infinities as ($\epsilon^{-8}, \epsilon^{-3}, \epsilon^{-1}$) respectively. 
This 
means ($g_s N, g_s M$) go to infinities as ($ \epsilon^{-11/2}, \epsilon^{-1/2}$) respectively, and 
($g_sN_f, g^2_sM N_f, g_sM^2/N$) 
go to zero as ($\epsilon^{3/2}, \epsilon, \epsilon^{9/2}$) respectively. 
This is one limit where we can have well defined UV completed 
gauge theories. Note however that for the kind of background that we have been studying one cannot make $N_f$ large 
because of the underlying F-theory constraints \cite{vafaF2}\cite{senF}\cite{DM1}. Since we only require $g_sN_f$ small, large or small 
$N_f$ choices do not change any of our results.}. 

The warp factor (\ref{logr}) has a good behavior at infinity and reproduces the ${\cal O}(g_sN_f)$ result locally.
Our conjecture then would be the complete form of the warp factor at large $r$
will be given by sum over $\alpha$ as in 
(\ref{epde}) but now $\alpha$ can take values $1\le \alpha \le \infty$. This conjecture in fact  justifies the 
holographic renormalizability of our boundary theory as we will find out shortly.

With the warp factor given by (\ref{logr}), we introduce perturbation $l_{\mu\nu}$ to background five dimensional metric
$g_{\mu\nu}$ given by (\ref{AdSmetric}). Recall that for non-AdS space the five dimensional effective action $S^{\rm eff}_5$
derived from ten dimensional action (\ref{action1}) takes the form 
\bg \label{nonAdSact}
S^{\rm eff}_5=\int d^5x \sqrt{G}\left(R+\Lambda(G_{\mu\nu})\right)
\nd
where $G_{\mu\nu}=g_{\mu\nu}+l_{\mu\nu}$ and the source $\Lambda(G_{\mu\nu})$ is in general a function of the metric. As
$l_{\mu\nu}$ is a perturbation, one can write the action (\ref{nonAdSact}) as a Taylor series in the perturbation.
Considering terms only up to quadratic order in $l_{\mu\nu}$ one obtains:
 
\bg \label{KS7c} 
&&{\cal S}^{(1)}[\Phi]~=~\int \frac{d^4q}{(2\pi)^4 \sqrt{g(r_c)}}\int dr
~\Bigg\{{1\over 2} A^{mn}_1(r,q)\Big[\Phi^{[1]}_m(r,q)\Phi''^{[1]}_n(r,-q) + 
\Phi''^{[1]}_m(r,q)\Phi^{[1]}_n(r,-q)\Big]\nonumber\\
&& + B^{mn}_1(r,q) \Phi'^{[1]}_m(r,q)\Phi'^{[1]}_n(r,-q)
+{1\over 2} C^{mn}_1(r,q) \Big[\Phi'^{[1]}_m(r,q)\Phi^{[1]}_n(r,-q) + \Phi^{[1]}_m(r,q)\Phi'^{[1]}_n(r,-q)\Big]\nonumber\\
&& + D^{mn}_1(r,q)\Phi^{[1]}_m(r,q)\Phi^{[1]}_n(r,-q)+{\cal T}^m_1(r,q)\Phi^{[1]}_m(r,q)
+E^m_1\Phi'^{[1]}_m(r,q)+F^m_1\Phi''^{[1]}_m(r,q)\Bigg\}\nonumber\\
\nd
where $m,n=1,..., 5$, prime denotes differentiation with respect to $r$, the script $[1]$ denote the {\it total} background to 
${\cal O}(g_sN_f, g_sM^2/N)$; 
and the explicit expressions for 
$A^{mn}_1,B^{mn}_1,C^{mn}_1,F^m_1$ for a specific case are given in Appendix D of \cite{FEP}. 
We have also defined $\Phi^{[1]}_m(r, q)$ 
in the following way (with $q_0 \equiv \omega \sqrt{g(r_c)}$ as before): 
\bg\label{phim}
\Phi^{[1]}_m(r,q)=\int \frac{d^4x}{(2\pi)^4 \sqrt{g(r_c)}}\;e^{i(q_0 t-q_1x-q_2y-q_3z)}l_{mm}(t,r,x,y,z)
\nd 
 We will see that the effective four dimensional boundary action is independent of $D^{mn}_1$ and ${\cal T}^m_1$
and hence their explicit expressions do not appear in the appendix of \cite{FEP}. Furthermore, note that the 
derivative terms in (\ref{KS7c}) all come exclusively from $\sqrt{-G}R$, whereas fluxes contribute powers of 
$\Phi^{[1]}$ but no derivative interactions. 
In the following we keep up to quadratic orders, and therefore the 
contributions from the fluxes will appear in $D^{mn}_1$ and $E^m_1$.

 The equation of
motion for $\Phi_n^{[1]}(r, -q)$ is given by:
\bg\label{eomphin}
&&{1\over 2} \Big[A_1^{mn}(r, q)\Phi_n^{[1]}(r, -q)\Big]''~ -~ \Big[B_1^{mn}(r, q)\Phi_n'^{[1]}(r, -q)\Big]' ~- ~
{1\over 2} \Big[C_1^{mn}(r, q)\Phi_n^{[1]}(r, -q)\Big]' \nonumber\\
&& ~~+~ D_1^{mn}(r, q)\Phi_n^{[1]}(r, -q) ~+~ {1\over 2} A^{mn}_1(r, q) \Phi_n''^{[1]}(r, -q)
~-~ {1\over 2} C_1^{mn}(r, q) \Phi_n'^{[1]}(r, -q)\nonumber\\ 
&&~~~~~~~~~~~~~~~~ + ~{\cal T}_1^m(r, q) ~-~ E_1'^m(r, q) ~+~ F''^m_1(r, q) = 0
\nd
The next few steps are rather standard and so we will quote the results. The variation of the action (\ref{KS7c}) 
can be written in terms of the variations $\delta \Phi_m^{[1]}(r, q)$ and $\delta \Phi_n^{[1]}(r, -q)$ 
in the following way\footnote{Henceforth, unless mentioned otherwise, $\Phi_m^{[1]}, \Phi_n^{[1]}$ 
will always mean $\Phi_m^{[1]}(r, q)$ and 
$\Phi_n^{[1]}(r, -q)$ respectively. Similar definitions go for the variations $\delta\Phi_m^{[1]}$ 
and $\delta\Phi_n^{[1]}$.}:
\bg\label{varofac}
&&\delta{\cal S}^{(1)} = {1\over 2}\int \frac{d^4q}{(2\pi)^4 \sqrt{g(r_c)}}\int_{r_h}^{r_c} dr\Bigg\{
\Big[(A_1^{mn}\Phi_m^{[1]})'' - (2B_1^{mn}\Phi_m'^{[1]})' + C_1^{mn}\Phi_m'^{[1]} + 2D_1^{mn}\Phi_m^{[1]} \nonumber\\ 
&&+ A^{mn}_1 \Phi_m''^{[1]} - (C_1^{mn}\Phi_m^{[1]})'\Big]\delta\Phi_n^{[1]}
+ \Big[(A_1^{mn}\Phi_n^{[1]})'' - (2B_1^{mn}\Phi_n'^{[1]})' + C_1^{mn}\Phi_n'^{[1]} + 2D_1^{mn}\Phi_n^{[1]} \nonumber\\ 
&& + A^{mn}_1 \Phi_n''^{[1]} - (C_1^{mn}\Phi_n^{[1]})'\Big]\delta\Phi_m^{[1]} + 
2\left({\cal T}_1^m - E_1'^m + F''^m_1\right)\delta\Phi_m^{[1]}\nonumber\\
&&\partial_r\Big[A_1^{mn}\Phi_m^{[1]} \delta\Phi_n'^{[1]} - (A_1^{mn}\Phi_m^{[1]})'\delta\Phi_n^{[1]}
 + 2B_1^{mn}\Phi_m'^{[1]}\delta\Phi_n^{[1]} + 
C_1^{mn} \Phi_m \delta\Phi_n^{[1]} + 2B_1^{mn}\Phi_n'^{[1]} \delta\Phi_m^{[1]}\nonumber\\
&& +C_1^{mn}\Phi_n^{[1]}\delta\Phi_m^{[1]} + 2E_1^m \delta\Phi_m^{[1]} + 2F_1^m\delta\Phi_m'^{[1]} - 
2F_1'^m \delta\Phi_m^{[1]} + 
A_1^{mn} \Phi_n^{[1]} \delta\Phi_m'^{[1]} - (A_1^{mn}\Phi_n^{[1]})'\delta\Phi_m^{[1]}\Big]\Bigg\}\nonumber\\
\nd
which includes the equations of motion as well as the boundary term. We can then write the variation of the 
action $\delta{\cal S}^{(1)}$ in the following way: 
\bg \label{KS7c1}
&&\delta{\cal S}^{(1)}~ = ~ \int \frac{d^4q}{(2\pi)^4 \sqrt{g(r_c)}}
\Bigg\{ \int_{r_h}^{r_c} dr \Big[\left({\rm EOM \;for\; \Phi_n^{[1]}}\right)\delta \Phi^{[1]}_m 
+ \left({\rm EOM \;for\; \Phi_m^{[1]}}\right)\delta \Phi^{[1]}_n\Big] \nonumber\\
&&+ {1\over 2}\Big[(2B_1^{mn} - A_1^{mn}) (\Phi_m'^{[1]}\delta\Phi_n^{[1]} 
+ \Phi_n'^{[1]}\delta\Phi_m) + (C_1^{mn} - A_1'^{mn})
(\Phi_m^{[1]}\delta\Phi_n^{[1]} + \Phi_n^{[1]}\delta\Phi_m^{[1]}) \nonumber\\
&& + 2(E_1^{m} - F_1'^{m})\delta\Phi_m^{[1]} + A_1^{mn}\Phi_m^{[1]}\delta\Phi_n'^{[1]} 
+ A_1^{mn}\Phi_n^{[1]}\delta\Phi_m'^{[1]} + 2F_1^{m}\delta\Phi_m'^{[1]}\Big]_{\rm boundary}\Bigg\}\nonumber\\
\nd 
where by an abuse of notation by
the ``boundary'' here, and the next couple of pages (unless mentioned otherwise), 
we mean that the functions are all measured at $r_h$ and $r_c$ i.e the horizon and the 
cut-off respectively\footnote{Note that we have not carefully described the degrees of freedom at the boundary as yet.
For large enough $r_c$ we expect large degrees of freedom at the UV. This would mean that the 
contributions to various gauge theories from these degrees of freedom would go like $e^{-{\cal N}_{\rm eff}}$, which 
would be negligible. Thus unless we cut-off the geometry at $r = r_c$ and add UV caps with specified degrees of freedom
we are in principle only describing the parent cascading theory. For this theory of course ${\cal N}_{\rm eff}$ is 
infinite at the boundary, which amounts to saying that UV degrees of freedom don't contribute anything here.
We will, however, give a more precise description a little later.}. 
It is now easy to see why $D_1^{mn}$, ${\cal T}_1^{mn}$ and $E_0^{mn}$ etc do not 
appear in the boundary action. Finally, we need to add another boundary term to (\ref{KS7c1}) to cancel of the 
term proportional to $\delta\Phi'_n$. This
is precisely the Gibbons-Hawking term \cite{Gibbons-Hawking}: 
\bg \label{KS7c2} 
{\cal K}_1 =- {1\over 2}\int \frac{d^4q}{(2\pi)^4 \sqrt{g(r_c)}}\Big(A^{mn}_1\Phi^{[1]}_m\Phi'^{[1]}_n + 
A^{mn}_1\Phi^{[1]}_n\Phi'^{[1]}_m 
+2 F^n_1\Phi'^{[1]}_n\Big)\Bigg{|}_{{\rm boundary}}
\nd
Taking the variation of (\ref{KS7c2}) $\delta{\cal K}_1$ we get terms proportional to $\delta\Phi'^{[1]}$ as well as 
$\delta\Phi^{[1]}$. Adding $\delta{\cal K}_1$ to $\delta{\cal S}^{(1)}$ we can get rid of all the $\delta\Phi'^{[1]}$
terms from (\ref{KS7c1}). This means we can alternately state that the boundary theory should have the following 
constraints\footnote{One can impose similar constraints at the horizon also.}:
\bg\label{conts}
 \delta\Phi_m'^{[1]}(r_c, q) ~=~ \delta\Phi_n'^{[1]}(r_c, -q)~ =~ 0
\nd

With all the above considerations we can present our final result for the boundary action. Putting the 
equations of motion constraints on (\ref{KS7c1}), as well as the derivative constraints (\ref{conts}), we can 
show that the variation (\ref{KS7c1}) can come from the following boundary $3+1$ dimensional action:
\bg \label{action2}
&& {\cal S}^{(1)}~=~\int \frac{d^4q}{(2\pi)^4 \sqrt{g(r_{\rm max})}}
\Bigg\{\Big[ C^{mn}_1(r,q)-A'^{mn}_1(r,q)\Big] \Phi^{[1]}_m(r, q) \Phi^{[1]}_n(r, -q)\nonumber\\
&&~~~~~~+\Big[B^{mn}_1(r,q)-A^{mn}_1(r,q)\Big] \Big[\Phi'^{[1]}_m(r, q) \Phi^{[1]}_n(r, -q)+\Phi^{[1]}_m(r, q)
\Phi'^{[1]}_n(r, -q)\Big]\nonumber\\
&&~~~~~~~~~~~ +\Big(E^m_1 -F'^m_1\Big)\Phi^{[1]}_m(r, q)\Bigg\}\Bigg{|}_{r_h}^{r_c}
\nd
However the above action diverges, as one can easily check 
from the explicit expressions for 
$A^{mn}_1,B^{mn}_1,C^{mn}_1,E^m_1$ and $F^m_1$ for the specific case worked in Appendix D of \cite{FEP}. 
Indeed, comparing it to the known AdS results, we
observe that from the boundary there are terms proportional to $r_c^4$ in each of 
$C^{mn}_1,A'^{mn}_1,E^m_1$ and $F'^{m}_1$ and proportional to $r_c^5$ in $A^{mn}_1, B^{mn}_1$. 
As $r_c \to \infty$ the action diverges so as it stands $r_c$ cannot 
completely specify the UV degrees of freedom at the 
scale $\Lambda_c$. 
Thus we need to
regularize/renormalize
it before taking functional derivative of it. This renormalization procedure will give us a finite 
boundary theory from which one could get meaningful results of the dual gauge theory.

Once we express the warp factors in terms of power series in $r_{(\alpha)}$ (\ref{logr})
the renormalizability procedure becomes 
much simpler. 
Note however that this renormalization is only 
in classical sense, as the procedure will involve removing the infinities in (\ref{action2}) by adding counter-terms
to it. Comparing with the known AdS results, and the specific example presented in Appendix C of \cite{FEP}, 
one can argue that
the infinities in (\ref{action2}) arise from the following three sources:
\bg\label{infinity}
&&1.~~~ C^{mn}_1(r_c,q)-A'^{mn}_1(r_c,q) ~ = ~ \sum_{\alpha} H_{\vert\alpha\vert}^{mn}(q)~ r_{c(\alpha)}^4 
~+~ {\rm finite~terms}\nonumber\\
&&2.~~~ B^{mn}_1(r_c,q)-A^{mn}_1(r_c,q)  ~ = ~ \sum_{\alpha} K_{\vert\alpha\vert}^{mn}(q)~ r_{c(\alpha)}^5 
~+~ {\rm finite~terms}\nonumber\\
&&3.~~~ E^m_1(r_c,q) -F'^m_1(r_c,q)  ~ = ~ \sum_{\alpha} I_{\vert\alpha\vert}^{m}(q) ~r_{c(\alpha)}^4 
~+ ~ {\rm finite~terms}
\nd
where in the above expressions we are keeping $\alpha$ arbitrary so that it can in general take both positive and 
negative values; and 
the finite terms above are of the form $r_{c(\alpha)}^{-n}$ with $n \ge 1$. Therefore
to regularize, first we write the metric perturbation also as a series in $1/r_{(\alpha)}$:
\bg \label{mpert1}
\Phi^{[1]}_n=\sum_{k=0}^{\infty} \sum_{\alpha}~{s_{nn}^{(k)[\alpha]}\over r_{(\alpha)}^k} 
\nd 
where 
the above relation could be easily derived using (\ref{phim}), taking the background warp factor
correctly.

Plugging in (\ref{mpert1}) and (\ref{infinity}) in (\ref{action2}) we can easily extract the divergent parts of it
. Thus the counter-terms are given by:  
\bg\label{cterms}
&& {\cal S}^{(1)}_{\rm counter}=\int \frac{d^4q}{2(2\pi)^4 \sqrt{g(r_{\rm max})}}
\sum_{\alpha, \beta, \gamma} \Big\{{H}_{\vert\alpha\vert}^{mn}
\Big[s_{mm}^{(0)[\beta]}s_{nn}^{(0)[\gamma]}~r_{(\alpha)}^4 + 
\Big(s_{mm}^{(1)[\beta]}s_{nn}^{(0)[\gamma]}r_{(\alpha_1)}^3\nonumber\\ 
&& + s_{mm}^{(0)[\beta]}s_{nn}^{(1)[\gamma]}r_{(\alpha_2)}^3\Big) 
+ \Big(s_{mm}^{(2)[\beta]}s_{nn}^{(0)[\gamma]}r_{(\alpha_3)}^2 
+ s_{mm}^{(0)[\beta]}s_{nn}^{(2)[\gamma]}r_{(\alpha_4)}^2 + 
s_{mm}^{(1)[\beta]}s_{nn}^{(1)[\gamma]}r_{(\alpha_5)}^2\Big)\nonumber\\ 
&&+\Big(s_{mm}^{(3)[\beta]}s_{nn}^{(0)[\gamma]}r_{(\alpha_6)} + 
s_{mm}^{(0)[\beta]}s_{nn}^{(3)[\gamma]}r_{(\alpha_7)}
+ ~ s_{mm}^{(2)[\beta]}s_{nn}^{(1)[\gamma]}r_{(\alpha_8)} + 
s_{mm}^{(1)[\beta]}s_{nn}^{(2)[\gamma]}\Big)r_{(\alpha_9)} \Big]\nonumber\\
&& + {K}_{\vert\alpha\vert}^{mn}
\Big[-\big(s_{mm}^{(0)[\beta]}s_{nn}^{(1)[\gamma]}r_{(\alpha_{10})}^3+s_{mm}^{(1)[\beta]}s_{nn}^{(0)[\gamma]}r_{(\alpha_{11})}^3\big)
- \Big(2s_{mm}^{(1)[\beta]}s_{nn}^{(1)[\gamma]}r_{(\alpha_{12})}^2 \nonumber\\
&& +2 s_{mm}^{(0)[\beta]}s_{nn}^{(2)[\gamma]}r_{(\alpha_{13})}^2
+2 s_{nn}^{(0)[\beta]}s_{mm}^{(2)[\gamma]}r_{(\alpha_{14})}^2\Big) 
-~\Big(2s_{mm}^{(1)[\beta]}s_{nn}^{(2)[\gamma]}r_{(\alpha_{15})} 
+ s_{mm}^{(2)[\beta]}s_{nn}^{(1)[\gamma]}r_{(\alpha_{16})}\nonumber\\
&&+ 3s_{mm}^{(0)[\beta]}s_{nn}^{(3)[\gamma]}r_{(\alpha_{17})} 
+2s_{nn}^{(1)[\beta]}s_{mm}^{(2)[\gamma]}r_{(\alpha_{18})} 
+ s_{nn}^{(2)[\beta]}s_{mm}^{(1)[\gamma]}r_{(\alpha_{19})} + 
3s_{nn}^{(0)[\beta]}s_{mm}^{(3)[\gamma]}r_{(\alpha_{20})}\Big)\Big] \nonumber\\
&& ~~~~~~~~~ +I_{\vert\alpha\vert}^m \theta(r_0-r)\Big(s_{mm}^{(0)[\beta]} r_{(\alpha)}^4
+ s_{mm}^{(1)[\beta]} r_{(\alpha_1)}^3 + s_{mm}^{(2)[\beta]} r_{(\alpha_3)}^2 
+ s_{mm}^{(3)[\beta]} r_{(\alpha_6)}\Big)\Big\}
\nd
with an equal set of terms with $r_{(-\alpha_i)}$. In the above expression 
$r_{(\alpha_i)} \equiv r^{1- \epsilon_{(\alpha_i)}}$;
and as before, the integrand is defined at the horizon $r_h$ and the cutoff $r_c$. The other variables namely, 
$s_{mm}^{(k)[\beta]}, {H}_{\vert\alpha\vert}^{mn},K_{\vert\alpha\vert}^{mn}$ and ${I}_{\vert\alpha\vert}^m$ 
are independent of $r$ but functions of $q^i$. 
For one specific case their values 
are given in Appendix C of \cite{FEP}. Finally the $\epsilon_{(\alpha_i)}$ can be defined by the following procedure. 
Lets
start with the expression:
\bg\label{ster}
&&H^{mn}_{\vert\alpha\vert} s_{mm}^{(a)[\beta]}s_{nn}^{(b)[\gamma]}~r_{(\alpha_k)}^p ~\equiv~ 
H^{mn}_{\vert\alpha\vert} s_{mm}^{(a)[\beta]}s_{nn}^{(b)[\gamma]}~ 
{r^4_{(\alpha)}\over r^a_{(\beta)}r^b_{(\gamma)}}\nonumber\\
&&K^{mn}_{\vert\alpha\vert} s_{mm}^{(c)[\beta]}s_{nn}^{(d)[\gamma]}~r_{(\alpha_l)}^q ~\equiv~
K^{mn}_{\vert\alpha\vert} s_{mm}^{(c)[\beta]}s_{nn}^{(d)[\gamma]}~ 
{r^5_{(\alpha)}\over r^c_{(\beta)}r^d_{(\gamma)}}
\nd
{}from where one can easily infer:
\bg\label{finfer}
&&p ~=~ 4-a-b,~~~~ \epsilon_{(\alpha_k)} ~=~ {4\epsilon_{(\alpha)} - a \epsilon_{(\beta)} -b\epsilon_{(\gamma)}\over 
4-a-b}\nonumber\\
&&q ~=~ 5-c-d,~~~~ \epsilon_{(\alpha_l)} ~=~ {5\epsilon_{(\alpha)} - c \epsilon_{(\beta)} -d\epsilon_{(\gamma)}\over 
5-c-d}
\nd
Using this procedure we can determine all the $r_{(\alpha_i)}$ in the counterterm expression (\ref{cterms}). 

\noindent At this point the analysis of the theory falls into two possible classes. 

\noindent $\bullet$ The first class 
is to analyze the theory right at the 
usual boundary 
where $r_c \to \infty$. This is the standard picture where there are infinite degrees of freedom at the boundary, and 
the theory has a smooth RG flow from UV to IR till it confines (at least from the weakly coupled gravity dual). The action
obtained by setting $r_c=\infty$ describes the `parent cascading theory'. 

\noindent $\bullet$ The 
second class is to analyze the theory by specifying the 
degrees of freedom at generic energy scale given by $r_c=\Lambda_c$ and then defining the theories at the boundary by adding
appropriate irrelevant and marginal operators for scales greater than $\Lambda_c$. By adding different operator for scales
$\Lambda>\Lambda_c$, one gets different theories.  All these different branches meet the parent cascading 
 theory at some scale $r_c=\Lambda_c$, as depicted in Fig {\bf 2.13}.
 The gravity duals of these theories are the usual 
{\it deformed} conifold 
geometries cutoff at various $r_c$ with appropriate UV caps added (of course for $r < r_c$ the geometries change 
accordingly). More details on  UV caps to geometries will be discussed in the next subsection.  

The first class of theories is more relevant for the 
pure AdS/CFT case whereas the latter is more relevant for the non-AdS case\footnote{In both cases of course we need to 
add appropriate number of seven branes to get the finite F-theory picture. The holographic 
renormalization procedure remains unchanged and the far IR physics remains unaltered. The UV caps affect mostly geometries
close to $r_c$, as expected.}.For the pure AdS/CFT case without flavors $\epsilon_{(\alpha_i)} = 0$ 
(so that the subscript $\alpha_i$'s can be ignored from 
all variables), 
we can subtract the counter-terms (\ref{cterms}) from the action (\ref{action2}) to get 
the following renormalized action:
\bg \label{actionren}
 {\cal S}^{(1)}_{\rm ren}&=&{\cal S}^{(1)} - {\cal S}^{(1)}_{\rm counter}\nonumber\\
&=&\int \frac{d^4q}{(2\pi)^4}
\Big[H^{mn}\Big(s_{mm}^{(4)}s_{nn}^{(0)} + s_{mm}^{(3)}s_{nn}^{(1)} + 
s_{mm}^{(2)}s_{nn}^{(2)}+ s_{mm}^{(1)}s_{nn}^{(3)} \nonumber\\
&&+ s_{mm}^{(0)}s_{nn}^{(4)} \Big) - 
K^{mn}\Big(4s_{mm}^{(0)}s_{nn}^{(4)} + 3s_{mm}^{(1)}s_{nn}^{(3)} + 
4s_{mm}^{(2)}s_{nn}^{(2)}+s_{mm}^{(3)}s_{nn}^{(1)}\nonumber\\
&& 4s_{nn}^{(0)}s_{mm}^{(4)} + 3s_{nn}^{(1)}s_{mm}^{(3)} +s_{nn}^{(3)}s_{mm}^{(1)} \Big)
+I^m s_{mm}^{(4)}\Big]
\nd
where we have made all the ${\cal O}(1/r_c)$ terms vanishing, and in the limit $r_h$ small the small shifts to 
$s_{nn}^{(j)}$ given by $s_{nn}^{(3)} + {\cal O}(r_h^4)$ can also be ignored. Observe that 
we can reinterpret the renormalized action 
(\ref{actionren}) as the following new action:
\bg\label{newren}
{\cal S}^{(1)}_{\rm ren} &=&  \int \frac{d^4q}{(2\pi)^4} \Big[Z^{mn} \Phi_m(q) \Phi_n(-q)+
U^{mn} \big(\Phi_m(q) \Phi'_n(-q)+ \Phi'_m(q) \Phi_n(-q)\big)\nonumber\\
&+& Y^m \Phi_m(q) 
+ Y^n \Phi_n(-q) +V^m \Phi'_m(q) 
+ V^n \Phi'_n(-q) + X\Big] \nonumber\\
\nd
where $\Phi_m(q)=\Phi_m(q,r=\infty)$ , $\Phi'_m(q)=d\Phi_m(q,r)/dr|_{r=\infty}$ and
$X, Y, Z,U,V$ could be functions of r and  $\overrightarrow{q}$ but evaluated for fixed $r=\infty$.
We can determine their functional form 
by comparing (\ref{newren}) with (\ref{actionren}). For us however the most relevant part is the energy momentum
tensors which we could determine from \ref{newren} by finding the coefficients $Y^m$ and $Y^n$. One can 
easily show that, up to a possible additive constant, $Y^m, Y^n$ are given by:
\bg\label{ymyn}
Y^m &=& H^{mn} s_{nn}^{(4)}-4K^{mn} s_{nn}^{(4)}, 
~~~~~ Y^n = H^{mn} s_{mm}^{(4)}-4K^{mn} s_{mm}^{(4)}\nonumber\\
V^m &=& K^{mn} s_{nn}^{(5)}, ~~~~~ V^n = K^{mn} s_{mm}^{(5)}
\nd

Now let us come to second class of theories 
 wherein we take any arbitrary $r = r_c$, with appropriate UV degrees of freedom such that they 
have good boundary descriptions satisfying all the necessary constraints. For these cases, once we 
subtract the counter-terms (\ref{cterms}), the renormalized action (specified by $r$) takes the following form:
\bg \label{actionren2}
 {\cal S}^{(1)}_{\rm ren}&=&{\cal S}^{(1)} - {\cal S}^{(1)}_{\rm counter}\nonumber\\
&=&\int \frac{d^4q}{2(2\pi)^4}\sum_{\alpha, \beta, \gamma}
\Big\{H_{\vert\alpha\vert}^{mn}\Big(s_{mm}^{(4)[\beta]}s_{nn}^{(0)[\gamma]} r^{4\epsilon_{(\beta)} - 
4\epsilon_{(\alpha)}}
+ s_{mm}^{(3)[\beta]}s_{nn}^{(1)[\gamma]} r^{3\epsilon_{(\beta)} + \epsilon_{(\gamma)} - 4\epsilon_{(\alpha)}}\nonumber\\
&&+s_{mm}^{(2)[\beta]}s_{nn}^{(2)[\gamma]} r^{2\epsilon_{(\beta)}+ 2 \epsilon_{(\gamma)} - 4\epsilon_{(\alpha)}}
+ s_{mm}^{(1)[\beta]}s_{nn}^{(3)[\gamma]} r^{\epsilon_{(\beta)}+ 3\epsilon_{(\gamma)} -4\epsilon_{(\alpha)}}
+ s_{mm}^{(0)[\beta]}s_{nn}^{(4)[\gamma]}r^{4\epsilon_{(\gamma)} - 4\epsilon_{(\alpha)}} \Big)\nonumber\\ 
&&-4K_{\vert\alpha\vert}^{mn}\Big(s_{mm}^{(0)[\beta]}s_{nn}^{(4)[\gamma]}r^{5\epsilon_{(\gamma)} - 
5\epsilon_{(\alpha)}}
+ s_{mm}^{(1)[\beta]}s_{nn}^{(3)[\gamma]}\big[r^{\epsilon_{(\beta)} + 4\epsilon_{(\gamma)} -5\epsilon_{(\alpha)}}  
+ r^{2\epsilon_{(\beta)} + 3\epsilon_{(\gamma)} -5\epsilon_{(\alpha)}}\big]\nonumber\\
&& + 
4s_{mm}^{(2)[\beta]}s_{nn}^{(2)[\gamma]}\big[r^{2\epsilon_{(\beta)} + 3\epsilon_{(\gamma)} -5\epsilon_{(\alpha)}}
+ r^{3\epsilon_{(\beta)} + 2\epsilon_{(\gamma)} -5\epsilon_{(\alpha)}}\big]
+ 4s_{nn}^{(0)[\beta]}s_{mm}^{(4)[\gamma]} r^{5\epsilon_{(\beta)} -5\epsilon_{(\alpha)}} \nonumber\\
&& +s_{mm}^{(3)[\beta]}s_{nn}^{(1)[\gamma]}\big[r^{4\epsilon_{(\beta)} + \epsilon_{(\gamma)} -5\epsilon_{(\alpha)}} +
r^{3\epsilon_{(\beta)} + 2\epsilon_{(\gamma)} -5\epsilon_{(\alpha)}}\big]\Big)
+I_{\vert\alpha\vert}^m s_{mm}^{(4)[\beta]} r^{5\epsilon_{(\beta)} - 5\epsilon_{(\alpha)}}\Big\}
\nd
evaluated at the cut-off $r_c$ and the horizon radii $r_h$ as usual. Notice now the appearance of 
$r^{m\epsilon_{(\alpha)} + n\epsilon_{(\beta)} + p\epsilon_{(\gamma)}}$ factors. One can easily show that:
\bg\label{shth}
{1\over 2}\big[r^{m\epsilon_{(\alpha)} + n\epsilon_{(\beta)} + p\epsilon_{(\gamma)}} + 
r^{-m\epsilon_{(\alpha)} -n\epsilon_{(\beta)} - p\epsilon_{(\gamma)}}\big]~ = ~ 1 ~ + ~ 
{\cal O}\big[\epsilon_{(\alpha, \beta, \gamma)}\big]^2
\nd
Since the warp factor $h$ is defined only for small values of 
$g_sN_f, g_sM^2/N, g^2_s N_fM^2/N$ we don't know the background (and 
hence the warp factor) for finite values of these quantities. 
Therefore for our case we can put ${\cal O}\big[\epsilon_{(\alpha, \beta, \gamma)}\big]^2$ 
to zero
so that the value in (\ref{shth}) is identically 1. For finite values of these 
quantities both the warp factor and the background 
would change drastically and so new analysis need to be performed to holographically renormalize the theory. 
Our conjecture would be that once we know the background for finite 
values of $g_sN_f, g_sM^2/N$, the terms like 
(\ref{actionren2}) would come out automatically renormalized by choice of our counterterms. 

Once this is settled, the renormalized action at the cut-off radius $r_c$ would only go as powers of 
$r^{-1}_{c(\alpha)}$. Thus we can express the total action as: 
\bg\label{renacto}
&&{\cal S}^{(1)}_{\rm ren} = \int \frac{d^4q}{(2\pi)^4} 
\sum_{\alpha, \beta}\Bigg\{\left(\sum_{j=0}^{\infty}{\widetilde{a}^{(\alpha)}_{mn(j)}\over r_{c(\alpha)}^j}\right) 
\widetilde{G}^{mn} \Phi_m(q) \Phi_n(-q) +
\left(\sum_{j=0}^{\infty}{\widetilde{e}^{(\alpha)}_{mn(j)}\over r_{c(\alpha)}^j}\right) 
\widetilde{M}^{mn} (\Phi_m(q) \Phi'_n(-q)\nonumber
\nd
\bg
&& +\Phi'_m(q) \Phi_n(-q)) 
+{H}_{\vert\alpha\vert}^{mn}\Big[s_{nn}^{(4)[\beta]}\Phi_m(q) + 
s_{mm}^{(4)[\beta]}\Phi_n\Big]+{K}_{\vert\alpha\vert}^{mn}\Big[-4s_{nn}^{(4)[\beta]}\Phi_m(q) 
-4 s_{mm}^{(4)[\beta]}\Phi_n(q)\nonumber\\
&&+s_{nn}^{(5)[\beta]}\Phi'_m(q) +s_{mm}^{(5)[\beta]}\Phi'_n(q)\Big]
+ \left(\sum_{j=0}^{\infty}\frac{\widetilde{b}^{(\alpha)}_{m(j)}}{r_{(\alpha)}^j}\right) 
\widetilde{J}^m\Phi_m(q) + X[r_{c(\alpha)}]\Bigg\}
\left[1-{r_h^4\over r_c^4}\right]^{-{1\over 2}}
\nd
where $\Phi_n$ are as before and $r_{c(\alpha)}^j= r_{(\alpha)}^j|_{r=r_c}$. Here we have ignored the terms evaluated at the
horizon as they are not proportional to the sources $\Phi_n$ and hence will not contribute to the expectation value. The 
explicit expressions for the other coefficients
listed above, namely,
$\widetilde{G}^{mn},\widetilde{M}^{mn} ,\widetilde{a}^{(\alpha)}_{mn(j)}, \widetilde{J}^m,
\widetilde{e}^{(\alpha)}_{mn(j)}$ 
and $\widetilde{b}^{(\alpha)}_{m(j)}$ can be worked out easily from our earlier analysis (see
Appendix C of \cite{FEP} for one specific example). Note
that $X[r_{(\alpha)}]$ is a function independent of $\Phi^{[0]}_m$ and appears for generic renormalized action.

Now the generic form for the energy momentum tensor is evident from looking at the linear terms in the 
above action (\ref{renacto}). This is then given by:
\bg \label{wake}
&&T_0^{mm} \equiv 
\int \frac{d^4q}{(2\pi)^4}
\Bigg[({H}_{\vert\alpha\vert}^{mn}+ {H}_{\vert\alpha\vert}^{nm})s_{nn}^{(4)[\beta]} 
-4({K}_{\vert\alpha\vert}^{mn}+ {K}_{\vert\alpha\vert}^{nm})s_{nn}^{(4)[\beta]}+({K}_{\vert\alpha\vert}^{mn}+ 
{K}_{\vert\alpha\vert}^{nm})s_{nn}^{(5)[\beta]}\nonumber\\
&&~~~~~~~~~~~~~~~~~~~~~~~~~~~ + \left(\sum_{j=0}^{\infty}\frac{\widetilde{b}^{(\alpha)}_{n(j)}}
{r_{c(\alpha)}^j}\right)
\widetilde{J}^n \delta_{nm}\Bigg]\left(1- {r_h^4\over r_c^4}\right)^{-{1\over 2}}
\nd
at $r = r_c$ (we ignore the result at the horizon) and sum over ($\alpha, \beta$) is implied. 
This result should be compared to the ones derived in 
 \cite{kostas1} \cite{kostas2}\cite{kostas3} \cite{kostas4}\cite{kostas5} \cite{ahabu} \cite{Yaffe-1} which don't have any $r_c$ dependence. 
The AdS/CFT result is of course the first line of the above result. Notice that the second line also has a $r_c$ independent
additive constant which are irrelevant for our purpose because the energy-momentum can always be 
shifted by a constant to absorb this factor. Now for non-AdS geometries, $r_c$ determines the number of degrees of
freedom at scale $\Lambda_c$ with the relation (\ref{Neff}) and inverting it with $r^4 h\sim {\rm log}(r)$ we get $r_c\sim
e^{ N_{\rm eff}}$ .  Thus once we specify the  effective degrees of freedom at scale
$\Lambda_c$, then
our result shows that the energy-momentum tensor not only
inherits the universal behavior of the parent cascading theory but there 
are additional corrections of ${\cal O}(e^{-jN_{\rm eff}})$ . But energy scale should
eventually be set to infinity, so how is keeping ${\cal O}(1/r_c)$ terms in (\ref{wake}) justified?   

To answer this, first consider a  daughter gauge
theory for which number of degrees of freedom do not change for scales $\Lambda>\Lambda_c$ and then $N_{\rm
eff}(\Lambda_c)=N_{\rm UV}(\infty)$. Taking the energy scale to infinity means replacing $r_c$ in (\ref{wake}) with
$e^{N_{\rm UV}(\infty)}$ as only gauge theory observables such as number of degrees of freedom should appear in the final
result for stress tensor. But as $N_{\rm
eff}(\Lambda_c)=N_{\rm UV}(\infty)$, for this particular theory, we can keep the corrections coming from
${\cal O} (1/r_c)$ as in (\ref{wake}). 

This argument of replacing $r_c\rightarrow {\cal O} (e^{N_{\rm UV}})$ is somewhat
naive and needs elaboration. If we cut off the geometry at $r=r_c$, and {\it do not} add any geometry from $r=r_c$ to
$r=\infty$, we are essentially ignoring gavitons that are present in the region $r\ge r_c$. This will in turn mean we are
ignoring UV modes in the dual field theory that have energies  $\Lambda \ge \Lambda_c$. Thus (\ref{wake}) is incomplete.
What we mean here is once we add gravitons from region $r\ge r_c$, their contribution to the final stress energy tensor will
can be accounted for by replacing $r_c$ in (\ref{wake})with ${\cal O} (e^{N_{\rm UV}})$.  

Note that for an AdS geometry, the dual gauge theory is conformal and do not change degrees of freedom with scale. Hence 
treating $N_{\rm
eff}(\Lambda_c)=N_{\rm UV}(\infty)$  amounts to adding an AdS UV cap to our dual geometry from $r=r_c$ to $r=\infty$. The
final result for stress tensor of a dual gauge theory with an AdS UV geometry will be of the form (\ref{wake}) with  
$r_c\rightarrow {\cal O} (e^{N_{\rm UV}})$ which may not be infinity. We will discuss this in detail in the following
section.

 In general, various gauge theories which are quite distinct at far UV scale can look identical in the IR. 
One can start with a given number of degrees of freedom at a scale $\Lambda_c$ and demand that it is the infrared
limit of several different UV theories. Finally specifying the UV degrees of freedom distinguishes the theories. 
From the gravity
side we can replace $r_c$ in (\ref{wake}) with diffent values of $e^{N_{\rm UV}}$ and obtain distinct UV gauge theories
which are identical at the infrared scale $\Lambda_c$.
In fact a more careful analysis shows that one can cut a bulk geometry at some $r_c$ and attach various UV geometries from
$r_c$ to $\infty$ to obtain distinct gauge theories at the boundary. We discuss this in some detail in the following section.
 
\subsection{UV Completion from Dual Gravity}      
The first important issue here is that we can study infinite number 
of UV completed theories in full F-theory setup. 
All of these theories have good boundary descriptions and have same degrees of 
freedom as the parent cascading theory at certain specified scales. The simplest UV complete theory  is of course the 
parent cascading theory of OKS model. Using (\ref{Neff}) with warp factor given by (\ref{hvalue}) one easily gets
\bg \label{Neff1}
N_{\rm eff}(\Lambda)\sim L^4 (1+{\cal O}({\rm log}(\Lambda)))
\nd
 which grows  with scale $\Lambda$ and diverges for $\Lambda\rightarrow \infty$.  
The question now is how to construct other possible theories by defining the degrees of freedom 
at scale $\Lambda\rightarrow \infty$ i.e. at the boundary of the geometry $r\rightarrow \infty$. The action for the  boundary theory  should be identified as:
\bg\label{bndiden}
\big[{\cal S}^{(1)}_{\rm ren}\big]_{r_h}^\infty ~ = ~ \big[{\cal S}^{(1)}_{\rm ren}\big]_{r_h}^{r_c} + 
\big[{\cal S}^{(1)}_{\rm ren}\big]_{r_c}^\infty
\nd 
where the boundary is at $r \to \infty$. For the boundary cascading theory the above expression simply means that 
\bg\label{casbnd}
 \big[{\cal S}^{(1)}_{\rm ren}\big]_{r_c}^\infty & = & - \int \frac{d^4q}{(2\pi)^4} \left(\sum_{j=0}^{\infty}
\frac{\widetilde{b}^{(\alpha)}_{n(j)}}{r_{c(\alpha)}^j}\right)
\widetilde{J}^n \Phi_n - \int \frac{d^4q}{(2\pi)^4} \left(\sum_{j=0}^{\infty} 
{B^{(\alpha)}_{n(j)}r_h^{4j} \over r_{c(\beta)}^j}\right)\Phi_n \nonumber\\
&& ~~~~~~~~~~~+~ {\{\widetilde{b}^{(\alpha)}_{n(j)}, {\cal O}(r_h^{4j})\}\over \infty}~{\rm factors} 
\nd
where the sign is crucial and sum over $\alpha$ is again implied (note (a) the cut-off dependence, and (b)
$r_{c(\beta)}$ is some function of $r_{c(\alpha)}$ that
one can determine easily). 
We now see that the contributions from the UV cap give the following values for $B_j$ for parent cascading theory:
\bg\label{cfuv} 
&& B^{(\alpha)}_{n(0)} ~ = ~ B^{(\alpha)}_{n(1)} ~ = ~ B^{(\alpha)}_{n(2)} ~ = ~ B^{(\alpha)}_{n(3)} ~ = ~ 0 \nonumber\\
&& B^{(\alpha)}_{n(4)} ~ = ~ {1\over 2}\bigg[\widetilde{b}^{(\alpha)}_{n(0)}
\widetilde{J}^n \theta(r_0-r_c) + {\cal O}(H_{\vert\alpha\vert}^{nm}, 
K_{\vert\alpha\vert}^{nm}, s_{nn}^{[\alpha]})\bigg],~ .....
\nd

The choice above is made precisely to {\it exactly} cancel the ${\cal O}(1/r_c)$ contributions 
coming in from the action measured from $r_h \le r \le r_c$. Finally the boundary energy-momentum tensor reads 
\bg\label{casenmo} 
\int \frac{d^4q}{(2\pi)^4} \sum_{\alpha, \beta}
\Big[({H}_{\vert\alpha\vert}^{mn}+ 
{H}_{\vert\alpha\vert}^{nm})s_{nn}^{(4)[\beta]} -4({K}_{\vert\alpha\vert}^{mn}+ 
{K}_{\vert\alpha\vert}^{nm})s_{nn}^{(4)[\beta]}+({K}_{\vert\alpha\vert}^{mn}+ 
{K}_{\vert\alpha\vert}^{nm})s_{nn}^{(5)[\beta]}\Big]\nonumber\\
\nd
which is the result derived in  \cite{kostas1}-\cite{kostas5} and \cite{Yaffe-1}.

The above way of reinterpreting the boundary contribution should tell us precisely how we could modify the boundary 
degrees of freedom to construct distinct UV completed theories. There are two possible ways we can achieve this:


\noindent $\bullet$ From the geometrical perspective
 we can cutoff the deformed conifold background at $r = r_c$ and attach an 
appropriate UV ``cap'' from $r = r_c$ to $r \to \infty$ by carefully modifying the 
geometry at the neighborhood of the junction point. As an example, this UV cap could as well 
be another AdS background
from $r_c$ to $r \to \infty$. There are of course numerous other choices available from the F-theory limit. 
Each of these caps would give 
rise to distinct UV completed gauge theories.

\noindent $\bullet$ From the action perspective we could specify  the value of the action measured from 
$r_c$ to $r \to \infty$, i.e $\big[{\cal S}^{(1)}_{\rm ren}\big]_{r_c}^\infty$. The simplest case where this is 
zero gives rise to a boundary theory for which $N_{\rm eff}(\Lambda_c)=N_{\rm UV}(\infty)$ as discussed before. To study more generic cases,
we need to see how much constraints we can put on our 
integral. One immediate constraint is the holographic renormalizability of our theory. This tells us that the 
value of the integral can only go as powers of $1/r_{c(\alpha)}$ otherwise we will not have finite actions. This in turn 
implies
\bg\label{sifrom}
&&\Big[{\cal S}^{(1)}_{\rm ren}\Big]_{r_c}^\infty = \int \frac{d^4q}{(2\pi)^4}\sum_{\alpha}
\Bigg\{\left(\sum_{j=0}^{\infty}{\widetilde{A}^{(\alpha)}_{mn(j)}\over r_{c(\alpha)}^j}\right) 
\widetilde{G}^{mn} \Phi_m \Phi_n +
\left(\sum_{j=0}^{\infty}{\widetilde{E}^{(\alpha)}_{mn(j)}\over r_{c(\alpha)}^j}\right) \widetilde{M}^{mn}\\ 
&&\times (\Phi_m \Phi'_n 
+\Phi'_m \Phi_n) + \left(\sum_{j=0}^{\infty}\frac{\widetilde{B}^{(\alpha)}_{m(j)}}
{r_{c(\alpha)}^j}\right) 
\widetilde{J}^m\Phi_m + X[r_{(\alpha)}]\Bigg\}\theta(r_0-r_c) + {\rm finite~ terms}\nonumber
\nd
where by specifying the coefficients $\widetilde{A}^{(\alpha)}_{mn(j)}, 
\widetilde{E}^{(\alpha)}_{mn(j)}$ and $\widetilde{B}^{(\alpha)}_{m(j)}$ 
we can specify the precise 
UV degrees of freedom! The finite terms are $r_c$ independent and therefore would only provide finite shifts 
to our observables. They could therefore be scaled to zero. Notice also that the contributions from (\ref{sifrom}) 
only renormalizes the coefficients $\widetilde{a}^{(\alpha)}_{mn(j)}, 
\widetilde{e}^{(\alpha)}_{mn(j)}$ and $\widetilde{b}^{(\alpha)}_{m(j)}$
in (\ref{renacto}), 
and 
therefore the final expressions for all the physical variables for various UV completed theories could be written
directly from (\ref{renacto}) simply by replacing the $1/r_c$ dependent coefficients by their renormalized values. 
This is thus our precise description of how to specify the UV degrees of freedom for various gauge theories 
in our setup (see Fig {\bf 2.15} below). 
\begin{figure}[htb]\label{cone}
		\begin{center}
\includegraphics[height=7cm, width=14cm]{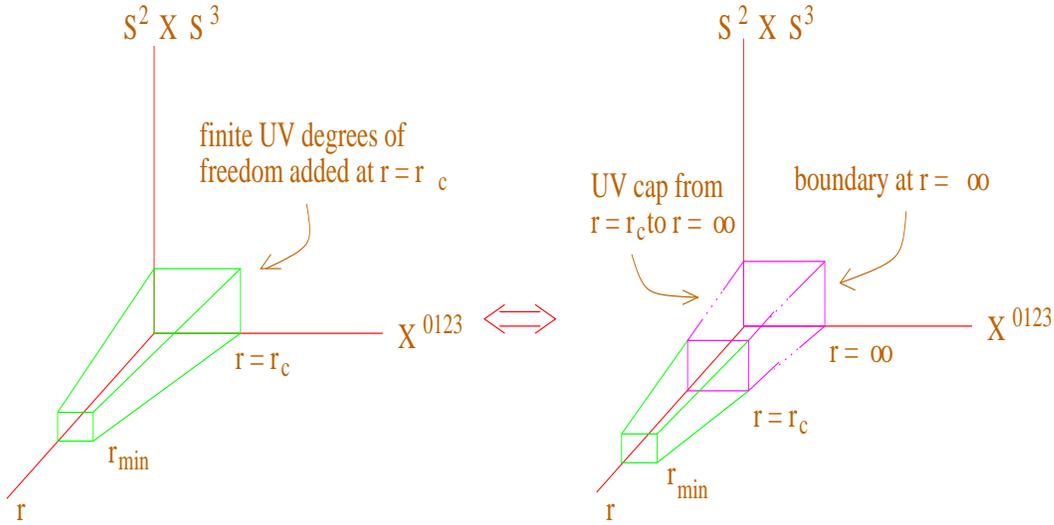}
		\caption{The equivalence between two different ways of viewing the boundary theory at zero temperature.
		  }
		\end{center}
		\end{figure}

For the left figure in {\bf 2.15} 
 we add finite UV degrees of freedom at $r = r_c$ of the deformed conifold geometry. Such a process is equivalent 
to the figure on the right where we cut-off the deformed conifold geometry at $r = r_c$ and
add a UV cap from $r = r_c$ to $r = \infty$. The boundary theory on the right 
has ${\cal N}_{\rm uv}$ degrees of freedom at $r = \infty$ and all physical quantities computed in either of these 
two pictures would only depend on ${\cal N}_{\rm uv}$ but not on $r = r_c$. At non-zero temperature the 
UV descriptions remain unchanged. 

\noindent Once the UV descriptions are properly laid out, we can determine the form for 
$\widetilde{A}^{(\alpha)}, 
\widetilde{B}^{(\alpha)}$ and $\widetilde{E}^{(\alpha)}$ by writing Callan-Symanzik type equations for them.  They tell us how 
$\widetilde{A}^{(\alpha)}, 
\widetilde{B}^{(\alpha)}$ and $\widetilde{E}^{(\alpha)}$ would behave with the scale $r_c$ or equivalently $\mu_c$. For
$\widetilde{A}$ the equation is\footnote{The following equation is derived from the scale-invariance of 
$\big[{\cal S}^{(1)}_{\rm ren}\big]_{r_h}^\infty$.}:
\bg\label{calsym}
\mu_c {\partial \widetilde{A}^{(\alpha)}_{mn(j)} \over \partial \mu_c} ~ = ~ j\left[1-\epsilon_{(\alpha)}\right]
\left[\widetilde{A}^{(\alpha)}_{mn(j)} + 
\widetilde{a}^{(\alpha)}_{mn(j)}\right]
\nd
with similar equations for $\widetilde{B}^{(\alpha)}$ 
and $\widetilde{E}^{(\alpha)}$. These equations tell us that physical quantities are
independent of scales. The  
parent cascading theory is defined as the scale-invariant limits of (\ref{calsym}), i.e:  
\bg\label{abe}
\widetilde{A}^{(\alpha)}_{mn(j)} ~= ~ - \widetilde{a}^{(\alpha)}_{mn(j)}, ~~~ \widetilde{E}^{(\alpha)}_{mn(j)} 
~= ~ - \widetilde{e}^{(\alpha)}_{mn(j)}, ~~~ 
\widetilde{B}^{(\alpha)}_{m(j)} ~= ~ - \widetilde{b}^{(\alpha)}_{m(j)} 
\nd
The above relation gives us a hint how to express $\widetilde{A}^{(\alpha)}, 
\widetilde{B}^{(\alpha)}$ and $\widetilde{E}^{(\alpha)}$ in terms of 
${\cal N}_{\rm eff}$, the effective degrees of freedom at $r = r_c$ and ${\cal N}_{\rm uv}$, the effective
degrees of freedom at $r = \infty$ 
i.e the boundary:
\bg\label{result1}
 \widetilde{A}^{(\alpha)}_{mn(j)} &~= ~& - \widetilde{a}^{(\alpha)}_{mn(j)}~ +~ 
\hat{a}^{(\alpha)}_{mn(j)} e^{-j\left[{\cal N}_{\rm uv} - (1-\epsilon_{(\alpha)})
{\cal N}_{\rm eff}\right]}\nonumber\\
\widetilde{E}^{(\alpha)}_{mn(j)} &~= ~& - \widetilde{e}^{(\alpha)}_{mn(j)}~ 
+~ \hat{e}^{(\alpha)}_{mn(j)} e^{-j\left[{\cal N}_{\rm uv} - (1-\epsilon_{(\alpha)})
{\cal N}_{\rm eff}\right]}\nonumber\\
 \widetilde{B}^{(\alpha)}_{m(j)} &~= ~& - \widetilde{b}^{(\alpha)}_{m(j)}~ 
+~ \hat{b}^{(\alpha)}_{m(j)} e^{-j\left[{\cal N}_{\rm uv} - (1-\epsilon_{(\alpha)}){\cal N}_{\rm eff}\right]}
\nd
where the actual boundary degrees of freedom are specified by knowing $\hat{a}_{mn(j)}, \hat{e}_{mn(j)}$ and  
$\hat{b}_{m(j)}$ as well as
 ${\cal N}_{\rm uv}$. Since $j$ goes from 0 to $\infty$, there are infinite possible UV complete 
boundary theories 
possible\footnote{The connection of $j$ with UV completions come from the coefficients $\hat a_{mn(j)}^{(\alpha)}, 
\hat e_{mn(j)}^{(\alpha)}$ and $\hat b_{mn(j)}^{(\alpha)}$ etc. that depend on $j$. For different choices of 
these coefficients we can have different UV completions. In this sense $j$ and 
UV completions are related.}. 
For very large ${\cal N}_{uv}$ (i.e ${\cal N}_{uv}\to \epsilon^{-n}, n >> 1$)
the boundary theories are similar to the original cascading 
theory. The various choices of 
($\hat{a}^{(\alpha)}_{mn(j)}(\overrightarrow{q}), \hat{e}^{(\alpha)}_{mn(j)}(\overrightarrow{q}), 
\hat{b}^{(\alpha)}_{m(j)}(\overrightarrow{q})$) 
tell us how the degrees of freedom
change from ${\cal N}_{\rm uv}$ to ${\cal N}_{\rm eff}$ under RG flow. The $\overrightarrow{q}$ dependence of all the 
quantities will tell us how the UV degrees of freedom affect IR physics. 
This is to be expected: addition of 
irrelevant operators do change IR physics, but not the far IR\footnote{The $\overrightarrow{q}$ dependences of the 
UV caps are also one-to-one correspondence to the changes in the local geometries near the cut-off radius 
$r_c$, as we discussed before. All in all this fits nicely with what one would have expected from UV degrees of 
freedom. Of course it still remains to verify the story from an actual supergravity calculation. We need to 
analyze the metric near the junction by studying the continuity and differentiability of the metric 
and see how far below $r = r_c$ we expect deformations from the UV caps. Various types of deformations will 
signal various sets of irrelevant operators. 
Needless to say, the 
far IR physics remain completely unaltered. In the following section, we will perform an exact calculation using appropriate sources in SUGRA action
accounting the deformation in the region near the junction.
  }.  
 
Therefore with this understanding of the boundary theories we can express the
energy-momentum tensor at the boundary with ${\cal N}_{uv}$ degrees of freedom at the boundary purely
in terms of gauge theory variables, as:
\bg \label{wakegt}
&&T_0^{mm} \equiv 
\int \frac{d^4q}{(2\pi)^4}
\Big[({H}_{\vert\alpha\vert}^{mn}+ {H}_{\vert\alpha\vert}^{nm})s_{nn}^{(4)[\beta]} 
-4({K}_{\vert\alpha\vert}^{mn}+ {K}_{\vert\alpha\vert}^{nm})s_{nn}^{(4)[\beta]}+({K}_{\vert\alpha\vert}^{mn}+ 
{K}_{\vert\alpha\vert}^{nm})s_{nn}^{(5)[\beta]}\nonumber\\
&&~~~~~~~~~~~~~~~~~~~~~~~~~~~~~~~~~~~ + \sum_{j=0}^{\infty}~\hat{b}^{(\alpha)}_{n(j)}(\overrightarrow{q}) 
\widetilde{J}^n  e^{-j{\cal N}_{\rm uv}}
\delta_{nm}\Big]
\nd
where sum over $\alpha$ is again implied, and
the first line is the universal property of the parent cascading theory inherited by our gauge theory. 
The second line specifies the precise 
degrees of freedom that we add at $r = r_c$ to describe the UV behavior of our theory at the boundary $r \to \infty$. 
Using this procedure, the final results of any physical quantities should be expressed only in terms of 
${\cal N}_{\rm uv}$ i.e the UV degrees of freedom\footnote{Restoring back the ${\cal O}(r_h)$ contributions would mean
that there should be an additional contribution to (\ref{wakegt}) of the form 
$\sum_{j=0}^{\infty}~ G(\hat b^{(\alpha)}_n, H^{mn}_{\vert\alpha\vert}, K^{mn}_{\vert\alpha\vert}, s_{nn}^{[\alpha]})
{\cal T}^{4j} e^{-j{\cal N}_{uv}}$ where $G$ is a function
whose functional form could be inferred from the UV integral (\ref{sifrom}).}.

\section{Towards Large N QCD from Dual Gravity}

From the above discussions we see how one obtains gauge theories which are quite distinct in the UV but have common IR dynamics. With a clear
understanding of how to UV complete a gauge theory with particular properties in the IR, we can construct a 
brane configuration that mimics
large N QCD. 

First observe from (\ref{Neff}) and (\ref{Neff1}) that   in OKS model the number of degrees of freedom keep on growing in the UV
. Furthermore with $B_2\sim {\rm log}(r)$ and axio-dilaton behaving as in (\ref{axfive}) we see that the gauge couplings run as 
$g_i\sim {\rm
log}(\Lambda)$ using (\ref{twocoup}). In fact in the far UV, at least one of the two gauge couplings blow up and we have Landau poles. Thus the UV of OKS model
has runaway behavior while QCD is asymptotically a free theory \cite{Gross-Will}\cite{Politzer}. However  the gauge couplings do run logarithmically and in the far IR using a
cascade of Seiberg dualities one obtains $SU(\bar{M})$ gauge group with fundamental matter. Hence the IR of OKS model has common features  
with large N QCD while the UV is quite different. But as discussed in the previous section, we can modify the UV of OKS model and build dual
gravity of a gauge theory that resembles large N QCD more closely. 

The ideal scenario would be to build a gauge theory which becomes asymptotically free and for scales where coupling is large, it has a
description in terms of weakly coupled classical gravity. Currently there are no brane setups that realizes such running of coupling and has
dual gravity in the appropriate regime. What we have been able to construct in \cite{LC} is a gauge theory that becomes asymptotically {\it conformal} in the
far UV and has logarithmic running  of couplings in the IR. The regime where the gauge theory has a classical supergravity description is
when gauge coupling is large and thus from UV to IR we have strong but running gauge coupling. This gauge theory has no Landau poles and the
number of degrees of freedom do not diverge as $\Lambda\rightarrow \infty$. In this section we will briefly describe the brane setup that
gives rise to such a gauge theory and construct its dual gravity. Details of our model can be found in \cite{LC}.

We start with the brane setup of Fig {\bf 2.11} i.e. the OKS model. The logarithmic running of the couplings come from the global logarithmic running
of  NS-NS two form $B_2\sim {\rm log}(r)$ and the { \it local} logarithmic running of axio-dilaton near a seven brane as in (\ref{axfive}). On the other hand
the  three form
fluxes run as $1/r$, i.e. similar to the scalar potential for a single charge located at $r=0$. By introducing anti charge at some location
$r\sim r_0$, we can get the total three form flux to behave like $1/r^2$ for large $r\gg r_0$ i.e. similar to scalar potential of a dipole.
This would result in total NS-NS two form $B_2\sim 1/r$ for $r\gg r_0$. We can also arrange 7 branes in such a way that F theory gives  
$\tau\sim {\cal
O}(1/r^n)$ for large $r\gg r_0$ with $n>0$. Details of the 7 brane embedding and the global behavior of $\tau$ will be discussed in section
{\bf 2.4.3}. The anti-charges we introduce are anti five branes in the neighborhood of $r\sim r_0$ . Of course anti branes want to slide down
to the tip of the conifold and we need to introduce sufficient flux to keep them at $r\sim r_0$. The resulting brane setup is depicted in 
Fig {\bf 2.16}.           
\begin{figure}[htb]\label{LCbranesetup}
		\begin{center}
\includegraphics[height=7cm]{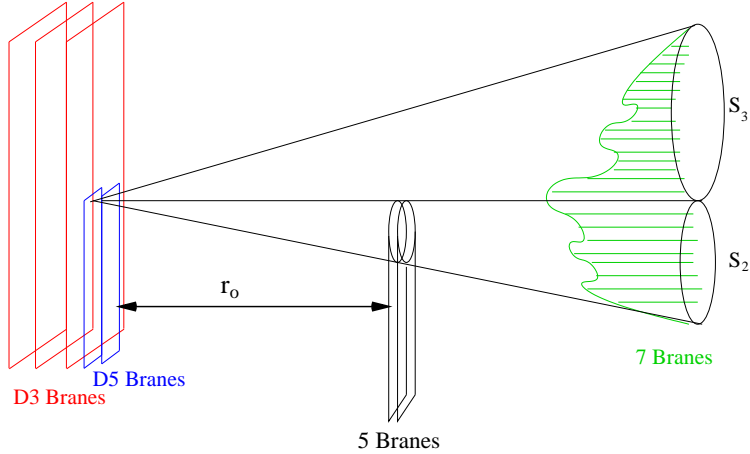}
		\caption{{Brane configuration for asymptotic conformal gauge theory.}}
		\end{center}
		\end{figure}

The dual gravity for the brane setup in Fig {\bf 2.16} considering non-zero temperature is sketched in Fig {\bf 2.17}.The geometry splits into
 three regions of interest: Regions 1, 2 and 3. The figure shows the various regions
of interest. As should be clear, most of the seven branes lie in Region 3, except for a small number of 
coincident seven branes that dip till $r_{\rm min}$ i.e Region 1. The interpolating region is Region 2.
\begin{figure}[htb]\label{sevenbraneconf4}
		\begin{center}
\includegraphics[height=6cm]{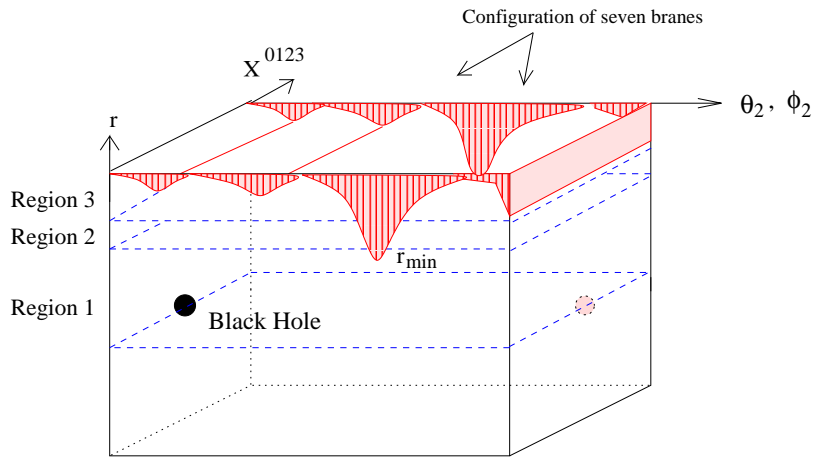}
		\caption{{Dual geometry for brane setup in Fig {\bf 2.16}}}
		\end{center}
		\end{figure} 
Region 1 is basically the one discussed in great details in \cite{FEP} and the previous sections. In this region there is one (or a coincident 
set of) seven brane(s) along with $B_2\sim {\rm log}(r)$. The logarithmic dependences of the warp factor, fluxes and 
subsequently of the gauge couplings originate from $B_2$ and these 
coincident 
(or single) seven branes . 

 The UV cap in the full F-theory framework is depicted as Region 3 in the above figure. 
In this region the seven branes are distributed so that axio-dilaton has the behavior i.e. $\tau\sim {\cal O}(1/r^n), n\ge 0$ while  
NS-NS two form $B_2$ go as ${\cal O }(1/r^m), m\ge 0$ . Using these precise forms in (\ref{twocoup}) one gets that the gauge couplings go as $g_i\sim
{\cal O}(1/r^m)={\cal O}(1/\Lambda^m)$, i.e.  we have 
asymptotic AdS space giving almost conformal field theory in the far UV. 

One big advantage 
about our UV cap is related to the issues raised in \cite{cschu}. Since the $H_{NS}$ and the 
axio-dilaton fields have  well defined 
behaviors at large $r$, there would be {\it no} UV divergences of the Wilson loops in our picture! 
Therefore our configuration can not only boast of  
holographic renormalizability, but also of the absence of Landau poles and the associated UV divergences
of the Wilson loops. We will have more to say about Wilson loops in section 3.3. 

The metric in all three regions can be written in the following form  \cite{LC}
\bg\label{bhmetko}
ds^2 = {1\over \sqrt{h}}
\Big[-g_1dt^2+dx^2+dy^2+dz^2\Big] +\sqrt{h}\Big[g_2^{-1}g_{rr}dr^2+ g_{mn}dx^m dx^n\Big]
\nd  
with $g_i$ being the Black-Hole factors and $h$ being the warp factor depends on all the internal 
coordinates ($r, \theta_i, \phi_i, \psi$).  To zeroth order in $g_sN_f$ and $g_s M$ we have our usual relations:
\bg\label{0gsnf}
h^{[0]} ~=~ {L^4\over r^4}, ~~~~g^{[0]} = 1-{r_h^4\over r^4}, ~~~~g^{[0]}_{rr} = 1, ~~~~
g^{[0]}_{mn}dx^m dx^n = ds^2_{T^{11}}
\nd
But in higher order in $g_sN_f,g_s M$, both the warp factor and the internal metric get modified because of the back-reactions
from the seven-branes, three form fluxes and the localized sources we embed. We can write this as:
\bg\label{wfmet}
h ~ = ~ h^{[0]} ~+~ h^{[1]}, ~~~~~~ g_{rr} ~=~ g_{rr}^{[0]} ~+~ g_{rr}^{[1]}, ~~~~~~ g_{mn} ~=~ g^{[0]}_{mn} ~+~ 
g^{[1]}_{mn}
\nd
where the superscripts denote the order of $g_sN_f$ and $g_sM$. Now observe that to linear order in $g_sN_f$ and $g_sM$,
$g_{rr}^{[1]}=g^{[1]}_{mn}=0$ and we recover OKS
metric (\ref{bhmet}) with warp factor given by (\ref{hvalue}) in Region 1. 
Thus only when  we include higher order terms, the  internal metric deforms from the metric of $T^{1,1}$. Whereas for a given
order in $g_sN_f,g_sM$ the warp factor $h$ behaves logarithmically for small $r$ and as $1/r^m$ for large $r$.

It is clear that one cannot jump from Region 1 to Region 3 abruptly. There should be an interpolating geometry 
where fluxes and the metric should have the necessary property of connecting the two solutions. 
This is Region 2 in our figure above.  

In the following, we briefly discuss the backgrounds for all the three regions . For a comprehensive analysis, please consult our paper
\cite{LC}. 

\subsection{Region 1: Fluxes, Metric and the Coupling Constants Flow}

 As already mentioned, the metric of the entire geometry has the form given in (\ref{bhmetko}) with warp factor $h$ 
 given by (\ref{hvalue}) in region 1.  The 
internal space retains its resolved-deformed conifold form up to ${\cal O}(g_sN_f)$. Beyond this order the 
internal space loses its simple form and becomes a complicated non-K\"ahler manifold.  
The background has {\it all} the type IIB fluxes 
switched on, with  the three-forms given by (\ref{3form}), (\ref{asymmetry}) while five-form and the axio-dilaton are 
given by (\ref{axfive}).

\subsection{Region 2: Interpolating Region and the Detailed Background}

To attach a UV cap that allows conformal  invariance i.e. vanishing beta function, we need at least a configuration of
vanishing 
NS three-form and axio-dilaton that becomes constant as $r\rightarrow \infty$.
This cannot be {\it abruptly} attached to Region 1: we need an interpolating region. This region, which we 
will call Region 2,
should have the behavior that at the outermost boundary the three-forms vanish, while solving the equations of 
motion. The innermost boundary of Region 2 $-$ that also forms the outermost boundary of Region 1 $-$ will be determined 
by the scale associated with the mass of the lightest quark, $m_0$, in our system. In Fig {\bf 2.17}, this 
is given by region in the local neighborhood of $r_{\rm min} \equiv m_0 T_0^{-1} + r_h$, where $T_0$ and $r_h$ 
are the string tension and the horizon radius respectively.  

For $B_2$  to change its form  from ${\rm log}(r)$ to behave as constant as $r\rightarrow \infty$, we will need three form flux that behaves as 
$1/r$ for small $r$ to run as $1/r^n$, $n\ge 2$ for large $r$. To get such behavior, it is useful  
 to define two functions $f(r)$ and $M(r)$ as (see Fig {\bf 2.18}):
\bg\label{mdefo}
f(r) ~ \equiv ~ {e^{\alpha(r-r_0)}\over 1 + e^{\alpha(r - r_0)}}, ~~~~~~~ M(r) ~\equiv~ M [1-f(r)], ~~~~~ \alpha \gg 1
\nd  
where  $M$ is as before related to the effective number of five-branes (or the 
RR three-form charge) and $r_0$ is the location of the sources, which we will elaborate shortly. 
Note that for $r << r_0$, $f(r) \approx e^{r-r_0}$, whereas for 
$r > r_0$, $f(r) \approx 1$. Thus for $r$ smaller than the scale $r_0$, $f(r)$ is a very small quantity; whereas for
$r$ bigger than the scale $r_0$, $f(r)$ is identity. In terms of $M(r)$
this means that for $r < r_0$, $M(r) \approx M$ whereas for 
$r > r_0$, $M(r) \to 0$.
This will be useful below. 

Using these functions, we see that one way in which logarithmic behavior  
along the radial direction can go to inverse $r$ behavior, is when the warp factor takes the following form:
\bg\label{warpy}
h ~ = ~ { c_0 + c_1 f(r) + c_2f^2(r)\over r^4} \sum_{\alpha} 
~{L_{\alpha} \over r^{\epsilon_{(\alpha)}}}  
\nd
where $c_i$ are constant numbers, and the denominator can be mapped to  
$r_{(\alpha)}$ defined in (\ref{logr}) with $\epsilon_{(\alpha)}$ functions of $g_sN_f, M, N$ and the resolution 
parameter $a$. 
$L_{\alpha}$'s are functions of the angular 
coordinates ($\theta_i, \phi_i, \psi$). For other details see \cite{FEP}. 
The warp factor $h$  
has the required logarithmic behavior as long as 
the exponents of $r$ are small and fractional numbers, and indeed switches to the inverse $r$ behavior as soon as 
the exponents become integers. The question now is what are the background sources that give rise to a warp factor of 
the form (\ref{warpy}).
All the elements of OKS model only give rise to the logarithmic warp factor and  
 it turns out, as we will discuss in some detail bellow, that we need to add sources (i.e. anti five branes) at the 
outermost boundary of region 2 along with additional background fluxes. The specific point in the radial direction 
beyond which Region 3 starts tells us exactly {\it where} to add the sources and the AdS cap. 

The demarcation point can be found by looking at the behavior of $H_{NS}$ and $H_{RR}$. For this we need to 
use the functions (\ref{mdefo}) to write the 
RR three-form. Our ansatz for ${\widetilde F}_3$ then is:
\begin{eqnarray}
&&{\widetilde F}_3  = \left({a}_o - {3 \over 2\pi r^{g_sN_f}} \right)
\sum_\alpha{2M(r)c_\alpha\over r^{\epsilon_{(\alpha)}}} 
\left({\rm sin}~\theta_1~ d\theta_1 \wedge d\phi_1- 
\sum_\alpha{f_\alpha \over r^{\epsilon_{(\alpha)}}}~{\rm sin}~\theta_2~ d\theta_2 \wedge
d\phi_2\right)\nonumber\\
&&~~ \wedge~ {e_\psi\over 2}-\sum_\alpha{3g_s M(r)N_f d_\alpha\over 4\pi r^{\epsilon_{(\alpha)}}}   
~{dr}\wedge e_\psi \wedge \left({\rm cot}~{\theta_2 \over 2}~{\rm sin}~\theta_2 ~d\phi_2 
- \sum_\alpha{g_\alpha \over r^{\epsilon_{(\alpha)}}}~ 
{\rm cot}~{\theta_1 \over 2}~{\rm sin}~\theta_1 ~d\phi_1\right)\nonumber \\
&& -\sum_\alpha{3g_s M(r) N_f e_\alpha\over 8\pi r^{\epsilon_{(\alpha)}}}
~{\rm sin}~\theta_1 ~{\rm sin}~\theta_2 \left({\rm cot}~{\theta_2 \over 2}~d\theta_1 +
\sum_\alpha{h_\alpha \over r^{\epsilon_{(\alpha)}}}~ 
{\rm cot}~{\theta_1 \over 2}~d\theta_2\right)\wedge d\phi_1 \wedge d\phi_2\label{brend}
\end{eqnarray}
where $a_o = 1 + {3\over 2\pi}$ and ($c_\alpha, ..., h_\alpha$) are constants. 
One may also notice three things: first, 
how the internal forms get deformed near the innermost boundary of the region, second, how the function $f(r)$ appears
for all the components, and finally, how
$N_f$ is not a constant but a delocalized function\footnote{We will soon see that $N_f$ in fact 
is the effective number of seven-branes.}. 
The function $f(r)$ becomes identity for 
$r > r_0$ and therefore ${\widetilde F}_3 \to 0$ for $r > r_0$. For $r < r_0$, the corrections coming from
$f(r)$ is exponentially small. Integrating ${\widetilde F}_3$ over the topologically non-trivial three-cycle:
\bg\label{3cycle} 
{1\over 2}{e_\psi} \wedge
\left({\rm sin}~\theta_1~ d\theta_1 \wedge d\phi_1- 
\sum_\alpha{f_\alpha \over r^{\epsilon_{(\alpha)}}}~{\rm sin}~\theta_2~ d\theta_2 \wedge
d\phi_2\right)
\nd
we find that the number of units of RR flux vary in the following way with respect to the radial coordinate $r$:
\bg\label{fvary}
M_{\rm tot}(r) =  M(r) \left(1 + {3\over 2\pi} - {3 \over 2\pi r^{g_sN_f}} \right)
\sum_\alpha{c_\alpha\over r^{\epsilon_{(\alpha)}}}
\nd
which is perfectly consistent with the RG flow, because for $r < r_0$, and $r \to r e^{-{2\pi\over 3 g_s M}}$,
$M_{\rm tot}$ decreases precisely as $M - N_f$ as the correction factor $e^{r-r_0}$ coming from $f(r)$ is 
negligible. For $r > r_0$, $M_{\rm tot}$ shuts off completely. This also means that below $r_0$, the total colors 
$N$ decrease by $M_{\rm tot}$ exactly as one would have expected for the RG flow with $N_f$ flavors. 
\begin{figure}[htb]\label{ffunction}
		\begin{center}
\includegraphics[height=10cm,width=8cm,angle=-90]{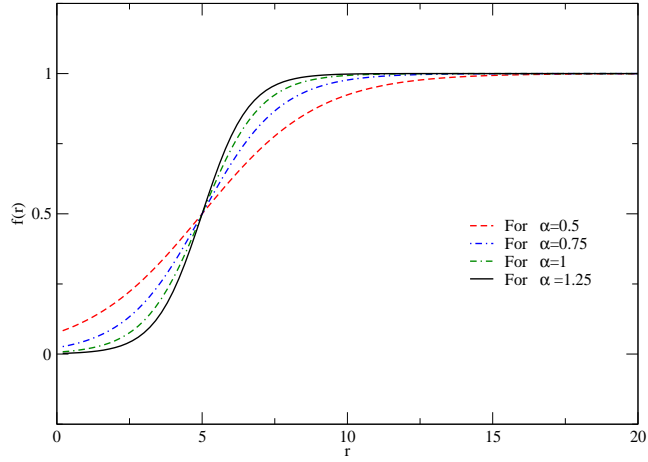}
		\caption{A plot of the $f(r)$ function for $r_0 = 5$ in appropriate units, and
various choices of $\alpha$. Observe that for large 
$\alpha$ the function quickly approaches 1 for $r > r_0$.}
	        \end{center}
		\end{figure}  
Using similar deformed internal forms, one can
also write down the ansatz for the NS three-form. The result is:
\begin{eqnarray}
&&H_3 =  \sum_\alpha {6g_s M(r) k_\alpha \over r^{\epsilon_{(\alpha)}}}\Bigg[1+\frac{1}{2\pi} - 
\frac{\left({\rm cosec}~\frac{\theta_1}{2}~{\rm cosec}~\frac{\theta_2}{2}\right)^{g_sN_f}}{2\pi r^{{9g_sN_f\over 2}}}
\Bigg]~ dr \nonumber\\
&&\wedge \frac{1}{2}\Bigg({\rm sin}~\theta_1~ d\theta_1 \wedge d\phi_1
-\sum_\alpha{p_\alpha \over r^{\epsilon_{(\alpha)}}} ~{\rm sin}~\theta_2~ d\theta_2 \wedge d\phi_2\Bigg)
+\sum_\alpha \frac{3g^2_s M(r) N_f l_\alpha}{8\pi r^{\epsilon_{(\alpha)}}}
\Bigg(\frac{dr}{r}\wedge e_\psi -\frac{1}{2}de_\psi \Bigg)\nonumber\\
&& \wedge \Bigg({\rm cot}~\frac{\theta_2}{2}~d\theta_2 
-\sum_\alpha{q_\alpha \over r^{\epsilon_{(\alpha)}}}~{\rm cot}~\frac{\theta_1}{2} 
~d\theta_1\Bigg) + 
g_s {dM(r) \over dr}
\left(b_1(r)\cot\frac{\theta_1}{2}\,d\theta_1+b_2(r)\cot\frac{\theta_2}{2}\,d\theta_2\right)\nonumber\\
&&\wedge e_\psi
\wedge dr
+{3g_s\over 4\pi} {dM(r) \over dr}\left[\left(1+g_sN_f -{1\over r^{2g_sN_f}} + {9a^2g_sN_f\over r^2}\right)
\log\left(\sin\frac{\theta_1}{2}\sin\frac{\theta_2}{2}\right) + b_3(r)\right]\nonumber\\
&& \sin\theta_1\,d\theta_1\wedge d\phi_1\wedge dr
-{g_s\over 12\pi}{dM(r) \over dr} \Bigg(2 -{36a^2g_sN_f\over r^2} + 9g_sN_f -{1\over r^{16g_sN_f}} - 
{1\over r^{2g_sN_f}} + {9a^2g_sN_f\over r^2}\Bigg)\nonumber\\ 
&& ~~~~~~~~~~~~~~~~ \sin\theta_2\,d\theta_2\wedge d\phi_2\wedge dr - 
{g_sb_4(r)\over 12\pi}{dM(r) \over dr}~ \sin\theta_2\,d\theta_2\wedge d\phi_2\wedge dr
\label{brend2}
\end{eqnarray} 
with ($k_\alpha, ..., q_\alpha$) being constants and $b_n = \sum_m {a_{nm}\over r^{m + \widetilde{\epsilon}_m}}$ 
where $a_{nm} \equiv a_{nm}(a^2, g_sN_f)$ and $\widetilde{\epsilon}_m \equiv \widetilde{\epsilon}_m(g_sN_f)$. 
The way we constructed the three-forms implies that $H_3$ is closed. In fact the 
${\cal O}(\partial f)$ terms that we added to (\ref{brend2}) ensures that. However $F_3$ is not closed. 
We can use the non-closure of $F_3$ to analyze {\it sources} that we need to add for consistency. 
These sources should in general be ($p, q$) five-branes, with ($p, q$) negative,
 so that they could influence both the 
three-forms and 
since the Imaginary Self Duality (ISD)  property of the three-forms is satisfied near $r = r_{\rm min}$ the sources should be 
close to the other boundary. A simplest choice could probably just be anti five-branes because adding anti D5-branes
would change ${\widetilde F}_3$, and to preserve the ISD condition, $H_3$ would have to change accordingly. 
Furthermore, as we mentioned before,
as $r \to r_0$, both $H_3 = {\widetilde F}_3 \to 0$. Therefore 
$r = r_0$ is where Region 2 ends and Region 3 begins, and we can put the sources there. They could be oriented along
the space time directions, located around the local neighborhood of $r = r_0$ and wrap the internal two-sphere 
($\theta_1, \phi_1$) so that they are parallel to the seven-branes.
However, putting in anti D5-branes near $r= r_0$ would imply non-trivial forces between the five-branes 
and seven-branes as well as five-branes themselves. Therefore if we keep, in general, the ($p, q$) 
five-branes close to say one
of the seven-branes then they could get {\it dissolved} in the seven-brane as electric and magnetic
gauge fluxes $\ast F^{(1)}$ and  $F^{(1)}$ respectively. Here, as before, $\ast$ is the hodge star operator and $\ast
F^{(1)}$ is the hodge dual $F^{(1)}$.  In general a configuration of $Dp$ branes is equivalent to a
configuration of $Dp+2$ brane with fluxes. This equivalence is due to M theory where a brane can be viewed as collapsed
version of another higher dimensional brane. In type IIA theory, this was first shown by  
\cite{susyrest1}\cite{susyrest2} where
a $D0$ brane is considered as a tubular $D2$ brane with electric and magnetic fields on the $D2$ brane world volume.
 Thus we may treat the five brane charges being soaked in by  
seven-branes, which in turn would mean that ${\widetilde F}_3$ in (\ref{brend}) 
and $H_3$ in (\ref{brend2}) will satisfy the following EOMs:
\bg\label{eombrend}
d{\widetilde F}_3~&=&~ F^{(1)} \wedge \Delta_2(z) - d\left({\bf Re}~\tau\right) \wedge H_3\nonumber\\
d\ast H_3 ~&=&~ \ast F^{(1)} \wedge \Delta_2(z) - d(C_4 \wedge F_3)
\nd
where the tension of the seven-brane is absorbed in $\Delta_2(z)$, which 
is the term that measures the delocalization of the seven branes (for localized 
seven branes this would be copies of
the two-dimensional delta functions) and $\tau$ is the axio-dilaton that we will determine below. In addition to 
that $d\ast F_3$ will satisfy its usual EOM. 

However the above set of equations (\ref{eombrend}) is still not the full story. Due to the anti GSO projections 
between anti-D5 and D7-brane, there should be tachyon between them. It turns out that the tachyon can be 
removed (or made massless) by switching on additional electric and magnetic fluxes on D7
along, say, ($r, \psi$)
directions! This would at least kill the instability due to the tachyon, although susy 
may not be restored. For details on the precise 
mechanism on how brane-anti brane configurations are stabilized , one may refer to \cite{susyrest1}-\cite{susyrest6}. But switching on gauge fluxes on D7 would generate 
extra D5 charges and switching on gauge fluxes on anti-D5s will generate extra D3 charges. This is one reason 
why we write ($N, N_f, M$) as effective charges. This way a stable system of anti-D5s and D7 could be constructed.  


To complete the rest of the 
story we need the axio-dilaton $\tau$ and the five-form. The five-form is easy to determine
from the warp factor $h$  using (\ref{axfive}). The total five-form charge should have contribution 
from the gauge fluxes also, which in turn would affect the warp factor.  
For regions close to $r_{\rm min}$ it is clear 
that $\tau$ goes as $z^{-g_sN_f}$ where $z$ is the embedding (\ref{seven}). More generically and for the whole of 
Region 2, looking at the warp factor and the three-form fluxes, we expect the axio-dilaton to go 
as\footnote{One may use this value of axio-dilaton and the three-form NS fluxes (\ref{brend2}) to determine the 
beta function from the relations (\ref{twocoup}). To lowest order in $g_sN_f$ we will reproduce the  beta functions
presented in section {\bf 2.2.1}. Notice that for $r > r_0$ the beta function {\it does not} vanish and both the gauge 
groups flow at the same rate. This will be crucial for our discussion in the following subsection.}:
\bg\label{axdilato}
\tau ~=~ [b_0 + b_1 f(r)]\sum_\alpha {C_\alpha\over r^{\epsilon_{(\alpha)}}}
\nd
where $b_i$ are constants and $C_\alpha$ are functions of the internal coordinates and are complex. These
$C_\alpha$ and the constants $b_i$ are determined from the dilaton equation of motion \cite{DRS, GKP}:
\bg\label{dieq}
{\widetilde\nabla}^2~\tau = {{\widetilde\nabla}\tau\cdot {\widetilde\nabla}\tau\over i{\rm Im}~\tau} - 
{4\kappa_{10}^2 ({\rm Im}~\tau)^2\over \sqrt{-g}} {\delta S_{\rm D7}\over \delta\bar\tau} + (p,q)~ 
{\rm sources}
\nd
where tilde denote the unwarped internal metric $g_{mn}$, and $S_{\rm D7}$ is the action for the {\it delocalized} 
seven branes. The $f(r)$ term in the axio-dilaton come from the ($p, q$) sources that are absorbed as gauge fluxes 
on the seven-branes\footnote{The $r^{-\epsilon_{(\alpha)}}$ behavior stems from additional anti seven-branes that we 
need to add to the existing system to allow for the required UV behavior from the F-theory completion. 
The full picture will become clearer in the next sub-section when we 
analyze the system in Region 3.}.
Because of this behavior of axio-dilaton we don't expect the unwarped metric to remain Ricci-flat 
to the lowest order in $g_sN_f$. The Ricci tensor becomes:
\bg\label{rten}
{\widetilde{\cal R}}_{mn} = \kappa^2_{10} {\partial_{(m}\partial_{n)}\tau\over 4({\rm Im}~\tau)^2} + 
\kappa_{10}^2 \left({\widetilde T}^{\rm D7}_{mn} - {1\over 8}{\widetilde g}_{mn}{\widetilde T}^{\rm D7}\right)
+ \kappa_{10}^2 \left({\widetilde T}^{(p,q)5-{\rm brane}}_{mn} - 
{1\over 4}{\widetilde g}_{mn}{\widetilde T}^{(p,q)5-{\rm brane}}\right)\nonumber\\
\nd 
where we see that ${\widetilde{\cal R}}_{rr}$ picks up terms proportional to $\epsilon^2_{(\alpha)}$ 
and derivatives of $f(r), N_f(r)$, implying that to zeroth order in $g_sN_f$ the interpolating region 
may not remain Ricci-flat. However since the coefficients 
are small, the deviation from Ricci-flatness is consequently small.

Finally the warp factor can be obtained using the five-form equation of motion:
\bg\label{5form}
d\ast d h^{-1} = H_3 \wedge {\widetilde F}_3 + \kappa_{10}^2 ~{\rm tr} 
\left(F^{(1)}\wedge F^{(1)} - {\cal R}\wedge {\cal R}\right)
\Delta_2(z) + \kappa_{10}^2 ~{\rm tr}~F^{(2)} {\widetilde \Delta}_4({\cal S})\nonumber\\
\nd
where $F^{(1)}$ is the seven-brane gauge fields that we discussed earlier, 
$F^{(2)}$ is the ($p, q$) five-brane gauge fields required for stabilization of five brane anti-five brane configuration and 
they also give proper interpretation of the colors in the 
gauge theory side\footnote{In fact one should view the gauge fluxes on the seven-branes and the five-branes as the 
total gauge fluxes that are needed to stabilize the system. We will see in the next subsection that the full stabilization
would require additional fluxes, but the structure would remain the same.}. Here 
${\cal R}$ is the pull-back of the Riemann two-form,
and ${\widetilde \Delta}_4({\cal S})$ is the term that 
measures the delocalization of the dissolved ($p, q$) five-branes over the space ${\cal S}$ embedded in the seven-brane
(again for localized five-branes there would be copies of four-dimensional
delta functions). The $H_3 \wedge {\widetilde F}_3$ term in (\ref{5form}) is proportional to ${M^2(r)\over 
r^{2\epsilon_{(\alpha)}}}$. This is precisely the form for the warp factor ansatz (\ref{warpy}) with the 
$f^2(r)$ term there accounting for the $M^2(r)$ term above. This way
with the warp factor (\ref{warpy}) and the 
three-forms (\ref{brend}) and (\ref{brend2}) we can satisfy (\ref{5form}) by switching on small gauge fluxes on the
seven-branes and five-branes. 

Therefore combining 
(\ref{warpy}), (\ref{brend}), (\ref{brend2}), (\ref{axdilato}) and the five-form, we can pretty much determine the 
supergravity background for the interpolating region $r_{\rm min} < r \le r_0$. At the outermost boundary of Region 2
we therefore
only have the metric and the axio-dilaton. Both the three-forms exponentially decay away fast, 
giving us a way to attach an
AdS cap there.    

\subsection{Region 3: Seven Branes, F-Theory and UV Completions} 

The interpolating region, Region 2, that we derived above can be interpreted alternatively as the {\it deformation} 
of the neighboring geometry once we attach an AdS cap to the OKS-BH geometry. The OKS-BH geometry is the range 
$r_h \le r \le r_{\rm min}$ and the AdS cap is the range $r > r_0$. The geometry in the 
range $r_{\rm min} \le r \le r_0$ is the deformation. Such deformations should be expected for all other UV caps 
advocated in \cite{FEP}. In this section we will complete the rest of the picture by elucidating the background 
from $r > r_0$ in the AdS cap. But before that let us give a brief gauge theory interpretation of 
background. 

For the UV region $r > r_0$ we expect the dual gauge theory to be $SU(N + M) \times SU(N + M)$ with fundamental 
flavors coming from the seven-branes. This is because addition of ($p, q$) branes at the junction, or more appropriately 
anti five-branes at the junction with gauge fluxes on its world-volume, tell us that the number of three-branes
degrees of freedom are $N + M$, with the $M$ factor coming from five-branes anti-five-branes pairs. As mentioned in the
previous section, one comes to this conclusion using M theory, as five anti-five brane configuration with flux is
 equivalent to three brane charge. 
Furthermore, the dual gauge theory for AdS conifold geometry is the Klenabov-Witten theory with gauge group
$SU(\tilde{N})\times SU(\tilde{N})$ for some $\tilde{N}$. Thus the gauge theory dual to region 3 is indeed  
$SU(N + M) \times
SU(N + M)$, but has RG flows because of the 
fundamental flavors (This RG flow is the remnant of the flow that we saw in the 
previous subsection. We will determine this in more details below).

At the scale $r = r_0$ we expect one of the gauge groups 
to be Higgsed, so that we are left with $SU(N + M) \times SU(N)$. This Higgsing is justified as follows:
 the five and anti-five branes
are separated by distance $r_0$, which means the gauge bosons that arise due to strings stretching between five and anti-five
branes have length $\sim r_0$. Thus these bosons are massive with mass ${\cal O}(r_0=\Lambda_0) $. For scales less than
$r_0=\Lambda_0$, these bosons are not produced and we only have the gauge theory arising from $D3, D5$ and seven branes
i.e. OKS-BH type theory with gauge group $SU(N + M) \times SU(N)$. Now both the gauge fields flow at different rates 
and give rise to the cascade that is slowed down  by the $N_f$ flavors. In the end, at far IR, we expect 
confinement at zero temperature.

The few tests that we did above, namely, (a) the flow of $N$ and $M$ colors, 
(b) the RG flows, (c) the decay of the 
three-forms, and (d) the behavior of the dual gravity background, all point to the gauge theory interpretation that we
gave above. What we haven't been able to demonstrate in \cite{LC} is the precise Higgsing that takes us from Klebanov-Witten
type gauge theory to Klebanov-Strassler type cascading  
picture. From the gravity side its clear how this could be interpreted. From the gauge theory side it would be 
interesting to demonstrate this. 

Coming back to the analysis of region 3, we see that 
in the region $r > r_0$ we do not expect three-forms but we do expect non-zero axio-dilaton. These non-zero axio-dilaton
come from the the seven branes that are present in region 3. Of course the complete set of seven-branes should be determined
from the F-theory picture \cite{vafaF1}\cite{vafaF2} to capture the full non-perturbative corrections. This is now subtle because the 
seven-branes are embedded non-trivially here (see (\ref{seven})). A two-dimensional base, parametrized by a 
complex coordinate $z$, 
on which we 
can have a torus fibration:
\bg\label{torus}
y^2 = x^3 + x F(z) + G(z)
\nd
can be identified with the $z$ coordinate of (\ref{seven}). This way vanishing discriminant 
$\Delta$ of (\ref{torus}) i.e $\Delta \equiv 4F^3 + 27G^2 = 0$, will specify the positions of the seven-branes exactly as
(\ref{seven}). Here we have taken $F(z)$ as a degree eight polynomial in $z$ and $G(z)$ as a degree 12 polynomial 
in $z$. 

As is well known, embedding of seven-branes in F-theory also tells us that we can have $SL(2, {\bf Z})$ jumps of 
the axio-dilaton \cite{senF}\cite{vafaF2}. We can define the axio-dilaton $\tau = C_0 + i e^{-\phi}$ as the modular parameter of a torus 
${\bf T}^2$ fibered over the base parametrized by the coordinate $z$. The holomorphic map\footnote{Holomorphic in 
$\tau$, the modular parameter.} 
from the fundamental domain
of the torus to the complex plane is given by the famous $j$-function:
\bg \label{axdil1}
j(\tau) ~\equiv ~ \frac{\left[\Theta_1^8(\tau)+\Theta_2^8(\tau)+\Theta_3^8(\tau)\right]^3}{\eta^{24}(\tau)} ~= ~
\frac{4(24{F}(z))^3}{27{G}^2(z)+4{F}^3(z)}
\nd
where $\Theta_i, i = 1, 2, 3$ are the well known Jacobi Theta-functions and $\eta$ is the Dedekind 
$\eta$-function:
\bg\label{dedekind}
\eta(\tau) ~=~ q^{1\over 24}\prod_n (1 - q^n),~~~~~~~~~ q ~= ~ e^{2\pi i \tau}
\nd
For our purpose, we can write the discriminant $\Delta(z)$ and the polynomial $F(z)$ generically as:
\bg\label{delF}
\Delta(z) ~=~ 4F^3 + 27G^2 ~=~ a \prod_{j =1}^{24} (z - {\widetilde z}_j), ~~~~~~ F(z) ~=~ b \prod_{i = 1}^8 (z - z_i)
\nd
where $\tilde{z}_j$ are (like $\mu$ in (\ref{seven})) complex constants and the $j$th seven brane is 
located at $z=\tilde{z}_j$. When we have weak type IIB coupling i.e $\tau = C_0 + i\infty$, $j(\tau) \approx e^{-2\pi i \tau}$ and 
using (\ref{axdil1})
the modular parameter can be mapped to the embedding coordinate $z$ as:
\bg\label{modmap}
\tau ~ &=& ~ {i\over g_s} ~+~
{i\over 2\pi} ~{\rm log}~(55926 ab^{-1}) - {i\over 2\pi} \sum_{n = 1}^\infty \left[{1\over nz^n} 
\left(\sum_{i=1}^8 3 z_i^n -  \sum_{j=1}^{24} {\widetilde z}_j^n\right)\right] \nonumber\\
&=&~ \sum_{n = 0}^\infty {{\cal C}_n + i{\cal D}_n\over {\widetilde r}^n}
\nd
where ${\cal C}_n \equiv {\cal C}_n(\theta_i, \phi_i, \psi)$ and ${\cal D}_n \equiv {\cal D}_n(\theta_i, \phi_i, \psi)$
are real functions and ${\widetilde r} = r^{3/2}$. To avoid cluttering
of formulae, we will use $r$ instead of ${\widetilde r}$ henceforth in this section unless mentioned otherwise. So the 
coordinate $r$ will parameterize Region 3, and $\tau = \sum {{\cal C}_n + i{\cal D}_n\over r^n}$.

The above computation was done
assuming that
$z > (z_i, {\widetilde z}_j)$, which at this stage can be guaranteed if we take $\theta_{1,2}$ small. 
This gives rise to special
set of configurations of seven-branes where they are distributed along other angular directions. 
However one might get a little worried if there exists some ${\widetilde z}_j \equiv {\widetilde z}_o$ related 
to the {\it farthest} seven-brane(s) where the above approximation fails to hold. This can potentially happen 
when we try to compute the mass of the heaviest quark in our theory. The question is whether we can still use the 
$\tau$ derived in (\ref{modmap}), or we need to modify the whole picture. 

Before we go into answering this question, the choice of $z$ bigger than ($z_i, {\widetilde z}_j$) 
already needs more
convincing
elaboration because allowing $\theta_{1,2}$ small is a rather naive argument. The situation at hand is more subtle 
than that and, as we will argue below, the picture that we have right now is incomplete.
 
To get the full picture, observe first that 
$z$ being defined by equation (\ref{seven}) means that if we want (\ref{modmap}) to hold  
in Region 3, we need to specify the condition $r > r_0$ in (\ref{seven}). This way a given $z$ will 
{always} imply points in Region 3 for varying choices of the angular coordinates ($\theta_i, \phi_i, \psi$). 
However a particular choice of ($z_i, {\widetilde z}_j$) may imply very large $r$ with small 
angular choices or small $r$ with large angular choices. Thus analyzing the system 
only in terms of the $r$ coordinate is tricky. In terms of the full complex coordinates, 
$z > (z_i, {\widetilde z}_j)$ would mean that we are always looking at 
points away from the surfaces given by $z = z_i$ and $z = {\widetilde z}_j$. 

What happens when we touch the $z = z_i$ surfaces? 
For these cases $F(z_i) \to 0$ and therefore we are no longer
in the weak coupling regime. For all $F(z_i) = 0$ imply $j(\tau) \to 0$ which in turn means 
$\tau = {\rm exp}~(i\pi/3)$ on these surfaces. 
These are the constant coupling regimes of \cite{DM1}-\cite{DM3} where the string couplings on these surfaces are 
 {\it not} weak. On the other hand, 
near any one of the seven-branes $z = {\widetilde z}_j$ we are in the 
weak coupling regimes and so (\ref{modmap}) will imply 
\bg\label{taulog}
\tau(z) = {1\over 2\pi i} ~{\rm log}~(z - {\widetilde z}_j) ~\to~i\infty
\nd
which of course is 
expected but nevertheless problematic for us. This is because we need logarithmic behavior 
of axio-dilaton in Region 2, but not in Region 3. For a good UV behavior, we need axio-dilaton to behave like 
(\ref{modmap}) everywhere in Region 3. 

In addition to that there is also the issue of the heaviest 
quarks creating additional log divergences that we mentioned earlier. These seven branes are located at 
$z = {\widetilde z}_j \equiv {\widetilde z}_o$, and therefore
if we can make the axio-dilaton independent of the coordinates
 ${\widetilde z}_o$ then at least we won't get any divergences from these seven-branes. 
It turns out that there are configurations (or rearrangements) of seven-brane(s) that allow us to do exactly 
that. To see one such configuration, let us define $F(z), G(z)$ and $\Delta(z)$ in 
(\ref{delF}) in the following way:
\bg\label{delFnow}
&&F(z) ~ = ~ (z - {\widetilde z}_o)\prod_{i = 1}^7 (z - z_i), ~~~~~~~~~
G(z) ~=~ (z- {\widetilde z}_o)^2 \prod_{i =1}^{10} 
(z - {\hat z}_i)\nonumber\\ 
&&\Delta(z) ~ = ~ (z - {\widetilde z}_o)^3 \prod_{j = 1}^{21} (z - {\widetilde z}_j) 
\nd
which means that we are stacking a bunch of {\it three} seven-branes at the point $z = {\widetilde z}_o$, and 
\bg\label{deldefn}
\prod_{j = 1}^{21} (z - {\widetilde z}_j) ~ \equiv ~ 
4\prod_{i = 1}^7 (z - z_i)^3 ~ + ~ 27 (z - {\widetilde z}_o) \prod_{i = 1}^{10} 
(z - {\hat z}_i)^2
\nd
implying that the axio-dilaton $\tau$ becomes independent of ${\widetilde z}_o$ and behaves
exactly as in (\ref{modmap}) with ($i, j$) in (\ref{modmap}) varying 
up to (7, 21) respectively. 

The situation is now getting better. We have managed to control a subset of log divergences. 
To get rid of the other set of log divergences that appear on the remaining 
twenty-one surfaces, one possible way would be  
to modify the embedding (\ref{seven}). In fact a change in the 
embedding equation will also explain the axio-dilaton choice (\ref{axdilato}) of Region 2. 
To change the embedding equation (\ref{seven}) we will 
use similar trick that we used to kill off the three-form fluxes, namely, attach anti-branes. These
anti seven-branes\footnote{They involve both local and non-local anti seven-branes.} 
are embedded via the following equation:
\bg \label{ABembedding}
r^{3/2} e^{i(\psi-\phi_1-\phi_2)}{\rm sin}~\frac{\theta_1}{2}~{\rm sin}~\frac{\theta_2}{2}~=~r_0 e^{i\Theta}
\nd
where $\Theta$ is some angular parameter, and could vary for different anti seven-branes. The above embedding 
will imply that their overlaps with 
the corresponding seven-branes are only partial\footnote{For example if we have a seven-brane at $z = {\widetilde z}_1$ 
such that lowest point of the seven brane is $r = \vert {\widetilde z}_1\vert^{2/3} < r_o$, 
then the corresponding anti-brane 
has only partial overlap with this.}. And since we require
$$ \vert{\widetilde z}_j\vert^{2/3} ~ < ~ r_o$$
it will appear effectively that we can only have seven-branes in 
Regions 1 and 2, and {\it bound} states of seven-branes and anti seven-branes in Region 3.\footnote{Of course this 
effective description is only in terms of the axio-dilaton charges. In terms of the embedding equation for the 
seven-branes (\ref{seven}) this would imply that we can define 
$z$ with $r > r_o$ and ${\widetilde z}_j$ with $r < r_o$.}
This way 
the axio-dilaton in Region 3 will indeed behave as (\ref{modmap}), with  the seven values of 
$z_i$ in (\ref{deldefn}) chosen to be in region 1 and 2. 

There are two loose ends that we need to tie up to complete this side of the story. The first one is the issue of 
Gauss' law, or more appropriately, charge conservation. The original configuration of 
24 seven branes had zero global charge, but now with  
the addition of anti seven-branes charge conservation seems to be problematic. There are a few ways to resolve this issue.
First, we can assume that that branes wrap topologically trivial cycles, much like the ones of \cite{4}. Then 
charge conservation is automatic. The second alternative is to isolate six seven-branes using some 
appropriate $F$ and $G$ functions, so that they are charge neutral. This is of course one part of the constant 
coupling scenario of \cite{senF}. Now if we make the ($\theta_2, \phi_2$) directions non-compact then we can 
put in a configuration of 18 seven-branes and anti seven-branes pairs together using the embeddings (\ref{seven}) and 
(\ref{ABembedding}) respectively. The system would look effectively like what we discussed above. Since the whole 
system is now charge neutral, compactification shouldn't be an issue here. 

The second loose end is the issue of tachyons between the seven-brane and anti seven-brane pairs. Again, as for the 
anti-D5 branes and D7-brane case \cite{susyrest1}-\cite{susyrest6}, 
switching on appropriate electric and magnetic fluxes will make the tachyon massless! 
Therefore the system will be stable and would behave exactly as we wanted, namely, the axio-dilaton will not have the 
log divergences over any slices in Region 3.  

This behavior of axio-dilaton justifies the $r^{-\epsilon_{(\alpha)}}$ in (\ref{axdilato}) 
in Region 2. So the 
full picture would be a set of seven-branes with electric and magnetic fluxes embedded via (\ref{seven})
and another set of anti seven-branes embedded via (\ref{ABembedding}) lying completely in Region 3.
 
Thus in Region 3 both the three-forms vanish and therefore $g_1 = g_2 = g_{\rm YM}$ with $g_1, g_2$ being the 
couplings for $SU(N+M), SU(N+M)$. From (\ref{twocoup}) we can compute the $\beta$-function for $g_{\rm YM}$ as:
\bg\label{betym}
\beta(g_{\rm YM}) ~\equiv~ {\partial g_{\rm YM}\over \partial {\rm log}~\Lambda} 
~ = ~ {g^3_{\rm YM}\over 16 \pi}~ \sum_{n = 1}^\infty ~ 
{n{\cal D}_n \over \Lambda^n}
\nd
where $\Lambda$ is the usual RG scale related to the radial coordinate in the supergravity approximation. For
$\Lambda \to \infty$, $\beta(g_{\rm YM}) \to 0$ implying a conformal theory in the far UV. We can fix the 
't Hooft coupling to 
be strong to allow for the supergravity approximation to hold consistently at least for all points away from the 
$z = z_i, i= 1,..., 7$ surfaces.

Existence of axio-dilaton $\tau$ of the form (\ref{modmap}) and the seven-brane sources will tell us, from (\ref{rten}), 
that the unwarped metric may not remain Ricci flat. For example it is easy to see that 
\bg\label{rrr}
\widetilde{\cal R}_{rr} = {{\cal A}_{\cal D}\over r^2{\cal D}_0^2} \sum_{n,m =1}^{\infty} nm 
{({\cal C}_n + i{\cal D}_n)({\cal C}_m - i{\cal D}_m)\over r^{n+m}} 
+ {\cal O}\left({1\over r^n}\right)
\nd
where the last term should come from the seven-brane sources and, because of these sources, we don't expect 
$\widetilde{\cal R}_{rr}$ to vanish to lowest order in $g_sN_f$.\footnote{Although, as discussed before, the deviation
from Ricci flatness will be very small.}
 The term ${\cal A}_{\cal D}$ is given by the 
following infinite series:
\bg\label{adddef}
{\cal A}_{\cal D} ~ = ~ 1-\sum_{k,l=1}^\infty {{\cal D}_k {\cal D}_l
{\cal D}_0^{-2}\over r^{k+l}} 
+ \sum_{k,l,p,q=1}^\infty {{\cal D}_k{\cal D}_l{\cal D}_p{\cal D}_q {\cal D}_0^{-2}\over r^{k+l+p+q}} + ...
\nd
Similarly one can show that 
\bg\label{rab}
\widetilde{\cal R}_{ab} = {{\cal A}_{\cal D}\over {\cal D}_0^2}
\sum_{n,m=0}^\infty {(\partial_a{\cal C}_n + i\partial_a{\cal D}_n)
(\partial_b{\cal C}_m - i\partial_b{\cal D}_m)\over r^{n+m}}
+ {\cal O}\left({1\over r^n}\right)
\nd
for ($a, b$) $\ne r$. For $\widetilde{\cal R}_{rb}$ similar inverse $r$ dependence can be worked out. In the far UV 
we expect the 
unwarped curvatures should be equal to the AdS curvatures. The warp factor $h$ on the other hand can be determined 
from the following variant of (\ref{5form}):
\bg\label{wfac}
d\ast d h^{-1} = \kappa_{10}^2 ~{\rm tr} 
\left(F^{(1)}\wedge F^{(1)} - {\cal R}\wedge {\cal R}\right)
\Delta_2(z) + ...
\nd
because we expect no non-zero three-forms in Region 3. The dotted terms are the non-abelian corrections from the 
seven-branes. As $r$ is increased i.e $r >> r_0$, 
we expect $F^{(1)}$ to fall-off (recall that they appear from the anti (1,1) five-branes located in the neighborhood of 
$r = r_0$) and therefore can be absorbed in ${\cal R}$. 
Once we embed the seven-brane gauge connection in some part of spin-connection, we expect 
\bg\label{boxh}
\Box~h^{-1} ~ = ~ {\cal O}\left({1\over r^n}\right)
\nd
where the box represents combinations of differential operators that arise from (\ref{wfac}). Solving this will reproduce the generic form for $h$:
\bg\label{heaft}
h ~= ~ \frac{L^4}{r^4}\left[1+\sum_{i=1}^\infty\frac{a_i(\psi,\theta_i, \phi_i)}{r^i}\right]
\nd
with a constant $L^4$ and $a_i$'s are suppressed by powers of $g_sN_f$. 
More details on this is given in the Appendix A and {\bf B} of \cite{LC}. At far UV we recover the AdS picture 
implying a strongly coupled conformal behavior in the dual gauge theory.

To summarize, we have obtained the dual gravity of a thermal field theory with matter in the fundamental representation . The gauge theory becomes
almost conformal in the UV with massive Higgs-like gauge bosons with  matter transforming under $SU(N+M)\times SU(N+M)$ gauge symmetry.  
In the IR the massive gauge bosons are Higgsed away and we end up with a $SU(N+M)\times SU(N)$ gauge theory. The dual
geometry has the metric of the form (\ref{bhmetko}) with a warp factor given by (\ref{hvalue}) for $r\ll r_0$, by
(\ref{warpy}) for $r\sim r_0$, and of the form (\ref{heaft}) for $r\gg r_0$.   
\begin{figure}[htb]\label{LCbranesetup}
		\begin{center}
\includegraphics[height=10cm, width=9cm]{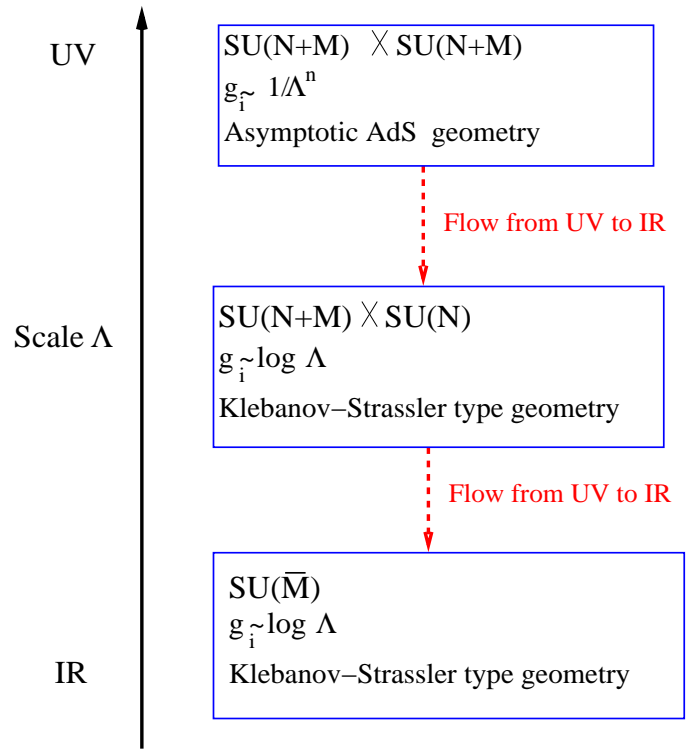}
		\caption{{Schematic depiction of flow in the gauge theory}}
		\end{center}
		\end{figure}  
Thus in UV we have
Klebanov-Witten type field theory while in the IR we have Klebanov-Strassler model with fundamental matter and temperature. Once we reach the
Klebanov-Strassler type description of our field theory, we expect Seiberg duality cascade to occur as we go further down in energy scale. At
the bottom of the cascade we end up with strongly coupled $SU(\bar{M})$ gauge theory with $N_f$ fundamental flavors and logarithmic running of
coupling. The scenario is sketched in Fig {\bf 2.19}. 

\chapter{Application to Thermal QCD}
Having discussed in some detail the gauge theories that arise from various brane setups and their dual geometries, now we attempt to make
connections with thermal QCD in the large N limit. The gauge theory arising from  the brane setup  in section {\bf 2.4} 
is obtained by UV completing OKS-BH geometry and  is
asymptotically conformal with logarithmic running of couplings in the IR. To our knowledge it is the brane setup that resembles large N QCD
the most in the `top down' approach where gauge/gravity duality is exactly derivable. Unlike the `bottom up' approach where we do not know the
exact brane configuration in higher dimensions or the precise origin of the gauge theory, our approach is based on open/closed string duality
and in some sense the correspondence is more theoretically complete. 

As sketched in Fig {\bf 2.19}, although the gauge theory has a very rich structure, it is of course not exactly QCD. However, the benefit of this 
construction is
at large coupling it has an exact classical supergravity description. Certain gauge theory observables which are extremely difficult  
to obtain using conventional field theory techniques become rather simple to calculate using dual gravity. The  field theory lives in  four 
flat spatial dimensions and incorporates matter in fundamental representation. We will refer to the fundamental matter as `quarks' of our
theory. The theory has no UV Landau poles while effective degrees of freedom converge in the far UV. As the theory reaches
conformal fixed point at large energy scales, our construction maybe viewed as the UV completion of Klebanov-Strassler model which is most
relevant for QCD.
      
Although the field content  is somewhat different from large N QCD, our theory has several common features to it and 
in fact
becomes almost large N QCD in the IR. Thus the physics extracted from our gauge theory could be very similar to that 
of QCD and it is worth
analyzing the thermodynamic properties of the field theory plasma. As this may turn out to be crucial in understanding 
the collective excitations
of QCD plasma, in this chapter we will examine the dynamics of `quarks', transport coefficients of the medium and the 
confining nature of
the our gauge theory.

\section{Quark Dynamics}
Strings in ten dimensional dual gravity with endpoint on a D7 brane corresponds to fundamental matter i.e. quarks in four dimensional
Minkowski space time. As the D7 brane fills the four dimensional Minkowski space,  the endpoint of the string  is a point there and the quark
is localized at the endpoint. Energy of the string gives energy of the quark and minimal energy of the static string configuration in the bulk
geometry gives the mass of the quark. If we consider a string traveling through the bulk geometry, it  corresponds to a 
quark moving through gauge theory plasma.
In the following subsections we will use this key concept to compute thermal mass of a quark and the drag it experiences 
as it traverses a medium.
We will also quantify the wake it leaves behind in the medium along with transverse momentum broadening of a fast moving quark.
  
\subsection{Thermal Mass and Drag of a Quark}
We start with the action of a string in ten dimensional geometry with metric (\ref{bhmet}). If $X^{i}(\sigma,\tau)$ is a map from world sheet coordinates $\sigma,\tau$ to 10 dimensional space time, 
then string action or fundamental string Born Infeld action is (see for example \cite{leigh1}\cite{leigh2}):
\bg \label{KS3}
&& S_{\rm string}= T_0\int d\sigma d\tau \Big[\sqrt{-{\rm det}(f_{\alpha\beta} + \partial_\alpha\phi \partial_\beta\phi)} 
+ {1\over 2} \epsilon^{ab} B_{ab}
+ J(\phi)\nonumber\\
 && ~~~~~~~~+ \partial X^m \partial X^n ~\bar\Theta~ \Gamma_m \Gamma^{abc....} \Gamma_n ~\Theta
~F_{abc....} + {\cal O}(\Theta^4)\Big] =\int d^{10}x ~ {\cal L}_{\rm string}(x)\nonumber
\nd
where $J(\phi)$ is the additional coupling of the dilaton $\phi$ to the string world-sheet, $T_0$ is the string tension, 
$X^n$ are the ten bosonic coordinates, $\Theta$ is a 32 component spinor, $F_{abc...} = [dC]_{abc..}$ with 
$C_{abc..}$ being the background RR form potentials,
and
$f_{\gamma\delta}$ is world sheet metric, given by the standard pull-back of the space time metric on the 
world-sheet:

\begin{displaymath}\label{metricmatrix}
f=\left(\begin{array}{cc}
 \dot{X}\cdot \dot{X} ~~&  \dot{X} \cdot X' \\  
 \dot{X} \cdot X'~~ & X'\cdot X'
\end{array} \right) =\left(\begin{array}{cc}
 \frac{\dot{X}^2}{\sqrt{h}} - \frac{g_1}{\sqrt{h}} & \frac{ \dot{X} X'}{\sqrt{h}}\\
\frac{\dot{X} X'}{\sqrt{h}} & \frac{\sqrt{h}}{g_2} + \frac{X'^2}{\sqrt{h}}  
\end{array} \right)
\end{displaymath}
where we have taken the background metric of the form (\ref{bhmetko}) ignoring $g_s^2N_f^2$ terms which means $g_{mn}$
is the metric of $T^{1,1}$ in (\ref{bhmetko}). Here $\gamma,\delta=0,1$ with parametrization $\eta^0=\tau=t$ and
$\eta^1=\sigma=r$.
  
In the ensuing analysis we will keep $B_{ab}$, $J(\phi)$ as well as $\partial_a\phi$
zero and the justification will be given shortly. The interesting thing however 
is to do with the background RR forms. Note that the RR forms {\it always} couple to the 32 component spinor. Therefore
once we switch-off the fermionic parts in (\ref{KS3}), the fundamental string is completely unaffected by the background 
RR forms\footnote{This is of course the familiar statement that the RR fields do not couple 
in a simple way to the fundamental string.}. Thus in the following analysis, for the mass and drag of the quark, we can 
safely ignore the RR fields.    
We have also defined:
\bg \label{KS4}
&&{\rm det}~f=-G_{ij}\dot{X}^{i}X'^{j}+(G_{ij}X'^{i}X'^{j})(G_{kl}\dot{X}^{k}\dot{X}^{l})\nonumber\\  
&& X'^{i}=\frac{\partial X^i}{\partial \sigma}, ~~~~~~
\dot{X}^{i}=\frac{\partial X^i}{\partial \tau}  
\nd
where $G_{ij}$ is more generic than the background metric, and could involve the back reaction of the 
fundamental string on the geometry. The analysis is very similar to the AdS case discussed in \cite{HK1}\cite{HK2}, however, 
since our background geometry in \cite{FEP} involves running couplings, the results will 
differ from the ones of \cite{HK1}\cite{HK2}. 

As mentioned earlier, a fundamental quark will be a string starting from  the D7 brane and ending on a D3 or fractional D3
(which is a D5 brane)
brane, giving the quark color. At non zero temperature, the dual geometry has a black hole and  quark in the
thermal medium is represented by a string stretching between seven brane  and  the horizon $r_h$ 
of the black hole.   
For simplicity of the calculation, we will then restrict to the case when 
\bg\label{lcone} \nonumber
&&X^0=t,~~X^4=r,~~X^1=x(\sigma,\tau),~~X^k=0~(k=2,3,5,6,7)\nonumber\\
&& (X^8, X^9) = (\theta_1, \theta_2) =\pi, ~~~~ \Theta = \bar\Theta = 0 
\nd
and we choose parametrization $\tau=t,\sigma=r$ also known as the static gauge. Thus
we are only considering the case when the string extends in the $r$ direction, does not interact with the RR fields, 
and moves in the $x$ direction of 
our
manifold. More general string profile, while being 
computationally challenging, does not introduce any new physics and hence
our simplification is a reasonable one.

Before moving further, let us consider two points. First is the effect of the black hole on the {\it shape} 
of the D7 brane. We expect due to gravitational effects the D7 brane will sag towards the black hole and eventually the 
string would come very close to the horizon. In fact putting a point charge on the D-brane tends
to create a long thin tube on the D-brane that in general extends to infinity. The end point of the string being a 
source of point charge should show similar effects (see \cite{anirban1} for a discussion of a somewhat similar 
scenario)\footnote{Even in the supersymmetric case, putting a point charge on a D-brane tends to create a long 
tube that extends to infinity. One can then view an open string to lie at the end of the thin tube. This effect 
is somewhat similar to the one discussed in \cite{calmal}.}. 
For our analysis here we will ignore this effect altogether and we hope to address the issue in our future work. 

The second point is to see how the background varying dilaton and NS-NS two form effect the string. With the axion and
dilaton behaving as (\ref{axfive}) near the 
location of D7 brane and as ${\cal O}(g_s N_f/r^n)$ globally, $\partial_\alpha \phi \sim {\cal O}( g_s N_f/r^m)$ for all $r$
at the location of the string. 
Note that $g_s N_f\ll 1$ and as the string extends from $r=r_0$ to $r=r_h$, for
$r_h$ large i.e. high temperature, ${\cal O}(g_s N_f /r^m)$ is negligible at the location of the string. This means we can
ignore
the contribution from $\partial_\alpha \phi$ in the string action. Similarly we can ignore effects of $B_2$ as it gives 
${\cal O}(g_s M{\rm log}r )$ for 
small $r$ and ${\cal
O}(g_s M/r^m)$ for large $r$, both of which can be ignored at the location of the string. This is because for reasonably large temperature, the IR
divergence of ${\rm log}(r)$ is absent (temperature which is of ${\cal O}(r_h)$ is the IR cutoff) and at the UV 
we can ignore $1/r^n$ terms. On the other hand, the dilaton additionally couples to the string world sheet through 
$J(\phi)$
term which is proportional to the Ricci scalar $R_{(2)}$ of the world sheet metric, $f_{\gamma \delta}$ and under a reparametrization of the
world sheet metric, we can make $R_{(2)}$ small enough as there are no divergences in the background metric.
 This way we can ignore $J(\phi)$ near the location of the string.
However in section {\bf 3.3} we will consider the effects of $B_2$ and dilaton on the string world sheet 
and we will see that these effects
although rather technical to track, do not change the overall physics. 
Thus we can simply set $J(\phi)=B_2=\partial_\mu \phi=0$ in the following analysis.     

Getting back to the string action, with
our choice of parametrization and string profile we get: 
\bg \label{KS5}
-{\rm det}~f=\frac{g_1(r)}{g_2(r)}+\frac{g_1(r)}{h(r,\pi,\pi)}x'^2-g_1(r)^{-1}\dot{x}^2
\nd
where the warp factor $h(r, \theta_1, \theta_2) = h(r, \pi, \pi)$ , evaluated at the location of the string.  With this, the rest of the analysis is a straightforward extension of \cite{HK1}\cite{HK2}. 
The Euler-Lagrangian equation for $X^1=x(t,r)$ derived from the action (\ref{KS3}) and the associated canonical 
momenta are:
\begin{eqnarray} \label{KS6} \nonumber 
&& \frac{1}{g_2}\frac{d}{dt}\Big(\frac{\dot{x}}{\sqrt{-{\rm det}~f}}\Big)+\frac{d}{dr}\Big(\frac{g_1x'}
{h\sqrt{-{\rm det}~f}}\Big)=0 
\\\nonumber
&& \Pi_i^0= -T_0 G_{ij}\frac{(\dot{X}\cdot X')(X^{j})'-(X')^2(\dot{X}^j)}{\sqrt{-{\rm det}~f}}\nonumber\\
&& \Pi_i^1 = -T_0 G_{ij}\frac{(\dot{X}\cdot X')(\dot{X}^{j})-(\dot{X})^2(X^j)'}{\sqrt{-{\rm det}~f}}
\end{eqnarray}
If we consider a static string configuration, i.e. $x(\sigma,\tau)=b={\rm constant}$, then energy can be interpreted as
the thermal mass of the quark in the dual gauge theory. Using the static solution in (\ref{KS6}), we obtain the thermal mass $m({\cal T})$
using $E = -\int d\sigma ~\Pi^0_t$, as
\bg \label{KS15b}
m({\cal T})=T_0(r_0-r_h)=T_0\left(\vert\mu\vert^{2/3}- {\cal T}\right) 
\nd 

Observe that the thermal mass decreases as temperature is increased. In our analysis, $\mu^{2/3}=r_0>{\cal T}$ and
$\mu^{2/3}$ is proportional to zero temperature mass. This means, we are dealing with heavy quarks where the temperature is
always less than the mass of the quark.On the other hand, our predictions for the thermal mass gets exact in the limit where number of
colors is infinite. Thus our results suggest that at large N and small temperatures, the heaviest quarks get less massive as
temperature is increased.

When a probe particle moves through a plasma, it interacts with the medium through collisions with the constituents of the medium and if it is
charged, it also radiates. Overall because of the interaction with the medium, the probe experiences drag force and this drag is a key 
characteristic of the plasma. As a moving string in the bulk corresponds to a moving quark in the medium, we can compute the drag 
coefficient
by considering a string moving with some velocity with the endpoint representing a quark which traverses the medium.  For the computation of 
drag, that is to analyze the response of the medium to a moving probe, it is enough to
consider a constant velocity quark and calculate the drag it experiences. 
Of course drag will try to slow down the quark 
and we need to apply force to 
keep it at constant speed. If there is no external force applied, the probe will slow down but the drag coefficient we extract considering
constant speed remains unchanged - as we can consider instantaneous velocity to be the constant velocity and the analysis stays the same. 

We consider the string profile (\ref{lcone}) with    
\bg \label{KS10}
x(t,r)=\bar{x}(r)+vt
\nd
 
Then from (\ref{KS6}), noting that $f$ is independent of time, we can solve the equation of motion to get: 
\bg \label{KS12}
\bar{x}'^2= \frac{h^2C^2v^2}{g_1g_2}\cdot \frac{g_1-{v^2}}{g_1-{h C^2v^2}}
\nd
where $C$ is a constant of integration that can be determined by demanding that 
$-{\rm det}~f$ is always positive. Using the value of $\bar{x}'^2$ from (\ref{KS12}) we can give an explicit expression
for the determinant of $f$ as:
\bg \label{KS13}
-{\rm det}~f=\frac{g_1}{g_2}\cdot \frac{g_1-v^2}{g_1-h C^2v^2}
\nd
For $-{\rm det}~f$ to remain positive for all $r$, we need both 
numerator and denominator to change sign at same value of $r$. This is the same argument as in \cite{HK1} \cite{HK2}. 
The
numerator changes sign at\footnote{Note that by ${\cal O}(g_sN_f,g_sM)$ we will always mean 
${\cal O}(g_sN_f,g^2_sMN_f, g_sM^2/N)$ unless mentioned otherwise.}  
\bg \label{cvaluea}
r^2=\frac{r_h^2}{\sqrt{1-v^2}} + {\cal O}(g_sN_f,g_sM)
\nd
where we use $h$ near horizon to be of the form in (\ref{hvalue}). \footnote{This is justified as for all the geometries considered in section {\bf
2.2-2.4}, the near horizon $r\sim r_h$ warp factor is always of the form (\ref{hvalue}). We have various choices for the large $r$ behavior of the
geometry and subsequently for $h$, but the IR behavior remains the same as the OKS model.} Requiring that denominator  also change sign at that value fixes $C$ to be:
\bg \label{cvalue}
C =\frac{r_h^2 L^{-2}}{\sqrt{1-v^2}}\cdot
\frac{1}{\sqrt{1+\frac{3 g_s\bar{M}^2}{2\pi N}~{\rm log}\Big[\frac{r_h}{(1-v^2)^{{1}/{4}}}\Big]
\Big(1+\frac{3 g_s\bar{N}_f}{2\pi}~\Big\{{\rm
log}~\Big[\frac{r_h}{(1-v^2)^{{1}/{4}}}\Big]+\frac{1}{2}\Big\}\Big)}}
\nd
where $\bar{M}$ and $\bar{N}_f$ differs from $M,N_f$ due to the ${\cal O}(g_sN_f,g_sM)$ terms in (\ref{cvaluea}). 
The first part of $C$ is the one derived in \cite{HK1} \cite{HK2}. The next part is new. 
Now the rate at which momentum is lost to the black hole is given by the momentum density at horizon
\bg \label{KS15a}
\Pi_1^x(r=r_h)= - T_0 Cv
\nd
while the force  quark experiences due to friction with the  plasma is $\frac{dp}{dt}=-\nu p$ with 
$p=mv/\sqrt{1-v^2}$. To keep the quark moving at
constant velocity, an external field ${\cal E}_i$ does work and the equivalent energy is dumped into the 
medium \cite{HK1}\cite{HK2}. Thus 
the rate at which a quark dumps energy and momentum into the thermal medium is 
precisely the rate at which the string loses energy and momentum to the black hole. Thus up to 
${\cal O}(g_sN_f, g_s M)$ we have ${\nu m
v \over \sqrt{1-v^2}} = - \Pi_1^x(r=r_h)$ and 
\bg \label{KS15b} 
\nu &= & \frac{T_0 C \sqrt{1-v^2}}{m}\\
&=&\frac{T_0}{mL^2}\frac{{\cal T}^2} 
{\sqrt{1+\frac{3 g_s\bar{M}^2}{2\pi N}~{\rm log}\Big[\frac{{\cal T}}{(1-v^2)^{{1}/{4}}}\Big]
\Big(1+\frac{3 g_s\bar{N}_f}{2\pi}~\Big\{{\rm
log}~\Big[\frac{{\cal T}}{(1-v^2)^{{1}/{4}}}\Big]+\frac{1}{2}\Big\}\Big)}}\nonumber
\nd
which should now be compared with the AdS result \cite{HK1}\cite{HK2}.
In the AdS case the drag coefficient $\nu$ is 
proportional to ${\cal T}^2$. For our case, when we incorporate RG flow in the gravity dual, 
we obtain ${\cal O}\left(1/\sqrt{A ~{\rm log}~{\cal T}+B ~{\rm log}^2 {\cal T}}\right)$ 
correction to the drag coefficient computed
 using AdS/CFT correspondence \cite{HK1} \cite{HK2}.

\subsection{Transverse Momentum Broadening}
With the computation of thermal mass of the quark and the drag it experiences in the medium, we will now study the
diffusion process through which a probe particle transfers momentum with the plasma. The analysis is of particular interest
as it relates to the formation of the quark gluon plasma at the relativistic heavy ion colliders. To be more precise, at the
earliest stages of a central collision, energy densities are expected to be high enough to form the
quark gluon plasma and several observables have been proposed to probe the plasma. Strangeness enhancement \cite{senhance},
$J/\psi$ suppression \cite{Matsui-Satz-1}, electromagnetic radiation \cite{Gale:2001yh} are among the key candidates while the ones that
are produced at the earliest stages of the collision gather special attention as they directly interact with the plasma.
Hard scatterings where the partons transfer large momentum take place right after the collision as only
then there is enough energy and the resulting partons fragment into jets of hadrons with large transverse momentum $P_{T}$.
 The partons produced in the heavy ion collisions are expected to lose
energy in the medium \cite{jq2} which should result in a suppression of high $P_T$ hadrons \cite{jq3} when compared to the high $P_T$ 
hadrons produced in proton-proton collision. Experiments at RHIC have indeed observed this suppression
\cite{Adcox:2001jp}-\cite{Gyulassy:1993hr} and this phenomenon is known as  `jet quenching'. 

There has been a lot of effort  to model the energy loss mechanism \cite{jqb}-\cite{jqe} and gluon bremsstrahlung with Landau-Pomeranchuk-Migdal
(LPM) effect \cite{LPM} is thought to be the dominant process in jet quenching. As the medium formed in the collison
expands, one needs to model the time evolution of the system accounting the dynamics of the fluid and then compare with
experimental data of suppression. As the dual geometry we constructed is time independent, we cannot address the effects
due to the evolution of medium. Rather we will quantify the diffusion process by computing the mean square transverse
momentum transfer between the {\it non expanding} medium and a fast moving parton. In principle, one needs to construct a 
time dependent dual 
geometry by considering the collision of strings ending on D7 branes and computing their back reaction, similar
 in the spirit of \cite{Shuryak:2005ia}. Then using this time dependent background geometry to compute Wilson loops, 
one can calculate 
momentum distribution and finally quantify `jet quenching'. Constructing the dual geometry of a heavy ion collision is
rather challenging and we hope to address the issue in our future work.             

For our current analysis, consider a parton moving through a plasma in four dimensional Minkowski space time with the 
following world line

\bg \label{wl1}
x(t)&=&vt\nonumber\\
z(t)&=&y(t)\equiv \delta y(t)
\nd  

For a fast moving parton, we can always choose coordinates such that (\ref{wl1}) is the world line. Now if the plasma has
matter in fundamental representation, has logarithmic running of coupling in the IR but becomes asymptotically conformal, 
we can treat the ten dimensional geometry with metric of the form (\ref{bhmet2}) to be the
dual gravity of this gauge theory which lives in four dimensional flat space time.  

To obtain the momentum broadening of the
parton  we shall use the Wigner distribution function $f$ as defined in QCD kinetic theory \cite{KT}
\bg
f(X,r_\perp)&\equiv&<f_{cc}(X,r_\perp)>=\rm{Tr}\left[\rho Q_a^{\dagger}(X_-^\perp)U_{ab}(X_-^\perp,X_+^\perp)Q_a(X_+^\perp)\right]
\nd
where $X_-^\perp=X-r_\perp/2$, $X_+^\perp=X+r_\perp/2$, 
$X=(t,\bf{x})$ is the world line of the parton field $Q_a$ without any fluctuation, $U_{ab}$ is the link and $\rho$ is
the density matrix. Of course the index $ab$ refers to color and for our choice of the world line for the parton we have 
$r_\perp^2=2\delta y^2$. Now Fourier transforming the distribution functions, we can get  the average 
transverse momentum to be
\bg
<p^2_\perp>=\int d^3x\int\frac{d^2k_\perp}{(2\pi)^2}k^2_\perp f(X, k_\perp)
\nd
\vspace{1cm}
\begin{figure}[ht]
\begin{center}
\includegraphics[height=2cm, width=\textwidth]{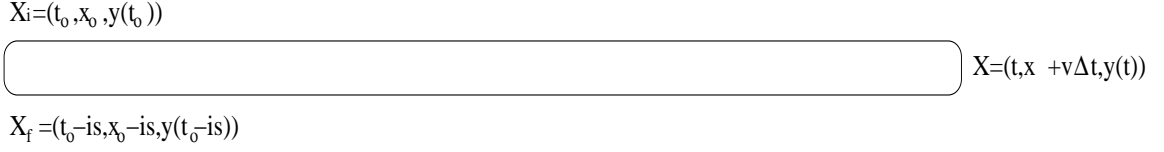}
\caption{{The contour C with $X$ denoting the coordinates of a point on the contour. The upper line with real time coordinate is the
world line of the heavy parton while the lower line has complex time coordinate.
 Here $s\rightarrow 0$ is real.}}
\label{charm_xsec}
\end{center}
\end{figure}

Now if we assume that the initial transverse momentum distribution is narrow, then we
have 
\bg
<p^2_\perp>&=&2\kappa_T {\cal T}\nonumber\\
\nd
with ${\cal T}$ some large time
interval, and $\kappa_T$ is the diffusion coefficients.

Following the arguments in \cite{DTCS},  one can write the diffusion coefficients solely in terms of functional derivative of
Wilson loops.  The final result is 
\bg
\kappa_T&=&\lim_{\omega\rightarrow 0}\frac{1}{4}
\int dt \;e^{i\omega t}\;(iG_{11}^y(t, 0)+iG_{22}^y(t, 0)+iG_{12}^y(t, 0)+iG_{21}^y(t, 0))
\nd
 where the Greens functions are 
\bg \label{prop}
G_{11}^y(t,t')&=& \frac{1}{\rm{tr}\rho^0 W_C[0,0]}\langle \rm{tr}\rho^0 \frac{\delta^2 W_C[\delta y_1,0]}{\delta
y_1(t)\delta
y_1(t')}\rangle\nonumber\\
G_{22}^y(t,t')&=& \frac{1}{\rm{tr}\rho^0 W_C[0,0]}\langle \rm{tr}\rho^0 
\frac{\delta^2 W_C[0,\delta y_2]}{\delta y_2(t)\delta
y_2(t')}\rangle\nonumber\\
G_{12}^y(t,t')&=& \frac{1}{\rm{tr}\rho^0 W_C[0,0]}\langle \rm{tr}\rho^0 \frac{\delta^2 W_C[\delta y_1,\delta y_2]}{\delta
y_1(t)\delta
y_2(t')}\rangle\nonumber\\
G_{21}^y(t,t')&=& \frac{1}{\rm{tr}\rho^0 W_C[0,0]}\langle \rm{tr}\rho^0 \frac{\delta^2
W_C[y_2(t),\zeta_2(t')]}{\delta y_1(t)\delta
y_2(t')}\rangle
\nd
 where  $t,t'$ are the real part of complex time $t_C,t_C'$ on contour $C$ (Fig
 {\bf 3.1}) .
  Here we have introduced type `1' and `2' fields ($\delta y_i,i=1,2$) of thermal field theory in real time formalism 
  \cite{TFTtext1}\cite{Gale-Kapusta} 
 evaluated with real and complex time coordinates - hence evaluated at the upper and lower horizontal line of contour C  .  
 We denote by $W_C[\delta y_1,\delta y_2]$ the Wilson loop with deformation $\delta y_1$ and  $\delta y_2$ on
 the upper and lower line of $C$.

We will now compute the Wilson loop at strong coupling by using holography, that is we  identify 
\bg
<{\rm tr}\rho^0W_C>=e^{iS_{\rm NG}}
\nd 
where $\rho^0$ is the density matrix \cite{DTCS}, $S_{\rm NG}$ being the Nambu-Goto action, with the boundary 
of the string world sheet being  the curve $C$. That is the string world sheet ends on the world line (\ref{wl1}) of the  heavy parton.
 If $X^\mu:(\sigma, \tau)\rightarrow
(t, \zeta, x, y, z,\psi,\phi_1,\phi_2,\theta_1,\theta_2)$ is a mapping from string world sheet to ten-dimensional geometry 
given by (\ref{bhmet}) with $\zeta=1/r$, then with parametrization
$\sigma=\zeta, t=\tau$, we have 
\bg
x(t, \zeta)=vt+\bar{x}(\zeta)\nonumber\\
z(t,\zeta)=y(t, \zeta)=\delta y(t,\zeta)
\nd
where $\bar{x}$ is the unperturbed solution for the mapping. Now using background metric of the form (\ref{bhmetko}) with
$g_{rr}=H$, the Nambu-Goto action up to quadratic order in the
 perturbation $\delta y$ gives 
 \bg \label{NG1}
 S_{\rm NG}=\frac{1}{2\pi \alpha'}\int d\zeta dt\sqrt{\frac{H}{\zeta^4}+\frac{g}{h}\left[(\bar{x}')^2+2\delta
 y'^2\right]-\frac{H}{g\zeta^4}\left[v^2+2\delta \dot{y}^2\right]}
 \nd
 where we approximated $g_i=g$ and prime means derivative with respect to $\zeta$ while dot means derivative with respect to `t', $\dot{A}\equiv
 \frac{dA}{dt},A'\equiv \frac{dA}{d\zeta}$. 
 Minimizing this action with respect to $\bar{x}$ with ignoring ${\cal O}(\delta y)$ gives the solution 
 \bg \label{xbar}
 \bar{x}'^2=\frac{h^2C^2v^2H(g-v^2)}{g^2\zeta^4(g-C^2v^2h)}
 \nd
 where $C$ is a constant of integration. 
 
Now to solve for the
 transverse fluctuation $\delta y$,  
note that  the pullback metric $f_{\alpha \beta}=\partial_\alpha X^\mu \partial_\beta X^\nu G_{\mu\nu}$ is off diagonal 
with components 
\bg \label{pullback1}
f_{tt}&=&h(\zeta)^{-1/2}\left(-\frac{1}{\gamma^2}+\zeta^4 / \zeta_h^4
+\delta\dot{y}^2\right)\nonumber\\
f_{t\zeta} &=& \frac{1}{\sqrt{h}}(v\bar{x}')+\frac{1}{\sqrt{h}}\delta \dot{y}\delta y'\nonumber\\
f_{\zeta \zeta} &=& \frac{H\sqrt{h(\zeta)}}
{\zeta^4 g(\zeta)}+\frac{1}{\sqrt{h}}(\bar{x}'^2+\delta y'^2)
\nd
with $\delta \dot{y}\equiv d\delta y / dt, \delta y'\equiv d\delta y/d\zeta$, 
$g(\alpha)=1- \alpha^4/\zeta_{h}^4$ for any $\alpha$ and $\gamma=1/\sqrt{1-v^2}$ . 
This pullback metric can be
diagonalized at zeroth order in $y$ with  the reparametrization

\bg
\hat{t}&=& \frac{1}{\sqrt{\gamma}} \left( t + F(\zeta) \right)\nonumber\\
\hat{\zeta}&=&\sqrt{\gamma}\zeta\nonumber\\
\frac{dF}{d\zeta}&=&-\frac{v\bar{x}'}{g(\zeta)-v^2}
\nd   
The resulting pullback metric is

\begin{eqnarray}
f_{\hat{t}\hat{t}}&=&\frac{1}{\sqrt{h(\zeta)}\gamma} \left(-g(\hat{\zeta}) + \delta \dot{\hat{y}}^2 \right)
 \nonumber\\
f_{\hat{t}\hat{\zeta}}&=&{\cal O}(\delta y^2);\nonumber\\
f_{\hat{\zeta}\hat{\zeta}}&=&\frac{1}{\gamma}\Bigg(\frac{H(\zeta)\sqrt{h(\zeta)}}{\zeta^4(g(\zeta)-C^2v^2h(\zeta)}
+\frac{1}{\sqrt{h(\zeta)}}\delta \hat{y}'^2\nonumber\\
&-&\frac{\delta
\dot{y}^2v^4h(\zeta)^2C^2H(\zeta)}{g(\zeta)^2\sqrt{h(\zeta)}\zeta^4(g(\zeta)-v^2)(g(\zeta)-C^2v^2h(\zeta))}\Bigg)
\end{eqnarray} 
Here $\delta \hat{y}=\sqrt{\gamma}\delta y$, $\delta \dot{\hat{y}}=\frac{d \delta \hat{y}}{d\hat{t}}$ and 
$\delta \hat{y}'=\frac{d\delta\hat{y}}{d \hat{\zeta}}$. Observe that the pullback metric has a horizon at
$\hat{\zeta}=\zeta_h$ which means at $\zeta=\zeta_h/\sqrt{\gamma}<\zeta_h$ and thus the pullback has larger horizon radius
than the space-time metric. Thus the world sheet only extends in the radial direction from $\zeta=0$ to
$\zeta=\zeta_h/\sqrt{\gamma}$. Also observe that at the horizon $g-C^2v^2h=0$ by our choice of C and thus indeed
$f_{\hat{\zeta}\hat{\zeta}}=\infty$ at the horizon. 
Now with this reparametrization , the terms quadratic order in $\delta \hat{y}$ in 
Nambu-Goto action  become 
\begin{eqnarray}
S_{NG,\delta y}^{[2]}&=&\frac{1}{2\pi \alpha'}\int d\hat{\zeta}d\hat{t}\left[
\frac{g(\hat{\zeta})\hat{\zeta}^2}{2\gamma^2h(\zeta)\sqrt{1+{\cal
A}}}\delta\hat{y}'^2-\frac{1+\gamma^2h(\zeta)v^4C^2}{2\hat{\zeta}^2g(\hat{\zeta})}\delta\dot{\hat{y}}^2\right]\nonumber\\
{\cal A}&=&\frac{g(\hat{\zeta})H(\zeta)}{g-C^2v^2h(\zeta)}-1
\end{eqnarray}  
 If we denote $\delta\hat{y}(\hat{t}, \hat{\zeta})=\int
d\hat{\omega} \;e^{i\hat{\omega} \hat{t}}\; \delta \hat{y}(\hat{\omega})\hat{Y}(\hat{\zeta})$, then the equation of motion can be written as  
\begin{eqnarray} \label{EOM}
&&\hat{Y}''+\frac{B'}{B}\hat{Y}'-\frac{D}{B}\hat{Y}=0\nonumber\\
&&B=\frac{g(\hat{\zeta})\hat{\zeta}^2}{\gamma^2h(\zeta)\sqrt{1+{\cal
A}}}\nonumber\\
&&D=-\frac{\hat{\omega}^2(1+\gamma^2h(\zeta)v^4C^2)}{2\hat{\zeta}^2g(\hat{\zeta})}
\end{eqnarray} 
 We try  solution to
(\ref{EOM}) of the form 
\bg \label{sol1}
\hat{Y}=g(\hat{\zeta})^{\beta}{\cal F}\nonumber\\
{\cal F}=(1+\beta {\cal H})
\nd
with $\beta$ being a constant and we
have written ${\cal F}$ only up to linear order in $\beta$ which is sufficient for what is to follow. Now from
equation (\ref{EOM}) near horizon $\hat{\zeta}\rightarrow \zeta_h$ and only considering up to quadratic order in $\beta$, 
we can isolate the most divergent terms to obtain

\bg \label{b1}
\beta^2&=&-\frac{\hat{\omega}^2\bar{h}\sqrt{1+\tilde{{\cal A}}}
(1+\zeta_h^4\bar{h}v^4C^2)}{g'(\zeta_h)}\nonumber\\
\Rightarrow \beta&=&\pm i \frac{\hat{\omega}}{4\pi T}(1+{\cal E})
\nd
where $\tilde{{\cal A}}={\cal A}(\hat{\zeta}=\zeta_h)$, ${\cal E}$ is of ${\cal O}(g_sM_{\rm eff}^2/N,g_s^2(N_f^{\rm eff}))^2$ 
and $\bar{h}=h(\zeta=\zeta_h/\sqrt{\gamma})$ while
$T=\frac{\zeta_h}{\pi h(\zeta_h)}$ is the temperature associated with the space time black-hole (not the world sheet
black hole). This determines our constant $\beta$ and now we can solve (\ref{EOM}) order by order in $\hat{\omega}$. For the purpose of
calculating $\kappa_T$, we will 
eventually take the zero frequency $\omega\rightarrow 0$ limit (albeit after dividing by $\omega$), 
thus it is sufficient to
solve (\ref{EOM}) only up to linear order in $\omega$. With the form of the solution as in (\ref{sol1}), the equation of
motion (\ref{EOM}) gives the following equation for ${\cal H}$,

\bg\label{EOM2}
{\cal H}''+\frac{B'}{B}{\cal H}'+\frac{g(\hat{\zeta})''}{g(\hat{\zeta})}-\frac{\tilde{h}'g(\hat{\zeta})'}{\tilde{h}g(\hat{\zeta})}=0
\nd
where $\tilde h=2\gamma^2h(\zeta)\frac{\sqrt{1+{\cal A}}}{\hat{\zeta}^2}$ and prime denotes a
derivative with respect to $\hat{\zeta}$.
From the above form of the equation, we observe with  $W={\cal H}'$ we can write it as 
\bg \label{EOM2a}
[WB]'&=&-B\left(\frac{g(\hat{\zeta})''}{g(\hat{\zeta})}-\frac{\tilde{h}'g(\hat{\zeta})'}{\tilde{h}g(\hat{\zeta})}\right)\nonumber\\
\Rightarrow {\cal H}'&=&-\frac{1}{B}\int d\hat{\zeta} ~~B\left(\frac{g(\hat{\zeta})''}{g(\hat{\zeta})}-\frac{\tilde{h}'g(\hat{\zeta})'}{\tilde{h}g(\hat{\zeta})}\right)
\nd
We will impose the boundary condition $Y(0)=1$ which implies ${\cal H}(0)=0$. Now using the solution for $\hat{Y}$ and 
taking appropriate linear 
combinations to build the type `1' and `2' fields
$\delta y_1, \delta y_2$ as in \cite{DTCS}, we can write the boundary action after integrating  the Nambu-Goto action 
and the result is
equation (3.51) of 
 \cite{DTCS} but $\hat{Y}$ is replaced with our solution, $\hat{\omega}, \omega$ replaced by $\hat{\omega}/\pi T, \omega/\pi T$ and
 $R=1$.
 Finally from the boundary action we can obtain the Greens function $G_{ij}$ and the result
 for the diffusion coefficient is
\begin{eqnarray} \label{NG1}
\kappa_T=\sqrt{\gamma g_s\bar{N}_{\rm eff}}\pi T^3 (1+{\cal B})
\end{eqnarray}            
where $\cal{B}$ is of ${\cal O}(g_sM_{\rm eff}^2/N,g_s^2(N_f^{\rm eff}))^2$ and higher, while $\bar{N}_{\rm eff}$ is the number of effective degrees of freedom for 
the boundary  gauge theory
\begin{eqnarray} \label{Neff}
\bar{N}_{\rm eff}&=&N\Bigg(1+\frac{27g_s^2M^2_{\rm eff}N_f^{\rm eff}}{32\pi^2 N}-\frac{3g_sM_{\rm eff}^2}{4\pi N}+
\left[\frac{3g_sM_{\rm eff}^2}{4\pi
N}-\frac{9g_s^2M_{\rm eff}^2N_f^{\rm eff}}{16\pi^2 N}\right]{\rm log}r_0
\nonumber\\
&+&\frac{9g_s^2M_{\rm eff}^2N_f^{\rm eff}}{8\pi^2N}{\rm log}^2 r_0\Bigg)
\end{eqnarray} 
where $r_0$ is as in section {\bf 2.4} i.e. it is the scale where warp factor changes to inverse power series from
logarithm. Note for duality to hold, $N$ must be quite large, making $N_{\rm eff}$ quite large. For $N_f^{\rm eff}=M_{\rm
eff}=0$, we get
back the value of $\kappa_T$ as computed in \cite{DTCS}. However for a non conformal field theory with fundamental matter
- which is more relevant for QCD - $M_{\rm eff}\neq 0, N_f^{\rm eff} \neq 0$, and our analysis thus generates a correction to the AdS/CFT result.

\subsection{Wake }
In the previous sections we computed the drag force on the quark along with the broadening of its momentum. 
Clearly a moving quark should leave some disturbance 
in the surrounding media. This disturbance is called the {\it wake} of the 
quark.
In order to quantify the wake left behind by a fast moving quark in the Quark Gluon Plasma, we need to compute 
the stress tensor $T^{pq}$, $p,q=0,1,2,3$ of the entire system. We expect a cone like disturbance in the medium with the quark located at 
the tip of the cone and this cone becomes apparent when one plots the stress tensor as a function of location of the fast quark.  We would
like to compute the energy momentum tensor the medium including the fast quark, then subtract the contribution from the quark to obtain the
disturbance left behind in the medium. Our goal therefore would be to 
compute:
\bg\label{goal}
T^{pq}_{{\rm medium} + {\rm quark}} ~- ~ T^{pq}_{\rm quark}
\nd
where the first term is basically the energy-momentum tensor obtained from dual geometry considering back reaction of 
a moving string  i.e
$T^{pq}_{{\rm background} + {\rm string}}$
. Similarly the second term is the energy momentum tensor of the string 
i.e $T^{pq}_{\rm string}$ restricted to 
four-dimensional space-time. This is similar to the 
analysis done in \cite{Yaffe-1} for the AdS case. For our case
the above idea, although very simple to state, will be rather technical because of the underlying RG 
flow in the dual gauge theory side. Our second goal would then be to see how much we differ from the AdS results 
once we go from CFT to theories with running coupling constants.   

For a strongly coupled QGP, we will apply the gauge/gravity duality
to compute $T^{pq}$ of QGP using the supergravity action. In the ten dimensional bulk geometry,  we introduce an additional string moving with some velocity and this string is dual to the fast
 parton creating the wake. The total  
metric  takes the following form:
\bg\label{metchange}
G_{ij}~&= & ~g_{ij}+\kappa l_{ij}\nonumber\\
 l_{ij}~ & \equiv & ~ {l}_{ij}(r,x,y,z,t)
\nd
where $g_{ij}$ is the background metric  and  $l_{ij}$ ($i,j=0,..,9$)
denote the perturbation from the moving string source (with $\kappa \to 0$).    
In order to compute $T^{pq}$, we need to write the supergravity action as a functional 
of the perturbation ${l}_{pq}$. We have ${\cal O}=T^{pq}$ in (\ref{O})  and
thus $\phi_0=\kappa {l}_{pq}$ is the source in the partition function (\ref{KS16}). It follows that  
\bg \label{KS17}
\langle T^{pq}\rangle ~= ~ {1\over \kappa} 
\frac{\delta S}{\delta {l}_{pq}}\Bigg{|}_{\kappa {l}_{pq}=0}
\nd 
where $S \equiv S_{\rm total} + S_{\rm GH} + S_{\rm counterterm} $ as discussed in section {\bf 2.3.1} where now $S_{\rm total}$ includes the
DBI action for the string.  

 Minimizing $S_{\rm total}$, we obtain  the equation of motion for $G_{ij}$ : 
\bg \label{KS7a}
R_{ij}\left(g_{\alpha\beta} + \kappa l_{\alpha\beta}\right) - {1\over 2}
\left(g_{ij} + \kappa l_{ij}\right) R\left(g_{\alpha\beta} + \kappa l_{\alpha\beta}\right) = 
T_{ij}^{\rm string} + T_{ij}^{\rm fluxes} + T_{ij}^{(p, q)7}
\nd
where the $T_{ij}^{\rm fluxes}$ come from the five-form fluxes $F_{(5)}$ (that give rise to the $AdS_5$ part) and the 
remnant of the $H_{NS}, H_{RR}$ and the axio-dilaton along the radial $r$ direction (that give rise to the 
deformation of the $AdS_5$ part). . The effect of $T_{ij}^{(p, q)7}$ will not be substantial if we take it as a probe in this 
background whereas the strings stress tensor is given explicitly by 
 \bg \label{KS8}
T^{ij}_{\rm string}(x)&=& \frac{\delta {S}_{\rm string}}{\delta G_{ij}} \\
&=& \int d\sigma d\tau ~\Bigg(\frac{2 \dot{X}\cdot X' \dot{X}^iX'^j-X'^iX'^j \dot{X}^2-\dot{X}^i\dot{X}^jX'^2}{2\sqrt{-{\rm det}f}}\Bigg)\;\delta^{10}(X-x)\nonumber
\nd
where $X$ is the mapping from string world sheet to space time, and we can consider the string profile given by (\ref{lcone}).

The Einstein equations can be worked out if one considers the effects of all the background fluxes in our theory. The 
result of such an analysis can be presented in powers of $\kappa$. For our case we are only interested in 
back reactions that are linear in $\kappa$. To this order 
the equation of motion satisfied by $l_{\alpha\beta}$ is determined 
by expanding (\ref{KS7a}) in the following way:
\bg\label{lalbe}
\kappa\Big(\triangle_{ij}^{\alpha\beta} - {\cal B}_{ij}^{\alpha\beta} - {\cal A}_{ij}^{\alpha\beta}\Big)
l_{\alpha\beta} = T_{ij}^{\rm string}
\nd
where $\triangle_{ij}^{\alpha\beta}, \alpha,\beta=0,..,9$ is an operator whereas ${\cal B}_{ij}^{\alpha\beta}$ and 
${\cal A}_{ij}^{\alpha\beta}$ are functions of $r$, the radial coordinate\footnote{There will be another 
contribution from the ($p, q$) seven branes in the background, although for small $g_s N_f$ these are sub leading.}.  
We have been able to determine 
the form for the operator $\triangle_{ij}^{\alpha\beta}$
 for any generic perturbation $l_{\alpha\beta}$ in five dimensions that is by setting $l_{ij}=0$ for $i,j=5,..,9$. The resulting equations are rather long and 
involved; and we give them in the Appendix B of \cite{FEP}. For the functions 
${\cal A}_{ij}^{\alpha\beta}$
and ${\cal B}_{ij}^{\alpha\beta}$
we have worked out a toy example in 
Appendix C of \cite{FEP} with only diagonal perturbations. For off diagonal perturbations we need
to take an inverse of a $5\times 5$ matrix to determine the functional form. We shall provide details of this in the 
following. The variables defined in (\ref{lalbe}) are given as:
\bg\label{listman}
&& \triangle_{ij}^{\alpha\beta} = 
\left({\delta R_{ij}\over \delta g_{\alpha\beta}}\right) - {1\over 2} ~g_{ij} 
\left({\delta R\over \delta g_{\alpha\beta}}\right) - {1\over 2}~R~\delta_{\mu\alpha} \delta_{\nu \beta}\nonumber\\ 
&& {\cal A}_{ij}^{\alpha\beta} = 5\sum_{b,c,d, ..} F_{(5)\mu bcd a}F_{(5)\nu b'c'd'a'}g^{bb'}g^{cc'}g^{dd'} 
g^{a\alpha}g^{a'\beta}\nonumber\\
&& {\cal B}_{ij}^{\alpha\beta} = -{5\over 8} \sum_{a,b,c,d, ..} F_{(5)n abcd} F_{(5)n' a'b'c'd'} g^{aa'} g^{bb'}
g^{cc'}g^{dd'} g^{nn'}
(g_{ij}g^{\alpha\beta} - \delta_\mu^\alpha\delta_\nu^\beta)\nonumber\\
&&~~~~~~~~ -{1\over 4}\sum_{i=1}^4 g_{ij} F_a^{(i)} F_b^{(i)} g^{a\alpha}g^{b\beta} + 
\sum_{i=1}^4 F_r^{(i)} F_r^{(i)} g^{rr} \delta_{\mu}^\alpha \delta_{\nu}^\beta 
\nd
where we have given the most generic form in (\ref{listman}) above for a five dimensional perturbation. Furthermore, 
${\delta R_{ij}\over \delta g_{\alpha\beta}}$ and ${\delta R\over \delta g_{\alpha\beta}}$ are operators
and not functions.

Once (\ref{lalbe}) is solved, with the exact solutions $l_{ij}$, we can write write the action as a functional of $l_{ij}$. We can then
integrate the radial direction along with the internal directions and add appropriate Gibbons-Hawkings terms along with counterterms to
renormalize the action if necessary. The renormalization and calculation of the stress tensor was described in some detail in section {\bf
2.3.1} and {\bf 2.3.2}.  The final result for the stress tensor $T^{pq}$ with $\kappa l_{ij}=\phi_m$ is given by (\ref{wakegt}). Knowing all
the solutions $l_{ij}$, one can easily evaluate the stress tensor and finally plot the wake. Here we have outlined the procedure and an exact
calculation  will be done in our future work.

\section{Transport Coefficients}
At low energies, an effective theory of fluids is hydrodynamics which describes the kinematics of the fluid and 
uniquely determines its  stress energy tensor. As dissipation is allowed, the theory is not described in terms of action but in terms
of equation of motion, namely the equation which guarantees conservation of energy-momentum tensor
\bg \label{stress-tensor}
\nabla_{\mu} T^{\mu\nu}=0
\nd
where $\mu,\nu=0,1,2,3$. There are four equations above and they can be solved by four independent variables
\cite{Gale-Kapusta}\cite{Son-Starinets}\cite{Forster}. If we assume local thermal equilibrium then  the state of the system can be uniquely
determined by  the local fluid velocity $u^{\mu}(x^\mu)$ and local temperature $T(x^{\mu})$.
There are actually three independent components of the four velocity $u^{\mu}$ as $u^{\mu}u_{\mu}=-1$ fixes one of the components and the
fourth independent variable is the temperature. Using these four independent variables, we can write the expression for stress tensor that
satisfies  equation (\ref{stress-tensor}) and at zeroth order in derivatives of $u^{\mu}$ we obtain the stress tensor of ideal fluid:
\bg
T_{\mu\nu}^{[0]}=(\epsilon +P)u_\mu u_\nu+ Pg_{\mu\nu}
\nd  
where $\epsilon$ is the energy, $P$ is the pressure and $g_{\mu\nu}$ is the metric of four dimensional space time where the fluid lies. 

Now at linear order in derivatives, the allowed terms are restricted by rotational symmetry and only nonzero components are $T_{ij}^{[1]},
i,j=1,2,3$  
\bg \label{TCoeff}
T_{ij}^{[1]} 
=P_{i\alpha}P_{j\beta}\left[\eta\left(\nabla^\alpha u^\beta+\nabla^\beta u^\alpha\right)+\left(\zeta-\frac{2}{3}\eta\right)g^{\alpha\beta} \nabla \cdot u\right]
\nd
where $P_{i\alpha}=g_{i\alpha}-u_{i}u_{\alpha}$ and the coefficients of the two terms determine the transport coefficients shear viscosity $\eta$ and bulk viscosity $\zeta$
\cite{Son-Starinets}. Thus
(\ref{TCoeff}) is the defining equation for viscosity. 

Physically shear viscosity measures the mixing between two layers of a fluid \cite{FEP}. More viscous the fluid, the faster momentum can be
transferred from a layer to the next. Somewhat counter intuitive is the fact that the stronger the
coupling in the fluid, the less the shear viscosity. This is because the rate of mixing is controlled by the mean free path.
When the mean free path is small compared to the
flow velocity variation  of the   
two laminas, the layers 
cannot easily mix since the exchange of particles
is limited to the small volume near the interface of the two laminas: Most 
particles in the fluid just flows
along as if there is no other layers nearby.
On the other hand, if the mean free path is comparable to the typical size
of the flow velocity variation, then mixing between different layers
can proceed relatively quickly.
 For an ideal fluid, laminas of fluid do not
interact at all and we have zero viscosity. 

Transport coefficient such as shear viscosity becomes crucial in understanding physics of quark gluon plasma. In particular, as already
mentioned in the introduction, in a heavy ion
collision the overlapping region of the two nuclei is of elliptical shape with different short and long axes. Thus the plasma formed 
undergoes elliptic flow due to difference in pressure along the long and short axes. On the other hand, a large viscosity would mean greater
interaction between the layers of fluid which would quickly equilibrate the system and one would observe very little elliptic flow. But the
data from RHIC is well described by ideal hydrodynamics with zero shear viscosity and shows  strong elliptic flow
\cite{RHIC-Eflow-1}-\cite{RHIC-Eflow-4}. 
This means the fluid
created in heavy ion collisions has small viscosity and thus is strongly coupled. Thus one cannot apply conventional perturbative field
theory techniques to compute the transport coefficients of the plasma. 

However in the regime of strong coupling, certain gauge theories can be
described by  dual supergravity. In particular for the gauge theories described in section {\bf 2.2-2.4}, which have several features common to QCD,
one can compute the transport coefficients at strong coupling using dual geometry. The result we obtain can then be suggestive of the
viscosities of QGP. In the following sections we compute shear viscosity $\eta$ for  gauge theories using dual supergravity and the ratio
$\eta/s$ which appears in various experimental observations.        
\subsection{Shear Viscosity }
In this section we will present our calculation \cite{FEP} of shear viscosity of the four dimensional theory
following some of the recent works \cite{Kats}\cite{buchel-1}. The shear viscosity described earlier can be obtained 
from correlation functions
using the Kubo formula \cite{Gale-Kapusta}:
\bg \label{SV-1}
\eta=\lim_{\omega\rightarrow 0}~\frac{1}{2\omega}\int dt d^3x\; e^{i\omega
t} \langle\left[ T_{23}(x), T_{23}(0)\right]\rangle =-\lim_{\omega\rightarrow
0}~\frac{{\rm Im}~G^R(\omega,0)}{\omega}
\nd
where $G^R(\omega,\overrightarrow{q})$ is the momentum space retarded propagator for
the operator $T_{23}$ at finite
temperature, defined by
\bg \label{SV-2}
{G}^R(\omega,\overrightarrow{q})= -i \int dt d^3x \; e^{i(\omega t-\overrightarrow{x}\cdot \overrightarrow{q})}
\theta(t) \langle\left[ T_{23}(x), T_{23}(0)\right]\rangle
\nd
In the following,
we will compute the Minkowski propagator following the conjecture made in \cite{S+Starinets1}\cite{S+Starinets2} 
using dual gravity. 
Note that our prescription (\ref{KS16}) computes a path integral 
with a classical action $S_{\rm SUGRA}$, unaware of the ordering of the operators whose expectation value is being computed. Therefore 
computing any commutator is subtle here. 
Hence, we compute only the correlator  
$\langle T_{23}(\tau,\overrightarrow{x}) T_{23}(0,\overrightarrow{x})\rangle$ 
and using the conjecture in \cite{S+Starinets1},
 relate this to the retarded Greens function.   
However before we compute this explicitly, let us evaluate the higher order corrections 
to the effective action from the wrapped D7 brane in our theory. 

In the case of a single D7 brane, the disc level action contains the term \cite{DJM1}\cite{DJM2}:
\bg\label{disclevel}
S_{\rm D7}^{\rm disc} = {1\over 192\pi g_s} \cdot {1\over (4\pi \alpha')^2} \int_{M^8} \left[ C_4 \wedge {\rm tr}~
(R \wedge R) - e^{-\phi}{\rm tr}~(R \wedge \ast R)\right]
\nd
where $C_4$ is the four-form, $R$ is the curvature two-form,
$\phi$ is the dilaton and $M^8$ is a non-trivial eight manifold which is the world-volume
of the D7 brane. 
The action is $SL(2, {\bf Z})$ invariant which was shown  by 
doing an explicit analysis \cite{DJM1}\cite{DJM2}. Since for our case the D7 wrap a non-trivial four-cycle, we can 
dimensionally reduce it over the four-cycle and obtain  the following action:
\bg\label{dimred}
S_{\rm D7}^{\rm disc} = {1\over 16\pi^2} \int_{M^4} {\rm Re}\left[ {\rm log}~\eta(\tau) ~{\rm tr}\left(R \wedge 
\ast R - i R \wedge R\right)\right]
\nd
where $\eta(\tau)$ is the Dedekind function, and $\tau$ is the modular parameter defined as follows:
\bg\label{taudef}
\tau = {1\over g_s(4\pi\alpha')^2} \left(\int_{S^4} C_4 + i {\cal V}_4\right) \equiv {1\over g_s} (\tau_1 + i \tau_2)
\nd
with ${\cal V}_4$ being the volume of the four-cycle on which we have the wrapped D7 brane.
In fact the above action can be {\it derived} from the following action that has two parts,  CP-even
and  CP-odd \cite{BBG}:
\bg\label{cpevodd}
{1\over 32\pi^2} \int_{M^4} {\rm log} \vert \eta(\tau)\vert^2 {\rm tr}~(R \wedge \ast R) - {i \over 32\pi^2} 
\int_{M^4} {\rm log} ~{\eta(\tau) \over \eta(\bar\tau)}~{\rm tr}~(R \wedge R)
\nd
where the first part is CP-even and the second part is CP-odd. To compare (\ref{cpevodd}) with (\ref{dimred}) 
note that the Dedekind $\eta$ function has the following expansion with $q \equiv e^{2\pi i \tau}$:
\bg\label{etaexpan}
&& {\rm log} \vert \eta(\tau)\vert^2 = - {\pi \over 6} \tau_2 - \left[ q + {3q^2 \over 2} + {4q^3\over 3} + ...
+ {\rm c.c}\right]\nonumber\\
&& {\rm log} ~{\eta(\tau) \over \eta(\bar\tau)} = + {i\pi\over 6}\tau_1 - \left[q + {3q^2 \over 2} + {4q^3\over 3} + ...
- {\rm c.c}\right]
\nd
Combining everything we see that, up to powers of $q$ (\ref{cpevodd}) and (\ref{dimred}) are 
equivalent. 
However writing the action in terms of (\ref{cpevodd}) instead of (\ref{dimred}) has the following advantage: from D7 
point of view (\ref{cpevodd}) captures the D3 instanton corrections in the system \cite{BBG}-\cite{oog4}. But 
a deeper reason for writing the action as (\ref{cpevodd}) is that  the CP-even and CP-odd terms can be expanded further 
to account  Gauss-Bonet type interactions \cite{BBG}:
\bg\label{cpeven}
S_{\rm CP-even} = -\alpha_1 \int_{M^8} e^{-\phi} {\cal L}_{\rm GB} - T_7 \int_{M^8} e^{-\phi}\left[\sqrt{G} - 
{(4\pi^2\alpha')^2 \over 24} ~{\cal L}_R + {\cal O}(\alpha^{'4})\right]
\nd
where $\alpha_1$ is a constant, and ${\cal L}_{\rm GB}$ and ${\cal L}_R$ are respectively the Gauss-Bonnet and the 
curvature terms defined in the following 
way:
\bg\label{GBR}
&&{\cal L}_{\rm GB} = {\sqrt{G}\over 32\pi^2} \Big(R_{\alpha\beta\gamma\delta}R^{\alpha\beta\gamma\delta} - 
4 R_{\alpha\beta}R^{\alpha\beta} + R^2\Big)\nonumber\\
&& {\cal L}_R = {\sqrt{G}\over 32\pi^2}
\Big(R_{\alpha\beta\gamma\delta}R^{\alpha\beta\gamma\delta} - 2R_{\alpha\beta}R^{\alpha\beta} - 
R_{ab\gamma\delta}R^{ab\gamma\delta} + 2 R_{ab}R^{ab}\Big)
\nd
In the above note that the three curvature terms $R_{\alpha\beta}R^{\alpha\beta}$, $R_{ab}R^{ab}$ and $R^2$ are 
{\it not} the pull-backs of the bulk Ricci tensor.  We have also used the notations ($\alpha, \beta$) to 
denote the world-volume coordinates, and ($a, b$) to denote the normal bundle. 

From the CP-even terms, the coefficient of $R_{\alpha\beta\gamma\delta}R^{\alpha\beta\gamma\delta}$ is 
given by \cite{FEP}:
\bg\label{coeffeven}
c_3 \equiv {e^{-\phi}\sqrt{G} \over 32\pi^2}\left({4\pi^4 \alpha^{'2}\over 3} - \alpha_1\right)
\nd
which has an overall plus sign because $\alpha_1$ in many cases is zero (see \cite{BBG} for 
a discussion on this).
However in general for certain exotic compactifications we can have 
$\alpha_1 << {4\pi^4 \alpha^{'2}\over 3}$. If we now compare this to \cite{Kats} we see that $c_3$, which
is the coefficient of $R_{\alpha\beta\gamma\delta}R^{\alpha\beta\gamma\delta}$ in \cite{Kats}, is indeed positive. 
This would 
clearly mean that adding fundamental flavors lowers the viscosity to entropy bound!\footnote{The analysis here was
motivated by discussions with Aninda Sinha. His paper \cite{anindapaper} dealing with the 
violation of viscosity to entropy bound  appeared recently and has some overlap with our analysis.}     

The CP-odd term on the other hand has a standard expansion 
\cite{GHM1}-\cite{GHM4},\cite{DJM1}-\cite{BBG}:
\bg\label{cpodd}
S_{\rm CP-odd} = T_7 \int_{M^8} \left(C_8 + {\pi^2 \alpha^{'2}\over 24} C_4 \wedge {\rm tr}~R \wedge R\right)
\nd
where the first term gives the  dual axionic charge of the D7 brane. Combining (\ref{cpeven}) and 
(\ref{cpodd}) we
get the full back reactions of the D7 brane up to ${\cal O}(\alpha^{'2})$. 

Having computed the back reactions of the embedded D7 brane, we can use this result to compute the shear viscosity.
To start the analysis, we need the correlation function of $T_{23}(x)$ and $T_{23}(0)$ to use it in the 
Kubo formula (\ref{SV-1}). From gauge/gravity duality we know that switching on $T_{23}$ in the gauge theory is 
equivalent to considering graviton modes along $x^2 = x$ and $x^3 = y$ directions. 
In the ten dimensional dual geometry with the metric of
the form (\ref{bhmetko}), we introduce graviton perturbations in the $xy$ direction. We start with the ten dimensional SUGRA action with the
perturbed metric and then integrate over the five internal compact directions to obtain an effective five dimensional action . 
The resulting five dimensional metric $ds^2_{5}=\tilde{g}_{\mu\nu} dx^\mu dx^\nu$ which minimizes the five dimensional action takes the following form:
\begin{displaymath}\label{metricmatrix}
\left(\begin{array}{ccccc}
\tilde{g}_{00}~~& \tilde{g}_{0x}~~ & \tilde{g}_{0y}~~& \tilde{g}_{0z}~~ & \tilde{g}_{0r}\\
\tilde{g}_{x0}~~& \tilde{g}_{xx}~~ & \tilde{g}_{xy}~~& \tilde{g}_{xz}~~ & \tilde{g}_{xr} \\
\tilde{g}_{y0}~~& \tilde{g}_{yx}~~ & \tilde{g}_{yy}~~& \tilde{g}_{yz}~~ & \tilde{g}_{yr}\\
\tilde{g}_{z0}~~& \tilde{g}_{zx}~~ & \tilde{g}_{zy}~~& \tilde{g}_{zz}~~ & \tilde{g}_{zr} \\
\tilde{g}_{r0}~~& \tilde{g}_{rx}~~ & \tilde{g}_{ry}~~& \tilde{g}_{rz}~~ & \tilde{g}_{rr} 
\end{array} \right) =\frac{1}{\sqrt{\bar{h}(r)}}\left(\begin{array}{ccccc}
-g(r)~~& 0~~ & 0~~& 0~~ & 0\\
0~~& 1~~ & \phi(r,t)~~& 0~~ & 0 \\
0~~& \phi(r,t)~~ & 1~~& 0~~ & 0\\
0~~& 0~~ & 0~~& 1~~ & 0 \\
0~~& 0~~ & 0~~& 0~~ & \frac{\bar{h}(r)}{g(r)} 
\end{array} \right)
\end{displaymath}
where we have ignored ${\cal O}(g_s^kN_f^k), k\ge 2$ terms (which is equivalent to setting $g_{rr}=1$ in (\ref{bhmetko}) and thus in five dimensions
the effective $g_{rr}$ is 1) and $\bar{h}(r)$ is now only a function the the radial coordinate. Note that $\bar{h}$   is obtained from  $h$ 
appearing in (\ref{bhmetko}) by integrating over the internal coordinates on which $h$ depends. When $h$ is independent of any of the angles
of internal space, like the case in AdS space, then $\bar{h}=h$. We have also taken the approximation
$g_1=g_2=g(r)$ for simplicity and this approximation becomes exact if we ignore ${\cal O}(g_sN_f,g_sM^2/N)$ corrections to the black hole
factors $g_i$.

Since our goal is to compute the Fourier transform of 
$\langle T_{23}(t,\overrightarrow{x}) T_{23}(0,\overrightarrow{x})\rangle$, we can do this by first
writing the supergravity action in momentum space, treating it  as a functional of Fourier modes for
$\phi(r, t)$ where: 
\bg \label{SV-7a}
&& \phi(r,t)=\widetilde{\phi}(r, t)\bar{\phi}(t) 
\equiv \int d\omega\; e^{-i\omega \tau} \phi(r,\omega) = \int d\omega\; e^{-i\omega \sqrt{g}t} \phi(r,\omega) \nonumber\\
&& \phi(r,\omega)= \widetilde{\phi}(r, \vert\omega\vert)\bar{\phi}(\omega) 
 \nd 
where as before, we defined the Fourier transform using the curved space time $\tau \equiv \sqrt{g(r_c)} ~t$ and 
not simply $t$. Although this definition is precise for the theory at the cut-off $r = r_c$ only, we will use
it also for any  $r$ because in the end we will only provide description at the boundary (i.e $r \to \infty$)
where the results would 
be independent of the choice of the cut-off.

The way we proceed now is the following\footnote{This is similar to the procedure of \cite{Kats}. 
Notice however that the theory considered by \cite{Kats} has no running 
but contains higher curvature-squared corrections.}. 
We consider the metric fluctuation as in (\ref{metricmatrix}) and plug this in the five dimensional effective action.
 Finally, 
we will call this resulting action as $S_{\rm SG}^{(2)}$ where 
the subscript (2) involves writing the action in terms of quadratic $\phi(r, \omega)$. 
We do this as  there exists a very useful relation for computing the shear viscosity (see for example \cite{S+Starinets1}\cite{S+Starinets2}):
\bg \label{SV-7a2}
\lim_{\omega\rightarrow 0}
{\rm Im}~{G}_{11}^{\rm SK}(\omega,\overrightarrow{0})~ = ~
\lim_{\omega\rightarrow 0} 
\frac{2T}{\omega}~ {\rm Im}~G^{R}(\omega,\overrightarrow{0})  
\nd 
where ${G}_{ij}^{\rm SK}$ is the Schwinger-Keldysh propagator \cite{S+Starinets1}-\cite{H+Son}. Comparing this 
with our earlier Kubo formula (\ref{SV-1}), we get the following expression for shear viscosity:
\bg\label{skbyt}
\eta ~ = ~-{1\over 2T}
\lim_{\omega\rightarrow 0}
~ {{\rm Im}~{G}_{11}^{\rm SK}(\omega,\overrightarrow{0}) }
\nd
Thus if we can write our effective supergravity action in the following way:
\bg \label{SV-7}
S_{\rm SG}^{(2)}[{\phi}(r_b, \omega)]~= ~ \frac{1}{2}\int \frac{d\omega d^3q}{(2\pi)^4}  ~ {\phi}_i(r_b, \omega)
{G}^{\rm SK}_{ij}(\vert\omega\vert,\overrightarrow{q}){\phi}_j(r_b, -\omega)
\nd 
where $r_b$ is a specified point on the boundary. 
Then taking the ${G}_{11}^{\rm SK}(\omega,\overrightarrow{0})$ part 
and using (\ref{skbyt}) we can easily obtain the shear viscosity\footnote{Notice that there would be an overall volume 
factor of $T^{1,1}$ that would appear 
with the effective action. This factor  just modifies the Newton's constant in five 
dimensions and does will not effect dimensionless ratios like $\eta/s$, as we shall find out.}.  
In other words, 
 we will be taking two functional derivatives of
$S_{\rm SG}^{(2)}[{\phi}(r_b, \omega)]$ with respect to ${\phi}(r_b, \omega)$ 
and thus are  interested about terms quadratic in
${\phi}(r_b, \omega)$ in the action. Of course, 
in real time formalism, we are concerned with the Schwinger-Keldysh propagator ${G}_{ij}^{\rm SK}$ of the
doublet fields $\phi_i(r, t),\phi_j(r, t)$. 
In the context of gauge/gravity duality, we follow the procedure outlined by \cite{H+Son} for
AdS/CFT correspondence and treat 
$\phi_1(r, t),\phi_2(r, t)$ as the perturbation $\phi(r, t)$
 and its doublet in the four dimensional Minkowski space\footnote{ Although  our background
is a deformation of the AdS space,  the arguments of \cite{H+Son} also apply here. We 
can still consider the $\phi_1$ perturbations to compute the Schwinger-Keldysh propagator because by definition
a propagator  is what appears sandwiched between the fields. See also \cite{sken2} for a generic 
approach.}.  
In ten dimensional gravity theory, $\phi_1(r)=\phi(r)$ is the 
 field in the R quadrant of the Penrose diagram and $\phi_2(r)$ is the field in the L quadrant. 
 For more details see \cite{H+Son}-\cite{sken2}.

To be more  precise, our aim is to get the effective action in the form (\ref{SV-7}). To this 
effect we take our metric (\ref{metricmatrix}) and plug it in the five dimensional effective action with  
net result:
\bg \label{SV-8}
S_{\rm SG}^{(2)}&=&\frac{1}{8\pi G_N\sqrt{g(r_c)}}\int \frac{d\omega d^3q}{(2\pi)^4}\int_{r_h}^{r_c} dr
\Bigg[ A(r)\phi(r,-\omega) \phi''(r,\omega)+B(r)\phi'(r,-\omega) \phi'(r,\omega)\nonumber\\
&+&C(r)\phi(r,-\omega) \phi'(r,\omega)+D(r)\phi(r,-\omega) \phi(r,\omega)\Bigg] 
\nd
 where prime denotes derivative with respect to $r$ and the explicit expressions for 
 $A,B,C,D$ are given in Appendix E of \cite{FEP}. The five dimensional Newton's constant is given by:
\bg\label{nc}
G_N \equiv {\kappa_{10}^2 L^5 \over 4\pi V_{T^{1,1}}}
\nd 
 where volume of $T^{1,1}$ i.e $V_{T^{1,1}}$ is dimensionful  
and $\kappa_{10}$ is proportional to ten dimensional Newton's 
constant. 

The fluctuation $\phi(r, \omega)$  satisfies the following Euler-Lagrange 
equation of motion:
\bg \label{SV-11a}
\phi''(r,\omega)+ \frac{A'(r)-B'(r)}{A(r)-B(r)}\phi'(r,\omega)+
\frac{2D(r)-C'(r)+A''(r)}{2\left[A(r)-B(r)\right]}\phi(r,\omega)=0
\nd
which we can derive from \ref{SV-8} by minimizing it. Once we plug in the values of $A, B$ etc., the above Euler-Lagrange
equation takes the following form:
\bg \label{SV-11b}
&& \phi''(r,\omega)+ \left[\frac{g'(r)}{g(r)}+\frac{5}{r} +{\cal M}(r)\right]\phi'(r,\omega)
+\left[\frac{\omega^2 g(r_c)\bar{h}(r)}{g(r)^2}+{\cal J}(r)\right]\phi(r,\omega)=0\nonumber\\
&&\bar{h}(r)\equiv \frac{L^4}{r^4}\Bigg\{1+\frac{3g_s N_f^2}{2\pi N}\left[1+\frac{3g_s N_f}{2\pi}
\left({\rm log}r+\frac{1}{2}\right)-\frac{g_s N_f}{4\pi}\right]
{\rm log}r\Bigg\}\nonumber\\
\nd
where ${\cal J}(r)$ and ${\cal M}(r)$ appear due to seven branes and fluxes in the geometry \footnote{In 
special cases we expect ${\cal J}(r)$ and ${\cal M}$ to 
vanish (see for example \cite{buch04}). 
However when this is not the case ,as  possible for non-trivial UV completions of our model, we could 
expect a non-minimally coupled scalar field.}. 
As before, primes in (\ref{SV-11a}) and (\ref{SV-11b})
denote derivatives with respect to the five dimensional radial coordinate $r$. 

Now as we mentioned above in (\ref{SV-7a}), $\phi(r, \omega)$ can be decomposed in terms of 
$\widetilde\phi(r, \vert\omega\vert)$ and 
$\bar\phi(\omega)$. Then 
as a trial solution, just like in \cite{Kats}, we first try $\widetilde{\phi}(r, \omega)= g(r)^{\gamma}$ and look at 
(\ref{SV-11b}) for $r$ near the horizon $r_h$ where $g(r)\rightarrow 0$. Plugging this in (\ref{SV-11b}) 
with $g(r) = 0$ we 
obtain :
\bg \label{SV-12} 
\gamma&=&\pm i\vert\omega\vert \sqrt{\frac{\bar{h}(r_h)g(r_c)}{16}}r_h\nonumber\\
&=&\pm i \frac{\vert\omega\vert}{4\pi T_c}
\nd
where 
in the last step we have used the definition of  
temperature $T_c$ as in (\ref{Temp4}). 

To get the solution with for general $r$, where $g(r) \ne 0$,  
we propose the following ansatz 
for the solution to (\ref{SV-11b}):
\bg\label{ansatsol}
\phi(r,\omega)~ = ~ g(r)^{\pm i \frac{\vert\omega\vert}{4 \pi T_c}}F(r, \vert\omega\vert)
\bar{\phi}(\omega)
\nd
Plugging this in (\ref{SV-11b}) we see that the 
equation satisfied by $F(r, \vert\omega\vert)$ can be expressed in terms of $\gamma$ and $\gamma^2$ in the following way: 
\bg \label{SV-13}
&& F''(r, \vert\omega\vert)+\left({g'(r) \over g(r)}+ {5 \over r} + {\cal M}(r) \right)F'(r, \vert\omega\vert)+
\left({\vert\omega\vert^2 g(r_c) {\bar h} \over
g^2(r)} + {\cal J}(r)\right) F(r, \vert\omega\vert)\\
&& + ~\gamma \Bigg\{{2g'(r)\over g(r)} F'(r, \vert\omega\vert) + 
\left[{g''(r) \over g(r)} + \left({5\over r} + {\cal M}\right){g'(r) \over 
g(r)}\right]F(r, \vert\omega\vert)\Bigg\} + \gamma^2 {g'^{2}(r) \over g^2(r)} F(r, \vert\omega\vert) = 0\nonumber
\nd
where the $\gamma^2$ terms come from both the last term in the above equation as well as the 
$\vert\omega\vert^2$ term above.
Furthermore,
note that the source ${\cal J}(r)\sim {\cal O}(g_s)+{\cal O}(g_s^2)$, so in the limit $g_s\rightarrow 0$ we find that 
(\ref{SV-13})
has a solution of the form $F(r, \vert\omega\vert) =c_1+c_2 g(r)^{-2\gamma}$ with $c_1, c_2$ constants. 
Then we expect the complete solution for $g_s\neq 0$ to be $F(r, \vert\omega\vert)=c_1+c_2
g(r)^{-2\gamma}+f(r, \vert\omega\vert)$. 
Demanding that $F(r, \vert\omega\vert)$ be regular at the horizon $r=r_h$ forces 
$c_2=0$ as $g(r_h)=0$. We choose $c_1=1$ and
$f ={\cal G}+\gamma{\cal H} + \gamma^2 {\cal K} + ...$ as a series solution in $\gamma$. Then our ansatz
for the solution to (\ref{SV-13}) becomes 
\bg\label{sersolo}
F(r, \vert\omega\vert) =1+{\cal G}(r) +\gamma{\cal H}(r) + \gamma^2 {\cal K}(r) + ...
\nd 
Once we 
plug in the ansatz (\ref{sersolo}) in (\ref{SV-13}) we see that the resulting equation 
can be expressed as a series in $\gamma$:  
\bg\label{seringama}
&& ~~~~~~~ {\cal G}'' + \left({g'\over g} + {5\over r} + {\cal M} \right) G' + {\cal J}(1 + {\cal G}) \nonumber\\ 
&& +~\gamma \Bigg\{{\cal H}'' + \left({g'\over g} + {5\over r} + {\cal M} \right){\cal H}' + {\cal J} {\cal H} + 
{2g'\over g} {\cal G}' + \left[{g''\over g} + \left({5\over r} + {\cal M} \right){g'\over g}\right]
(1 + {\cal G})\Bigg\}\nonumber\\
&& +~\gamma^2 \Bigg\{{\cal K}'' +  \left({g'\over g} + {5\over r} + {\cal M} \right){\cal K}' + {\cal J} {\cal K} + 
{2g'\over g} {\cal H}' + \left[{g''\over g} + \left({5\over r} + {\cal M} \right){g'\over g}\right]{\cal H}\nonumber\\
&& ~~~~~~~~~~~~~~ + ~\left(\kappa_0 + {g'^{2}\over g^2}\right) (1 + {\cal G}) \Bigg\}\nonumber\\ 
&& + ~ \gamma^3 \Bigg\{{2g'\over g} {\cal K}' + 
\left[{g''\over g} + \left({5\over r} + {\cal M} \right){g'\over g}\right]{\cal K} + 
\left(\kappa_0 + {g'^{2}\over g^2}\right) {\cal H} + ...... \Bigg\} + {\cal O}(\gamma^4) ~ = ~ 0\nonumber\\
\nd
where we have avoided showing the explicit $r$ dependences of the various parameters to avoid clutter. We have also 
defined $\kappa_0$ in terms of the variables of (\ref{SV-12}) in the following way:
\bg\label{kappa0}
\kappa_0 ~ \equiv~ - {16 \over {\cal T}^2 g^2}
\nd
Although the above equation (\ref{seringama}) may look formidable there is one immediate simplification that could 
be imposed, namely, putting the coefficients of $\gamma^0, \gamma, \gamma^2, ...$ individually to zero. This is 
possible because one can view $\gamma$ to be an arbitrary parameter that can be tuned by choosing the graviton 
energy $\omega$ or the temperature $T_c$. This means that the zeroth order in $\gamma$ we will have the following 
equation: 
\bg \label{SV-13a}
{\cal G}''(r)+\left[{g'(r)\over g(r)}+ {5\over r} + {\cal M}(r) \right]{\cal G}'(r)
+{\cal J}(r)[1+{\cal G}(r)]~= ~ 0
\nd
In the above equation observe that 
the source ${\cal J}(r)$, for cases where it is non-zero, has a complicated structure with logarithms and powers of $r$. 
To simplify the 
subsequent expressions, let us choose to work near the cut-off $r = r_c$. This is similar to the spirit of the 
previous section where we eventually analyzed the system from the boundary point of view. Then
to solve (\ref{SV-13a}) near $r\sim r_c$ we can
switch to following coordinate system
\bg \label{SV-13a1}
r=r_c(1- {\zeta}) 
\nd 
Taylor expanding all the terms ${\cal J}(r),g(r),{1\over r_c^n(1-{\zeta})^n}$  in (\ref{SV-13a}) 
about ${\zeta}=0$, 
we obtain a power series solution for ${\cal G}$ as: 
\bg \label{SV-13a2}
{\cal G}(r) ~ = ~ \sum_{\alpha} \sum_{i=0}^{\infty}~ {{\tilde a}_i^{(\alpha)} \over r_{c(\alpha)}^{4i}(1-\zeta)^{4i}} ~ 
\equiv ~  \sum_{i=0}^{\infty}~a_i {\zeta}^i
\nd
Since (\ref{SV-13a}) is a second order differential equation, 
we can fix two coefficients and we choose $a_0=a_1=0$. Then the rest of $a_i's$ are
determined by equating coefficients of ${\zeta}^i$ on both sides of equation (\ref{SV-13a}). 
The exact solutions are listed in Appendix E of \cite{FEP}. Note that all $a_i$ are proportional to $g_s$ and in the limit
$g_s\rightarrow 0$, ${\cal G}\rightarrow 0$. 

To next order in $\gamma$ we have an equation for ${\cal H}$ that also depends on the solution that we got for 
${\cal G}$. 
The equation for ${\cal H}(r)$ can be taken from (\ref{seringama}) as:
\bg \label{SV-13c}
&& {\cal H}''(r) + \left[{g'(r)\over g(r)} + {5\over r} + {\cal M}(r) \right]{\cal H}'(r) + {\cal J}(r) {\cal H}(r) =  
-{2g'(r)\over g(r)} {\cal G}'(r)\nonumber\\
&& ~~~~ - \left\{{g''(r)\over g(r)} + \left[{5\over r} + {\cal M}(r) \right]
{g'(r)\over g(r)}\right\}
\left[1 + {\cal G}(r)\right]
\nd 
To solve this we make 
the coordinate transformation (\ref{SV-13a1}) and plug in the series solution for ${\cal G}(r)$ given above. 
The final 
result for ${\cal H}$ can again be expressed as a series solution in ${\zeta}$ in the following way:
\bg \label{SV-13c1}
{\cal H}(r)~ = ~ \sum_{\alpha}\sum_{i=0}^{\infty}~ {{\tilde b}_i^{(\alpha)} \over r_{c(\alpha)}^{4i}(1-\zeta)^{4i}} ~
\equiv ~ \sum_{i=0}^{\infty}b_i {\zeta}^i  
\nd 
We again set $b_0=b_1=0$ and following similar ideas used to solve for ${\cal G}$, 
we determine all $b_i's$ by equating coefficients in(\ref{SV-13c}). The exact
solution is given in Appendix E of \cite{FEP}. Again note that all $b_i$ are of 
at least ${ \cal O}(g_s)$ and thus with $g_s\rightarrow
0$, ${\cal H}\rightarrow 0$.

Finally the second order in $\gamma$ is a much more involved equation that uses results of the previous two equations 
to determine ${\cal K}$. This is given by:
\bg\label{SV-13c2} 
&& {\cal K}''(r) +  \left({g'(r)\over g(r)} + {5\over r} + {\cal M}(r) \right){\cal K}'(r) + {\cal J}(r) {\cal K}(r)  =  
- {2g'(r)\over g(r)} {\cal H}'(r)\\
&& ~~~ - \left[{g''(r)\over g(r)} + \left({5\over r} + {\cal M}(r) \right){g'(r)\over g(r)}\right]{\cal H}(r)
 - \left(\kappa_0 + {g'^{2}(r)\over g^2(r)}\right) \left[1 + {\cal G}(r)\right]\nonumber
\nd
which could also be solved using another series expansion in ${\zeta}^i$ (we haven't attempted it here). 
Therefore combining 
(\ref{SV-13a2}) and (\ref{SV-13c1}) we finally have the
solution for the metric perturbation: 
\bg \label{SV-13e}
\widetilde{\phi}(r, \vert\omega\vert)_\pm= g(r)^{\pm i\frac{\vert\omega\vert }{4\pi T_c}}
\left[1+{\cal G}(r)\pm i{\vert\omega\vert  \over 4{\pi T_c}} {\cal H}(r) 
- {\vert\omega\vert^2 \over 16 \pi^2 T_c^2} 
{\cal K}(r) + ....\right]
\nd     
We can analyze this in the regime where the gravitons have very small energy, i.e $\omega \to 0$ or equivalently 
$\gamma \to 0$. In this limit 
we can Taylor expand  $\widetilde{\phi}(r, \vert\omega\vert)$ about $\gamma = 0$ to give us the two possible solutions: 
\bg \label{SV-13f} 
\widetilde{\phi}(r, \vert\omega\vert)_\pm &=& 1+{\cal G}(r)\pm i\frac{\vert\omega\vert }{4\pi T_c}
\Big\{{\cal H} (r) + [1+{\cal G}(r)] {\rm log}~g(r)\Big\}\\
&& - {\vert\omega\vert^2 \over 16 \pi^2 T_c^2}\Big\{{\cal K}(r) + {\cal H}(r) ~{\rm log}~g(r) + [1 + {\cal G}(r)]
{\rm log}^2 g(r)\Big\} + {\cal O}(\vert\omega\vert^3)\nonumber
\nd
which 
consequently means that to the first order in $\omega$ the off diagonal gravitational perturbation at low energy is 
given by two possible solutions corresponding to positive and negative frequencies as: 
\bg \label{SV-13g}
\phi(r,\omega)_\pm = \left[1+{\cal G}(r)\right] {\bar \phi}(\omega) \pm i\frac{\vert\omega\vert } {4 \pi T_c} 
\Big\{{\cal H} (r) + [1+{\cal G}(r)] {\rm log}~g(r)\Big\} \bar{\phi}(\omega) 
\nd
As is well known, following \cite{Unruh} \cite{H+Son}, we can define field on the right 
${\bf R}$ and left ${\bf L}$ quadrant of the 
Kruskal plane in terms of $\phi_+(r, \omega)$ and $\phi_-(r, \omega)$ in the following way:
\bg \label{SV-13g1}
\phi_{{\bf R},\pm}(\omega,r)&=& \phi_{\pm}(\omega,r) ~~~ {\rm in ~~ {\bf R}} \nonumber\\ 
 &=&  0 ~~~ {\rm in ~~ {\bf L}} \nonumber\\
 \phi_{{\bf L},\pm}(\omega,r)&=& \phi_{\pm}(\omega,r) ~~~ {\rm in ~~ {\bf L}}\nonumber\\
&=& 0 ~~~ {\rm in ~~ {\bf R}} 
\nd  
Now  $\phi_{{\bf R},\pm},\phi_{{\bf L},\pm}$ contain positive and negative frequency modes but a 
certain linear combination of  
$\phi_{{\bf R},\pm},\phi_{{\bf L},\pm}$
 gives purely positive or purely negative frequency modes in the entire Kruskal plane \cite{Unruh} \cite{H+Son}. 
Furthermore imposing that positive frequency modes are
 in falling at the horizon in ${\bf R}$ quadrant and negative frequency modes are outgoing at the horizon in ${\bf R}$ 
fixes two combinations :
\bg \label{SV-13g2}
\phi_{\rm pos}&=&e^{\omega/T_c}\phi_{{\bf R},-}(\omega,r) +e^{\omega/2T_c}\phi_{{\bf L},-}(\omega,r)\nonumber\\
\phi_{\rm neg}&=&\phi_{{\bf R},+}(\omega,r) +e^{\omega/2T_c}\phi_{{\bf L},+}(\omega,r)
\nd  
With (\ref{SV-13g2}) we see that we can define fields in ${\bf R}({\bf L})$ 
quadrant as linear combination of positive and negative frequency modes 
\bg \label{SV-13g3}
\phi_{\bf R}(\omega,r)&\equiv& {{\tilde a}_0}\big[\phi_{{\bf R},+}(\omega,r) 
- e^{\omega/T_c}\phi_{{\bf R},-}(\omega,r)\big]\equiv\phi_1\nonumber\\
\phi_{\bf L}(\omega,r) &\equiv&{{\tilde a}_0}e^{\omega/2T_c}\big[\phi_{{\bf L},+}(\omega,r) 
- \phi_{{\bf L},-}(\omega,r)\big]\equiv\phi_2
\nd
where we have identified $\phi_{\bf R}(\phi_{\bf L})$ with the thermal field $\phi_1(\phi_2)$ 
defined on the complex time contour which familiarly
appears in the Schwinger-Keldysh propagators of real time thermal field theory. Here ${{\tilde a}_0}$ is a constant.  
The final physical quantity that we
will extract from here will only depend on ${\cal T}$, as we will show soon. 

Having got the graviton fluctuations $\phi(r, \omega) \equiv \phi_{\bf R}(\omega,r)$, 
we are almost there to compute the viscosity $\eta$ using 
(\ref{skbyt}). Our next step would be to compute the Schwinger-Keldysh propagator 
$G^{\rm SK}_{11}(0, \overrightarrow{q})$. All we now need is to write the action (\ref{SV-8}) as (\ref{SV-7}) and 
from there extract the Schwinger-Keldysh propagator. This analysis is similar to the one that we 
did in the previous section, so we could be brief (see also \cite{buchel-1}). The action (\ref{SV-8}) can be used to 
get the boundary action once we shift $\phi(r, \omega)$ to $\phi(r, \omega) + \delta \phi(r, \omega)$ in the 
following way:
\bg\label{boundary}
S_{\rm SG}^{(2)}(\phi &+&\delta\phi) =\frac{g(r_c)^{-1/2}}{8\pi G_N}\int \frac{d\omega d^3q}{(2\pi)^4}\int_{r_h}^{r_c} dr
\Big\{A(r)\phi(r,-\omega) \phi''(r,\omega)+B(r)\phi'(r,-\omega) \phi'(r,\omega)\nonumber\\
&&+C(r)\phi(r,-\omega) \phi'(r,\omega)+D(r)\phi(r,-\omega) \phi(r,\omega) + \Big[2A(r) \phi''(r, \omega)\\
&&- 2B(r) \phi''(r, \omega) - 2 B'(r) \phi'(r, \omega) - C'(r) \phi(r, \omega) + 2D(r) \phi(r, \omega) \nonumber\\
&&+ A''(r) \phi(r, \omega) + 2 A'(r) \phi'(r, \omega)\Big]\delta\phi(r, -\omega) + 
\partial_r \Big[2B(r) \phi'(r, \omega)\delta\phi(r, -\omega) \nonumber\\
&&+ C(r) \phi(r, \omega) \delta\phi(r, -\omega) + A(r) \phi(r, \omega) \delta\phi'(r, -\omega) 
- \partial_r\left(A(r) \phi(r, \omega)\right)\delta\phi(r, -\omega)\Big]\Big\}\nonumber
\nd
Plugging in the background value of $\phi(r, \omega)$ will tell us that only the boundary term survives. And as before,
 to cancel the $A(r) \phi(r, \omega) \delta\phi'(r, -\omega)$ we will have to add the Gibbons-Hawking term to the 
action \cite{Gibbons-Hawking}. The net result is the following boundary action:
 \bg \label{SV-9}
S_{\rm SG}^{(2)}&=&{g(r_c)^{-1/2}\over 8\pi G_N}\int \frac{d\omega d^3q}{(2\pi)^4}
\phi(r,-\omega) \Bigg\{\frac{1}{2}\Big[C(r)-A'(r)\Big]
+\Big[ B(r)-A(r)\Big] {\phi'(r,-\omega)\over \phi(r,-\omega)}\Bigg\} \phi(r,\omega)\Bigg\vert_{r_h}^{r_c}\nonumber\\
&\equiv& {1\over 8\pi G_N\sqrt{g(r_c)}}\int \frac{d\omega d^3q}{(2\pi)^4}{\cal F}(\omega,r)\Bigg{|}_{r_h}^{r_c} 
\nd 
Now
comparing (\ref{SV-7}) with \ref{SV-9} we see that the terms between the braces combine to give us the required 
Schwinger-Keldysh propagator:
\bg \label{SV-10}
{G}^{\rm SK}_{11}(0, \overrightarrow{q})&=& \lim_{\omega \to 0}~{1\over 4\pi G_N\sqrt{g(r_c)}}
\frac{{\cal F}(\omega,r)}{\phi_1(r,\omega)\phi_1(r,-\omega)}
\Bigg{|}_{r_h}^{r_c}\\
& = & \lim_{\omega \to 0} ~{1\over 4\pi G_N\sqrt{g(r_c)}}\Bigg\{
\frac{1}{2}\Big[C(r)-A'(r)\Big] +\Big[ B(r)-A(r)\Big] {\phi_1'(r,-\omega)\over \phi_1(r,-\omega)}\Bigg\}
\Bigg{|}_{r_h}^{r_c}\nonumber
\nd 
where we assume\footnote{At this point one might worry that the solution for $\phi_1$ is only known around 
$r_c$. That this is not the case can be seen in the following way:
Integration by parts gives (\ref{SV-10}) which says one only needs to
know the value of the field $\phi_1$ at $r_c $ and $r_h$. The solution for
$\tilde{\phi_1} ={\phi_1\over {\bar{\phi_1}}}$ is given in (\ref{SV-13e}) from which it is
clear that $\phi_1(r_h)=0$ as $g(r_h)=0$.  Furthermore to know $\eta$ we only
need to know the imaginary part of (\ref{SV-10}), which is evaluated using (\ref{vislim})
in (\ref{SV-10}) and using boundary values of $\phi_1(r_c)$ and $\phi_1(r_h)$.}
that $\phi_1(r_h, \omega) = 
{\tilde a}_0\big[\phi_{{\bf R},+}(r_h, \omega) - e^{\omega/T_c}\phi_{{\bf R},-}(r_h, \omega)\big]$. Now
to evaluate the shear viscosity from the above result we need to perform two more steps:

\noindent $\bullet$ Evaluate the contributions from the UV cap that we attach from $r = r_c$ to 
$r = \infty$.  

\noindent $\bullet$ Take the imaginary part of the resulting {\it total} Schwinger-Keldysh propagator. This should 
give us result independent of the cut-off. 

\noindent To evaluate the first step i.e
contributions from the UV cap, we need to see precisely the singularity structure of 
$S_{\rm SG}^{(2)}$. The second step would then be to extract the imaginary part of SK propagator from there. Since the 
imaginary part can only come from the second term of (\ref{SV-9}), we only need to evaluate:
\bg\label{visuv}
\lim_{\omega \to 0} ~{1\over 4\pi G_N \sqrt{g(r_c)}}
~\Big[ B(r)-A(r)\Big] {\phi_1'(r,-\omega)\over \phi_1(r,-\omega)}
\Bigg{|}_{r_c}^{\infty}
\nd 
with $\phi(r, -\omega)$ being the graviton fluctuation in the regime $r > r_c$. To analyze this let us first 
consider a case where $g_s \to 0$ and (${\cal G}(r), {\cal H}(r), {\cal K}(r), ...$) $\to 0$. In this limit we 
expect for $r_h \le r \le r_c$:
\bg\label{vislim}
&& B(r) - A(r) ~ =  -{1\over 2 g_s^2} ~g(r) r^5 ~ + ~ {\cal O}(g_s N_f) \\
&& \phi_1(r, \omega) ~ = ~ {\tilde a}_0\Bigg[-{\omega\over T_c}\left(1+{\cal G} - i{\vert\omega\vert \over 4\pi T_c} 
{\cal H}\right) + i{\vert\omega\vert \over 2\pi T_c}{\cal H} + i{\vert \omega \vert \over 2\pi T_c} {\rm log}~g (1 + 
{\cal G})\Bigg]\nonumber\\
&& {\phi_1'(r, -\omega)\over \phi_1(r, -\omega)} = {g'(r)\over g(r)} \left({2\pi\over 4\pi^2 + {\rm log}^2 g(r)}\right)
\nonumber
\nd
The above considerations would mean that the contribution to the viscosity, $\eta_1$, 
for this simple case without incorporating
the UV cap will be:
\bg\label{visfirst}
\eta_1 & = & {r_h^4 \over 2\pi T_c g_s^2 G_N \sqrt{g(r_c)}} \Bigg({1\over 4\pi + {1\over \pi} ~{\rm log}^2 
g(r_c)}\Bigg) = {{\cal T}^3 L^2 \over 2 g_s^2 G_N}\left({1\over 4\pi + {1\over \pi} ~{\rm log}^2 g(r_c)}\right)\nonumber\\
\nd
where we have used the relations $\pi T_c \sqrt{g(r_c)} = \left[r_h \sqrt{{\bar h}(r_h)}\right]^{-1}$ and 
${\bar h}(r_h) \approx {L^4 \over r_h^4}$ in this limit. This helps us to write everything in terms of ${\cal T}$ and 
not the scale dependent temperature $T_c$. In fact as we show below, once we incorporate the contributions from the 
UV cap, the $r_c$ dependence of the above formula will also go away and the final result will be completely independent 
of the cut-off. Note that in the limit $r_c \to \infty$ we recover the result for the cascading theory.    

Combining all the ingredients together, the contribution to the viscosity in the limit where 
(${\cal G}(r), {\cal H}(r), {\cal K}(r), ...$) etc are non-zero can now be presented succinctly as (although 
$\eta_1$ below doesn't have any real meaning on the gauge theory side as this is an intermediate quantity):
\bg\label{vishu} 
\eta_1 = {r_h^5 \sqrt{{\bar h}(r_h)}\over 2 g_s^2 G_N} \left\{ {1 + {r_c^5 g(r_c) \over 4 r_h^4} 
\left[{{\cal H}' \over 1 + {\cal G}} - {{\cal H}{\cal G}' \over (1+{\cal G})^2}\right]\over 
4\pi + {1\over \pi} \left[{\rm log}~g(r_c) + {{\cal H} \over 1 + {\cal G}}\right]^2}\right\}
\nd
Note that the above expression is exact for our background at least in the limit 
where we take the leading order $r^5$ singularity 
of the background. This is motivated from our detailed discussion that we gave in the previous section. Note that 
the second term in the action (\ref{SV-9}) is exactly the second equation of the set (\ref{infinity}) whose 
singularity structure has been shown to be renormalizable. Thus taking the leading order singularity $r^5$ instead of 
the actual $r_{(\alpha)}^5$ will not change anything if we carefully compensate the coefficients with appropriate
$g_sN_f, g_sM^2/N$ factors!  
 
But this is still not the complete expression as we haven't 
added the contributions from the UV cap. Before we do that, we want to re-address
the singularity structure of the
above expression. The worrisome aspect is the existence of $r_c^5$ factor in (\ref{vishu}). Does that create a problem for
our case? 

The answer turns out to be miraculously no, because of the form of ${\cal H}$ and ${\cal G}$ given in (\ref{SV-13c1}) 
and (\ref{SV-13a2}). This, taking only the leading powers of $r_c$, yields:
\bg\label{sigcal}
{\cal H}' ~ = ~ -{4{\tilde b}_1 \over r_c^5} -{8{\tilde b}_2 \over r_c^9} + ...., 
~~~~{\cal G}' ~ = ~ -{4{\tilde a}_1 \over r_c^5} -{8{\tilde a}_2 \over r_c^9} + .... 
\nd
killing the $r_c^5$ dependence in (\ref{vishu})\footnote{It is now easy to see why ${\tilde b}_1 = {\tilde a}_1 = 0$ is
consistent. For non-zero ${\tilde b}_1, {\tilde a}_1$ there would have been additional ${\rm log}~r$ terms from 
$r_{(\alpha)}^5$. These would have made the theory non-renormalizable. Thus holographic renormalizability would 
demand ${\tilde b}_1 = {\tilde a}_1 = 0$ from the very beginning $-$ consistent with what we choose earlier.}. 
This would make $\eta_1$ completely finite and all the $r_c$ 
dependences would go as ${\cal O}(1/r_c)$. Therefore we expect the contribution to the viscosity from the UV cap to 
go like:
\bg\label{eta1n}
\eta_2 ~ \equiv ~ \eta\big\vert_{r_c}^\infty ~ = ~ \sum_{i = 0}^\infty ~{G_i \over r_c^{4i}}
\nd
where the total viscosity will be defined as $\eta \equiv \eta_1 + \eta_2$. As this is a physical quantity  
we expect it to be independent of the scale. Therefore 
\bg\label{pq}
 {\partial \eta\over \partial r_c} ~ = ~ 0
\nd
which will give us similar Callan-Symanzik type equations, as discussed in the previous section,
from where we could derive the precise forms for 
$G_i$ in \ref{eta1n}. Finally when the dust settles, the result for shear viscosity can be expressed as:
\bg\label{shearvo} 
\eta ~ = ~ {{\cal T}^5 \sqrt{{\bar h}({\cal T})}\over 2 g_s^2 G_N} \Bigg[{1 + \sum_{k = 1}^\infty
\alpha_k e^{-4k {\cal N}_{\rm uv}} \over 4\pi + {1\over \pi}~{\rm log}^2 \left(1 - 
{\cal T}^4 e^{-4{\cal N}_{uv}}\right)}\Bigg]
\nd
where $\alpha_k$ are functions of ${\cal T}$ that can be easily determined from the coefficients 
(${\bar a}_i, {\bar b}_i$) in (\ref{SV-13a2}) and (\ref{SV-13c1}) or ($a_i, b_i$) worked out in Appendix E of \cite{FEP}; and
${\bar h}({\cal T}) \equiv {L^4\over {\cal T}^4} + {\cal O}(g_s, N_f, M)$.  
Observe that the final result for shear viscosity is completely independent of $r_c$ and $T_c$; and only depend 
on ${\cal T}$ and the degrees of freedom at the UV i.e through $e^{-{\cal N}_{uv}}$. Needless to say, for large enough 
${\cal N}_{uv}$ (which is always the case for our case because ${\cal N}_{uv} \to \epsilon^{-n}, n \ge 1$), 
the shear viscosity is only sensitive to the characteristic 
temperature ${\cal T}$ of the cascading theory. The interesting thing however is that the shear viscosity with finite
but large enough ${\cal N}_{uv}$ can be {\it smaller} than or {\it equal to}
the shear viscosity with ${\cal N}_{uv} \to \epsilon^{-n}, n >> 1$ i.e for the 
parent cascading theory provided:
\bg\label{smlim}
\alpha_k ~\le ~ {1\over 4\pi^2} \sum_{n \in {\bf Z}} ~{{\cal T}^{4k}\over n(k-n)}, ~~~~~ n \le k, ~~ k \in {\bf Z}
\nd
in the limit of small characteristic temperature ${\cal T}$. 
This will have effect on the viscosity to entropy ratio, to which we turn next.

\subsection{$\eta/s$}

Going back to the stress tensor of fluid described by hydrodynamics, one can obtain \cite{FEP}   
\be
\ave{\delta T_{ij}} 
=
-{\eta\over \varepsilon + P}
\left(\nabla_i\ave{T_j^0} + \nabla_j\ave{T_i^0} - {2\over
3}\delta_{ij}\nabla_l \ave{T^{l0}}\right)
-{\zeta\over \varepsilon + P}
\delta_{ij} \nabla_l \ave{T^{l0}}
\ee
where $\delta T^{ij}$ is the deviation from the ideal fluid stress
tensor $T^{ij}$ in the fluid rest frame and $\varepsilon$ and $P$ are the local
energy density and the pressure, respectively.
Using the thermodynamic identity
$Ts = \varepsilon + P$ where $s$ is the entropy density,
the two coefficients can be also written as 
$\eta/Ts$ and $\zeta/Ts$.
Since the temperature is the only relevant energy scale in the highly
relativistic fluid, one can easily see that the importance of the viscous
terms depends on the size of the dimensionless ratios $\eta/s$ and $\zeta/s$. Thus not the absolute value of shear viscosity rather the ratio
$\eta/s$ is what is relevant in  viscous hydrodynamics. 

Furthermore, the dimensionless number $\eta/s$ has been shown to be universal for a
large class of gauge theories where the dual graviton perturbation is a scalar field \cite{Nabil-1} and 
even more strikingly $\eta/s=1/4\pi$ has been conjectured to be 
its lowest possible value \cite{kovtun-1} for {\it any} fluid using arguments of \cite{PSS}. Although it has been shown by 
\cite{FEP}\cite{Kats}-\cite{etasv7} 
that the lower bound can be violated for gauge theories which have higher derivative terms (i.e. curvature square terms and beyond) in their
dual gravity, these theories may violate other constraints and may not be most relevant for physics of fluids. 
In any case, the ratio $\eta/s$ becomes crucial in describing fluid dynamics and in this section we will compute this ratio using dual
gravity.
     
We will calculate the ratio for two cases i.e one with 
only RG flow, and the other with both RG flow and curvature squared corrections. As usual the former is easier to handle
so we discuss this first. 

Starting with the type IIB supergravity action (\ref{action1}) in ten dimension,  
the entropy is given by Wald's formula \cite{wald1},\cite{wald2},\cite{wald3},\cite{wald4}
\bg \label{e2}
&& {\cal S}=-2\pi \oint dxdydz d^5{\cal M} \sqrt{{\cal P}}\frac{\partial{{\cal L}_{10}}}{\partial {R}_{abcd}}\epsilon^{ab}
\epsilon^{cd}
\nd
where the integral is over the eight dimensional surface of the horizon at $r=r_h,{\cal L}_{10}$ is the Lagrangian density of the action in
(\ref{action1}),
${\cal P}_{ab},a,b=1..8$ 
 is the induced $8\times 8$ metric at horizon, $\epsilon_{ab}$ is the bi normal normalized to  
$\epsilon_{ab}\epsilon^{ab}=-2$. Finally using
 explicit expression for the metric (\ref{bhmetko}) with warp factor given by  (\ref{hvalue}) near horizon, we have
 \bg \label{e2a}
 s &= &
{{\cal S}\over V_3} = -\frac{\pi r_h^5}{108 V_3\kappa_{10}^2} \oint dxdydz ~d^5{\cal M}~
{\rm sin}~\theta_1~{\rm sin}~\theta_2 ~\sqrt{h(r_h,\theta_1,\theta_2)}~\frac{\partial{{\cal L}_{10}}}{\partial {R}_{abcd}}\epsilon^{ab}
\epsilon^{cd}\nonumber\\
&= & \frac{r_h^3 L^2}{2 g_s^2 G_N} \Bigg\{1+\frac{3g_s M_{\rm eff}^2}{2\pi N}\left[1+\frac{3g_s N_f^{\rm eff}}{2\pi}
\left({\rm log}~r_h+\frac{1}{2}\right)-\frac{g_s N_f^{\rm eff}}{4\pi}\right]
{\rm log}~r_h\Bigg\}^{1/2}
\nd
where $V_3$ is the infinite three dimensional volume 
and we have used the definition of five dimensional Newton's constant $G_N$ introduced in (\ref{nc}).
 
Once we replace $r_h$ by the characteristic temperature ${\cal T}$, we see that the entropy is only sensitive to the 
temperature and is independent of any other scale of the theory. Since the above result is also independent of 
${\cal N}_{uv}$ it would seem that the Wald formula only gives the entropy for the theory 
with ${\cal N}_{uv} = \infty$ i.e for the parent 
cascading theory\footnote{This can be argued by observing that fact that in 
a renormalizable theory, like ours, the 
dependences on degrees of freedom go like ${\cal O}\left(e^{-{\cal N}_{uv}}\right)$ corrections as we saw in 
the previous sections.}. The interesting question now would be to ask what is the entropy for the theory whose UV 
description is different from the parent cascading theory? In other words, what is the effect of the UV cap attached 
at $r = r_c$ on the entropy? 

To evaluate this, observe first that in finite temperature 
gauge theory, entropy density of a thermalized medium having stress tensor 
$\langle T^{\mu\nu}\rangle ={\rm diagonal} (\epsilon,P,P,P)$ is given by
\bg \label{entropyC1}
s=\frac{\epsilon +P}{T}
\nd 
where $\epsilon$ is the energy density, $P \equiv P_x=P_y=P_z$ is the pressure of the medium and 
T being the temperature. With our gravity dual
we can compute the stress tensor $\langle T^{pq}_{\rm med}\rangle$
(and thus the energy $\epsilon= \langle T_{\rm med}^{00}\rangle$ and 
the pressure $P= \langle T^{11}_{\rm med}\rangle$) of the medium through equation of the form (\ref{KS17}), i.e
\bg \label{entropyC2}
\langle T^{pq}_{\rm med}\rangle = \frac{\delta_b {\bf S}_{\rm total}}{\delta_b {\bf g}_{pq}}
\nd 
where again $p,q=0,1,2,3$ and ${\bf g}_{pq}$ is the four dimensional metric obtained from  
the ten dimensional OKS-BH metric
$g_{ij},i,j=0,1..,9$; and $\delta_b$ operation has been defined earlier. 
There are two ways by which we could get a four-dimensional metric from the corresponding 
ten-dimensional one. The first way is to integrate out the $\theta_i, \phi_i$ directions to get the four-dimensional 
effective theory. This is because the warp factor for our case is dependent on the $\theta_i$ directions. The second way 
is to work on a slice in the internal space. The slice is coordinated by choosing some specific values for the internal
angular coordinates. Such a choice is of course ambiguous, and we can only rely on it if the physical quantities that we 
want to extract from our theory is not very sensitive to the choice of the slice. Clearly the first way is much more 
robust but unfortunately not very easy to implement. We will therefore follow the second way by choosing the  
the five dimensional slice 
as $\theta_1=\theta_2=\pi, \psi=\phi_1=\phi_2=0$
and thus obtaining 
\bg \label{entropyC2}
{\bf g}_{\mu\nu}~\equiv~ g_{\mu\nu}(\theta_i=\pi,\psi=\phi_i=0)
\nd 
with $\mu,\nu=0,1,2,3,4$. The next step would be to evaluate all the fluxes and the axio-dilaton on the slice. To do this 
we define:
\bg \label{entropyC3}
&&|{\bf H}_3|^2~=~ |H_3|^2(\theta_i=\pi,\psi=\phi_i=0);~~~~~~|{\bf F}_3|^2~ = ~|\widetilde{F}_3|^2(\theta_i=\pi,\psi=\phi_i=0)\nonumber\\
&& |{{\bf F}}_5|^2~ =~ |\widetilde{F}_5|^2(\theta_i=\pi,\psi=\phi_i=0);~~~~~~~|{\bf F}_1|^2~=~|F_1|^2(\theta_i=\pi,\psi=\phi_i=0)\nonumber\\
&&{\bf \Phi}~=~\Phi(\theta_i=\pi,\psi=\phi_i=0)
\nd
Once the fluxes have been defined, we need the description for ${\bf S}_{\rm total}$ which gives rise to (\ref{entropyC2}). This is easily
obtained from (\ref{action1}) as:
\bg \label{entropyC3}
{\bf S}_{\rm total}&=&\frac{1}{2\kappa_{5}^2}\int d^{5}x~e^{-2{\bf \Phi}} \sqrt{-{\bf g}}
\Bigg({\bf R}-4\partial_i {\bf \Phi} \partial^j {\bf \Phi}-\frac{1}{2}|{\bf H}_3|^2\Bigg)\nonumber\\
&-&\frac{1}{2\kappa^2_5} \int d^{5}x
\sqrt{-{\bf g}}\Bigg(|{\bf F}_1|^2+|{{\bf F}}_3|^2+\frac{1}{2}|{\bf F}_5|^2\Bigg)
\nd
 with ${\bf R}$ being the Ricci-scalar for ${\bf g}_{\mu\nu}$ and 
${\bf g}={\rm det}~{\bf g}_{\mu\nu}$.
Note that in the definition for the {\it slice} sources ${\bf H}_3, {\bf F}_1, {\bf F}_3,
{\bf F}_5$ and ${\bf R}$, 
we still have $g_{ij}, i,j\geq 5$ which we evaluate at 
$\theta_i=\pi, \psi=\phi_i=0$, treating them simply as functions and not metric degrees of freedom. 

To complete the background we need the line element. Here we will encounter some subtleties regarding the choice of the 
black-hole factors and the corresponding $g_s N_f$ type corrections to them. 
With the definition of ${\bf g}_{\mu\nu}$ the line element is:
\bg \label{entropyC5}
&& ds^2~=~-\frac{\bar{g}_1(r)}{\sqrt{h(r,\pi,\pi)}}dt^2+\frac{\sqrt{h(r,\pi,\pi)}}{\bar{g}_2(r)}dr^2+\frac{1}{\sqrt{h(r,\pi,\pi)}}d\overrightarrow{x}^2\\
&&\bar{g}_1(r)~=~g_1(r,\theta_1=\pi,\theta_2=\pi)=1-\frac{r_h^4}{r^4}+\sum_{i,j=0}^{\infty}\alpha_{ij}\frac{{\rm log}^i(r)}{r^j} = 1 + \sum_{j, \alpha} {\sigma_j^{(\alpha)}\over r^j_{(\alpha)}}\nonumber\\
&&\bar{g}_2(r)~=~g_2(r,\theta_1=\pi,\theta_2=\pi)=1-\frac{r_h^4}{r^4}+\sum_{i,j=0}^{\infty}\beta_{ij}\frac{{\rm log}^i(r)}{r^j}= 1 + \sum_{j, \alpha} {\kappa_j^{(\alpha)}\over r^j_{(\alpha)}}\nonumber
\nd
where $\alpha_{ij},\beta_{ij}$ are all of ${\cal O}(g_sN_f,g_sM)$ and only involve the parameters of the theory 
namely, $r_h, L$ and $\mu$ from the
embedding equation (\ref{seven}) and guarantees that $\frac{\alpha_{ij}}{r^j},\frac{\beta_{ij}}{r^j}$ are 
dimensionless. On the other hand $\sigma_j^{(\alpha)}, \kappa_j^{(\alpha)}$ can incorporate zeroth orders in $g_sN_f$. 
However note that so far we have been assuming $g_1(r) \approx g_2(r) = g(r)$, ignoring their 
inherent $\theta_i$ dependences, and also the inequality stemming from the choices of $\alpha_{ij}$ and $\beta_{ij}$. 
This will be crucial in what follows, so we will try to keep the black hole factors unequal. These considerations do 
not change any of our previous results of course. 


Now looking at the form of the metric, knowing the warp factor $h(r,\pi,\pi)$ and $\bar{g}_i(r)$, just like before we 
can expand the line
element as $AdS_5$ line element plus ${\cal O}(g_sN_f,g_sM)$ corrections. We can then rewrite the line element 
(\ref{entropyC5}) as: 
\bg \label{entropyC6}
&&ds^2~=~ -\frac{r^2}{L^2}\left[g(r)+l_1\right]dt^2+\frac{\sqrt{h(r,\pi,\pi)}}
{\bar{g}_2(r)}dr^2+\frac{r^2}{L^2}\left(1+l_2\right) d\overrightarrow{x}^2\nonumber\\
&&l_1(r)~=~\sum_{i,j=0}^{\infty}\gamma_{ij}\frac{{\rm log}^i(r)}{r^j}\nonumber\\
&&l_2(r)~=~\sum_{i,j=0}^{\infty}\zeta_{ij}\frac{{\rm log}^i(r)}{r^j}
\nd 
where again $\gamma_{ij},\zeta_{ij}$ are of ${\cal O}(g_sN_f,g_sM)$ and we are taking $h(r, \pi, \pi) =  
{L^4\over r^4} + {\cal O}(g_sN_f,g_sM)$. Such a way of writing the local line element tells us that there are two 
induced four-dimensional metrics at any point $r$ along the radial direction:
\bg\label{indme}
{\bf g}^{(0)}_{pq} ~\equiv~ {\rm diagonal}~(-g(r),1, 1, 1), ~~~~~~~ 
{\bf g}^{(1)}_{pq} ~\equiv~ {\rm diagonal}~(-l_1, l_2, l_2, l_2)
\nd
where we haven't shown the $r^2/L^2$ dependences. The reason for specifically isolating the four-dimensional part is to 
show that we can
study the system from boundary point of view where the dynamics will be governed by our choice of the boundary degrees 
of freedom. It should also be clear, from four-dimensional point of view, the metric choice ${\bf g}^{(0)}_{pq}$ 
is directly related to the AdS geometry whereas the other choice ${\bf g}^{(1)}_{pq}$ is the deformation due to 
extra fluxes and seven branes. 

The above decomposition also has the effect of simplifying our calculations of the energy momentum tensor 
$\langle T^{pq}_{\rm med}\rangle$. We can rewrite the total energy momentum tensor as the sum of two parts, one coming
from the AdS space and the other coming from the deformations, in the following way: 
\bg \label{entropyC7}
\langle T^{pq}_{\rm med} \rangle &=&\frac{\delta_b {\bf S^{[0]}}_{\rm total}}{\delta_b {\bf g}_{pq}^0}
~+ ~ \frac{\delta_b {\bf S}^{[1]}_{\rm total}}{\delta_b{\bf g}_{pq}^1}\nonumber\\
&\equiv& \langle T^{pq}_{\rm med}\rangle_{\rm AdS}~+ ~ \langle T^{pq}_{\rm med}\rangle_{\rm def} \nonumber\\
{\bf S} _{\rm total}&=&{\bf S}^{[0]}_{\rm total}~ + ~ {\bf S}^{[1]}_{\rm total}
\nd 
where ${\bf S}^{[0]}_{\rm total}$ is zeroth order in $g_sN_f,g_sM$ and  ${\bf S}^{[1]}_{\rm total}$ is higher order in 
$g_sN_f,g_sM$.
Note that $\langle T^{pq}_{\rm med}\rangle_{\rm AdS} =\frac{\delta_b {\bf S}^{[0]}_{\rm total}}
{\delta_b {\bf g}_{pq}^0}$ 
is the well known AdS/CFT result obtained from the analysis of \cite{kostas1}-\cite{kostas5}
in the limit $r_c \to \infty$. 
With the
${\cal O}(1/r)$ series expansion of our metric ${\bf g}_{00}^0=1-{r_h^4}/{r^4}, {\bf g}_{11}^0= {\bf g}_{22}^0
= {\bf g}_{33}^0=1$,
the result at the boundary is  
\bg \label{entropyC8}
\langle T^{00}_{\rm med}\rangle_{\rm AdS} &=& \frac{r_h^4}{2g_s^2 G_N}=\frac{{\cal T}^4}{2 g_s^2 G_N}\nonumber\\
\langle T^{mn}_{\rm med}\rangle_{\rm AdS} &=&0~~~~~~~~~~~~~m,n=1,2,3
\nd
This only gives the CFT stress tensor as we evaluate the tensor on the AdS boundary at infinity, reproducing the expected
first term of (\ref{e2a}). 
How do we then 
evaluate the ${\cal O}(g_s N_f, g_s M)$
contributions from the deformed AdS part i.e the energy momentum tensor 
$\langle T^{pq}_{\rm med}\rangle_{\rm def}$ at any $r = r_c$ cut-off in the geometry? 

In fact the procedure to evaluate exactly such a result has already been discussed in section {\bf 2.3}. 
Therefore without going into any details, the final answer
after integrating by parts, adding appropriate Gibbons-Hawkings terms and then using the equation of motion for
 ${\bf g}_{pq}^{[1]}$, we have
\bg \label{action1a} 
&& {\bf S}^{[1]}_{\rm total}~=~{1\over 8\pi G_N}\int \frac{d^4q}{(2\pi)^4 \sqrt{g(r_c)}}
\Bigg\{\Big[ \bar{C}^{mn}_1(r,q)-\bar{A}^{'mn}_1(r,q)\Big] \Phi^{[1]}_m(r, q) \Phi^{[1]}_n(r, -q)\nonumber\\
&&~~~~~~+\Big[\bar{B}^{mn}_1(r,q)-\bar{A}^{mn}_1(r,q)\Big] \Big[\Phi'^{[1]}_m(r, q) \Phi^{[1]}_n(r, -q)+\Phi^{[1]}_m(r, q)
\Phi'^{[1]}_n(r, -q)\Big]\nonumber\\
&&~~~~~~~~~~~~~~~~ +\Big(\bar{E}^m_1 -\bar{F}'^m_1\Big)\Phi^{[1]}_m(r, q)\Bigg\}\Bigg{|}_{r_h}^{r_c}
\nd
 The values of the coefficients are given in Appendix F of \cite{FEP}. 
The above form is
exactly as we had before, and so all we now need is to get the mode expansion for $\Phi^{[1]}_m$. Note however that 
the subscript $m$ can take only two values, namely 
$m =0,1$ as there are only two distinct fields ${\bf g}_{00}^{[1]}$ and 
${\bf g}_{11}^{[1]}= {\bf g}_{22}^{[1]}={\bf g}_{33}^{[1]}$. Therefore our proposed mode expansion is:
\bg\label{mexp}
\Phi^{[1]}_m~ =~ {\bf g}_{mm}^{[1]}~= ~\sum_{\alpha}\sum_{i=0}^{\infty}
\frac{{\bf s}_{mm}^{(i)[\alpha]}}{r_{c(\alpha)}^i}
\nd
Just like our analysis in section {\bf 2.4}, the action in (\ref{action1a}) is
divergent due to terms of ${\cal O}(r_c^4), {\cal O}(r_c^3)$ and hence we need to renormalize the action. 
The equations for renormalization are identical to the set of equations
(\ref{cterms})$-$(\ref{ymyn}), and therefore we analogously subtract 
the counter terms to obtain the following renormalized action:
\bg \label{actionren1a}
&& {\bf S}^{[1]}_{\rm ren}= {1\over 8\pi G_N} 
\int \frac{d^4q}{(2\pi)^4}\left[1-{r_h^4\over r_c^4}\right]^{-{1\over 2}} \sum_{\alpha, \beta}
\Bigg\{\left(\sum_{i=0}^{\infty}{\widetilde{{\cal A}}^{(\alpha)}_{mn(i)[1]}\over r_{(\alpha)}^i}\right) 
\widetilde{{\cal G}}^{mn[1]} \Phi_m
\Phi_n \nonumber\\ && + X[r_{(\alpha)}]
+ \left(\sum_{i=0}^{\infty}{\widetilde{{\cal E}}^{(\alpha)}_{mn(i)[1]}\over r_{(\alpha)}^i}\right) 
\widetilde{{\cal M}}^{mn[1]} (\Phi_m
\Phi'_n+\Phi'_m \Phi_n) 
+{H}_{\vert\alpha\vert}^{mn[1]}\Big[s_{nn}^{(4)[\beta]}\Phi_m +
{s}_{mm}^{(4)[\beta]}\Phi_n\Big] \nonumber\\
&& +{K}_{\vert\alpha\vert}^{mn[1]}\Big[-4\tilde{s}_{nn}^{(4)[\beta]}\Phi_m 
-4 {s}_{mm}^{(4)[\beta]}\Phi_n+{s}_{nn}^{(5)[\beta]}\Phi'_m+{s}_{mm}^{(5)[\beta]}\Phi'_n\Big]
+ \left(\sum_{i=0}^{\infty}\frac{\widetilde{b}^{(\alpha)}_{m(i)[1]}}{r_{(\alpha)}^i}\right) 
\Phi_m\Bigg\}\nonumber\\
\nd
where the radial coordinate is measured at the two boundaries $r_h$ and $r_c$ and $\Phi_m$ are independent 
of $r$ as before. Note
that $X[r_{(\alpha)}]$ is a function independent of $\Phi_m$ and appears for generic renormalized action.

Now the generic form for the energy momentum tensor is evident from looking at the linear terms in the 
above action (\ref{actionren1a}). This is again the same as before. However now we also need the entropy from the 
energy-momentum tensor as in (\ref{entropyC1}). The result for the energy-momentum tensor   
at $r = r_c$ is given by:
\bg\label{energymeda}
&&{\langle T^{mm}_{\rm med}\rangle_{\rm def}}\equiv 
{1\over 8\pi G_N} \int \frac{d^4q}{(2\pi)^4}{1\over \sqrt{g(r_c)}}\sum_{\alpha, \beta} 
\Bigg[({H}_{\vert\alpha\vert}^{mn[1]}+ {H}_{\vert\alpha\vert}^{nm[1]}){s}_{nn}^{(4)[\beta]} 
-4({K}_{\vert\alpha\vert}^{mn[1]} \nonumber\\
&&+{K}_{\vert\alpha\vert}^{nm[1]}){s}_{nn}^{(4)[\beta]}+({K}_{\vert\alpha\vert}^{mn[1]}+
{K}_{\vert\alpha\vert}^{nm[1]}){s}_{nn}^{(5)[\beta]}
+\left(\sum_{i=0}^{\infty}\frac{\widetilde{b}^{(\alpha)}_{n(i)[1]}}{r_{c(\alpha)}^i}\right)
 \delta_{nm}\Bigg]
\nd
The  explicit expressions for the coefficients
listed above, namely,
${H}_{\vert\alpha\vert}^{mn[1]}, 
{K}_{\vert\alpha\vert}^{mn[1]},\widetilde{b}^{(\alpha)}_{n(i)[1]}$ and ${s}_{nn}^{(i)[1]}$ are given
in Appendix F of \cite{FEP}.

To complete the story we need the contribution from the UV cap. This is similar to our earlier results. The final 
expression for the ratio of the energy-momentum tensor to the temperature takes the simple form:
\bg\label{finent} 
&&{\langle T^{mm}_{\rm med}\rangle_{\rm def}\over T_b} \equiv 
{\pi {\cal T}\sqrt{h({\cal T})}\over 8\pi G_N}\int \frac{d^4q}{(2\pi)^4}\sum_{\alpha, \beta} 
\Big[({H}_{\vert\alpha\vert}^{mn[1]}+ 
{H}_{\vert\alpha\vert}^{nm[1]}){s}_{nn}^{(4)[\beta]} -4({K}_{\vert\alpha\vert}^{mn[1]} \nonumber\\
&&+{K}_{\vert\alpha\vert}^{nm[1]}){s}_{nn}^{(4)[\beta]}+({K}_{\vert\alpha\vert}^{mn[1]}+
{K}_{\vert\alpha\vert}^{nm[1]}){s}_{nn}^{(5)[\beta]}
+ \sum_{j=0}^{\infty}~{\widetilde{b}^{(\alpha)}_{n(j)[1]}}\delta_{nm} {e^{-j{\cal N}_{uv}}} 
\Big]
\nd
We would like to make a few comments here: First, observe that the final result is independent of our choice of 
cut-off. Secondly, in the string frame there should be a $1/g_s^2$ dependence. Finally, we can pull out a 
${\cal T}^4$ term because the coefficients have an explicit $r_h^4$ dependences (see Appendix F of \cite{FEP}). This means that 
both from the AdS and the deformed calculations performed above we can show that the entropy is of the form:
\bg\label{entad}
s ~ = ~ {{\cal T}^5 \sqrt{h({\cal T})}\over 2 g_s^2 G_N}
\left[1 + {\cal O}\left(g_sN_f, g_s M, e^{-{\cal N}_{uv}}\right)\right]
\nd 
where the first part is from (\ref{entropyC8}) and the second part is from (\ref{finent}). The result for the parent 
cascading theory is (\ref{e2a}), and so we should regard (\ref{entad}) as the entropy for the theory with 
${\cal N}_{uv}$ degrees of freedom at the boundary. Of course in the limit ${\cal N}_{uv} \to \epsilon^{-n}, n >> 1$ 
we should recover
the entropy formula (\ref{e2a}) for the parent theory. All in all we see that the correction due to 
${\cal N}_{uv}$ degrees of freedom only goes as $e^{-{\cal N}_{uv}}$, so in practice this is always small for the 
type of ${\cal N}_{uv}$ that we consider here.  
This means that we can use the entropy for the parent cascading theory to estimate the viscosity by entropy ratio 
for a system with ${\cal N}_{uv}$ degrees of freedom at the UV as:
\bg \label{final} 
\frac{\eta}{s}~=~ {\left[{1 + \sum_{k = 1}^\infty
\alpha_k e^{-4k {\cal N}_{\rm uv}} \over 4\pi + {1\over \pi}~{\rm log}^2 \left(1 - 
{\cal T}^4 e^{-4{\cal N}_{uv}}\right)}\right]}
\nd
\noindent where we see that the boundary entropy term (\ref{e2a}) neatly cancels the ${\cal T}^3$ coefficient in the 
viscosity (\ref{shearvo}) to give us the precise bound of ${1\over 4\pi}$ when ${\cal N}_{\rm uv} \to \infty$. Of 
course from our other analysis (\ref{entad}) we might expect a
${\cal O}\left(g_sN_f, g_s M, e^{-{\cal N}_{uv}}\right)$ contribution that 
would make 
(\ref{final}) saturate the celebrated bound ${1\over 4\pi}$ if the total entropy density factors 
compensate the factors coming from the viscosity. This would seem consistent with, for example,  
\cite{Nabil-1}\footnote{Provided of course if we assume that $\alpha_k$'s are more general now, being functions of 
${\cal T}, g_sM, g_sN_f$. This way even for non-zero $M, N_f$, whenever we have ${\cal N}_{uv} \to \infty$ the 
bound is exactly ${1\over 4\pi}$.}. 
In fact our conjecture would be for non-zero $M, N_f$ and 
${\cal N}_{uv} \to \epsilon^{-n}, n >> 1$, the bound is exactly saturated\footnote{Note that any possible deviations
from ${1\over 4\pi}$ due to (\ref{sigcal}) in (\ref{vishu}) {\it cannot} happen because the underlying holographic 
renormalizability will make ${\tilde a}_1 = {\tilde b}_1 =0$, as discussed earlier. Thus the bound in itself is a rather 
strong result.}  
i.e ${\eta\over s} = {1\over 4\pi}$.

Our second and final step would be to 
incorporate both the RG flow as well as curvature square corrections. As we 
discussed before the curvature squared corrections are typically of the form
$c_3 R_{\mu\nu\rho\sigma}R^{\mu\nu\rho\sigma}$ with $c_3$ being the coefficient (\ref{coeffeven}) 
that we computed before. 

The crucial point here is that (see \cite{Kats} where this has also been recently emphasized) in the presence of 
curvature squared corrections the five dimensional metric itself changes to:  
\bg \label{metricR^2}
ds^2=\frac{-g_1(r)}{\sqrt{h(r, \pi, \pi)}}dt^2+\frac{\sqrt{h(r, \pi, \pi)}}{g_2(r)}dr^2+\frac{d\overrightarrow{x}^2}{\sqrt{h(r, \pi, \pi)}}
\nd 
where the black hole factors $g_i$ are no longer given by (\ref{grdef}). 
They take the following forms:
\bg\label{grdefnow}
g_1(r)&=&1-\frac{r_h^4}{r^4}+\alpha + \gamma \frac{r_h^8}{r^4}+\widetilde{\alpha}_{mn}\frac{{\rm log}^m r}{r^n}\nonumber\\
g_2(r)&=&1-\frac{r_h^4}{r^4}+\alpha+\gamma\frac{r_h^8}{r^4}+ \widetilde{\beta}_{mn}\frac{{\rm log}^m r}{r^n}
\nd
where $\widetilde{\alpha}_{mn},\widetilde{\beta}_{mn}$ are all of ${\cal O}(g_sM,g_sN_f)$ and 
can be worked out with some effort (we will not derive their explicit forms here). Similarly we could also 
express (\ref{grdefnow}) in terms of inverse powers of $r$ to have good asymptotic behavior.    
Observe that we can still impose $g_1 \approx g_2$ because the corrections are to ${\cal O}(g_sN_f, g_s M)$, although 
all our previous analysis have to be changed in the 
presence of curvature corrections because the explicit values of $g_i(r)$ have changed.  
We will address these issues in our future work. Finally ($\alpha, \gamma$) 
are given by 
\bg\label{alga}
\alpha&=&\frac{4c_3\kappa}{3L^2};~~~~~~~~~\gamma=\frac{4c_3\kappa}{L^2} 
\nd
At this point one might get worried that the metric perturbation on this background would become very complicated. 
On the contrary our analysis becomes rather simple once we 
ignore terms of ${\cal O}(c_3g_sM, c_3g_sN_f)$ (which is a valid approximation with $c_3<<1$). In this limit 
the metric perturbation can be written simply as a {\it linear} combination of  the
terms proportional to $c_3$ in $\Phi$ which appears in \cite{Kats} and our solution
 (\ref{SV-13e}), (\ref{SV-13f}) derived for RG flow. The final result is:
 \bg \label{SV-13f1} 
\tilde{\phi}(r, \vert\omega\vert)_{\pm,R^2} &=& 1\pm i\frac{\vert\omega\vert}{4\pi T_c}
\Bigg\{{\cal H} (r) + [1+{\cal G}(r)] {\rm log}~g(r) + \frac{\alpha r^8+\gamma r_h^8}{r^8 g(r)}-\alpha+4\gamma
\frac{r_h^4}{r^4}\Bigg\}\nonumber\\
&+ & {\cal G}(r) - {\vert\omega\vert^2 \over 16 \pi^2 T^2_c}\Big\{{\cal K}(r) 
+ {\cal H}(r) ~{\rm log}~g(r) + [1 + {\cal G}(r)]
{\rm log}^2 g(r)\Big\}\nonumber\\
&&~~~~~~~~~~ + {\cal O}(\vert\omega\vert^3) +  {\cal O}(c_3 g_s N_f) + {\cal O}(c_3 g_s M)\nd
where we have written \ref{grdef} as;
\bg\label{mother}
&& g_1(r) ~=~ g(r) ~+~ {\cal O}(c_3 g_s N_f)~ +~ {\cal O}(c_3 g_s M)\nonumber\\ 
&& g_2(r) ~=~ g(r) ~+~ {\cal O}(c_3 g_s N_f) ~+~ {\cal O}(c_3 g_s M)
\nd
with $g(r)=1-\frac{r_h^4}{r^4}$ being the usual black hole factor. Of course as emphasized above, this is 
valid only in the limit $c_3 << 1$, which at least for our background seems to be the case (see \ref{coeffeven}).
 
The above corrections are not the only changes. The  
entropy computed earlier
also gets corrected and therefore the horizon can no longer be at $r=r_h$. To evaluate the 
correction to entropy we again ignore the terms of 
${\cal O}(c_3g_sM,c_3g_sN_f)$. In this limit the correction terms are precisely given by the analysis of \cite{Kats} and
are proportional to the $c_3$ factor (\ref{coeffeven}) as expected. 
This means that the final result for 
$\eta/s$ including all the ingredients i.e RG flows,
Riemann square corrections as well as the contributions from the UV caps; is given by:   
\bg \label{finala} 
\frac{\eta}{s} &=&~ {\left[{1 + \sum_{k = 1}^\infty
\alpha_k e^{-4k {\cal N}_{\rm uv}} \over 4\pi + {1\over \pi}~{\rm log}^2 \left(1 - 
{\cal T}^4 e^{-4{\cal N}_{uv}}\right)}\right]}\nonumber\\
&-&\frac{c_3\kappa}{3 L^2 \left(1- {\cal T}^4 e^{-4{\cal N}_{uv}}\right)^{3/2}}
 \left[\frac{{B_o}(4\pi^2-{\rm log}^2 ~C_o)+4\pi{A_o}~{\rm log}~C_o}{\Big(4\pi^2-{\rm
log}^2~C_o\Big)^2+16\pi^2~{\rm
log}^2~C_o}\right]
\nd
where we see two things: one, the bound is completely independent of the cut-off $r = r_c$ in the geometry, and two, 
the bound {\it decreases} in the presence of curvature square corrections {even} when 
${\cal N}_{uv} \to \epsilon^{-n}$ with $n = {\cal O}(1)$.\footnote{In \cite{Cherman:2007fj}, 
non-relativistic systems that appear to have no lower bound
were constructed.
However, these are systems which necessarily require large chemical
potentials and low temperature.
In highly relativistic system created at high energy colliders such as the
RHIC or the LHC,
chemical potentials are small and temperature is high. Our discussion here
assumes that the system under
discussion has such properties so that the use of thermodynamic identity
$\varepsilon + P = Ts$ is valid.
Hence, our discussion here is in no direct conflict with the models
constructed in \cite{Cherman:2007fj}.}
The constants appearing in \ref{finala} are defined as:
\bg\label{constants}  
{C_o}&=&1-{{\cal T}^4}{e^{-4{\cal N}_{uv}}}\nonumber\\
{A_o}&=&-18 {\cal T}^8 e^{-8{\cal N}_{uv}}+\left(3{\cal T}^8 e^{-8{\cal N}_{uv}}
- 47 {\cal T}^4 e^{-4{\cal N}_{uv}}\right){\rm log}~{C_o}+ 26 {\cal T}^4 e^{-4{\cal N}_{uv}} \nonumber\\
&+&24 \left(1 + {\cal T}^2 e^{-2{\cal N}_{uv}}\right){\rm log}~{C_o} \nonumber\\
{B_o}&=&- 88 \pi {\cal T}^8 e^{-8{\cal N}_{uv}}+ 48 \pi {\cal T}^4 e^{-4{\cal N}_{uv}} +48  
\nd
This 
is consistent with \cite{Kats}, and the only violation of $\eta/s$ may be entirely from the $c_3$ factor provided the 
increase in bound 
from the first term of (\ref{finala}) is negligible, as we discussed 
earlier for (\ref{final}). This means in particular:
\bg\label{jolba}
{\eta\over s} ~ = ~ {1\over 4\pi} ~-~ n_b c_3 ~+~ {\cal O}\left({\cal T}e^{-{\cal N}_{uv}}\right)
\nd
where $n_b$ can be extracted from (\ref{finala}).  
Here we will not 
study the subsequent implication of this result, for example
whether there exists a causality violation in our theory due to the curvature corrections as in \cite{etasv1}-\cite{etasv7}. 
We hope to 
address this in our future work.

\section{Confinement in QCD}
At low temperatures only color neutral states exist and there are no free color charges i.e. quarks in QCD at temperatures
below some critical value. The mechanism which may explain this confinement of quarks was first proposed by Wilson back in
the early 70's \cite{Wilson-1} where the only requirement is the existence of abelian or non-abelian gauge fields which are
strongly coupled to the matter fields that confine. The strong coupling calculation can be done using lattice gauge theory
and it turns out that at low temperatures one can show that the free energy of a couple of charges grow linearly with the
distance between them \cite{Susskind}\cite{Polyakov}. This means it takes infinite amount of energy to separate two color charges 
which are
strongly coupled to gauge fields and thus only color singlet combinations of charges have finite energy at low temperatures.
The linear behavior of free energy as a function of inter charge separation is referred  as  linear confinement. At high
temperatures the free energy takes the form of Coulomb potential \cite{Susskind}\cite{Polyakov} and thus the interaction between 
the 
charges is negligible for large distances. This results in a deconfined phase of matter and gauge fields at sufficiently 
large temperatures. 

 The linear confinement of quarks at large separation and low
temperatures is a strong coupling phenomenon and one uses lattice QCD  to compute the free energy of the bound state of
quarks. On the other hand at high enough temperatures, if the gauge coupling is weak, one can compute the free energy using
perturbative QCD and obtains Coulombic interactions between the quarks. 
Whether using perturbative QCD at weak coupling or 
lattice QCD and effective field theory techniques at strong couplings, one finds that at high temperatures, the Coulomb
potential is Debye screened  \cite{Gale-kapustaVqq}\cite{Susskind}.  

The study of heavy quark potential gathers special attention as it is linked to a possible signal from Quark Gluon
Plasma. In particular, heavy quarks are formed during early stages of a heavy ion collision as only then there exists enough
 energy for their formation . On the other hand right after the collision  QGP is formed as only then the temperature is  
 high enough for quarks to be deconfined.  As temperature goes down, heavy quarks form bound states and as these states are 
 very massive, 
they are formed at temperatures higher than deconfinement value.  This  means heavy quark bound states like  $J/\psi$ can 
coexist with QGP and can act as probes to the medium. 

In particular one can study the $J/\psi$ bound state formed in
proton-proton collisions and compare with heavy ion collision where a medium is formed. The medium will screen the
$c\bar{c}$ interaction making them less bound and eventually resulting in a suppression  of $J/\psi$ production. This
phenomenon is known as the $J/\psi$ suppression and considered as a signal of QGP formation \cite{Matsui-Satz-1}.                 

In order to quantify quarkonium suppression and analyze features of plasma that causes the suppression, one has to compute
free energy of $J/\psi$. At large couplings, lattice QCD calculations are most reliable
and there has been extensive studies quantifying the potential for heavy quarks \cite{Fkarsch-1}-\cite{HQG-4} . 
The effective potential between the quark anti-quark pairs separated by a distance $d$
at temperature ${\cal T}$
can then be expressed 
succinctly in terms of the free energy $F(d, {\cal T})$, which generically takes the following form:
\bg\label{freeenergy}
F(d, {\cal T}) = \sigma d ~f_s(d, {\cal T}) - {\alpha\over d} f_c(d, {\cal T})
\nd
where $\sigma$ is the string tension, $\alpha$ is the gauge coupling and $f_c$ and $f_s$ are the screening 
functions\footnote{We expect the screening functions $f_s, f_c$ to equal identity when the temperature goes to 
zero. This gives the zero temperature Cornell potential.} 
(see for example \cite{Fkarsch-1}-\cite{Fkarsch-5} and references therein). At zero temperature free energy is the potential energy of the pair.
On the other hand free energy or potential energy of quarkonium is related to the Wilson loop which we will elaborate here. 

Consider the Wilson loop of a rectangular path ${\cal C} $ with space like width $d$ and time like length $T$.
The time like
paths can be thought of as world lines of pair of quarks $Q\bar{Q}$ separated by a spatial distance $d$. 
Studying the expectation value of the Wilson loop in the
limit $T\rightarrow \infty$, one can show that it behaves as 
\bg \label{WL-1}
\langle W({\cal C})\rangle ~\sim~ {\rm exp}(-T E_{Q\bar{Q}})
\nd   
where $E_{Q\bar{Q}}$ is the energy of the $Q\bar{Q}$ pair which we can identify with their 
potential energy $V_{Q\bar{Q}}(d)$ as the quarks are static. At this
point we can use the principle of holography \cite{Maldacena-1} \cite{Witten-2} \cite{Mal-2} 
and identify the expectation value of the Wilson loop with 
the exponential of the {\it renormalized} Nambu-Goto action,
\bg \label{Holo-1}
 \langle W({\cal C})\rangle ~\sim ~ {\rm  exp}(-S^{\rm ren}_{{\rm NG}}) 
\nd
with the understanding that ${\cal C}$ is now the boundary of string world sheet. 
Note that we are computing Wilson loop of gauge theory living on
flat four dimensional space-time $x^{0, 1, 2, 3}$. Whereas the string world sheet is embedded in curved five-dimensional 
manifold with coordinates
$x^{0, 1, 2, 3}$ and $r$. We will identify the five-dimensional manifold with Region 3 that we discussed in section {\bf
2.4}. For the
correspondence in (\ref{Holo-1}) to be valid, we need the t'Hooft  coupling which is the gauge coupling in the theory to be
large. On the other hand as discussed before, it is in this regime of strong coupling that linear confinement is realized in
gauge theories. Thus using gauge/gravity duality is most appropriate in computing quarkonium potential.  
 
To be consistent with the recipe in \cite{Witten-2}, we need to make sure that the induced four dimensional metric 
at the boundary
of the string world sheet ${\cal C}$ is flat. For an AdS space, this is guaranteed as long as the world sheet ends on 
boundary of AdS space
where the induced four dimensional metric can indeed be written as $\eta_{\mu\nu}$. 
Using the geometry  constructed in section {\bf 2.4}, with metric of the form (\ref{bhmetko}) and warp factor given by
(\ref{heaft}) for large $r$,
we see that the metric is asymptotically AdS and therefore induces a flat Minkowski metric at the boundary via:
\bg\label{induce}
\lim_{u \rightarrow 0}~ u^2 g_{\mu\nu}~ = ~ \eta_{\mu\nu}
\nd
where $u = r^{-1}$ and $g_{\mu\nu}$ is the full metric (including the warp factor) in Region 3. 
Thus we can
make the identification (\ref{Holo-1}). Once this subtlety is resolved, comparing (\ref{WL-1}) and (\ref{Holo-1}) 
we can read off the 
potential  
\bg \label{Vqq}
V_{Q\bar{Q}}~ = ~ \lim_{T \to \infty} \frac{S^{\rm ren}_{{\rm NG}}}{T}
\nd      
Thus knowing the renormalized 
string world sheet action, we can compute $V_{Q\bar{Q}}$ for a strongly coupled gauge theory.

For non-zero temperature, the free energy is related to  the Wilson lines $W\left(\pm {d\over 2}\right)$
 via:
\bg\label{wlfe}
{\rm exp}\left[-{F(d, {\cal T})\over {\cal T}}\right] ~ = ~ 
{\langle W^\dagger\left(+{d\over 2}\right) W\left(- {d\over 2}\right)\rangle \over 
\langle W^\dagger\left(+{d\over 2}\right)\rangle \langle W\left(-{d\over 2}\right)\rangle}
\nd
In terms of Wilson loop, the free energy (\ref{freeenergy}) is now related to the renormalized Nambu-Goto 
action for the string on a background with a black-hole\footnote{There is a big literature on the 
subject where quark anti-quark potential has been computed using various different approaches like 
pNRQCD \cite{HQG-1}$-$\cite{HQG-4}, hard wall AdS/CFT \cite{polstrass}$-$\cite{boschi4} and other 
techniques \cite{reyyee}$-$ \cite{cotrone2}. Its reassuring
to note that the results that we get using our newly constructed background matches very well with the results
presented in the above references. This tells us that despite the large $N$ nature there is an underlying 
universal behavior of the confining potential.}. One may also note that the theory we get is a four-dimensional theory 
{\it compactified} on a circle in Euclideanised version and not a three-dimensional theory.   

\subsection{Computing the Nambu-Goto Action: Zero Temperature} 
Our first attempt to compute the NG action would be to consider the zero temperature for the field theory. This means that we take
the black hole factors $g_i$ in (\ref{bhmetko}) to be identity. The string configuration we take to compute the action    
is 
shown in Fig {\bf 3.2}. Note that we are considering the case when the string is exclusively in 
region 3 of geometry shown in Fig {\bf 2.17} and the justification for this will be given shortly. Even if the string enters region 2 and 1, 
the arguments for linear confinement
still holds provided the warp factor is monotonic and satisfies a certain equation. We will discuss it later in this section. 
For the time being observe that the configuration in 
Fig {\bf 3.2} has one distinct advantage over all other configurations studied in the literature, namely, that 
because of the absence of three-forms in region 3, we will not have the UV divergence of the Wilson loop 
due to the logarithmically varying $B$ field \cite{cschu}. 

\begin{figure}[htb]\label{wilsonloop}
		\begin{center}
\includegraphics[height=6cm]{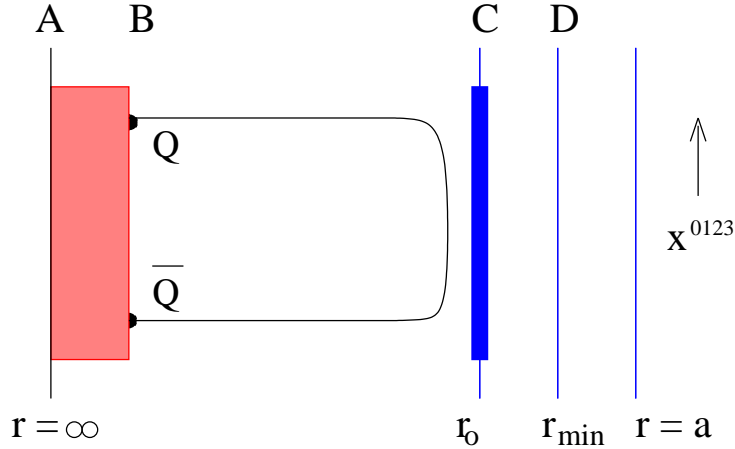}
		\caption{{The string configuration that we will use to evaluate the Wilson loop in the dual 
gauge theory. The line $A$ determines the actual boundary, with the line $B$ denoting the extent of the seven brane.
We will assume that line $B$ is very close to the line $A$. The line $C$ at $r = r_o$ denotes the boundary between 
Region 3 and Region 2. Region 2 is the interpolating region that ends at $r = r_{\rm min}$. At the far IR the geometry is
cut-off at $r = a$ from the blown-up $S^3$.}}
		\end{center}
		\end{figure} 

\noindent As the system is not dynamical, the world line for the static
 $Q\bar{Q}$ can be chosen to be 
\bg \label{qline}
x^1~=~\pm \frac{d}{2},~~~~~ x^2 ~= ~ x^3 ~= ~0
\nd
and using $u \equiv 1/r$ we can rewrite the metric in region 3 as\footnote{We  use the Einstein summation 
convention henceforth unless mentioned otherwise.}:
\bg \label{reg3met}
ds^2&=& g_{\mu\nu} dX^\mu dX^\nu ~=~
{\cal A}_n(\psi,\theta_i,\phi_i)u^{n-2}\left[-g(u)dt^2+d\overrightarrow{x}^2\right]\nonumber\\
&+&\frac{
{\cal B}_l(\psi,\theta_i,\phi_i)u^{l}}{
{\cal A}_m(\psi,\theta_i,\phi_i)u^{m+2}g(u)}du^2+\frac{1}
{{\cal A}_n(\psi,\theta_i,\phi_i)u^{n}}~ds^2_{{\cal M}_5}
\nd
where ${\cal A}_n$ are the coefficients that can be extracted from the $a_i$ in (\ref{heaft}), the 
black hole factor $g(u) = 1$ for the zero-temperature case, 
and 
$ds^2_{{\cal M}_5}$ is the metric of the internal space that includes the corrections given in  
(\ref{rab}). This can be 
made precise as
\bg\label{cala}
{1\over \sqrt{h}}~ = ~ {1\over L^2 u^2 \sqrt{a_i u^i}} \equiv {\cal A}_n u^{n-2} ~=~ {1\over L^2 u^2}\left[a_0  
-{a_1u\over 2} + \left({3a_1^2\over 8a_0} - {a_2\over 2}\right)u^2 + ...\right]
\nd
giving  ${\cal A}_0 = {a_0\over L^2}, {\cal A}_1 = -{a_1\over 2L^2}, 
{\cal A}_2 = {1\over L^2}\left({3a_1^2\over 8a_0} - {a_2\over 2}\right)$ and so on. Observe that since $a_i$, $i \ge 1$ 
are of ${\cal O}(g_sN_f)$ and $L^2 \propto \sqrt{g_sN}$, all ${\cal A}_i$ are very small.  
The $r^{-n}$ corrections along the radial direction given in (\ref{rrr}) are accommodated above 
through ${\cal B}_l u^l$ series.   

Now suppose $X^\mu:(\sigma,\tau)\rightarrow (x^{0123}, u,\psi,\phi_i,\theta_i)$ is a mapping from string 
world sheet to space-time. Choosing a parametrization $\tau= x^0 \equiv t,\sigma= x^1 \equiv x$ with the boundary of the 
world sheet overlapping with the world line of the $Q\bar{Q}$ pair, we see that we can have
\bg \label{ws-1}
&&X^0~= ~t,~~~ X^1~= ~x,~~~ X^2~ = ~ X^3 ~ = ~ 0, ~~~ X^7~=~u(x), ~~~X^6 ~=~\psi ~=~ 0 \nonumber\\
&&(X^4, X^5) ~ = ~ (\theta_1, \phi_1) ~ = ~ (\pi/2, 0), ~~~ 
(X^8, X^9) ~ = ~ (\theta_2, \phi_2) ~ = ~ (\pi/2, 0)
\nd
which is almost like the slice that we chose in \cite{FEP}. The advantage of such a choice is to 
get rid of the awkward angular variables that appear for our background geometry so that we will have only a $r$ (or $u$) 
dependent background. We also impose 
the boundary condition 
\bg \label{bc-1}
u(\pm d/2)~ = ~ u_\gamma ~\approx ~ 0
\nd
where $u_\gamma$ denote the position of the seven brane {\it closest} to the 
boundary $u=0$.  The Nambu-Goto action for the string connected to this seven brane is:
\bg \label{NG-1}
&& S_{\rm string}= {1\over 2\pi \alpha'}\int d\sigma d\tau 
\Big[\sqrt{-{\rm det}\left[(g_{\mu\nu} + \partial_\mu\phi\partial_\nu\phi)\partial_a X^\mu \partial_b X^\nu\right]} 
+ {1\over 2} \epsilon^{ab} B_{ab}
+ J(\phi)\nonumber\\
 && ~~~~~~~~+ \epsilon^{ab}
\partial_a X^m \partial_b X^n ~\bar\Theta~ \Gamma_m \Gamma^{abc....} \Gamma_n ~\Theta
~F_{abc....} + {\cal O}(\Theta^4)\Big]
\nd   
where $a, b=1,2$, $\partial_1\equiv \frac{\partial}{\partial \tau}$, $\partial_2\equiv \frac{\partial}{\partial \sigma}$. 
The other fields appearing in the action are the pull backs of the NS $B$ field $B_{ab}$, the dilaton coupling $J(\phi)$ 
and the RR field strengths
$F_{abc..}$. Its clear that if we switch off the fermions i.e $\Theta = \bar\Theta = 0$ the RR fields decouple.  
The $B_{NS}$ field do couple to the fundamental string but as we discussed before, in region 3 we don't expect to 
see any three-form field strengths. This is because the amount of $B_{\rm NS}$ that could leak out from region 2 to 
region 3 is:
\bg\label{leaking} 
B_{\rm NS} ~= ~ M {\cal S} [1-f(r)] ~ = ~ M{\cal S} ~e^{-\alpha(r - r_0)}, ~~~~ r > r_0
\nd
where ${\cal S}$ is the two-form:
\bg\label{sdefin}
&&{\cal S} = g_s 
\left(b_1(r)\cot\frac{\theta_1}{2}\,d\theta_1+b_2(r)\cot\frac{\theta_2}{2}\,d\theta_2\right)\wedge e_\psi
- {g_sb_4(r)\over 12\pi}\sin\theta_2\,d\theta_2\wedge d\phi_2\\
&& +{3g_s\over 4\pi}\left[\left(1+g_sN_f -{1\over r^{2g_sN_f}} + {9a^2g_sN_f\over r^2}\right)
\log\left(\sin\frac{\theta_1}{2}\sin\frac{\theta_2}{2}\right) + b_3(r)\right]
\sin\theta_1\,d\theta_1\wedge d\phi_1 \nonumber\\
&&-{g_s\over 12\pi}\Bigg(2 -{36a^2g_sN_f\over r^2} + 9g_sN_f -{1\over r^{16g_sN_f}} - 
{1\over r^{2g_sN_f}} + {9a^2g_sN_f\over r^2}\Bigg)\sin\theta_2\,d\theta_2\wedge d\phi_2\nonumber 
\nd
and $b_n$ have been defined in section {\bf 2.4.2}. We see that not only $B_{\rm NS}$ has an inverse $r$ fall off, but also has  
a strong exponential decay as $\alpha >> 1$. This is the main reason why there are no NS or RR three-forms in 
region 3, making our computation of the Wilson loop relatively easier compared to the pure Klebanov-Strassler model.

On the other hand the dilaton {\it will} couple {\it additionally} 
via the $J(\phi)$ term. Although this coupling of $\phi$ is not directly to the 
$X^\mu$, we can still control this coupling by arranging the other seven-branes such that:
\bg\label{sevbrarr}
{\rm {\bf Re}}\left(\sum_{i=1}^{n_1} {3z^n_i\over z^n} - \sum_{j=1}^{n_2} {{\widetilde z}^n_j \over z^n}\right) 
~ < ~ \epsilon ~~~~~ {\rm for}~~ 0\le n \le m_o
\nd
with $\epsilon$ very small and $m_o$ a sufficiently big number. Under this condition the dilaton will be 
essentially constant and the axio-dilaton $\tau$  
would behave as:
\bg\label{axdbel}
\tau ~ = ~ \tau_0 + \sum_{n = 1}^\infty {{\cal C}_n \over r^n} + 
i\sum_{n > m_o}^\infty {{\cal D}_n \over r^n}
\nd 
so that its contribution to NG action can be ignored although the ${\cal B}_l u^l$ contribution still remains, 
because the seven-branes continue to affect the geometry from their energy-momentum tensors and the axion 
charges. In this limit both string and Einstein frame metrics are identical and the background dilaton is 
\bg\label{bgdil}
\phi = {\rm log}~g_s - g_s{\cal D}_{n+m_o} u^{n+m_o} + {\cal O}(g_s^2)
\nd 
which, in the limit $g_s \to 0$, will be dominated by the 
constant term (note that $m_o$ is fixed). On the other hand, $J(\phi)\sim \phi R_{(2)}$, where $R_{(2)}$ is the Ricci scalar
for the world sheet metric. As the background space time metric run as ${\cal O}(1/r^n)$, we can make $R_{(2)}$
arbitrarily small by
a reparametrization of the world sheet. This means we can ignore $J(\phi)$ in the string action while
the NG string will see a slightly different background metric as 
evident from (\ref{NG-1}). 

Thus once all the effects have been accounted for, using the metric (\ref{reg3met})
with the embedding $X^\mu$ given by (\ref{ws-1}), one can easily show that at zero temperature the NG action is 
given by:
\bg \label{NG-2} 
S_{\rm{NG}}=\frac{\tilde{T}}{2\pi}\int_{-{d\over 2}}^{+{d\over 2}} \frac{dx}{u^2}\sqrt{\Big({\cal A}_n u^n\Big)^2
+ \Big[{\cal B}_m u^m + 2g_s^2 {\widetilde{\cal D}}_{n+m_o} {\widetilde{\cal D}}_{l+m_o} {\cal A}_k u^{n+l+k+2m_o} 
+ {\cal O}(g_s^4)\Big]
\left(\frac{\partial u}{\partial x}\right)^2 }\nonumber\\
\nd
where we have used $\int dt=T,\tilde{T}=T/\alpha'$, 
${\widetilde{\cal D}}_{n+m_o} = (n+m_o){\cal D}_{n+m_o}$; 
and ${\cal A}_n, {\cal B}_n$ and ${\cal D}_{n+m_o}$
are now defined for choices of the angular coordinates
given in (\ref{ws-1}).  
The above action can be condensed by redefining:
\bg\label{redefine}
{\cal B}_m u^m + 2g_s^2 {\widetilde{\cal D}}_{n+m_o} {\widetilde{\cal D}}_{l+m_o} {\cal A}_k u^{n+l+k+2m_o} 
+ {\cal O}(g_s^4) 
~ \equiv ~ {\cal G}_l u^l
\nd
which would mean that the constraint
equation i.e $\partial_1 T^1_1 = 0$, $T^1_1$ being the stress-tensor,
 for $u(x)$ derived from the action (\ref{NG-2}) using (\ref{redefine}) can be written as 
\bg \label{EL-1}
\frac{d}{dx}\left(\frac{\left({\cal A}_n\ u^n\right)^2}{u^2\sqrt{\left({\cal A}_m\ u^m\right)^2
+ {\cal G}_m u^m\left(\frac{\partial u}{\partial x}\right)^2 }}\right) ~= ~0
\nd
implying that:
\bg\label{feqn}
\frac{\left({\cal A}_n u^n\right)^2}{u^2\sqrt{\left({\cal A}_m u^m\right)^2
+ {\cal G}_m u^m ~u'(x)^2}} ~ = ~ C_o
\nd
where $C_o$ is a constant, and 
$u'(x)\equiv\frac{\partial u}{\partial x}$. 
This constant $C_o$ can be determined in the following way:
as we have the endpoints of the string at $x=\pm d/2$, 
by symmetry the string will be U shaped and if $u_{\rm max}$ is the maximum value of $u$, 
we can choose the $x$ coordinate such that $u(0)= u_{\rm
max}$ and $u'(x=0)=0$. Plugging this in (\ref{feqn}) we get:
\bg \label{C}
C_o ~= ~ \frac{{\cal A}_n u_{\rm max}^n }{u_{\rm max}^2}
\nd    
Once we have $C_o$, we can use (\ref{EL-1}) to get the following simple differential equation: 
\bg \label{EL-2}
{du\over dx} ~= ~ \pm \frac{1}{C_o\sqrt{{\cal G}_m u^m}}\left[\frac{\left({\cal A}_n u^n\right)^4}{u^4}
-C_o^2\left({\cal A}_m u^m\right)^2\right]^{1/2}
\nd
which in turn can be used to write $x(u)$ as:
\bg \label{EL-3}
x(u) ~= ~ C_o \int_{u_{\rm max}}^u dw \frac{w^2\sqrt{{\cal G}_m w^m}}{\left({\cal A}_n w^n\right)^2} 
\left[1-\frac{C_o^2 w^4}{\left({\cal A}_m
w^m\right)^2}\right]^{-1/2}
\nd 
where we have used $x(u_{\rm max})=0$. Now using the boundary condition given in (\ref{bc-1}) i.e $x(u=u_\gamma)=d/2$, 
and defining $w = u_{\rm max} v, \epsilon_o = {u_\gamma\over u_{\rm max}}$
we have
\bg \label{D-1}
d~ = ~2u_{\rm max} \int_{\epsilon_o}^{1} dv ~v^2\frac{\sqrt{{\cal G}_m u_{\rm max}^m v^m}
\left({\cal A}_n u_{\rm max}^n\right)}{\left({\cal A}_m u_{\rm max}^m v^m\right)^2} 
\left[1-v^4\left(\frac{{\cal A}_n u_{\rm max}^n}{{\cal A}_m u_{\rm max}^m v^m}\right)^2\right]^{-1/2}
\nd
At this stage we can assume all ${\cal A}_n > 0$. This is because
for ${\cal A}_n > 0$ we can clearly have degrees of freedom in the gauge theory growing towards UV, which is an 
expected property of models with RG flows. Of course this is done to simplify the subsequent analysis. Keeping 
${\cal A}_n$ arbitrary will also allow us to derive the linear confinement behavior, but this case will require a 
more careful analysis. We will get back to this in the section {\bf 3.3.3}. Note also that 
similar behavior is seen for the  
the Klebanov-Strassler model, and we have already discussed how degrees of freedom run in regions 2 and 3. 
Another obvious condition is that $d$, which is the distance between the quarks, cannot be imaginary. 
From
(\ref{D-1}) we can see that the integral becomes complex for 
\bg \label{real-1}
{\cal F}(v)~\equiv~ v^4\left(\frac{{\cal A}_n u_{\rm max}^n}{{\cal A}_m u_{\rm max}^m v^m}\right)^2~ > ~ 1
\nd
whereas for ${\cal F}(v)= 1$ the integral becomes singular. Then for $d$ to be always real we must have 
\bg\label{real-2}
{\cal F}(v)~ \leq ~ 1
\nd 
We will now set, without loss of generality, ${\cal A}_0 = 1$ and ${\cal A}_1 = 0$ . 
Such a choice is of course consistent 
with supergravity solution for our background (as evident from (\ref{cala})). This choice also defines our units in that we
are setting the AdS throat radius $L\equiv 1$ and this is only to make subsequent computations convenient. 
. Now   
analyzing the condition (\ref{real-2}), one easily finds that we must have
\bg\label{real-3}
{1\over 2}(m+1){\cal A}_{m + 3}\ u_{\rm max}^{m + 3} ~ \leq  ~ 1
\nd
for $d$ to be real. This condition puts an upper bound on $u_{\rm max}$ and we can use this to constrain the fundamental
string to lie completely in region 3. 
Observe that for AdS spaces, ${\cal A}_n=0$ for
$n>0$ and hence there is no upper bound for $u_{\rm max}$. This is also the main reason why we see confinement 
using our background but not from the AdS backgrounds. Furthermore one might mistakenly think that generic 
Klebanov-Strassler background should show confinement because the space is physically cut-off due to the presence of a 
blown-up $S^3$. Although such a scenario may imply an upper bound for $u_{\rm max}$, this doesn't naturally 
lead to confinement because due to the presence of 
logarithmically varying $B_{\rm NS}$ fields there are UV divergences of the Wilson loop. These divergences {\it cannot}
be removed by simple regularization schemes \cite{cschu} and creates problems in the interpretation of Wilson loop which 
describe the potential.    

Coming back to (\ref{NG-2}) we see that it can be further simplified.
Using (\ref{feqn}), (\ref{C}) and (\ref{EL-2}) in (\ref{NG-2}), 
we can write it as an integral over $u$:
\bg \label{NG-3}
S_{\rm NG}&=&~\frac{\tilde{T}}{\pi} \int_{u_\gamma}^{u_{\rm max}} {du\over u^2} \sqrt{{\cal G}_l u^l}
\left[1-\frac{C_o^2 u^4}{\left({\cal A}_m u^m\right)^2}\right]^{-1/2}\nonumber\\
&=&~\frac{\tilde{T}}{u_{\rm max}\pi}\int_{\epsilon_o}^1 {dv\over v^2} \sqrt{{\cal G}_m u_{\rm max}^m v^m}
\left[1-v^4\left(\frac{{\cal A}_n u_{\rm max}^n}{{\cal A}_m u^m_{\rm max}
v^m}\right)^2\right]^{-1/2} 
\nd
where in the second equality we have taken $v= u/u_{\rm max}$.

This simplified action (\ref{NG-3}) is however not the full story. It is also divergent in the limit $\epsilon_o \to 0$. 
But we can isolate the divergent part of the above integral (\ref{NG-3}) by first computing it as a function of $\epsilon_o$. The
result is
\bg
S_{\rm NG}&\equiv& S_{\rm NG}^{\rm I} + S_{\rm NG}^{\rm II} ~ = ~
\frac{\tilde{T}}{\pi}\frac{1}{u_{\rm max}}\int_{\epsilon_o}^1 {dv\over v^2} \sqrt{{\cal G}_m u_{\rm max}^m v^m}\nonumber\\
&+&\frac{\tilde{T}}{\pi}\frac{1}{u_{\rm max}}\int_{\epsilon_o}^1 {dv\over v^2} \sqrt{{\cal G}_m u_{\rm max}^m v^m}
\left\{\left[1-v^4\left(\frac{{\cal A}_n u_{\rm max}^n}{{\cal A}_m u^m_{\rm max}
v^m}\right)^2\right]^{-1/2}-1\right\}
\nd
Now by expanding $\sqrt{{\cal G}_m u_{\rm max}^m v^m}={\widetilde{\cal G}}_lv^l$ we can compute the first integral to be 
\bg \label{NG-3A}
S_{\rm NG}^{\rm I}&=&\frac{\tilde{T}}{\pi}\frac{1}{u_{\rm max}}\left(-{\widetilde{\cal G}}_0+\frac{{\widetilde{\cal G}}_0}
{\epsilon_o}+\sum_{l=2}\frac{{\widetilde{\cal G}}_l}{l-1}+{\cal O}(\epsilon_o)+..\right)
\nd
where ${\widetilde{\cal G}}_0 = {\cal G}_0, ~{\widetilde{\cal G}}_1 = {1\over 2} {\cal G}_1 u_{\rm max}$ 
and so on. The second integral becomes
\bg \label{NG-3B}
S_{\rm NG}^{\rm II}&=&\frac{\tilde{T}}{\pi}\frac{1}{u_{\rm max}}\int_0^1 {dv\over v^2} \sqrt{{\cal G}_m u_{\rm max}^m v^m}
\left\{\left[1-v^4\left(\frac{{\cal A}_n u_{\rm max}^n}{{\cal A}_m u^m_{\rm max}
v^m}\right)^2\right]^{-1/2}-1\right\}+{\cal O}(\epsilon_o^3)\nonumber\\
\nd
where the $\epsilon_o$ dependence here appears to 
${\cal O}(\epsilon^3_o)$; and we have set ${\cal G}_1=0$ without loss of generality. Now combining the
result in (\ref{NG-3A}) and (\ref{NG-3B}), we can obtain the renormalized action by subtracting the divergent term ${\cal
O}(1/\epsilon)$ in the limit $\epsilon_o \to 0$ and we obtain the following result 
\bg \label{NG-4}
S_{\rm NG}^{\rm ren}&=&\frac{\tilde{T}}{\pi}\frac{1}{u_{\rm max}}
\Bigg\{-{\widetilde{\cal G}}_0+ \sum_{l=2}\frac{{\widetilde{\cal G}}_l}{l-1} 
- \int_0^1 {dv\over v^2} \sqrt{{\cal G}_m u_{\rm max}^m v^m} + 
{\cal O}(g_s^2) \nonumber\\
&+& \int_0^1 {dv\over v^2} \sqrt{{\cal G}_m u_{\rm max}^m v^m}
\left[1-v^4\left(\frac{{\cal A}_n u_{\rm max}^n}{{\cal A}_m u^m_{\rm max}
v^m}\right)^2\right]^{-1/2} + {\cal O}(\epsilon_o)\Bigg\}
\nd
where the third term in (\ref{NG-4}), including the ${\cal O}(g_s^2)$ correction,  
is related to the action for a straight string in this background in the limit 
$g_s \to 0$. Our
subtraction scheme is more involved because the straight string sees a complicated metric due to the background 
dilaton and non-Ricci flat unwarped metric. This effect is {\it independent} of any choice of the warp factor and we 
expect this action to be finite in the limit $\epsilon_o \to 0$. 

Once we have the finite action,
we should use this to compute the $Q\bar{Q}$ potential through (\ref{Vqq}). 
Looking at (\ref{D-1}) we observe that the relation between $d$ and $u_{\rm max}$ is parametric and can be 
quite involved depending on the
coefficients ${\cal A}_n$ of the warp factor. 
If we have ${\cal A}_n=0, {\cal G}_n=0$ for $n>0$, 
we recover the well known AdS result, namely: $d\sim u_{\rm max}$ and $V_{Q\bar{Q}}\sim {1\over d}$. But in
general (\ref{D-1}) and (\ref{NG-4}) should be solved together to obtain the potential.  

As it stands, (\ref{D-1}) and (\ref{NG-4}) are both rather involved. So to find some correlation between them, we will observe 
the limiting behavior of $u_{\rm max}$. Therefore in the following,  
we will study the behavior of $d$
and $S_{\rm NG}^{\rm ren}$ for the cases where $u_{\rm max}$ is large and small.

\subsubsection{Quark-Anti quark potential for small $u_{\rm max}$}

Let us first consider the case where $u_{\rm max}$ is small.  In this limit 
we can ignore higher order terms in $u_{\rm max}$ and approximate
\bg\label{lbaz}
{\cal A}_n u^n_{\rm max} ~=~ {\cal A}_0 ~+~ {\cal A}_2 u^2_{\rm max} ~ \equiv~ 1 ~+~ \eta
\nd
where ${\cal A}_0 = 1$ and ${\cal A}_2 u^2_{\rm max}=\eta$. Using this we can 
 write both (\ref{D-1}) and (\ref{NG-4}) as Taylor series in $\eta$ around $\eta=0$. The result is
\bg \label{D-3}
&& d ~ = ~ \sqrt{\eta}\left[a_0 ~ + ~ a_1\eta + {\cal O}(\eta^2)\right]\nonumber\\
&& S_{NG}^{\rm ren} ~ = ~ {\tilde{T}\over \pi} \left[{b_0 + b_1\eta + {\cal O}(\eta^2)\over \sqrt{\eta}}\right]
\nd 
with $a_0, a_1, b_0, b_1$ are defined in the following way:
\bg\label{abdefn}
&&a_0 ~=~ {2\over \sqrt{{\cal A}_2}} \int_0^1 dv {v^2\over \sqrt{1-v^4}} ~=~ {1.1981\over \sqrt{{\cal A}_2}}\nonumber\\
&& a_1 ~=~ 
{2\over \sqrt{{\cal A}_2}} \int_0^1 dv {v^2\over \sqrt{1-v^4}}\left[{1-v^6\over 1-v^4} + \left({{\cal G}_2 - 
4{\cal A}_2\over 2{\cal A}_2}\right)v^2\right]\nonumber\\
&&b_0 ~ = ~ \sqrt{{\cal A}_2}\left[-1 +
\int_0^1 dv \left({1-\sqrt{1-v^4} \over v^2\sqrt{1-v^4}}\right)\right] ~ = ~ -0.62 \sqrt{{\cal A}_2}\nonumber\\
&& b_1 ~ = ~ {1\over 2\sqrt{{\cal A}_2}}\Bigg\{{\cal G}_2 +
\int_0^1 dv \left[{2{{\cal A}_2} v^4 + {\cal G}_2 v^2(1+v^2)(1-
\sqrt{1-v^4})\over v^2(1+v^2)\sqrt{1-v^4}}\right]\Bigg\}
\nd
where we have taken ${\cal G}_0 = 1$ and ${\cal G}_1 = 0$ without loss of generality.  
In this limit clearly increasing $\eta$ increases $d$, the distance between the quarks. For small $\eta$, 
$d = a_0\sqrt{\eta}$, and therefore the Nambu-Goto action will become:
\bg\label{NGac}
S_{\rm NG}^{\rm ren}  ~ = ~ \tilde{T}\left[-\left({a_0 \vert b_0\vert\over \pi}\right) {1\over d} 
~+~ \left({b_1\over \pi a_0}\right)d
+ {\cal O}(d^3)\right]
\nd
where all the constants have been defined in (\ref{abdefn}). Using (\ref{Vqq}) we can determine the short-distance 
potential to be :
\bg\label{sdpot}
V_{Q\bar{Q}} ~&& = ~ \frac{1}{\alpha'}\left[-\left({a_0\vert b_0\vert\over \pi}\right) {1\over d} ~+~ \left({b_1\over \pi a_0}\right)d
+ {\cal O}(d^3)\right]\nonumber\\
&&= ~ \sqrt{g_sN}\left[-{0.236\over d} ~+~\left(0.174{\cal G}_2 +0.095{\cal A}_2\right) d ~ + ~ {\cal
O}(d^3)\right]
\nd
where $N$ is the number of D3 branes in the gauge theory and  we have used string tension 
$1/\alpha'=\sqrt{g_s N}/L^2\equiv \sqrt{g_s N}$ as $L^2\equiv 1$ 
according to our choice of units. The potential is dominated by the inverse $d$ behavior, i.e 
the expected Coulombic behavior. Note that the coefficient of the 
Coulomb term which is a dimensionless number only depends on the number of D3 branes. Thus it is independent of the warp 
factor  and hence should be universal. 
This result, in appropriate units, 
is of the same order of magnitude as the real Coulombic term obtained by comparing with  Charmonium 
spectra as first modelled by \cite{charmonium} and subsequently by several authors \cite{Fkarsch-1}-\cite{Fkarsch-5},\cite{HQG-1}-\cite{HQG-4},\cite{boschi1}-\cite{boschi4}. This 
prediction, along with the overall minus sign, should be regarded as a success of our model (see also \cite{zakahrov}
where somewhat similar results have been derived in a string theory inspired model). The second term on the other 
hand is model dependent, and vanishes in the pure AdS background. 

Note also that the above computations are valid for infinitely massive quark-anti quark pair. For lighter quarks, we 
expect the results to differ. It would be interesting to compare these results with the ones where quarks are 
much lighter.  

\subsubsection{Quark-Anti quark potential for large $u_{\rm max}$}

Lets  analyze the integrals (\ref{D-1}) and (\ref{NG-4}) in the limit $u_{\rm max}$ is close to its upper bound set
by (\ref{real-3}) (see also \cite{zakahrov}). In particular if ${\bf u}_{\rm max}$ 
is the upper bound of $u_{\rm max}$, then it is found by solving
\bg \label{real-4}
{1\over 2}(m+1){\cal A}_{m + 3} {\bf u}_{\rm max}^{m + 3} ~ = ~ 1
\nd 
We observe that both the integrals (\ref{D-1}) and (\ref{NG-4}) are
dominated by $v\sim 1$ behavior of the integrands. Near $v=1$ and 
$u_{\rm max}\rightarrow {\bf u}_{\rm max}$ the distance $d$ between the quark and the anti quark can be written as:
\bg \label{D-4}
d&& =~~2\frac{\sqrt{{\cal G}_m {\bf u}_{\rm max}^m}{\bf u}_{\rm max}}{{\cal A}_n {\bf u}_{\rm max}^n} 
\int_0^{1} \frac{dv}{\sqrt{{\bf A} (1-v)+{\bf
B}(1-v)^2}}\nonumber\\
&& = -2 \frac{\sqrt{{\cal G}_m {\bf u}_{\rm max}^m}{\bf u}_{\rm max}}{{\cal A}_n {\bf u}_{\rm max}^n} 
\left[{{\rm log}{\bf A}-{\rm log}\left(2\sqrt{{\bf B}({\bf A}+{\bf B})}+2{\bf B}+{\bf A}\right)\over \sqrt{\bf B}}\right]
\nd
where note that we have taken the lower limit to 0. This will not change any of our conclusion as we would soon see. On  
the other hand, the renormalized Nambu-Goto action for the string now becomes:
\bg \label{NG-6}
S_{\rm NG}^{\rm ren}&&= ~ {\tilde{T}\over \pi}\frac{\sqrt{{\cal G}_m {\bf u}_{\rm max}^m}}{{\bf u}_{\rm max}}
\left[\int_0^{1} \frac{dv}{\sqrt{{\bf A} (1-v)+{\bf
B}(1-v)^2}} -1\right] - {\tilde{T}\over \pi {\bf u}_{\rm max}} + {\cal O}({\bf u}^2_{\rm max})\nonumber\\
&& = ~ -{\tilde{T}\over \pi}\frac{\sqrt{{\cal G}_m {\bf u}_{\rm max}^m}}{{\bf u}_{\rm max}}
\left[{{\rm log}{\bf A}-{\rm log}\left(2\sqrt{{\bf B}({\bf A}+{\bf B})}+2{\bf B}+{\bf A}\right)\over \sqrt{\bf B}} 
-1\right]\nonumber\\
&& ~~~~~~ - {\tilde{T}\over \pi {\bf u}_{\rm max}} + {\cal O}({\bf u}^2_{\rm max})
\nd
where ${\bf A}$ and ${\bf B}$ are defined as:
\bg \label{AB}
{\bf A}&&= ~ 4 - 2\frac{n{\cal A}_n {u}_{\rm max}^n}{{\cal A}_m {u}_{\rm max}^m}\\
{\bf B}&& = ~ 8\frac{n{\cal A}_n {u}_{\rm max}^n}{{\cal A}_m {u}_{\rm max}^m}
-3\left(\frac{n{\cal A}_n {u}_{\rm max}^n}{{\cal A}_m {u}_{\rm max}^m}\right)^2
+ \frac{(n^2-n){\cal A}_n {u}_{\rm max}^n}{{\cal A}_m{u}_{\rm max}^m} - 6\nonumber
\nd
Observe that in the integral (\ref{D-4}) and (\ref{NG-6}) we have to take the limit 
$u_{\rm max}\rightarrow {\bf u}_{\rm max}$. 
So ${\bf A},{\bf B}$ should be evaluated in the same limit. Interestingly, comparing (\ref{AB}) to (\ref{real-4})
we see that
\bg \label{AB1}
\lim_{u_{\rm max}\rightarrow {\bf u}_{\rm max}}~{\bf A}\rightarrow 0 
\nd
thus vanishes when computed exactly at ${\bf u}_{\rm max}$. The other quantity ${\bf B}$ remains finite at that point and
in fact behaves as:
\bg\label{bbeh}
{\bf B} = {n^2{\cal A}_n {\bf u}^n_{\rm max} \over {\cal A}_m {\bf u}^m_{\rm max}} - 4 ~ > ~0
\nd
Our above computation would  mean that the distance $d$ between the quark and the anti quark, and the Nambu-Goto 
action will have the following dominant behavior:
\bg\label{dNG}
d && = ~ \lim_{\epsilon\to 0}~
\frac{2\sqrt{{\cal G}_m {\bf u}_{\rm max}^m}{\bf u}_{\rm max}}{{\cal A}_n {\bf u}_{\rm max}^n} 
~ {{\rm log}~\epsilon \over \sqrt{\bf B}}\nonumber\\
S_{\rm NG}^{\rm ren}&& = ~ \lim_{\epsilon\to 0}~ 
{\tilde{T}\over \pi}\frac{\sqrt{{\cal G}_m {\bf u}_{\rm max}^m}}{{\bf u}_{\rm max}}~
 {{\rm log}~\epsilon \over \sqrt{\bf B}}
\nd
which means both of them have identical logarithmic divergences. Thus the {finite} quantity is the {\it ratio} between
the two terms in \ref{dNG}. This gives us:
\bg\label{ratio} 
{S_{\rm NG}^{\rm ren}\over d} ~ = ~ {\tilde{T}\over \pi}{{\cal A}_n {\bf u}_{\rm max}^n \over {\bf u}_{\rm max}^2} ~ = ~ 
T \times {\rm constant}
\nd
Now using the identity (\ref{Vqq}) and the above relation (\ref{ratio}) we get our final result:
\bg\label{lipo}
V_{Q\bar Q} ~ = ~ \left({{\cal A}_n {\bf u}_{\rm max}^n \over \pi {\bf u}_{\rm max}^2}\right) ~d/\alpha'
\nd
which is the required linear potential between the quark and the anti quark. 

Before we end this section one comment is in order. The result for linear confinement only depends on the existence
of ${\bf u}_{\rm max}$ which comes from the constraint equation (\ref{real-4}). We have constructed the background 
such that ${\bf u}_{\rm max}$ lies in region 3, although a more generic case is essentially doable albeit technically 
challenging without necessarily revealing new physics. 
For example when ${\bf u}^{-1}_{\rm max}$ is equal to the size of the blown up $S^3$ at the IR will require
us to consider a Wilson loop that goes all the way to region 1. The analysis remains similar to what we did 
before except that in regions 2 and 1 we have to additionally consider $B_{\rm NS}$ fields of the form 
$u^{\epsilon_{(\alpha)}}$ and ${\rm log}~u$ respectively. Of course both the metric and the dilaton will also have 
non-trivial $u$-dependences in these regions. One good thing however is that the Wilson loop computation have 
no UV or IR divergences whatsoever despite the fact that now the 
analysis is technically more challenging. Our expectation would be to get similar linear behavior as (\ref{lipo}) 
here too.

\subsection{Computing the Nambu-Goto Action: Non-Zero Temperature}

After studying the zero temperature behavior we will now discuss the case when we switch on a non-zero 
temperature i.e make $g(u) < 1$ or equivalently the inverse horizon radius, $u_h$ finite in (\ref{reg3met}), where 
\bg\label{g}
g(u)=1-{u^4\over u_h^4}
\nd 
Choosing the same quark world line (\ref{qline}) and the string embedding (\ref{ws-1}) with the same boundary condition
(\ref{bc-1}) but now in Euclidean space with compact time direction, the string action at 
finite temperature can be written as
\bg \label{NGfinT} 
S_{\rm{NG}}=\frac{\tilde{T}}{2\pi}\int_{-{d\over 2}}^{+{d\over 2}} \frac{dx}{u^2}\sqrt{g(u)\Big({\cal A}_n u^n\Big)^2
+ \Bigg[{\cal G}_m u^m - {2g_s^2 {\widetilde{\cal D}}_{n+m_o} {\widetilde{\cal D}}_{l+m_o} 
{\cal A}_k u^{4+n+l+k+2m_o} \over u_h^4}\Bigg]
\left(\frac{\partial u}{\partial x}\right)^2 }\nonumber\\
\nd
where ${\cal G}_m u^m$ is defined in (\ref{redefine}) and the correction to ${\cal G}_m u^m$ is suppressed by 
$g_s^2$ as well as $u^4/u_h^4$ because the background dilaton and non-zero temperature induces a slightly 
different world-sheet metric than what one would have naively taken. To avoid clutter, we will further redefine
these corrections as:
\bg\label{redefagain}
{\cal G}_m u^m - {2g_s^2 {\widetilde{\cal D}}_{n+m_o} {\widetilde{\cal D}}_{l+m_o} 
{\cal A}_k u^{4+n+l+k+2m_o} \over u_h^4} ~\equiv~ {\widetilde{\cal D}}_l u^l
\nd
Minimizing this action gives the equation of motion for $u(x)$ and using the exact same procedure 
as for zero temperature,
the corresponding equation for the distance between the quarks can be written as:
\bg \label{DT-1}
d&=&2u_{\rm max} \int_{0}^{1} dv \Bigg\{v^2 \sqrt{{\widetilde{\cal D}}_m u_{\rm max}^m v^m}
\frac{\sqrt{1-\frac{u_{\rm max}^4}{u_h^4}}{\cal A}_n u_{\rm max}^n}
{\left(1-\frac{v^4u_{\rm max}^4}{u_h^4}\right)\left({\cal A}_m u_{\rm max}^m v^m\right)^2}\nonumber\\
&& 
\left[1-v^4\frac{\left(1-\frac{u_{\rm max}^4}{u_h^4}\right)}{\left(1-\frac{v^4u_{\rm max}^4}{u_h^4}\right)}
\left(\frac{{\cal A}_n u_{\rm max}^n}{{\cal A}_m u_{\rm max}^m v^m}\right)^2\right]^{-1/2}\Bigg\}
\nd
Once we have $d$, the 
renormalized Nambu-Goto action can also be written following similar procedure. The result is
\bg \label{NGT-3}
S_{\rm NG}^{\rm ren}&=&\frac{\tilde{T}}{\pi}\frac{1}{u_{\rm max}}
\Bigg\{-{\widehat{\cal D}}_0+ \sum_{l=2}\frac{{\widehat{\cal D}}_l}{l-1} 
- \int_0^1 {dv\over v^2} \sqrt{{\widetilde{\cal D}}_m u_{\rm max}^m v^m} + 
{\cal O}(g_s^2)\\
&+& \int_0^1 {dv\over v^2} \sqrt{{\widetilde{\cal D}}_m u_{\rm max}^m v^m}
\left[1-v^4\frac{\left(1-\frac{u_{\rm max}^4}{u_h^4}\right)}{\left(1-\frac{v^4u_{\rm max}^4}{u_h^4}\right)}
\left(\frac{{\cal A}_n u_{\rm max}^n}{{\cal A}_m u^m_{\rm max}
v^m}\right)^2\right]^{-1/2} + {\cal O}(\epsilon_o)\Bigg\}\nonumber
\nd
which is somewhat similar in form with (\ref{NG-4}), which we reproduce in the limit $u_h \to \infty$. 
Also as in (\ref{NG-4}), we have defined 
$\sqrt{{\widetilde{\cal D}}_m u_{\rm max}^m v^m} \equiv {\widehat{\cal D}}_l v^l$. 

Now just like the zero temperature case, requiring that $d$ be real, 
sets an upper bound to $u_{\rm max}$, that we 
denote again by ${\bf u}_{\rm max}$, and is found by solving the following equation:
\bg\label{real-5}
{1\over 2} (m+ 1){\cal A}_{m+3} {\bf u}_{\rm max}^{m+3}
+ {1\over j!}\prod_{k=0}^{j-1}\left(k-\frac{1}{2}\right)\left(\frac{{\bf u}_{\rm max}^4}{u_h^4}\right)^j
\left[{\cal A}_{l}{\bf u}_{\rm max}^{l}\left({l\over 2}+ 2j - 1\right)\right] ~= ~1\nonumber\\
\nd
Once we fix $u_h$ and the coefficients of the warp factor ${\cal A}_n$, ${\bf u}_{\rm max}$ will be known. We will 
assume that ${\bf u}_{\rm max}$ lies in region 3. 

Rest of the analysis is very similar to the zero temperature case, although the final conclusions would be quite 
different. We proceed further by defining certain new variables in the following way:
\bg\label{newvar}
{\widetilde{\cal A}}_l ~& = & ~ \sum_m {{\cal A}_m\over u_h^{l-m}} {1\over \left({l-m\over 4}\right)!} 
\prod_{k=0}^{{l-m\over 4}-1} \left(k - {1\over 2}\right), ~~~~~ l-m \ge 4 \nonumber\\
&= & ~ 0 ~~~~~~~~~~~~~~~~~~~~~~~~~~~~~~~~~~~~~~~~~~~~~~~ l-m < 4\nonumber\\
& = & ~ {\cal A}_l ~~~~~~~~~~~~~~~~~~~~~~~~~~~~~~~~~~~~~~~~~~~~~ l - m = 0
\nd
As before, we observe that for $u_{\rm max}\rightarrow {\bf u}_{\rm max}$, 
both the integrals (\ref{DT-1}),(\ref{NGT-3}) are dominated by the behavior of the integrand near $v\sim 1$, 
where we can write 
\bg\label{dT}
d&& =~~2\frac{\sqrt{{\widetilde{\cal D}}_m {\bf u}_{\rm max}^m}{\bf u}_{\rm max}}
{\sqrt{1-{{\bf u}_{\rm max}^4\over {u}^4_{h}}}{\cal A}_n {\bf u}_{\rm max}^n} 
\int_0^{1} \frac{dv}{\sqrt{{\widetilde{\bf A}} (1-v)+ {\widetilde{\bf
B}}(1-v)^2}}\nonumber\\ 
&& = -2 \frac{\sqrt{{\widetilde{\cal D}}_m {\bf u}_{\rm max}^m}{\bf u}_{\rm max}}
{\sqrt{1-{{\bf u}_{\rm max}^4\over {u}^4_{h}}}{\cal A}_n {\bf u}_{\rm max}^n} 
\left[{{\rm log}{\widetilde{\bf A}}-{\rm log}\left(2\sqrt{{\widetilde{\bf B}}({\widetilde{\bf A}}+{\widetilde{\bf B}})}
+2{\widetilde{\bf B}}+{\widetilde{\bf A}}\right)\over \sqrt{\widetilde{\bf B}}}\right]
\nd
where taking the lower limit of the integral to 0 again do not change any of our conclusion. On  
the other hand, the renormalized Nambu-Goto action for the string now becomes:
\bg \label{NGT}
S_{\rm NG}^{\rm ren}&&= ~ {\tilde{T}\over \pi}\frac{\sqrt{{\widetilde{\cal D}}_m {\bf u}_{\rm max}^m}}{{\bf u}_{\rm max}}
\left[\int_0^{1} \frac{dv}{\sqrt{{\widetilde{\bf A}} (1-v)+ {\widetilde{\bf
B}}(1-v)^2}} -1\right] - {\tilde{T}\over \pi {\bf u}_{\rm max}} + {\cal O}({\bf u}^2_{\rm max})\nonumber\\
&& = ~ -{\tilde{T}\over \pi}\frac{\sqrt{{\widetilde{\cal D}}_m {\bf u}_{\rm max}^m}}{{\bf u}_{\rm max}}
\left[{{\rm log}{\widetilde{\bf A}}-{\rm log}\left(2\sqrt{{\widetilde{\bf B}}({\widetilde{\bf A}}+{\widetilde{\bf B}})}
+2{\widetilde{\bf B}}+{\widetilde{\bf A}}\right)\over \sqrt{\widetilde{\bf B}}} 
-1\right]\nonumber\\
&& ~~~~~~ - {\tilde{T}\over \pi {\bf u}_{\rm max}} + {\cal O}({\bf u}^2_{\rm max})
\nd
where ${\widetilde{\bf A}}$ and ${\widetilde{\bf B}}$ are defined exactly as in (\ref{AB}) but with ${\cal A}_n$ replaced
by ${\widetilde{\cal A}}_n$ given by (\ref{newvar}) above. It is also clear that:
\bg \label{ABnm}
\lim_{u_{\rm max}\rightarrow {\bf u}_{\rm max}}~{\widetilde{\bf A}}\rightarrow 0 
\nd
and so both (\ref{dT}) as well as (\ref{NGT}) have identical logarithmic divergences. This would imply that the 
finite quantity is the ratio between (\ref{NGT}) and (\ref{dT}):
\bg\label{rationow} 
{S_{\rm NG}^{\rm ren}\over d} ~ = ~ {\tilde{T}\over \pi}
\left({1-{{\bf u}_{\rm max}^4\over {u}^4_{h}}}\right)^{1\over 2}{{\cal A}_n {\bf u}_{\rm max}^n 
\over {\bf u}_{\rm max}^2}
\nd 
Now using the identity (\ref{wlfe}) and the above relation (\ref{ratio}) we get our final result:
\bg\label{liponow}
V_{Q\bar Q} ~ = ~ {\sqrt{1-{{\bf u}_{\rm max}^4\over {u}^4_{h}}}}\left({{\cal A}_n {\bf u}_{\rm max}^n 
\over \pi {\bf u}_{\rm max}^2}\right) ~d/\alpha'
\nd 
This gives linear potential at large distances and from the form of the potential above, we can see that the higher the temperature, i.e.
lower the $u_h$, the lower the slope of the linear term. In fact we will explicitly plot the potential at various temperature for certain 
choice of warp factors in section {\bf 3.3.4} and analyze this melting of the potential. 

\subsection{Linear Confinement from generic dual geometries}
Having argued for linear confinement for geometries which are like region 3 of Fig {\bf 2.17}, we will now consider a more
general warp factor choice of $h$ in dual geometry with metric (\ref{bhmetko}). Much of the discussion here closely follows 
 our recent work \cite{melting} and further details can be found there. 
We will restrict to  cascading gauge theories where the effective number of colors grows as scale grows. This property of a
gauge theory is most relevant for physical theories as new degrees of freedom emerge at UV and effective degrees of freedom
shrink in IR to form condensates at low energy \cite{Strassler:2005qs}. 
The number of colors at any scale $u = 1/r$ is given by (\ref{Neff})
and for the analysis given here, it is simpler to define 
\be
\calH(u) \equiv {u^2\over \sqrt{h}} = 
{\sqrt{N}\over L^2\sqrt{N_{\rm eff}}}
\label{eq:F_of_u}
\ee
instead of $N_{\rm eff}(u)$. 
The coefficients ${\cal A}_n$ in the previous section are related to $\calH(u)$ by
$\calH(u) = {\cal A}_n u^n$; and $h$ is the warp factor.
In terms of $\calH(u)$, the condition that $N_{\rm eff}(u)$ is 
a decreasing
function of $u=1/r$ becomes
\be
\calH'(u) > 0
\label{eq:fprime}
\ee
Combining Eqs.(\ref{eq:F_of_u}) and (\ref{eq:fprime})
yields the following condition 
\be
\calH(u) > {1\over L^2}
\ee
{}From now on, the value of $L$ is set to $1$ for the rest of this subsection, so that $\calH(u)> 1$. 

\subsubsection{Zero temperature}

Let $\umax$ be the maximum value of $u$ for the string between 
the quark and the anti-quark.
Then the relationship between $\umax$ and 
the distance between the quark and the anti-quark is given by \cite{LC}

\bg
d(\umax) =
2\umax
\calH(\umax)
\int_{\epsilon_0}^1dv\,
{v^2 \sqrt{\calG_m \umax^m v^m}
\over (\calH(\umax v))^2}
\left[1-v^4\left(\calH(\umax)\over \calH(\umax v)\right)^2\right]^{-1/2}
\label{eq:dumax}
\nd

For the above  expression to represent the physical
distance between a quark and an anti-quark in vacuum,  
the integral must be real. This is guaranteed if for all   $0 \le v \le 1$:
\bg \label{cond1}
W(v|\umax) \equiv v^2\left( {\calH(\umax)\over \calH(\umax v)}\right) \le
1
\nd
For AdS space, (\ref{cond1}) is automatic, as ${\cal H}=1$ and then $d$ is proportional to $\umax$ which results in 
only Coulomb potential. 
But for a generic warp factor, 
(\ref{cond1}) gives rise to an upper bound for $\umax$ as already discussed. 

To show confinement at large distances the potential between the quark and the
anti-quark
must be long ranged. That is, $d(\umax)$ must range from 0 to $\infty$ as
$\umax$ varies from 0 to its upper bound, say $\umax=\xmax$.
Since $\calH(u)> 1$, the only way to satisfy these conditions is via sufficiently fast vanishing of the square-root
in Eq.(\ref{eq:dumax}) as $v\to 1$ at 
$\umax=\xmax$.

For most $\umax$, $1-W(v|\umax)^2$ vanishes only linearly as $v$ approaches
1.
In this case, $d(\umax)$ is finite as the singularity in the integrand
behaves like $1/\sqrt{1-v}$ and hence it is integrable.
To make $d(\umax)$ diverge at $\umax=\xmax$,
$1-W(v|\xmax)^2$ must vanish quadratically 
as $v$ approaches 1 to make the integrand  sufficiently singular,
$1/\sqrt{1-W(v|\xmax)^2} \sim 1/|1-v|$.
Therefore, the function $W(v|\xmax)$ must have a maximum at $v=1$.

To determine the value of $\xmax$, consider
\bg
W'(v|\xmax)
& = &
2v \left( \calH(\xmax)\over \calH(\xmax v)\right)
\left(
1 - (\xmax v) {\calH'(\xmax v)\over 2\calH(\xmax v)}
\right)
\label{eq:Wprime}
\nd
For this to vanish at $v=1$, $\xmax$ must be 
the smallest positive solution of
\be
x\calH'(x) - 2\calH(x) = 0
\label{eq:zeroTcond}
\ee
With the definition $\calH(u) = {\cal A}_n u^n$, one can easily show that
this is equivalent to the condition (\ref{real-4}) which was originally
derived in \cite{LC}.
The allowed range of $\umax$ is then 
\be
0 \le \umax \le \xmax
\ee
and within this range, $d(\umax)$ varies from 0 to $\infty$.
How it varies will depend on the values of $\calG_m$ as well as
$\calH(u)$.

\subsubsection{Finite temperature}

At finite temperature, the relation between $\umax$ and the distance
between
the quark and the anti-quark is obtained by
replacing $\calH(u)$ with $\sqrt{1-u^4/u_h^4}\,\calH(u)$ in
Eq.(\ref{eq:dumax}):
\bg
d_T(\umax)&=&2\umax
\sqrt{1-\umax^4/u_h^4}\calH(\umax)
\int_{\epsilon_0}^1dv\,
{v^2 \sqrt{\calD_m \umax^m v^m}
\over (1-v^4\umax^4/u_h^4)(\calH(\umax v))^2}\nonumber\\ 
&&\times
\left[
1-v^4{(1-\umax^4/u_h^4)\over(1-v^4\umax^4/u_h^4)}
\left(\calH(\umax)\over \calH(\umax v)\right)^2
\right]^{-1/2}
\label{eq:dumaxT}
\nd
The explicit factor of $\umax$ makes $d_T(\umax)$ vanish at $\umax=0$
as in the $T=0$ case.
As $\umax$ approaches $u_h$, the integral near $v=1$ behaves like
\bg
d_T(\umax)\sim 
\int_0^1 dv
{\sqrt{1-\umax^4/u_h^4} \over \sqrt{(1-v)(1-v\umax/u_h)}}
\nd

which indicates that $d_T(\umax)$ goes to 0 as $\umax$ approaches $u_h$. 
Hence, at both $\umax=0$ and $\umax=u_h$, $d_T(\umax)$ vanishes.
Since $d_T(\umax)$ is positive in general,
there has to be a maximum between $\umax=0$ and $\umax=u_h$.
Whether the maximum value of $d_T(\umax)$ is infinite as in the $T=0$ case
depends on the temperature (equivalently, $u_h^{-1}$) as we now show.

The fact that the physical distance needs to be real
yields the following condition. For all $0 \le v \le 1$,
\be
W_T(v|\umax)\equiv
v^2\left(\calH(\umax)\over \calH(\umax v)\right)
\sqrt{1-\umax^4/u_h^4\over 1-\umax^4 v^4/u_h^4} \le 1
\ee
Taking the derivative gives
\bg
W_T'(v|\umax)&=&
{(1-\umax/u_h^4)^{1/2}\over(1-\umax^4 v^4/u_h^4)^{3/2}} 
{v\calH(\umax)\over\calH(\umax v)}
\nonumber\\
&&\times
\left[
-(\umax v)(1-(\umax v/u_h)^4)\calH'(\umax v) + 2 \calH(\umax v)
\right]
\non
\label{eq:Wprime}
\nd
Similarly to the $T=0$ case, $d_T(\umax)$ can have an infinite range
if the derivative vanishes at
$v=1$ for a certain value of $\umax$, say $\umax = \ymax$.
This value of $\ymax$ is determined by
the smallest positive solution of the following equation
\be
y{\bf \calH}'(y) - 2\calH(y) = (y/u_h)^4\,y\calH'(y) 
\label{eq:finiteTopt}
\ee
which then forces $W'_T(1|\ymax)$ to vanish.
Note that the left hand side is the same as the zero temperature
condition,
Eq.(\ref{eq:zeroTcond}). The right hand side is the temperature $(u_h)$
dependent part.
Using the facts that:
\bg\label{knownfact}
\prod_{k=0}^{j-1}\left(k-1/2\right)
= -(2j-3)!!/2^j,~~~ \sum_{j=1}^\infty x^j\, (2j-3)!!/2^j j!
= 1- \sqrt{1-x},
\nd
it can be readily
shown that Eq.(\ref{eq:finiteTopt}) is equivalent to Eq.(\ref{real-5}) as
long as $\umax < u_h$.
It is also clear that $y=u_h$ cannot be a solution of
Eq.(\ref{eq:finiteTopt}) because at $y=u_h$, the equation reduces to
$\calH(u_h) = 0$ which is inconsistent with the fact that $\calH(y)\ge 1$.

Recall that we are considering gauge theories for which $\calH(y)\ge 1$ and $\calH'(y)\ge 0$, and we
assume that the equation $y\calH'(y)-2\calH(y)=0$ has a real positive
solution $\xmax$ which gives confinement at zero temperature.
Hence as $y$ increases from 0 towards $\xmax$,
the left hand side of Eq.(\ref{eq:finiteTopt}) 
increases from $-2$ 
while the right hand side increases from 0.
The left hand side reaches 0 when $y = \xmax$
which is the point where the distance
$d(\umax)$ at $T=0$ becomes infinite.
At this point the right hand side of Eq.(\ref{eq:finiteTopt}) is
positive and has the value  
$(\xmax/u_h)^4\xmax\calH'(\xmax)$.
Hence the solution of Eq.(\ref{eq:finiteTopt}), if it exists,
must be larger than $\xmax$.

Consider first low enough temperatures so that $u_h \gg \xmax$.
For these low temperatures,
Eq.(\ref{eq:finiteTopt}) will have a solution, as the right hand side will be still small around $y =\xmax$.
This then implies that 
the linear potential at low temperature will have 
an infinite range if the zero temperature potential has an infinite range.

Now we  show that the infinite range potential cannot be maintained at all 
temperatures. We can have a black hole such that $u_h = \xmax$.
When the left hand side vanishes at $y=\xmax$,
the right hand side is 
$\xmax\calH'(\xmax) = 2\calH(\xmax)$
which is positive and finite.
For $y > \xmax$, the left hand side ($y\calH'(y)-2\calH(y)$)
may become positive, but it is always smaller than $y\calH'(y)$ 
since $\calH(y)$ is always positive.
But for the same $y$, the right hand side ($(y/u_h)^4y\calH'(y)$)
is always positive and necessarily larger than
$y\calH'(y)$ since $(y/u_h) > 1$. 
Hence, Eq.(\ref{eq:finiteTopt}) cannot have a real and positive
solution when $u_h = \xmax$. 
Therefore between $u_h = \infty$ and $u_h = \xmax$, there must be a point
when Eq.(\ref{eq:finiteTopt}) cease to have a positive solution.

When Eq.(\ref{eq:finiteTopt}) has no solution, then the expression for
$d_T(\umax)$, (\ref{eq:dumaxT}) will not diverge for any $\umax$ 
within $(0, u_h)$.  Furthermore, since the expression vanishes at both
ends,
there must be a maximum $d_T(\umax)$ at a non-zero $\umax$. 
When the distance between the quark and the anti-quark is greater than
this
maximum distance, there can no longer be a string connecting the quark and
the anti-quark. 

To summarize, we have just shown that if we start with a dual geometry that allows infinite range linear potential at zero 
temperature,
there exists some critical temperature above which the string connecting the quarks breaks. This shows that at high enough
temperatures quarkonium state melts and gives rise to 'free quarks'. In the following subsection, we will try to quantify
the melting temperature using geometries with exponential warp factors.

\subsection{Numerical analysis of melting temperatures}

After discussing the most general choice for warp factors that give rise to $y_{\rm max}$ and 
consequently linear potential,  we will now give specific examples of geometries that may
 arise as solutions to Einstein's equation.  We start with the
following ansatz for the metric:
\bg \label{metric}
ds^2&=&-\frac{g}{\sqrt{h}}dt^2+\frac{1}{\sqrt{h}}(dx^2+dy^2+dz^2)+\frac{\sqrt{h}}{u^2}\left(\frac{H}{gu^2}du^2+ds^2_{{\cal M}_5}\right)\nonumber\\
&\equiv&-\frac{g}{\sqrt{h}}dt^2+\frac{1}{\sqrt{h}}(dx^2+dy^2+dz^2)+\frac{\sqrt{h}}{u^2}\widetilde{g}_{mn}dx^m dx^n
\nd
where $h\equiv h(u,\theta_i,\phi_i,\psi), H\equiv H(u,\theta_i,\phi_i,\psi)$, and 
$g\equiv 1-u^4/u_h^4$; ${\cal M}_5$ is the
compact five dimensional manifold parametrized by
coordinates $(\theta_i,\phi_i,\psi)$ and can be thought of as a
perturbation over $T^{1,1}$. Here $u=0$ is the boundary and $u=u_h$ is the horizon. 
As discussed in \cite{LC}, the above metric arises in region 3 of \cite{LC} when one considers the 
running of axio-dilaton $\tau$, $D7$ brane local action and fluxes due to 
anti five-branes on a
geometry that deviates from the IR OKS-BH geometry from the back reactions of the above sources.
The three-form fluxes sourced by ($p,q$) anti-branes are proportional to $r^{-i}f(r)$ 
for some positive $i$ (see \cite{LC} for details about $f(r)$), where the function
 $f(r)\rightarrow 1$ as $r\rightarrow \infty$ and $f(r)\rightarrow 0$ as $r\rightarrow 0$. 
With the coordinate $u=1/r$, there is another function: $k(u) \equiv {\rm
 exp}(-u^{\cal A}), {\cal A}>0$, that also has somewhat similar behavior as $f(u)$
and may allow us to have a better analytic 
control on the background. 
With such a choice of $k(u)$, the total three form flux is proportional to
 $u^A M(u)$ with 
\bg
M(u) \equiv M[1-k(u)] = M\left[1- {\rm exp}(-u^{\cal A})\right]
\nd
where $M$ is the number of bi-fundamental flavors.  
Thus three-form fluxes are decaying fast as 
$M u^A \left[1-{\rm exp}\left(-u^{\cal A}\right)\right]$
and, as shown in \cite{LC}, the seven-branes could be arranged such that the axio-dilaton 
$\tau$ behaves typically as 
$\tau \sim u^B$.
This means that
from the behavior of the internal Riemann tensor one may conclude that the internal metric 
$\widetilde{g}_{mn}$ behaves as
$\widetilde{g}_{mn}\sim u^C {\rm exp} (c_ou^{\cal C})$ 
where $A, C, {\cal A}$ and ${\cal C}$ are all positive and $c_o$ could be positive or negative 
depending on the precise background informations. 

{}From the above discussions it should be clear that taking the three-forms and world-volume gauge 
fluxes to be exponentially decaying in the IR
(but axio-dilaton to be suppressed only as $u^B$) should solve all the equations
of motion, giving the following behavior for the warp factor $h$ and the internal metric $H$ in 
(\ref{metric}) \footnote{See also the interesting works of \cite{andreev} where exponential warp factors have 
been chosen.}:
\bg\label{warpy}
h~=~ L^4 u^4 {\rm exp}(-\alpha u^{\widetilde{\alpha}}), ~~~~~~ H~=~{\rm exp}(\beta u^{\widetilde{\beta}})
\nd    
where we are taking 
$\alpha, \widetilde{\alpha}, \beta, \widetilde{\beta}$ to be all positives with $\alpha, \beta$ to be 
functions of internal coordinates ($\theta_i,\phi_i,\psi$) and $L^4 = g_s N \alpha'^2$ to be the 
asymptotic AdS throat radius\footnote{Note that $\beta$ in (\ref{warpy}) could be considered negative so that $H$ would 
be decaying to zero in the IR. However since region 3 doesn't extend to the IR we don't have to worry about the far IR 
behavior of Eq.(\ref{warpy}).}.  

Motivated by the above arguments, we will consider Nambu-Goto action of the string in the geometry with 
$(\widetilde{\alpha},
\widetilde{\beta})=(3,3)$ and $(\widetilde{\alpha},\widetilde{\beta})=(4,4)$ at temperatures $T^{(1)}$ and $T^{(2)}$
respectively in Eq.(\ref{warpy}). 
As in \cite{FEP}\cite{LC} we consider mappings
$X^\mu(\sigma, \tau)$, which are points in the internal space, to lie on the slice:
\bg\label{slice}
\theta_1~=~\theta_2~=~\pi, ~~~~~~ \phi_i~=~ 0, ~~~~~~ \psi~=~0
\nd
so that on this slice $\alpha, \beta$ are fixed and we set it to
$(\alpha,\beta)=(0.1, 0.05)$ for both choices  $(\widetilde{\alpha}, \widetilde{\beta})$. 
(Such a choice of slice will also help us to ignore the three-form contributions to the Wilson loop.)
With these fixed choices for the warp factors, we plot
the inter quark separation $d$ as a function of $u_{\rm max}$  in Figures {\bf 3.3} and {\bf 3.4}
for various values of $T\equiv 1/u_h$.
\begin{figure}[htb]\label{dVSumax-1}
		\begin{center}
\includegraphics[width=0.45\textwidth,height=10cm,angle=-90]{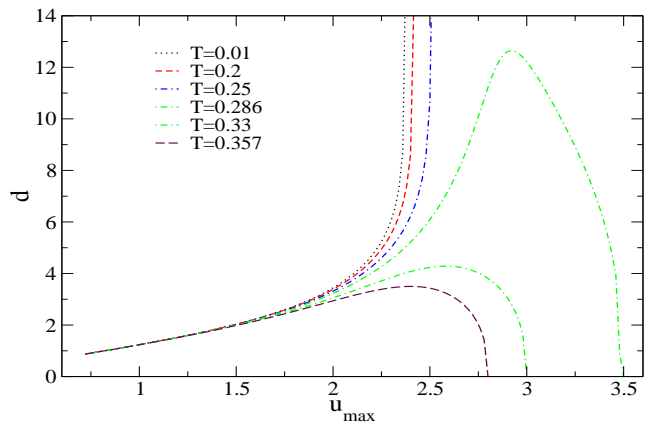}
		\caption{{Inter quark distance as a function of $u_{\rm max}$ for various temperatures and warp factor with 
		$(\alpha,\widetilde{\alpha},\beta,\widetilde{\beta})=(0.1,3,0.05,3)$ in the warp factor equation.}}
		\end{center}
		\end{figure}
		
\begin{figure}[htb]\label{dVSumax-2}
		\begin{center}
\includegraphics[width=0.45\textwidth,height=10cm,angle=-90]{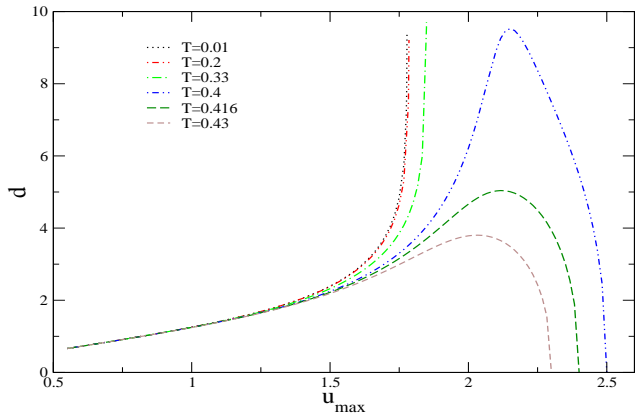}
		\caption{{Quark-anti quark distance as a function of $u_{\rm max}$ for various temperatures and warp factor with 
		$(\alpha,\widetilde{\alpha},\beta,\widetilde{\beta})=(0.1,4,0.05,4)$ in the warp factor equation.}}
		\end{center}
		\end{figure}
Note that for both choices of warp factors, for low enough temperatures, there exist $u_{\rm max}= y_{\rm max}$ where
$d\rightarrow \infty$. As the temperature is increased, $y_{\rm max}$ increases modestly. On the other hand from
figure {\bf 3.3}, one sees that when $T>T_c^{(1)}\sim 0.28$ there 
exists a $d_{\rm max}$ which is finite. This means for
inter quark distance $d > d_{\rm max}$, 
there is {\it no} string configuration with boundary condition $x(0)=\pm d/2$ implying that the
string attaching the quarks breaks
and we have two {\it free} partons for $d > d_{\rm max}$. Thus we can interpret $d_{\rm max}$
to be a ``screening length''. From Fig {\bf 3.4} we observe similar behavior but now $d_{\rm max}$ exists for
$T>T_c^{(2)}\sim 0.399$. 

\begin{figure}[htb]\label{dmaxvsT}
		\begin{center}
\includegraphics[width=0.45\textwidth,height=10cm,angle=-90]{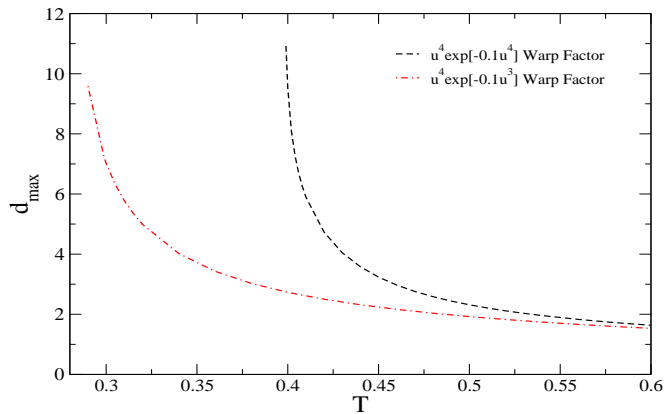}
		\caption{{Maximum inter quark separation $d_{\rm max}$ as a function of $T=1/u_h$ for both cubic and 
quartic warp factors.}}
		\end{center}
		\end{figure}
In figure {\bf 3.5}, $d_{\rm max}$ as a function of $T$ is plotted. We note that for a small change in the 
temperature near $T_c^{(1)}$ 
(or near $T_c^{(2)}$ equivalently) there is a sharp decrease in screening length 
$d_{\rm max}$, but for $T>> T_c^{(i)}$, $i = 1, 2$, 
 the screening length does not change much. In fact $d_{\rm max}$ behaves as
$C + {\rm exp}(-\gamma T)$ (where $C$ and $\gamma$ are constants) 
which in turn 
could be an indicative of a phase transition near $T_c^{(i)}$ for $i = 1,2$ i.e the two choices of warp factor. 
\begin{figure}[htb]\label{dmaxvsT}
		\begin{center}
\includegraphics[width=0.45\textwidth,height=10cm,angle=-90]{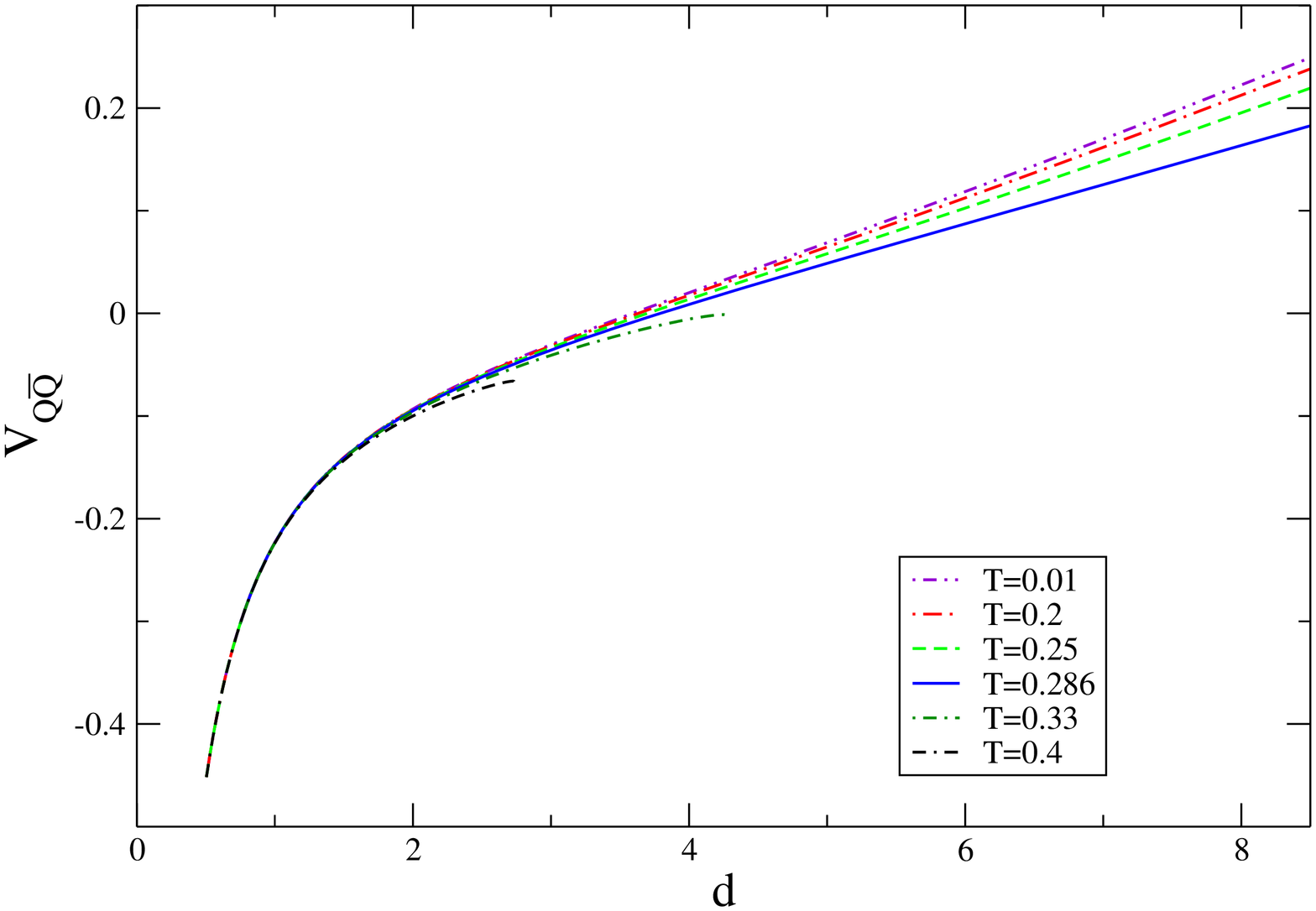}
		\caption{{Heavy quark potential $V_{Q\bar{Q}}$ as a function of quark separation $d$ with cubic warp 
factor, or equivalently,  
		$(\alpha,\widetilde{\alpha},\beta,\widetilde{\beta})=(0.1,3,0.05,3)$ in 
the warp factor equation for various temperatures.}}
		\end{center}
		\end{figure}
		
\begin{figure}[htb]\label{dmaxvsT}
		\begin{center}
\includegraphics[width=0.45\textwidth,height=10cm,angle=-90]{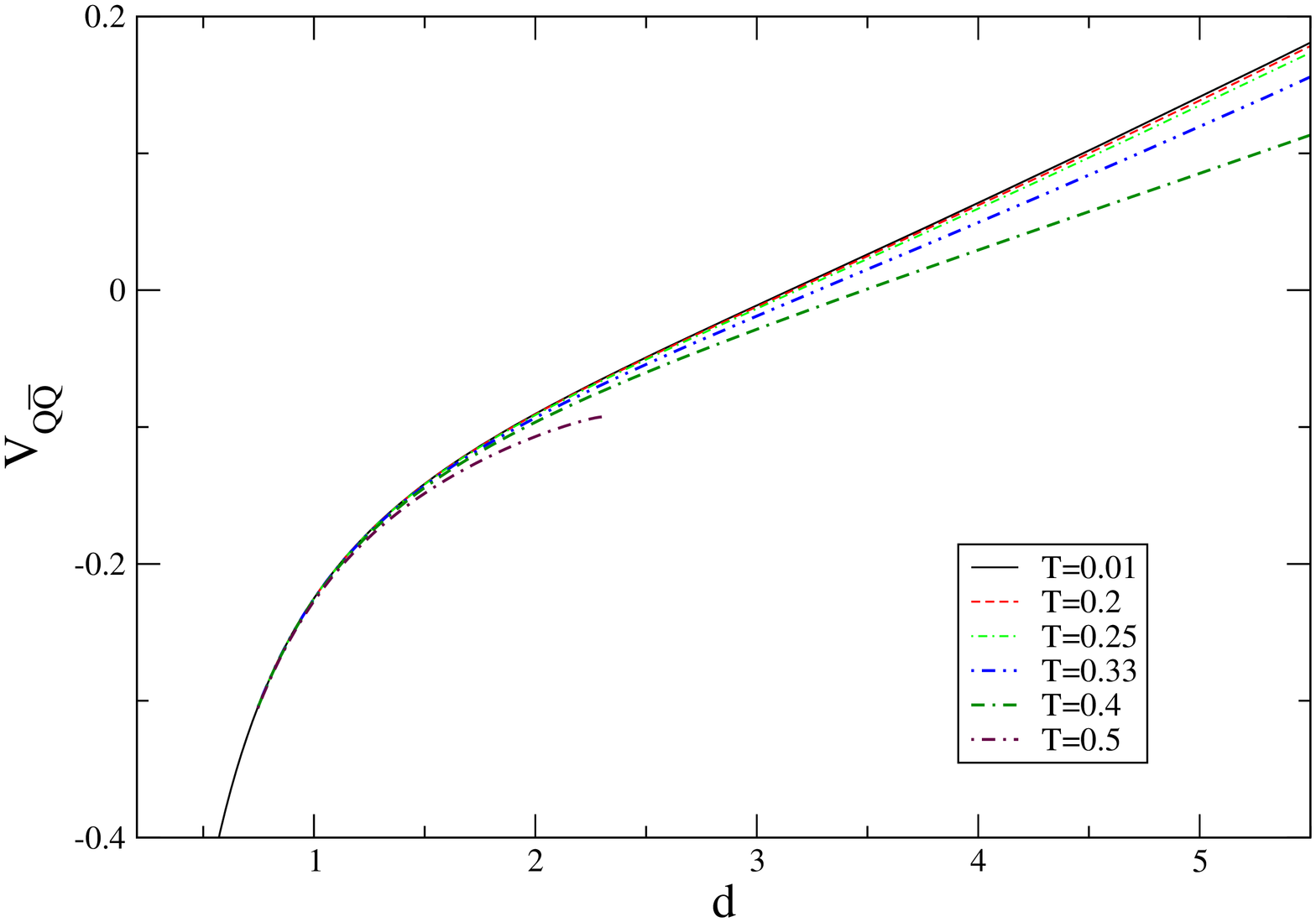}
		\caption{{Heavy quark potential $V_{Q\bar{Q}}$ as a function of quark separation $d$ with quartic warp 
factor, or equivalently,  
		$(\alpha,\widetilde{\alpha},\beta,\widetilde{\beta})=(0.1,4,0.05,4)$ in 
the warp factor equation for various temperatures.}}
		\end{center}
		\end{figure}
\begin{figure}[htb]
\begin{minipage}[htb]{0.5\linewidth}
\centering
\includegraphics[width=\textwidth, height=0.35\textheight]{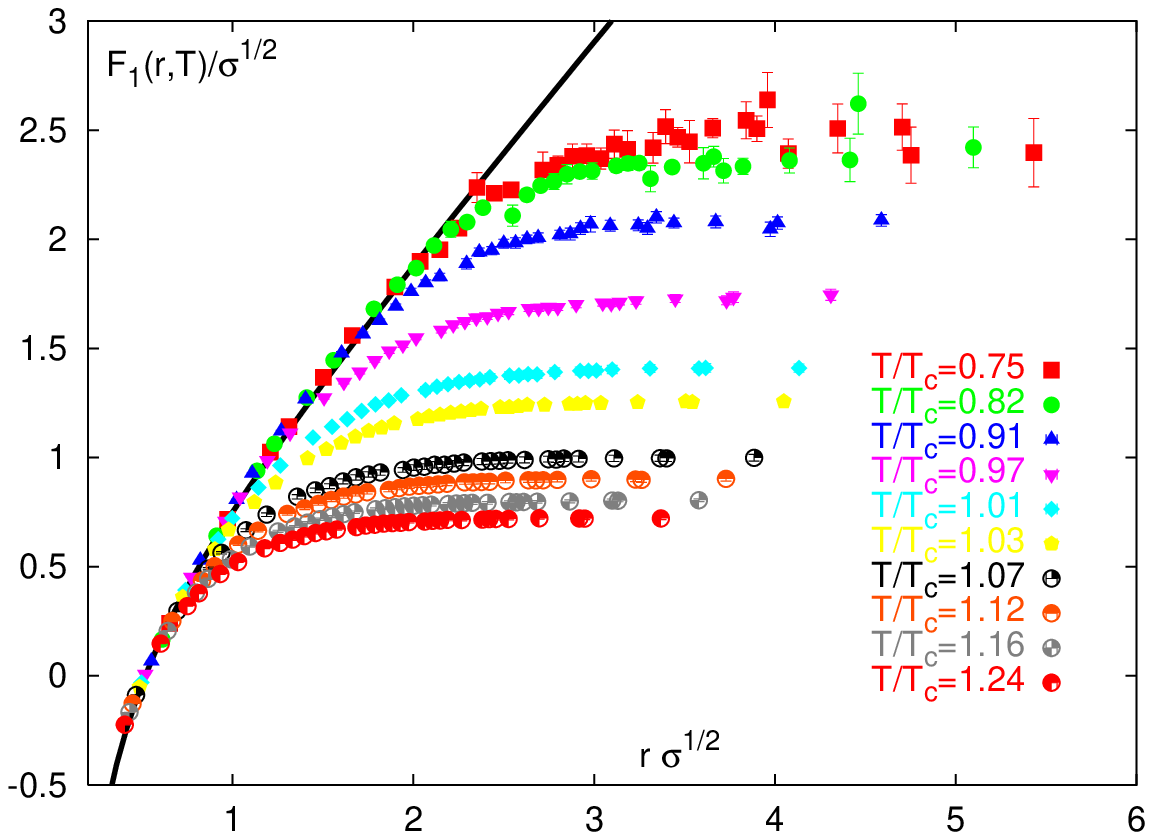}
\end{minipage}
\hspace{0.5cm}
\begin{minipage}[htb]{0.5\linewidth}
\centering
\includegraphics[width=1.3\textwidth, height=0.33\textheight, angle=270]{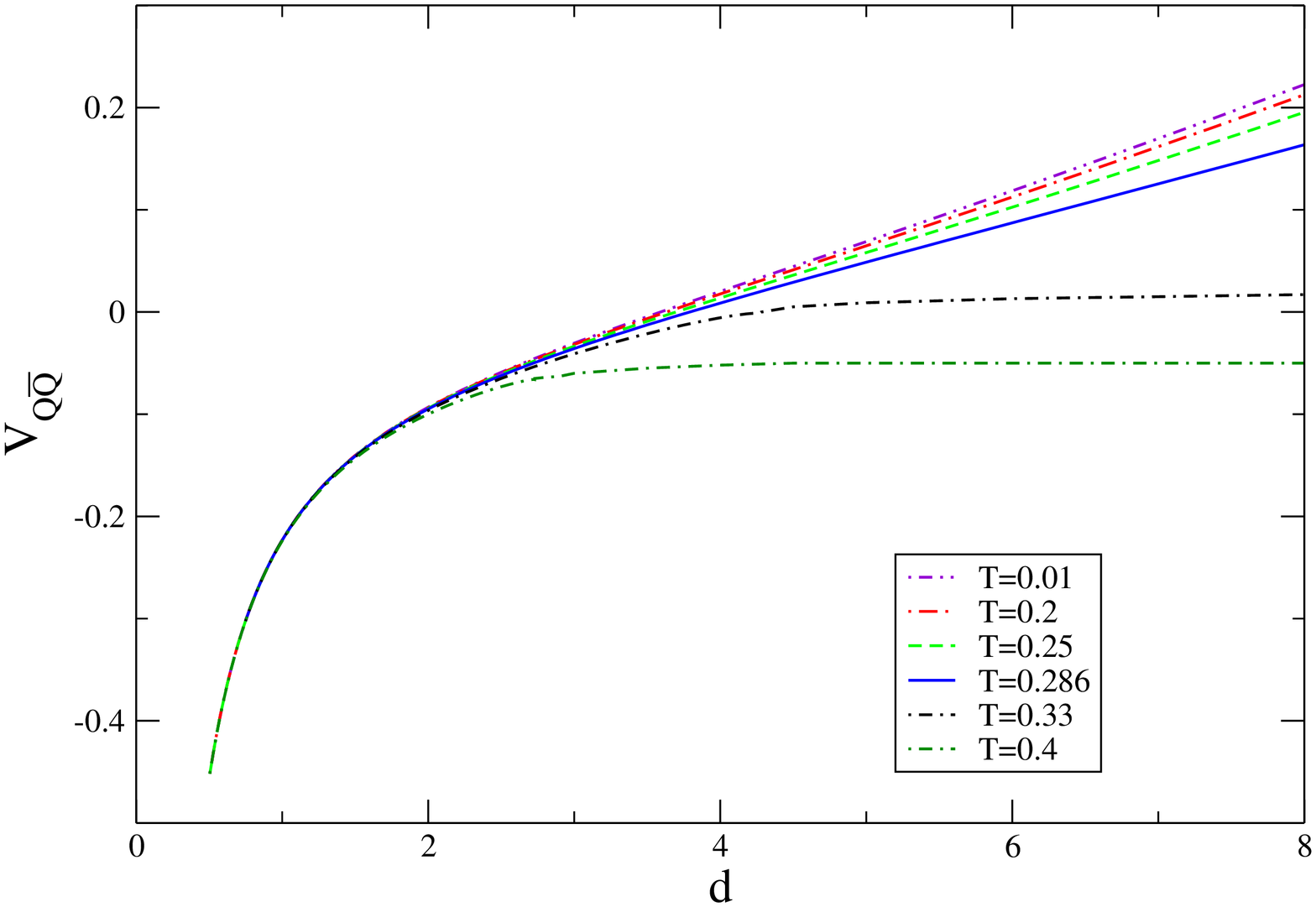}
\end{minipage}
\caption{{Comparison between lattice QCD results \cite{Kaczmarek} and our analysis. The left figure is the lattice plot
whereas the right figure is our calculation for the potential.}}
\end{figure}

Finally we plot the potential energy $V_{Q\bar{Q}}$ as a function of $d$ in Figures {\bf 3.6} and {\bf 3.7} for the two
choices of warp factor. For $T<T_c^{(1)}$ in Fig {\bf 3.6} and $T<T_c^{(2)}$ in Fig {\bf 3.7}, we have energies
linearly increasing with an arbitrarily large increment of the 
inter quark separations. Thus we have linear confinement of quarks for large
distances and small enough temperatures. For $T>T_c^{(i)}$, $i=1$ or $2$, there exists a $d_{\rm max}$ and for
all distances 
$d>d_{\rm max}$ there are no Nambu-Goto actions, $S_{\rm NG}$, for the
string attaching {\it both} the quarks. This means that we have free
quarks and $V_{Q\bar{Q}}$ is constant for $d>d_{\rm max}$. Of course looking at Fig {\bf 3.6} and {\bf 3.7} one shouldn't 
conclude that the free energy {\it stops} abruptly. 
What happens for those two cases is that the string joining the quarks 
breaks, and then the free energy is given by the sum of the 
energies of the two strings (from the tips of the seven-branes to the 
black-hole horizon) and the total energies of the small fluctuations on the world-volume of the strings. The latter 
contributions are non-trivial to compute and we will not address these in any more detail here,
but energy conservation should tell us how to extrapolate the curves in Fig {\bf 3.6} and {\bf 3.7}, 
beyond the points where the string
breaks, for all $T > T_c$. Of course after sufficiently long time the two strings would dissipate their energies 
associated with their world-volume fluctuations and settle down to their lowest energy states.

To compare with lattice QCD calculation of free energy \cite{Kaczmarek}, in Fig {\bf 3.8}, we plot side by side the lattice
results
and our calculation with $(\alpha,\widetilde{\alpha},\beta,\widetilde{\beta})=(0.1,3,0.05,3)$ where the potential energy has
been extrapolated to account for the energy of the disjoint strings i.e. the free quarks for $d>d_{\rm max}$. We observe the striking 
similarity between the shape of curves in the two plots. Two completely different
approaches yield very similar results which only strengthens the validity of applying gauge/gravity correspondence in the
study of strongly coupled QCD. Qualitatively the curves show how linear potential melts at high temperatures and there is
also evidence of screening of the Coulomb potential at small separation of the quarks.   
\begin{figure}[htb]\label{slope}
		\begin{center}
\includegraphics[height=9cm,angle=-90]{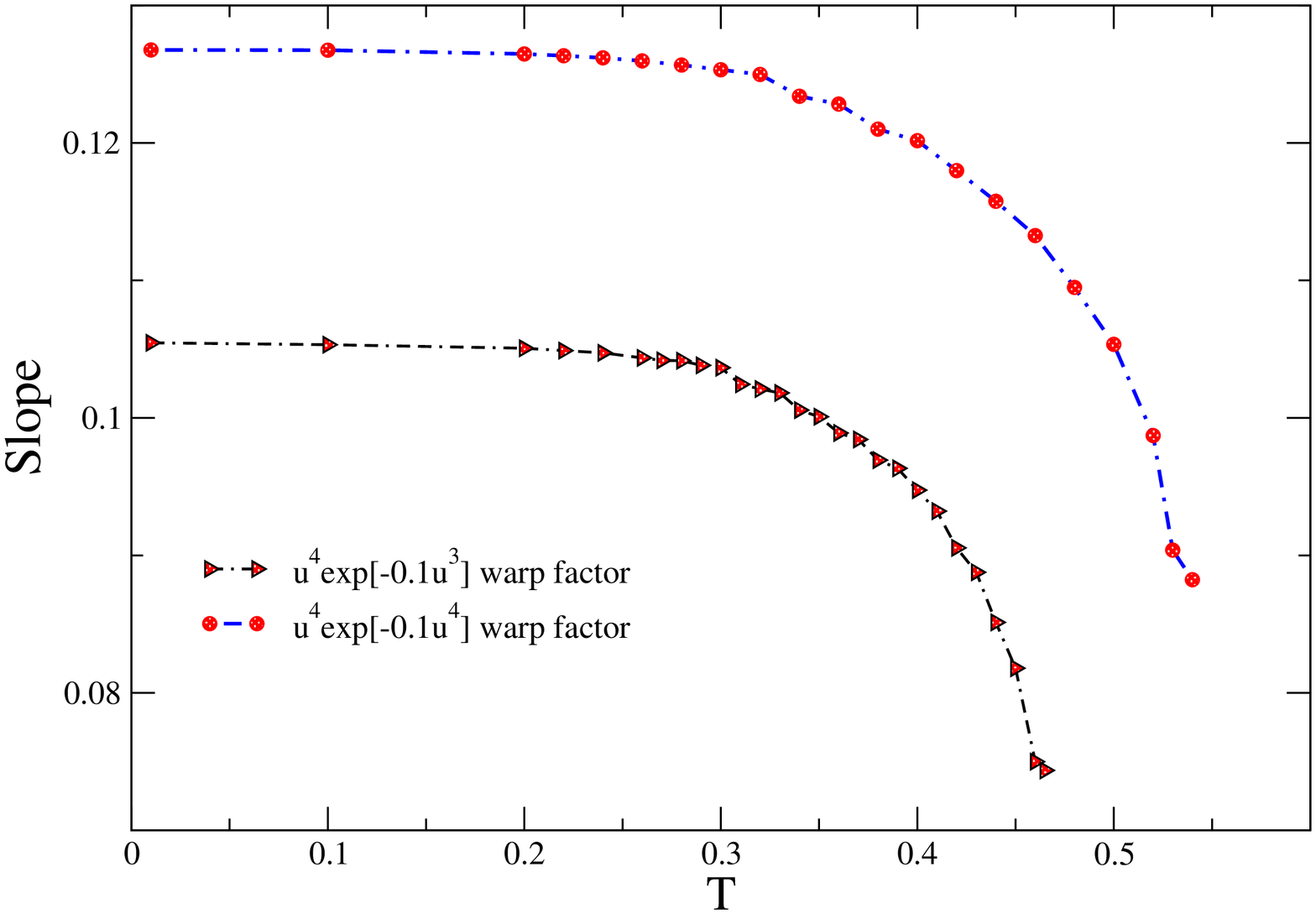}
		\caption{{Slope of linear potential as a function of T for both cubic and quartic warp factors. 
Note that in the figure the slopes have 
been computed as ${\Delta V\over \Delta d}$ with a range of $d$ from 1.6 to 1.7 in appropriate units.}}
		\end{center}
		\end{figure}

Observe that for a wide range of temperatures $0< T <T_c^{(i)}$, the
potential and thus the free energy hardly changes. But near a narrow range of temperatures $T^{(i)}_{c} - \epsilon
< T < T^{(i)}_{c} + \epsilon$ (where $\epsilon \sim 0.05$),
free energy changes significantly. For Fig {\bf 3.6} the change is more abrupt than Fig {\bf 3.7}. This means as we go for 
bigger  
values of $\widetilde{\alpha}$, the change in free energy is sharper.

In Fig {\bf 3.9} , we plot the slope of the linear potential as a function of $T$. 
Again for a wide range of temperatures, there are no significant changes in the slope
but near $T_c^{(i)}$, the change is more dramatic: the slope decreases sharply, indicating again
the possibility of a phase transition near $T_c^{(i)}$.
As we noticed before, 
here too bigger exponent $\widetilde{\alpha}$ gives a sharper 
decline in the slope hinting that when $\widetilde{\alpha}>>1$, the transition would be more manifest.

To conclude, the above numerical analysis suggest the presence of a deconfinement transition, where 
for a narrow range of temperatures $0.28 \le T_c  \le 0.39$ the free energy of 
$Q\bar{Q}$ pair shows a sharp decline. Interestingly, changing the powers of $u$ in the exponential 
changes the range of $T_c$ only by a small amount. So effectively $T_c$ lies in the range $0.2 \le T_c \le 0.4$. 
Putting back units, and defining the {\it boundary} temperature\footnote{See sec. (3.1) of \cite{FEP} for details.} 
${\cal T}$ as 
${\cal T} \equiv {g'(u_h)\over 4\pi\sqrt{h(u_h)}}$, 
our analysis reveal:
\bg
\frac{0.91}{L^2} ~\le ~{\cal T}_c ~\le ~ \frac{1.06}{L^2}
\nd
which is the range of the melting temperatures in these class of theories for heavy quarkonium states. Since the 
temperatures at both ends do not differ very much, this tells us that the melting temperature is inversely related to the 
asymptotic AdS radius in large $N$ thermal QCD.

\chapter{Conclusions}
In this thesis we have proposed the gravity dual for a non conformal finite temperature field theory with matter in fundamental representation. The
gauge couplings run logarithmically in the IR while in the UV they become almost constant and the theory approaches
conformal fixed point. To our knowledge, the brane construction and the dual geometry in section {\bf 2.4} is the first
attempt to UV complete a Klebanov-Strassler type gauge theory with asymptotically conformal field theory. 

Although our
construction is rather technical with the gauge group being of the form $SU(N+M)\times SU(N)$ in the IR, one can perform
a cascade of Seiberg dualities to obtain the group $SU(\bar{M})$ and identify this with strongly coupled QCD. One may
even interpret that the gauge groups depicted in Fig {\bf 2.19} contain strongly coupled large N QCD. This is indeed
consistent  as the coupling  $g_{N+M}$ of $SU(N+M)$
factor  in
$SU(N+M)\times SU(N+M)$ or in $SU(N+M)\times SU(N)$,  {\it always} decreases as scale is increased. To see this, observe
that in the IR, the coupling $g_{N+M}$ runs logarithmically with scale and gets stronger as scale is decreased. 
At the UV, we can arrange the
sources in the dual geometry such that $g_{N+M}$ decreases as the scale grows and runs as $g_{M+N}\sim a_k/\Lambda^k$. By 
demanding $-k a_k/\Lambda^{k-1}<0$, we see that $g_{N+M}$ indeed decreases as scale is increased. Thus from UV to IR
coupling always increases, just like QCD. 

Of course the 't Hooft coupling
$\lambda_{N+M}=(N+M)g_{N+M}$ for the group
$SU(N+M)$ is still large in the limit $N\rightarrow \infty$, even if $g_{N+M}$ has decreased to very small value.  This allows us to use the classical dual
gravity description for the gauge theory which has a coupling that shrinks in the UV, mimicking  QCD. For even higher
energies, 't Hooft coupling will eventually become too small for finite $N+M$ and  supergravity description will no longer
hold.  But we can use perturbative
methods to analyze the field theory in the highest energies. For finite $N+M$ and very high energies,
the gauge coupling may even vanish giving rise to asymptotically free theory. These arguments lead us to conclude that, 
in principle, 
the brane configuration we proposed can incorporate QCD and
in the large N limit, the gravity dual we constructed can capture features of QCD.         

Using the dual geometry, we have studied the dynamics of `quarks' in the gauge theory and computed shear viscosity $\eta$ 
and its
ratio to entropy $\eta/s$. The key to most of our analysis was the calculation of the stress tensor of gauge theory and 
we showed how different
UV completions contribute to its expectation value. Using the correlation function for the stress tensor, we computed the shear viscosity of
the medium while the computation of pressure and energy density allowed us to calculate the entropy of the system. 
Using a similar procedure to
calculate correlators of stress energy tensors, with introducing diagonal perturbations in the background metric, 
we can easily evaluate the 
bulk viscosity $\zeta$ of the non-conformal fluid. One can consider vector and tensor fields of higher rank to couple to the graviton
perturbations in the five dimensional effective theory and study how this coupling affects the bulk viscosity. The calculation is underway and 
we hope to report on it in the near future.  

All the calculations we performed regarding properties of the plasma did not account the effect of expansion of the medium
which is crucial in analyzing fluid dynamics. One possible improvement would be to construct a time dependent dual gravity which can describe
the expansion of the QGP formed in heavy ion collisions. A first attempt would be to consider collisions of open strings
ending on D7 branes and then compute their back-reactions on the geometry. The gravity waves associated with the collisions
evolve with time and from the induced boundary metric, one can compute the energy momentum tensor of the field theory. Analyzing the  time
dependence of this stress tensor, one can learn about the evolution of the medium and subsequently account for the effects 
it has on the quark
dynamics.    

We have computed the fluxes and the form of the warp factor for our static dual geometry, but did not give explicit 
expressions for the deformation of the internal five dimensional metric to all orders in $g_s N_f$. However, we have
explicitly shown the Einstein equations that determine the form of the internal metric and using our ansatz in \cite{LC},
one can in principle compute coefficients in the expansion of the internal metric  to all orders in $g_sN_f$. 
Even without a  precise knowledge of the 
internal geometry, we were able to extract crucial information about the dual gauge theory and formulated how the higher order
corrections may enter into our analysis. Most of the calculation only relied on the warp factor and with the knowledge of its
precise form, we were able to calculate thermal mass, drag and diffusion coefficients, $\eta/s$ and finally free energy of
$Q\bar{Q}$ pair. For completeness of the supergravity analysis, we hope to compute the exact solution for the 
internal metric using the ansatz of \cite{LC} in our future work. 

In our computation of the heavy quark potential, we classified the most general dual gravity that allows linear confinement of quarks at large
separation and small temperatures. We showed that if a gauge theory has dual gravity description and its effective degrees of freedom grows
 monotonically in the UV, it always shows linear confinement at large distances, as long as the dual warp factor satisfies a very simple relation given by
 (\ref{eq:finiteTopt}). Thus (\ref{eq:finiteTopt}) can be regarded as a sufficient condition for linear confinement of gauge theories with dual
 gravity. It would be interesting to study what are the general brane configurations that allow warp factors 
 which satisfy (\ref{eq:finiteTopt}) and thus give rise to confining gauge theories. 
 We leave it as a future direction to be explored.            

As the dual geometry incorporates features of Seiberg duality cascade, our construction is ideal for studying phase transitions. The gauge
theories we studied have description in terms of gauge groups of lower and lower rank. From the gravity dual analysis, by cutting the geometry
at certain radial location  and attaching another geometry up to infinity, we can construct gravity description for various phases of a gauge
theory. Each phase will have different dual geometries attached in the large $r$ region while the small $r$ region will be common to all the
theories. A flow from large $r$ to small $r$ geometry can be interpreted as `flow' from an effective theory in the UV to another in the IR.
From UV to IR, the different effective theories will describe different phases of the gauge theory and this can allow one to study the various
phases of dense matter. Thus our construction is not only useful to analyze strongly coupled gauge theory, but also has potential for studying
phase transitions in ultra dense medium and we hope to address this issue in the future.

\appendix



\end{document}